# Generalization of Eight Methods for Determining *R* in the Ideal Gas Law

## Donald B. Macnaughton*

The ideal gas law of physics and chemistry says that $PV = nRT$. This law is a statement of the relationship between four variables ($P$, $V$, $n$, and $T$) that reflect properties of a quantity of gas in a container. The law enables us to make accurate predictions of the value of any one of the four variables from the values of the other three. The symbol $R$ (called the "molar gas constant") is the sole parameter or constant of the law. $R$ stands for a fixed number that has been shown through experiments to equal approximately 8.314 472. Eight methods are available to analyze the data from a relevant experiment to determine the value of $R$. These methods are specific instances of eight general methods that scientists use to determine the value(s) of the parameter(s) of a model equation of a relationship between variables. Parameter estimation is one step in the study of a relationship between variables.

Part A of this tutorial paper discusses eight methods for determining $R$ (the molar gas constant) in the ideal gas law. Part B discusses how the ideas can be generalized. The generalization permits us to view many empirical research projects in terms of the same simple concepts. Ten appendices discuss some details behind the ideas.

### PART A: DETERMINING THE MOLAR GAS CONSTANT

#### 1. The Ideal Gas Law

The ideal gas law is a relationship between four variables that says that for a quantity of gas held in a container, $PV = nRT$. Here the variable $P$ reflects the pressure, $V$ reflects the volume, $n$ reflects the amount, and $T$ reflects the temperature of the gas. Many experiments have shown that this law is "true". That is, if we measure the numeric values of $P$, $V$, $n$, and $T$ in a "standard" situation in which a gas is held in a container, we find that $PV$ is almost exactly equal to $nRT$.

The ideal gas law is based on the work of eminent scientists in the 17th through 19th centuries: Boyle, Charles, Gay-Lussac, and Avogadro.

Following convention, this paper assumes that $P$, $V$, $n$, and $T$ are measured in the units of the International System of Units, known as the SI (from the French *Système International d'Unités*). The SI is universally accepted among scientists and engineers and is defined and discussed by the International Bureau of Weights and Measures (BIPM 2006). In SI units the pressure of a gas is measured in pascals, the volume is measured in cubic meters, the amount is measured in moles, and the temperature is measured in kelvins.

The equation $PV = nRT$ is called the "model equation" or "model" of the relationship between the four variables. Model equations are used to model relationships between variables throughout all branches of science.

The $R$ in the model equation of the ideal gas law is called the "molar gas constant". (It is also sometimes called the "universal gas constant," or simply the "gas constant".) $R$ is the sole parameter of the equation and is a fixed number that serves as a scaling factor for the relationship between the variables. That is, $R$ makes the equation consistent with the (SI) units of measurement of each of the variables in the equation.

As with almost all parameters in model equations, the value of $R$ must be determined through performing an appropriate empirical research project. Mohr, Taylor, and Newell (2008, p. 684) discuss how two recent experiments imply that the current "official" value of $R$ is 8.314 472.

The ideal gas law is important to theorists because it reflects a key relationship between variables in the theory of gases. In addition, the law is important to anyone working with gases because it gives us the valuable ability to accurately predict or control for a gas (the numeric value of) any of the four variables in the law by measuring or controlling (the values of) the other three.

The ideal gas law is highly accurate at predicting or controlling in many situations. Thus the law is regularly used in scientific work and in practical applications that require knowledge of the relationship between $P$, $V$, $n$, and $T$. However, as with most laws of science, the law isn't the final word on the relationship between the four variables. This is discussed further in Appendix A.

#### 2. Eight Methods for Determining *R*

If we wish to use the ideal gas law to predict or control, we need to know the value of $R$. Suppose that we are physical scientists and suppose that we wish to perform a research project to determine the precise value of $R$. How should we proceed?

A reasonable approach is to collect multiple sets of values of $P$, $V$, $n$, and $T$ for a quantity of gas under different conditions. For example, we might collect sets of values of *P*, *V*, *n*, and *T* for a fixed amount of a gas (*n*) as we vary the pressure, volume, and temperature of it. This would yield a data table like Table 1.

*Donald B. Macnaughton is president of MatStat Research Consulting Inc. Email: donmac@matstat.com

Table 1
Sample Data Table for Determining *R*

| P Pressure (Pa) | V Volume (m³) | n Amount (mole) | T Temperature (K) |
|---|---|---|---|
| 81,899 | 0.342 | 10.34 | 325.9 |
| 134,400 | 0.217 | 10.34 | 339.3 |
| 225,229 | 0.128 | 10.34 | 335.3 |
| (more rows of data would appear here) | | | |

Each row of numbers in the table provides a set of simultaneous values of *P*, *V*, *n*, and *T* for a quantity of gas held in a container. Thus each row provides information that we can use to determine the value of *R*, as we will see momentarily.

For simplicity, the preceding discussion speaks of *determining* the value of *R*. The verb "determine" suggests that an exact value of *R* is somehow obtained. But that isn't possible because the value of *R* is obtained by analyzing the measured values of *P*, *V*, *n*, and *T* (or related values) in a table like Table 1, and it is never possible to measure these values with perfect precision. Thus henceforth this paper speaks of *estimating* the value of *R*.

The following discussion refers to the "direction" of a variable. This concept is a straightforward generalization of the concept of 'direction' in the everyday physical world to the world of vector algebra. In the physical world the concept of 'direction' has three dimensions. We can describe any direction in this world by relating it to the three reference directions or dimensions of north-south, east-west, and up-down. In vector algebra the concept of 'direction' is generalized to worlds (vector spaces) with four or more dimensions. (The ideal gas law exists in a world of four dimensions, which are *P*, *V*, *n*, and *T*.) You don't need to understand the mathematical details of the vector-algebra concept of 'direction' to understand the key points in this paper. You need only understand that different "directions" exist in higher-dimensional worlds, and the different directions entail some of the different methods for estimating the value of *R*.

After we have a data table like Table 1 (perhaps with, say, 70 rows of data) we can use any of the following eight methods to estimate the value of *R*:

1. Solve the model equation for *R* to yield
$$R = PV/(nT).$$
Then, for each of the available sets of values of *P*, *V*, *n*, and *T* in the table, substitute the obtained numerical values into the right side of the equation and evaluate the expression to yield an estimate of *R*. Then compute the average of all the estimates to yield a final more precise estimate of *R*.

2a. Solve the model equation for *P* to yield
$$P = nRT/V.$$

Then use an appropriate general curve-fitting computer program (i.e., a linear or nonlinear regression program with the ability to omit the intercept term) to "fit" the equation to the data points (rows) in the data table with *P* as the "response" variable, and with *nT/V* as the "predictor" variable. In particular, determine the value of *R* so as to minimize the sum of the squared distances of the data points from the line that is fitted to the data points. (Here the value *nT/V* is viewed as a single variable as opposed to a composite of three variables.) Compute the distances between the points and the line in the direction of the response variable.

2b. Use Method 2a above, but use *V* as the response variable (and *nT/P* as the predictor variable).

2c. Use Method 2a above, but use *n* as the response variable.

2d. Use Method 2a above, but use *T* as the response variable.

3. Estimate *R* by fitting the equation $PV = nRT$ to the data points, but don't fit it to minimize the sum of the squared distances between each data point and the best-fitting line in the direction of one of the variables in the model equation (as is done above in Methods 2a through 2d). Instead, fit the equation to minimize the sum of the squared distances in the *scaled orthogonal* direction. (In this more general case we are actually fitting a surface to the data, not a line.) The orthogonal direction is the direction that is the "average" of the directions of all the variables, which results in each variable playing an equal role in the analysis. The scaling takes account of the differing standard error (standard uncertainty) of each variable, as discussed by Draper and Smith (1998, sec. 3.4) and Björck (1996, sec. 9.4).

4. Use one of the above methods, but instead of using the standard "least-squares" approach that is used above, choose the value of *R* so as to minimize (or perhaps maximize) the sum of the values of some other reasonable function of the individual distances between the points and the fitted line (or surface), as discussed by Huber (1964, 1981). In theory, we can use a decreasing function of distance, which gives more weight to points that are closer to the line or surface instead of giving more weight to points that are farther away. In this case we must choose the values of the parameters so as to *maximize* the sum of the values of the function. (In practice, the availability of useful functions is limited because many functions don't give stable or unique parameter estimates.)

5. Use one of the above methods, but instead of minimizing or maximizing the sum of the values of a function of the distances, use the "maximum-likelihood" method (Fisher 1922; Aldrich 1997) to estimate the value of *R*. This method provides the most "likely" estimate for the value of *R* given the data and given certain assumptions about the data. (In many standard situations the maximum-likelihood estimate of the value of a parameter is mathematically identical to the least-squares estimate.)

6. Use one of the above methods, but (if reasonable and feasible) use weighting to appropriately control the contribution that each data point makes to the analysis.

7. Use one of the above methods, but extend it with appropriate principles (e.g., principles of meta-analysis or Bayesian statistics) to take account of estimates of $R$ that were obtained in earlier research.

8. Use another method or combination of methods.

It is noteworthy that the actual use of one of the eight methods might be sensibly preceded by first "transforming" the values of one or more of the variables with mathematical functions. Such transformations are used transform the data into a form that will satisfy the underlying assumptions of the analysis approach being used or to provide another analysis perspective. For example, we might take the logarithm of each of the variables in the ideal gas law to turn the equation from a product of variables into a sum of variables. Often variables are transformed by taking powers, such as squares or square roots. However, any function is permissible if it assists the analysis. Transformations are discussed further by Draper and Smith (1998) and Chatterjee and Hadi (2006).

Methods 1 and 2 in the list are important methods for estimating the values of the parameters in a model equation and thus these methods deserve names. In the following discussion Method 1 is called the "parameter-focus" method, and Method 2 is called the "response-variable-focus" method. Methods 3 through 7 can be reasonably viewed as sensible extensions or modifications of the first two methods.

(Method 1 is actually a specific simple instance of the general parameter-focus approach—the general approach also includes Method 3. Method 2 is actually a specific instance of the general response-variable-focus approach—an instance that uses the method of least-squares.)

It is easily seen with real or simulated data that if we apply all the methods for estimating $R$ to data that have typical noise, we find (with the exception noted above) that for a given set of data *we generally obtain a slightly different estimated value of $R$ from each method.* This raises the important question of which method is best, which is discussed below in section 4.

### 3. The Official Estimate of $R$

As noted, Mohr, Taylor, and Newell (2008) report the current official estimate of $R$, which is 8.314 472. This value is based mainly on the results of an exemplary experiment by Moldover, Trusler, Edwards, Mehl, and Davis (1988), who used a necessarily complex extension of Method 1 above. This section discusses the Moldover et al. experiment and how its estimate of $R$ (which is 8.314 471) is the main determinant of the official estimate. The discussion is in terms of analyses of the actual Moldover et al. raw data, which Moldover et al. give in Appendices 1 through 3 of their article.

The Moldover et al. experiment is complex because it is known from earlier experiments that the measured value of $PV/(nT)$ depends slightly on the pressure, slightly on the temperature, and slightly on the type of gas being studied. Therefore, the accepted approach is to estimate $R$ as a "baseline" value—the (estimable) value that $PV/(nT)$ would have in a gas if the pressure were reduced to zero at a standard temperature. This value has the important property that it is (in theory and available practice) exactly the same value for all gases.

Since it is impossible to obtain a perfect vacuum (i.e., with the pressure equal to zero), and since a perfect vacuum would contain zero moles of gas, Moldover et al. couldn't directly estimate the value of $R$ at zero pressure. They circumvented this problem by estimating the value of $R$ indirectly. They did this by (in effect) measuring 70 instances of $PV/(nT)$ in the gas argon at 12 different nonzero pressures, and then "fitting a line" to the relationship between $PV/(nT)$ and pressure, and then extrapolating the line to the value that $PV/(nT)$ would have (according to the line) if the pressure were reduced to zero. If the pressure is plotted on the horizontal axis of a graph and if the measured value of $PV/(nT)$ is plotted on the vertical axis, then the estimate of $R$ is the *y*-intercept of the fitted line. Figure 1 shows what Moldover et al. effectively did.

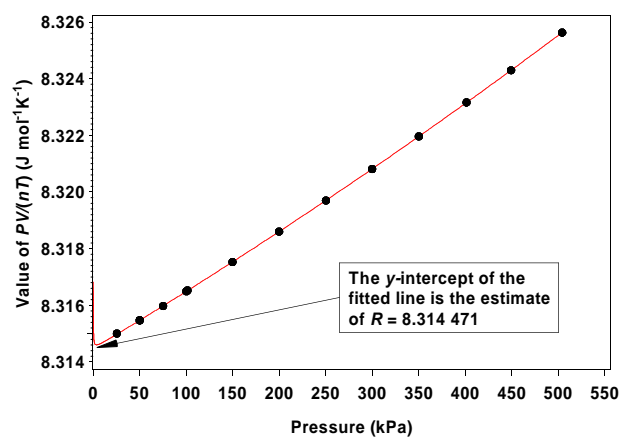

*Figure 1.* A graph illustrating the conceptual operation of the Moldover et al. experiment to estimate the value of the molar gas constant, $R$. The height of each black dot is the average of five or ten almost identical measurements of $PV/(nT)$ in argon at the indicated pressure. The red line is the best-fitting line for the 70 measurements of $PV/(nT)$ that lie behind the 12 black dots. The line was fitted using the Moldover et al. model equation of the relationship between $PV/(nT)$ and pressure. The almost-vertical direction of the line at the left edge of the graph is discussed in the text.

By convention, if a response-variable-focus method is used to study a relationship between variables, the *response variable* in the relationship is associated with the vertical axis of a graph or scatterplot illustrating the relationship. This can be contrasted with Figure 1 in which it appears that a *parameter*—i.e., $R_P = PV/(nT)$ at pressure $P$—is associated with the vertical axis. Under the approach used by Moldover et al. the value of $R_P$ (which is closely associated with parameter $R$) has become the response variable in terms of the mathematical role it plays in the analysis. This role switching is permissible because a variable can reflect any property that can be measured, including the varying results of the formula for a parameter in a model equation.

Of course, although $R_P$ is a variable, a parameter itself generally can't be a variable because a parameter doesn't vary—it is generally viewed as a fixed number. In the present situation the value of the parameter $R$ is the fixed numerical value of $R_P$ when the pressure $P$ is reduced to zero.

Figure 1 shows what Moldover et al. *effectively* did. However, the Moldover et al. experiment was additionally complex because it used an indirect way of measuring the values of some of the variables in order to increase the precision of the estimate of $R$. In what may initially seem odd, Moldover et al. give a careful argument (1988, sec. 1) to show how a very accurate and precise estimate of the value of $R$ can be obtained by measuring the *speed of sound* in a gas as a function of the *pressure* of the gas (at a fixed temperature).

Thus although Moldover et al. effectively fitted the line shown in Figure 1, they actually fitted a different line, although their line leads to exactly the same estimate of $R$. In particular, they used a standard (weighted least-squares) approach to fit a line based on a hybrid polynomial equation (discussed below) that models *the square of the speed of sound in argon* as a function of the *pressure* of the argon at a fixed temperature (1988, sec. 9.4). This line was derived from 70 carefully made measurements of the speed of sound in argon (5 or 10 measurements of the speed at each of the 12 pressures shown in Figure 1), and at a temperature of 273.16 kelvins (which is close to the freezing temperature of water). This line is shown in Figure 2.

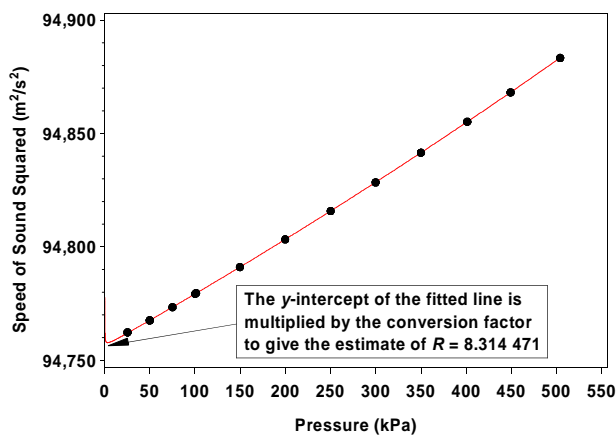

*Figure 2.* A graph illustrating the nominal operation of the Moldover et al. experiment to estimate the value of the molar gas constant, $R$. The height of each black dot is the average of five or ten almost identical measurements of the square of the speed of sound in argon at the indicated pressure and at a temperature of 273.16 kelvins. The red line is the best-fitting line for the 70 measurements of speed-squared that lie behind the 12 black dots. The line was fitted using the Moldover et al. model equation of the relationship between speed-squared and pressure.

Moldover et al. show that the *y*-intercept of the line in Figure 2 (i.e., speed-squared at zero pressure) can be multiplied by a known factor (approximately $8.77 \times 10^{-5}$) to yield a very accurate and precise estimate of $R$. This paper refers to the known factor as "the conversion factor"—the details of its computation are discussed at the end of Appendix G.11.

(Moldover et al. used argon as the gas in which they measured the speed of sound because argon is relatively easy to purify and handle and because it is relatively easy to compute a precise estimate of the value of the conversion factor for purified argon.)

Note the tight parallel between Figures 1 and 2—the points in Figure 1 are simply all the speed-of-sound-squared data values from Figure 2 multiplied by the conversion factor. Adopting the point of view of Figure 1 enables the vertical axis of the graph to be in the units of $R_P = PV/(nT)$ instead of units of speed-squared, which makes this somewhat complicated procedure easier to understand.

The data behind Figure 1 are speed-of-sound data. Thus the figure doesn't directly reflect measured values of $PV/(nT)$ at different non-zero pressures. Thus the *actual* measured values of $PV/(nT)$ in argon at the different pressures (except zero) at the fixed temperature might or might not be the same as those shown on Figure 1. However, that fact has no effect on the obtained estimate of $R$ because the mathematical parallelism between the approaches implies that regardless of whether we work in units of speed-squared or in units of $PV/(nT)$ we obtain exactly the same estimate of $R$, as illustrated in Appendix G.19.

If you are viewing this on a computer screen, you can zoom in on a small area of Figure 1 or Figure 2 by using the "zoom" tools in the reader software. If you zoom in on the fitted line in Figure 1 near the *y*-intercept, you will see the sharp upward turn of the line as the pressure approaches zero. Despite the upward turn, the computed *y*-intercept of the line is as indicated at 8.314 471, as discussed in Appendix G.11.

The estimated value of $R$ of 8.314 471 provided by Moldover et al. (1988) is different from the official value reported by Mohr, Taylor, and Newell (2008, p. 708), which is 8.314 472. This difference arises because the official value is the weighted average of the Moldover et al. value and the value obtained by Colclough, Quinn, and Chandler (1979), which is 8.314 504. The Moldover et al. value (being substantially more precise) received most of the weight.

As noted, Moldover et al. (1988) give their main raw data in their article. An analysis of these data suggests that Moldover et al. may "overfitted" the data, which leads to a slight but instructive error. If the overfitting aspect is removed from the analysis, the estimate of $R$ obtained from the Moldover et al. data changes from 8.314 471 to 8.314 477. These findings are discussed in Appendix G.

Appendix B discusses the benefits and risks of publishing research data.

## 4. What Is the Best Method to Estimate *R*?

The list of eight methods for estimating the value of *R* leads to the question of which of the methods is best. In particular, if we eliminate the overfitting problem, is the Moldover et al. approach the best approach, or might one of the other approaches be better?

None of the variables in the ideal gas law is generally viewed as deserving precedence as the sole response variable in the relationship. Therefore, Moldover et al.'s choice of an extension of the parameter-focus method (Method 1) is a reasonable choice for the best method.

The Moldover et al. experiment is an extension of the parameter-focus method because it has the following additional features: (a) indirect measurement, (b) least-squares fitting of speed-squared as a hybrid polynomial function of pressure instead of simple averaging at a fixed pressure, (c) extrapolation, and (d) weighting (Method 6 above). These points are discussed further in Appendix G.

The parameter-focus method itself doesn't fit a line, but simply computes the average of the values of the parameter estimate obtained from each row of data in the data table. However, because Moldover et al. used an extension of the parameter-focus method (due to the need for extrapolation to a pressure value of zero), their research requires fitting a line to a set of data—the line that is illustrated in Figure 2.

Thus although Moldover et al. used at a high level the parameter-focus method of Method 1, they are in effect again (recursively) able to visit the list above in section 2 to choose the method to perform the "second-level" fitting from among Methods 1 through 8. Moldover et al. chose to use Method 2, the response-variable-focus method, in which a best-fitting line is derived for the relationship between speed-squared and pressure with speed-squared in the role of the response variable.

Although Moldover et al. chose Method 2, it is also reasonable to consider any of Methods 3 through 8. (Method 1 can't be readily used in this situation because several parameters must be simultaneously estimated, and Method 1 can't readily estimate the value of more than one parameter at a time. The parameters being estimated are discussed in Appendix G.2)

Moldover et al. combined Method 2 with Method 6 (weighting), as discussed in Appendix G.10. The use of weighting took account of the varying precision of Moldover et al.'s measurements, which allowed computation of a more precise estimate of *R*.

Two other reasonable methods to fit the extrapolation line to the Moldover et al. data are the maximum-likelihood method (Method 5) and the equivalent scaled orthogonal method (Method 3). Moldover et al.'s use of Method 2 to perform this second-level fitting is equivalent to the special case of using Methods 5 and 3 in which the standard error (standard uncertainty) of the pressure values is zero.

Because all measured variables are subject to measurement error, the assumption that the standard error of a "continuous" variable such as pressure is zero is theoretically never reasonable. However, as a practical matter, analysis of the Moldover et al. data reveals that taking account of the measurement errors in the pressure values has no effect on the estimate of *R*. That is, analysis of the Moldover et al. data reveals that after rounding the estimates to the allowable seven significant digits we obtain (if the overfitting aspect is removed) exactly the same estimate of *R* from Methods 5 and 3 as we obtain from the Moldover et al. extended version of Method 1. The details of an analysis using Method 3 are discussed in Appendix G.20.

The different methods for estimating *R* yield estimates with slightly different standard errors and possibly slightly different biases in (a) the estimate of *R* and (b) the predictions of *P*, *V*, *n*, or *T* that are based on the estimate of *R*. Thus we can choose the "best" estimate of *R* to be the one that gives the lowest standard error and lowest bias in the estimate or predictions in some reasonable sense, perhaps a situation-relevant sense. However, this approach generally doesn't provide practical benefits because the different estimates are generally identical or very close after they are rounded to their allowable number of significant digits. Furthermore, defining *R* with a situation-relevant approach imposes a theoretical and mathematical burden because the relevant assumptions must be stated, and then somewhat complicated mathematical-statistical arguments must be developed. Things are simpler and neater if we have a single overall estimate of *R*. Thus the simple Method 1 (with the Moldover et al. extensions, but without overfitting, and possibly using the maximum-likelihood method for the second-level fitting) is a reasonable choice as the best method.

## PART B: GENERALIZATION

### 5. Generalization of Model Equations

Let us now consider some generalizations of the preceding ideas. These generalizations allow us to view many empirical research projects in terms of the same simple concepts.

Most empirical research projects can be sensibly viewed as studying variables and relationships between variables. This can be seen by noting that a central activity of most empirical research projects is to collect data obtained in the research in a data table. The columns of this table are usually viewed as reflecting different variables. And the rows of the data table represent different instances of measurement of the values of the variables. (Occasionally variables are reflected in the *rows* of a data table, or are reflected in a table in other reasonable senses.) A key part of the research project is an "analysis" of the data in the data table. Analysis of the data is sensibly viewed as mainly studying relationships between the variables.

(The branch of research called "qualitative research" is an exception to the points in the preceding paragraph. Qualitative research is used in some research in the social sciences. Qualitative research doesn't collect data in a data table and instead develops carefully written narrative dis-cussions of the entities [often people or social groups] that are under study.)

Most research projects that study relationships between variables have a response variable. If the response variable in a research project isn't explicitly identified, we can identify it by listing all the variables that are in the data table and asking whether the research is attempting to discover how to predict or control one of the variables on the basis of measuring or controlling the other(s). If the answer is yes, the variable being predicted or controlled is the response variable. The other variables (possibly with a few exceptions) are the predictor variables.

(A few research projects don't study relationships *between* variables but study instead the distribution of a single response variable, with no [explicit] predictor variables. For example, current experiments that look for evidence of the existence of the nuclear particle called the Higgs boson typically study the distributions of the masses of selected particles that emerge from collisions of very-high-energy particles. If the Higgs boson exists, the distribution of the particle masses should have a certain form, but otherwise the distribution should have a different form, as discussed by Lyons [2008].)

Many names are used for response variables and predictor variables in research, which is confusing for beginners. Response variables are sometimes called predicted or dependent variables. They are also called consequent, criterion, effect, outcome, output, or target variables, or *y*-variables. Predictor variables are sometimes called (with various shades of meaning) explanatory, independent, input, active, antecedent, carrier, cause, classification, concomitant, control, design, grouping, manipulated, predicated, regressor, stimulus, stratification, and treatment variables. Predictor variables are also sometimes called factors, covariates, covariables, regressors, risk factors, or *x*-variables. I contrast some of these terminologies in a paper (2002a, app. G).

To illustrate response and predictor variables, consider a simple medical experiment to determine if a new drug reduces headaches. Such an experiment typically has two key variables: The first reflects the reported severity of a headache in each person in the studied sample of people (say) 20 minutes after taking a pill to reduce the headache. The second key variable reflects the amount of the drug administered to the person to treat the headache, with two as-wide-apart-as-medically-sensible doses typically being used. (Thus one dose is zero, administered as a placebo. This dosing regimen generally maximizes the power of the experiment to detect the relationship between variables it is looking for.) The researcher wishes to learn how (if possible) to control headaches by controlling dose, and not the other way around. Therefore, severity of headaches is the response variable and amount of drug is the predictor variable. In this example we wish to learn how (if possible with this drug) to correctly control (i.e., reduce) the values of the response variable in people by manipulating the value of the predictor variable.

Often when we study a relationship between variables we use a model equation to express the known or postulated relationship between the variables. For example, the ideal gas law expresses the relationship between variables that reflect the pressure, volume, amount, and temperature of a gas.

Often model equations are presented in a standard form, with the chosen response variable on the left-hand side of the equals sign and with the predictor variable(s) combined in a mathematical expression on the right. This form is useful because if we have a response variable, we are generally interested in predicting or controlling the value of this variable on the basis of the values of the predictor variables. The standard form of the model equation simplifies this operation.

For example, if we wish to use the ideal gas law to predict the pressure of a gas, the appropriate standard form of the law is $P = nRT/V$. In contrast, if we wish to use the ideal gas law to predict the temperature of a gas, the appropriate standard form is $T = PV/nR$. In general, any model equation containing a response variable can be presented in the standard form by solving the equation for the response variable.

Consider the ideal gas law in the form

$$P = nRT/V. \tag{1}$$

This model equation can be generalized in the following five ways:

*Multiple Parameters.* Equation (1) has only a single parameter, $R$. However, a model equation can have more than one parameter. For example, here is a model equation of the relationship between $P$, $V$, $n$, and $T$ that has three parameters, which are $R$, $a$, and $b$:

$$P = \frac{nRT}{(V - bn)} - a\left(\frac{n}{V}\right)^2. \tag{2}$$

This equation is slightly more accurate than the ideal gas law and is discussed further in Appendix A.

*Multiple Terms.* Model equation (1) has only a single term on the right-hand side of the equals sign, but a model equation in the standard form can have more than one term on the right-hand side. For example, model equation (2) has two terms on the right-hand side of the equals sign. Similarly, here is the model equation of the relationship between variables that is illustrated in Figure 2:

$$c^2 = b_0 + b_1 P + b_2 P^2 + b_3 P^3 + b_4 P^{-1}. \tag{3}$$

This model equation has five terms on the right-hand side of the equals sign and is discussed further in Appendix G.2.

*Nonlinear Equations.* Model equation (1) is a linear model equation in which each parameter of the equation appears as a multiplier of one term in a sum of one or more terms on the right-hand side of the equals sign. [Equation (1) is a simple form of a linear equation because it has only one term on the right-hand side and thus has only a single parameter.] Model equation (3) is an example of a more complicated linear model equation. Linear equations are often used to model relationships between variables because they are versatile and well understood. However, sometimes in research the data or the underlying theory suggests that a nonlinear equation is more reasonable to model a relationship between variables than a linear equation. In such an equation one or more of the parameters of the equation don't appear as simple multipliers of functions of predictor variables in a linear combination of terms on the right-hand side of the equals sign, but appear in another role. For example, parameters may appear in exponents of variables of the equation.

Model equation (2) resembles a linear equation because parameter $R$ and parameter $a$ each multiply a separate term in the equation. However, the equation isn't a linear equation because parameter $b$ doesn't multiply its own term in a high-level sum of terms. Instead, parameter $b$ multiplies a term in a sum of terms in the denominator of another term. Thus (2) is a nonlinear equation.

*Discrete Variables.* All the variables in equations (1), (2), and (3) are continuous variables, which are the most commonly occurring type of variable in the physical sciences. A variable is a continuous variable if it is a numeric variable (or if it is expressible as a numeric variable) and if it is theoretically capable of having any intermediate value in its range of allowable numeric values. Otherwise the variable is a "discrete" variable. Examples of discrete variables are atomic number, blood type, and gender. One or more variables in a model equation can be discrete variables. Types of variables are discussed further in Appendix C.

*Vector-Valued Variables.* The variables in the model equations discussed above are all simple scalar variables. However, one or more of the variables in a model equation can be vectors, although this level of complexity is rare.

*Scope of the Preceding Generalizations.* The preceding five subsections discuss five generalizations of model equations of relationships between variables. Examination of model equations in empirical research suggests that most (all?) such model equations can be characterized as simple linear equations or as equations enhanced by one or more of the five generalizations.

## 6. The General Best Method for Parameter Estimation

Methods 2 through 7 in the list of methods in section 2 for estimating $R$ can be generalized in a straightforward manner to estimate (the values of) the parameters in model equations that are generalized in one or more of the ways discussed in section 5. McCullagh and Nelder (1989) give a unifying technical introduction to many of the methods.

In most cases, choosing the best method to estimate the value of one or more parameters in a model equation is straightforward, as follows: If (as usual) a response variable can be identified, the response-variable-focus method (Method 2) is usually chosen. This method is preferred because it is easy to understand and because it focuses on maximizing the accuracy and precision of the prediction or control of the values of the response variable by the model equation.

A few research projects simultaneously study *multiple* response variables, but these projects can often be conceptually broken into separate research projects, each with a single response variable. In a very few cases multiple response variables are sensibly analyzed simultaneously, as in factor analysis, cluster analysis, principal components analysis, and a few specialized cases with vector-valued response variables.

In a few empirical research projects a response variable *can't* be identified, such as in the Moldover et al. experiment to estimate $R$. In these research projects the parameters of the model equation can be estimated with the parameter-focus method (Method 1) or its generalization in the scaled orthogonal and maximum-likelihood methods (Methods 3 and some instances of Method 5). These methods are all reasonable here because they give no special status to any of the variables and because they focus on maximizing the accuracy and precision of the parameter estimates.

Thus Methods 2, 5, and 3 are the main methods for parameter estimation. The other methods in the list (4, 6, 7, and perhaps 8) are sometimes useful additions in the sense that they can sometimes provide increased accuracy or precision of the parameters estimates. (Method 1 isn't often used because it doesn't readily generalize.)

The least-squares method of parameter estimation is directly or indirectly associated with all of the first seven methods in section 2. Appendix D discusses why the least-squares method is viewed as being optimal.

Although choosing between the response-variable-focus and parameter-focus methods is straightforward, choosing the detailed approach to analyze a particular data table is surprisingly complicated. Some sources of the complexity are discussed in Appendix F.

## 7. A General Procedure for Studying a Relationship Between Variables

The preceding sections focus mainly on estimating the values of the parameters in a model equation. Parameter estimation is the tenth step in a set of twelve general steps that researchers often use (usually implicitly and sometimes omitting steps) to design and perform an empirical research project to study the relationship between a response variable and one or more predictor variables. Here are descriptions of the steps:

1. Specify the things (entities, instances) that are under study. These may be any type of thing or organism that can be observed. For example, physical scientists often study occurrences or trials of some physical phenomenon, and both medical scientists and social scientists often study people.
2. Specify the measurements of interest of the things of interest. These measurements are represented formally by variables.
3. Specify the response variable for the model equation. This is the variable that we would like to discover how to predict or control. For example, in the experiment to estimate $R$ the response variable is $R_P$. In a medical research project the response variable is often some measure of the extent of health or disease in a person.

4. Specify a set of one or more candidate predictor variables for the model equation. These are the variables that will be measured or controlled in the research to determine whether they can be used to predict or control the values of the response variable. For example, in the experiment to estimate $R$ the main predictor variable is pressure. In a medical research project the main predictor variable is often a measure of the amount or type of treatment given to a patient. For maximum efficiency a variable should be included in the set of candidate predictor variables if (a) there is *any* suspicion that the variable is related to the response variable, and (b) the value of the variable will vary (or can be varied) enough to have a noticeable effect during the research, and (c) the research budget can afford to measure the variable in the research.

5. Perform library research to identify and study all earlier work that investigated the relationship between variables of interest or that investigated related relationships. If this step is carefully done, it generally provides many helpful ideas and saves substantial time by eliminating unnecessary duplication of effort.

6. Identify the hypothesized most complicated possible form of the model equation by writing it down. Terms expressing interactions between groups of two or more predictor variables may be necessary in the equation, as discussed in data-analysis textbooks. In general, a term should be included in the equation if there is *any* suspicion that the term might belong. (This sixth step is sometimes omitted in research projects due to inexperience and because in many simpler cases modern data-analysis software for studying relationships between variables automatically adopts a sensible default model equation.)

7. Design a research project (experiment or observational study) to study the relationship between the response variable and the predictor variables as implied by the model equation. The research will study the relationship in a sample of instances that the researcher will select from the population of relevant instances. The research should be designed in a way that eliminates the possibility of reasonable alternative explanations arising of the results and in a way that has as good a chance as possible of finding the sought-after relationship between the variables if the relationship actually exists in nature. To enable generalizability, the sample must be properly representative of the population of interest. In theory, this implies that the sample should be a *random* sample from the population. However, in many cases in the physical sciences and in some cases in the biological sciences a set of consecutive independent instances or trials of the phenomenon of interest is found to be sufficient for proper generalizability. In other cases and often in the social sciences random sampling is necessary to ensure generalizability.

8. Perform the physical part of the research project and obtain a data table containing the relevant data—i.e., containing the measured value of the response variable and the measured value(s) of the candidate predictor variable(s) for each instance in the sample. Carefully check the data for errors, which occur surprisingly often and which can substantially weaken conclusions.

9. Choose terms for inclusion in the model equation (or otherwise choose the best functional form of the equation) through appropriate analyses of the data. This may involve using *p*-values as discussed in Appendix G or using other sensible methods.

10. Estimate the values of the parameters for the chosen terms, perhaps using the method of least squares or the method of maximum likelihood. (In practice, this step is usually performed concurrently with step 9.)

11. Check whether the assumptions underlying the analysis methods are adequately satisfied. If the assumptions aren't adequately satisfied, appropriately revise either (a) the model equation, (b) the analysis methods, or (c) the design of the research project so that the assumptions will (likely) be adequately satisfied and repeat the relevant steps in this procedure.

12. Draw conclusions from the results, being careful to consider reasonable alternative explanations. The conclusions will contain statements describing the relationships between variables that were studied and will often be presented graphically for ease of understanding. If the conclusions have useful theoretical or practical ramifications for the field of study, carefully consider the ramifications, which are a key output of the research. If the ramification are interesting, communicate the results and the ramifications to the relevant scientific community.

The preceding twelve steps are often used across the physical, biological, and social sciences to study relationships between variables. For a given relationship between variables the steps produce a simple model equation that makes accurate predictions.

Appendix E discusses the steps in more detail. I discuss some definitions of the concept of 'relationship between variables' in a Usenet post (2002b). I give a general discussion of these ideas in a paper (2002a). I discuss the distinction between the two main types of empirical research projects—experiments and observational research projects—in a Usenet post (2007).

## <u>SUMMARY</u>

Using the parameter $R$ of the ideal gas law as an example, this paper discusses how scientists estimate the values of the parameters of a model equation of a relationship between variables as a step in the study of a relationship between variables.

## APPENDIX A: EXTENSIONS TO THE MODEL EQUATION OF THE IDEAL GAS LAW

Various extended versions of the ideal gas law have been discovered. These are extended versions in the sense that the same basic model equation $PV = nRT$ is used, and the equation contains exactly the same parameter $R$. However, additional terms and additional parameters appear in the equation. [These extended versions are directly related to the fact that $PV/(nT)$ is related to pressure, as shown in Figure 1.] These extended versions of the law make more accurate predictions than the law itself can make.

For example, the model equation of the van der Waals extended version of the ideal gas law is given above in (2) and can also be written as

$$\left( P + a \left[ \frac{n}{V} \right]^2 \right) (V - bn) = nRT. \qquad (4)$$

Here $a$ and $b$ are two new parameters in the equation, each with a fixed numerical value, just like the parameter $R$. However, unlike $R$, parameters $a$ and $b$ have different fixed values for different gases. Note how if $a$ and $b$ are both zero, then (4) degenerates into $PV = nRT$. Sensible theoretical interpretations of the $a$ and $b$ terms in (4) are discussed by Vawter (2003) and Nave (2004).

The values of $a$ and $b$ in the van der Waals extended version of the ideal gas law have been estimated for various gases by fitting (4) to data obtained from experiments that measure $P$, $V$, $n$, and $T$ (or related variables) in these gases. Estimates of $a$ and $b$ for three well-known gases are given in Table 2.

Table 2
Experimentally Obtained Estimates of Parameters $a$ and $b$ in the van der Waals Extension of the Ideal Gas Law for Three Gases*

| Gas | Estimated Value of $a$ | Estimated Value of $b$ |
|---|---|---|
| Hydrogen | .0247 | $2.65 \times 10^{-5}$ |
| Air | .1358 | $3.64 \times 10^{-5}$ |
| Ammonia | .4233 | $3.73 \times 10^{-5}$ |

*Values are from Vawter (2003).

The $a$- and $b$-parameters are small in a relative sense, which tends to make the $a$ and $b$ terms quite close to zero relative to the terms $P$ and $V$ they are added to or subtracted from. Study of (4) indicates that the $a$ and $b$ terms will have a greater effect on the equation if $n$ is large relative to $V$, that is, if the gas is quite dense.

The van der Waals model equation (called an equation of state) has been superseded by other more complicated equations of state for the relationship between $P$, $V$, $n$, and $T$ that are even more accurate and that apply under a wider set of conditions. However, no currently known equation of state is capable of accurately predicting the relationship between $P$, $V$, $n$, and $T$ under all conditions (Wikipedia, 2010).

The preceding discussion raises the question whether the "final word" on a relationship between variables can exist. For example, in Einstein's equation does $E$ (in the appropriate context) really equal $mc^2$ exactly? Or might another very small term be involved in this model equation? It appears to be humanly impossible to tell if the final word exists for a given relationship between variables. This is because an extension to the relationship might be discovered tomorrow, as measurement methods improve and as new variables are discovered.

## APPENDIX B: BENEFITS AND RISKS OF PUBLISHING RESEARCH DATA

In keeping with the extreme care they exercised in their research, Moldover et al. (1988) published their main data in appendices in their article. Unfortunately, relatively few empirical research projects follow the useful practice of publishing the relevant data.

I recommend that all the data and the data analysis for an empirical research project be published either in a reliable archive on the Internet or in appendices in the relevant article or paper. This openness in publishing the data and openness in the analysis demonstrates the researcher's confidence in the conclusions and enables any interested reader to examine, verify, extend, or clone the analyses. This openness also enables beginning researchers and students to learn about data analysis with practical real data.

If information about a research project is published in an Internet archive, there is effectively no limit to the amount of information that can be stored in the archive. I recommend that all the following information be stored:

- a link (or links) to the relevant journal article(s) or papers about the research and possibly to other material associated with the research
- the complete set (or sets) of data used in the analysis in an easy-to-read and easy-to-import format
- a complete description of the organization of the data
- details of how each variable was measured
- the measurement-instrument-calibration data accompanied by the relevant other information categories in this list if the conclusions depend on proper instrument calibration
- a discussion of the sampling design of the research project (if not fully discussed in the associated journal article or paper)
- identification of the brand and version of the software used to do the analysis
- the integrated complete computer program used to do the analysis organized in a logical order of execution, with each main command or statement in the program documented with carefully written comments, with understandable variable names and variable labels, with sensible value labels for discrete variables, and with descriptive titles and subtitles in the output
- all the relevant output files from the analysis (e.g., listing, graphics, log, and data)
- dates on which the information was entered and updated in the database.

I have attempted to demonstrate some of these features with the analysis of the Moldover et al. data, as discussed below in Appendix G.

My recommendations are similar to the recommendations of the National Academy of Sciences for ensuring the integrity, accessibility, and stewardship of research data (2009). Schofield et al. (2010) discuss issues regarding biomedical data. Some journals now require that authors make the data behind their articles available to readers on request. I think this is an excellent first step. However, I hope that journals will soon require that their authors publish their data.

Some researchers are reluctant to voluntarily publish their data because they fear that this might reveal errors or oversights in their analyses. (Considerations of confidentiality are sometimes invoked as preventing publication of data although this issue can often be resolved by "anonymizing" the data.) In view of the complexity of data analysis (as discussed in below Appendix F), the fear in some researchers of exposure of errors or oversights isn't surprising. However, it is counterproductive to allow this fear to inhibit publication of data.

Kaiser (2008) notes that publishing research data may lead to incorrect reanalyses of the data by less experienced analysts. Such incorrect reanalyses can certainly occur. However, if an incorrect reanalysis is published, the original researcher(s) will generally respond with a careful rebuttal to protect their reputation. If the rebuttal appears correct, this will harm the reputation of the critic. Knowledge of this possibility will motivate thoughtful researchers doing reanalyses to proceed carefully, perhaps discussing contradictory findings with the original researchers before publishing the findings. In the end, if the research is important, the reanalyses and possible rebuttal will advance knowledge, regardless of who is correct. Thus it is sensible that the reanalyses be allowed to proceed.

Kaiser (2008) also notes that publishing original data may lead to novel analyses of the data being performed by new researchers and these analyses may pre-empt further analyses of the data by the original researchers. The original researchers can prevent this possibility by publishing only the variables they used in the analyses. (Hopefully, the original researchers used *all* the available relevant predictor variables because omitting relevant predictor variables from an analysis can cause important effects that would be visible in an analysis of the full data to be undetected.) Thus assuming the original researchers used the most appropriate analysis approach, no pre-empting analyses should be possible. On the other hand, if the original researchers didn't use the most appropriate approach, then other researchers deserve the chance to use that approach on the data.

Despite the preceding point, it is conceivable that a completely novel analysis is possible, perhaps with a different variable in the role of response variable. If so, and if the original researchers know about this analysis, it can be mentioned in the article accompanying the report of the analysis. This tends to give the original researchers the first right to analyze the data for this relationship if the analysis is done in reasonable time after the data are first published.

## APPENDIX C: TYPES OF VARIABLES

Section 5 of this paper introduced the distinction between continuous and discrete variables. Reasonable definitions of the two types are

• A variable is a *continuous* variable if it is theoretically capable of having any intermediate value in its range of allowable values. Continuous variables are usually numeric, but their values can also be represented by other continuous means, such as graphically by points on a line.

• If a variable isn't a continuous variable, it is a *discrete* variable.

Most variables in the physical sciences (e.g., mass, length, time, and temperature) are viewed as continuous variables, but a few are discrete, such as atomic number. Most discrete variables assume less than 30 distinct values, many assume less than 8 distinct values, and some assume only 2 distinct values (as in the case of gender). The values of discrete variables are often represented by words instead of numbers, such as "female" and "male" for the two values of gender

We can measure the degree of continuousness of a variable in terms of the number of distinguishable values in the variable's range. For example, a variable indicating the main computer operating system that a person uses can presently be sensibly defined with four possible values, which are Apple, Unix-family, Windows, and other. The small number of possible values and the lack of continuity of the values imply that this variable is a discrete variable. In contrast, a thermometer may have more than a thousand distinguishable values, and values *between* two measureable values on a thermometer are sensible, even though those values may not be resolvable with a given thermometer. Thus temperature variables are generally viewed as being continuous variables.

A second popular classification system for variables divides them into four types, which are (a) interval variables, (b) ordinal variables, (c) nominal variables, and (d) binary variables.

*Interval* variables predominate in the physical world. For example, height, weight, temperature, and speed are reasonably viewed as interval variables. By (problematic) definition, a variable is an interval variable if intervals of values of the variable of the same length in different parts of the permissible range of values are viewed as being "equivalent". For example, if we are measuring the height of a building in meters, then an interval of one meter near the ground is viewed as being equivalent to an interval of one meter near the top of the building. Thus height is an interval variable.

By definition, a variable is an *ordinal* variable if its values reflect an ordering, but with no requirement about equivalent intervals. Ordinal variables are often used in social research where highly precise direct measurement generally isn't possible. For example, an attitude test using a Likert scale presents respondents with a set of statements and asks them to rate each statement on typically a five- or seven-point agree-disagree scale. Thus on a test of political attitudes respondents may be presented with the statement "My political philosophy is more liberal than conservative". They may then be asked to indicate which of the following five statements best describes their attitude toward the statement: *strongly agree, agree, neither agree nor disagree, disagree,* or *strongly disagree.* These statements reflect an ordering of the values in terms of the amount of agreement with the statement and thus the variable is an ordinal variable.

If an ordinal variable has numeric values, it is technically possible to compute a measure of spread of the values, such as the standard deviation or the variance of the values. However, it isn't customary to compute a standard deviation or variance for the values of an ordinal variable because the measures of spread are based on the assumption of the equivalence of intervals across the range of values, and that assumption isn't necessarily satisfied for an ordinal variable.

The distinction between interval variables and ordinal variables is fuzzy because the distinction isn't strictly empirical because it generally isn't empirically possible to demonstrate "equivalence" of intervals. For example, after we have divided the scale on a temperature-measuring instrument into Celsius degrees, how could we demonstrate (apart from citing our instrument or the calibrating instrument) that a Celsius degree in the vicinity of 0 degrees Celsius is "equivalent" to a Celsius degree in the vicinity of 20 degrees Celsius? Similarly, although we can carry a ruler from the bottom of a building to the top, we can't demonstrate that the ruler didn't shrink or grow during the trip, so we can't empirically demonstrate that a meter near the ground is equivalent to a meter high above the ground.

However, although the distinction between interval and ordinal variables is fuzzy, it is useful because some variables that reflect an ordering carry more information in their values than others. In addition, some procedures for studying relationships between variables are characterized by the distinction between interval and ordinal variables, such as procedures of parametric and nonparametric correlation and regression.

Conover and Iman (1981) support the idea that the distinction between interval variables and ordinal variables is fuzzy by showing that statistical procedures that are designed for interval variables generally also work well with ordinal variables if the values of the ordinal variables are converted to ranks. Thus further research into the distinction may demonstrate that the distinction is unnecessary. (One eminent statistical dictionary appears to reject the distinction between interval and ordinal variables by omitting definitions of the concepts, Dodge, 2003.)

In contrast to interval and ordinal variables, the values of a *nominal* variable contain no implicit ordering, and the values serve only to identify (i.e., name, categorize) the different categories. For example, the variable nationality of people isn't generally viewed as reflecting an ordering of the different nationalities, but reflects only a categorization.

Finally, a *binary* variable can have only two values, which, for example, are sometimes Yes and No, or one and zero, or True and False, or female and male. Binary variables are reasonably viewed as a degenerate type of interval, ordinal, or nominal variable, so the binary category could be omitted. However, some data-analysis procedures (e.g., logistic regression analysis) are specifically designed for a binary response variable. Therefore, it seems reasonable to explicitly identify the type. Binary variables are sometimes called "dichotomous," "two-valued," or "two-point" variables.

The interval-ordinal-nominal typology was invented by Stevens (1946, 1951). Another type of variable invented by Stevens is a ratio variable. This type of variable has all the properties of an interval variable and it also has the property that ratios of its values are meaningful and it has the associated property that a zero value of the variable is a "true" zero. (The zero is a true zero in the sense that negative values aren't possible, and thus the value of the variable can never be lower than zero.) However, distinguishing this type of variable seems less important because practical situations in which we wish to study the ratios of values of a variable occur, but only rarely. Thus it is often sufficient to view a variable that qualifies as an interval variable *as* an interval variable, even if it also qualifies as a ratio variable.

The relationship between the continuous - discrete typology for variables and the interval - ordinal - nominal - binary typology is straightforward: Usually interval variables are viewed as continuous variables, and usually continuous variables are viewed as interval variables. Ordinal variables are usually viewed as discrete variables. Nominal and binary variables are always discrete variables.

The fact that the distinction between continuous and discrete variables is generally equivalent to the distinction between interval variables and the other three types suggests that the continuous - discrete typology might be abandoned. However, the continuous - discrete typology is useful if we don't need the detail of the three discrete categories, or if we wish to avoid the fuzzy "interval" concept.

The interval - ordinal - nominal - binary typology is useful because we can use it to assist in selecting an appropriate method to study a relationship between variables, as discussed below in Appendix F. Velleman and Wilkinson (1993) discuss problems with the typology. I believe that these problems deserve careful study. However, I don't think they are sufficient to eliminate the typology from being an important simplifying typology to help a researcher choose an appropriate statistical method from among the many available methods.

This paper introduces the typologies of variables because they assist us to choose appropriate statistical methods to study a relationship between the variables. However, it is statistically possible to ignore the typologies of the variables in a research project and to work at the more basic level of the distributional assumptions about the variables. Focusing on the distributional assumptions is statistically more elegant, but I think that the approach is harder for beginners to understand due to its heavier mathematical burden.

## APPENDIX D: THE OPTIMALITY OF THE LEAST-SQUARES APPROACH

The elegantly simple and widely used least-squares approach was independently invented by Adrien-Marie Legendre and Carl Friedrich Gauss, with important contributions by Pierre-Simon Laplace, as discussed by Plackett (1972) and Stigler (1977; 1981; 1986, ch. 1; 1999, ch. 17).

The Gauss-Markov theorem that is discussed in many mathematical statistics textbooks proves that the least-squares approach for estimating the values of the parameters of a standard (linear) statistical model equation is (under certain standard assumptions) optimal in the sense that (among standard linear estimators) the estimates are unbiased and have minimum variance. However, the theorem depends on how the concept of 'variance' is defined. If the definition of variance is changed, the theorem changes accordingly. Thus other reasonable methods for estimating the values of parameters (as given in the list in section 2) can be shown to be optimal if we simply give variance another (appropriately parallel) definition.

For example, we might define the population "new variance" as the average of the square roots of the absolute values (as opposed to the average of the squares of the values) of the deviations of the observed values from the expected values. This definition is sensible because it gives outliers relatively less weight than the squaring definition. Outliers arguably deserve less weight because their outlying status suggests that they are questionable values.

The preceding paragraph raises the idea that the statistical concept of 'variance' (and the associated concept of 'standard deviation') is arbitrary, and other reasonable measures of the spread of the values of a continuous variable can be defined. However, despite the arbitrariness of the least-squares / variance approach, it is a standard approach. This is because it leads to simpler mathematics and greater numerical stability than some (and perhaps all) other approaches. (To obtain estimates of the values of parameters with minimum 'variance' it is necessary to perform mathematical differentiation of the 'variance' with respect to each parameter. If 'variance' is computed in the standard way using the squares of the deviations of the original data values from the expected values, then the derivative of 'variance' with respect to a parameter is a function of the simple first power of the data values.) This choice of the simplest approach is supported by the principle of parsimony, which is discussed below in Appendix G.3. The simplicity and stability suggest that the least-squares / variance approach is "correct", even though it is arbitrary.

## APPENDIX E: NOTES ABOUT THE PROCEDURE FOR STUDYING A RELATIONSHIP BETWEEN VARIABLES

Section 7 of this paper discusses a twelve-step procedure that researchers often use to study a relationship between variables. This appendix discusses some ideas related to the steps in more detail.

### E.1 Choosing the Variables to Study

Choosing the variables to study in a research project helps to bring the main ideas of the research into sharp focus. Experienced researchers choose variables for which knowledge of relationships will yield theoretical or practical benefits. For example, if physics researchers can find a relationship between the direction in which light is traveling and the speed of the light, this will have important theoretical implications in physics. Similarly, if medical researchers can find a relationship between the amount of some drug administered to patients and the symptoms of the common cold in the patients, this relationship may have important implications for curing or reducing the common cold.

As noted in section 7, the more relevant variables that are chosen for study in a research project, the better. This is because the more variables that are studied, the more possible it is that subtle relationships will be found that wouldn't be found if fewer variables had been studied. (A proper data analysis will eliminate any predictor variables that aren't relevant, as illustrated below in Appendix G.)

### E.2 Specifying the Hypothesized Form of the Model Equation of the Relationship Between Variables

A key step in the study of a relationship between variables is to specify the "form" of the model equation of the relationship. Initially only a vague form may be specified. For example, the initial statement of the form may be that there is thought to be an "increasing" relationship between a particular pair of variables. That is, if one of the variables has a high value in an entity, the other variable will also tend to have a high value, and if one of the variables has a low value, the other will also tend to have a low value.

In some research projects the step of specifying the hypothesized form of the model equation comes before the variables are chosen. That is, a theory or a set of one or more research hypotheses or research questions may be stated. Then the ideas of the theory or the hypotheses are translated (perhaps implicitly) into a statement of one or more relationships between variables. This translation then implies the choice of the variables (measurements) that will be studied in the research and it helps to suggest possible forms of the model equation.

### E.3 Designing an Efficient Research Project

Some key research design goals are to design a research project that gives maximal relevant information about the relationship between the variables we are studying at minimal cost and with minimal chance of error or misunderstanding. Research designs that can satisfy these goals can be found to assist in the study of all relationships between variables, with the choice of the best design depending on details of the variables and other aspects of the research situation, as discussed below in Appendix F. These designs can be obtained from textbooks on research design. The ideal textbook contains a broad selection of modern examples that are taken from the researcher's own field of interest.

A key decision in designing an empirical research project is to choose how many rows of data (observations) to collect in the data table. This number is often represented by the symbol $N$ (or $n$). For example, in determining their estimate of $R$, Moldover et al. (1988) collected 70 rows of data in their data table. In general, choosing a higher value of $N$ leads to a greater likelihood of finding a relationship between the variables (if a relationship exists). Also, choosing a higher value of $N$ generally leads to a more detailed knowledge of the relationship between the variables (if a relationship exists), including more precise parameter estimates. However, choosing a higher value of $N$ generally also leads to a more expensive research project. Thus the researcher must make a tradeoff decision about what the value of $N$ will be. This choice is often made on the basis of experience and intuition. Efficient formal methods (statistical power analysis software and the theory of error propagation) are available to assist with the choice. Less experienced researchers sometimes choose too low a value for $N$, which causes them to miss discovering the phenomenon they had hoped to discover.

In performing a scientific research project the researcher can choose the statistical method to analyze the data either *before* or *after* the data have been collected. Less experienced researchers sometimes choose their data-analysis method after the data have been collected, perhaps postponing the choice because it is complicated, as discussed below in Appendix F. However, it is generally more efficient to choose and plan the data-analysis method during the design phase of the research because this often leads to a more efficient research design. This generally increases the chance that the research will find what it is looking for (if what it is looking for is there), and generally increases the precision of the parameter estimates, and generally reduces costs and errors.

I recommend that during the design phase of a research project less experienced researchers perform an analysis of realistic simulated data for their planned design as a way of confirming that the chosen data-analysis approach is viable and as a way of obtaining increased understanding of the research. I discuss methods for generating realistic simulated research data in a Usenet post (2007, app. B).

After a research project has been designed, but before the design is finalized, it is often helpful if the researcher is less experienced or if the research project is more complicated to send the written project plan to a research statistician or to an experienced researcher for review. For simpler research projects it generally takes between one and four hours of the reviewer's time to write comments about the plan and possible recommendations for improvements—a small price to pay for what often leads to substantial increases in research efficiency.

Many university statistics departments run a statistical consulting service that provides low-cost or free assistance for members of the university community and possibly for outside clients. Such a service will usually be pleased to review a research plan.

A good reviewer will frame his or her comments in the researcher's language (as opposed to the language of statistics). If statistical concepts are needed, they will be carefully explained. The reviewer will usually be able to make suggestions that will help to (a) maximize the power of the statistical tests, (b) maximize the prediction accuracy, (c) maximize the prediction precision, (d) maximize the accuracy and precision of the parameter estimates, (e) minimize the costs, and (f) possibly eliminate reasonable alternative explanations of the results. Even very-well-designed research projects can benefit from this service, if only to learn that the reviewer couldn't find any weaknesses in the research design.

### E.4 Performing the Physical Part of the Research Project and Collecting the Data in the Data Table

After we have designed an empirical research project, we can perform the physical part of it. Performing this step is an art that is learned through experience. In performing a research project it is important to carefully attend to details because key new facts and ideas are sometimes waiting to be discovered somewhere in the details.

As noted, the output from performing the physical part of a research project is a data table containing the collected data. Each column in the table represents one of the variables that were measured in the research project, and each row represents a set of "simultaneous" measurements of the values of the variables in an entity.

(Some beginners use a spreadsheet to capture data and mistakenly transpose the rows and columns in the data table, so that the variables are associated with the rows and the simultaneous measurements of the values are associated with the columns. This non-standard format leads to much confusion and frustration because all standard data-analysis software is designed for the standard data-table format, and the software can't be used if the data table is transposed from this format.)

### E.5 Identifying and Correcting Data-Table Errors

Experience reveals that many data tables contain errors. (Even very good typists make a surprising number of errors when entering data.) If these errors aren't corrected, they generally lessen the chance that the research will obtain correct and useful results. Thus all errors must be identified and corrected.

An efficient way to minimize errors in a data table is to collect the data with a computer that is programmed to automatically collect the data directly from the measuring instruments. This approach minimizes errors because computers (properly programmed, properly operating, and properly connected to properly operating instruments) don't make data entry errors.

However, in many research projects the data can't be collected by a computer and thus the data entry is done manually by transcribing the data from (typically) handwritten data collection forms into the computer. In this case, an effective way to identify data-entry errors is to enter the data into the computer *twice* and then to use the computer to compare the respective values in the two entries and to flag values that are different. These different values can then be corrected. If this approach is carefully used, virtually all data transcription errors can be identified and corrected. Better statistical software systems have procedures that can compare two data tables and list all the discrepancies between them.

After transcription errors have been corrected in a data table it is helpful to further examine the data for errors by examining a separate *summary of the values* of each variable in the table. All general statistical software systems have procedures to summarize the values of variables. Carefully labeled graphical summaries are often best. The best type of graphical summary for a variable depends on the type of the variable. Continuous variables can be summarized with dot plots or histograms (which must provide a clear indication of outliers). Discrete variables can be summarized with dot plots or bar charts. Any values that are discrepant from the other values or any other anomalies in the graphs should be followed up to confirm that the values behind the anomalies are correct.

If some data values are missing from a data table and can't be obtained, these values are often left as missing values in the table because modern statistical software can generally handle these missing values properly. Procedures are also available to "impute" missing values on the basis of the other values in the data table, which is sometimes useful. However, these procedures are complicated and are somewhat speculative and thus should be used with caution.

Generally data values should never be deleted from a data table merely because they *look* discrepant. Otherwise the selective deletion may bias the results. Values should only be deleted if they are somehow known to be incorrect.

Finally, before the main analysis of the data is begun it is sometimes helpful to study the relationships between various pairs (or larger groups) of variables because such study may reveal (as outliers) otherwise unidentified anomalous observations that should be checked to confirm that they are correct.

## E.6 Determining Whether a Relationship Exists Between or Among a Set of Variables

After we have collected the data for a research project and after any data-table errors have been corrected, an important next step is to analyze the data to determine whether the relationship of interest actually exists between or among the variables. This step is necessary before we can specify a final model equation of a relationship between variables because (following the principle of parsimony) we should only include a variable in a model equation if we have good empirical evidence that the variable *belongs* in the model equation. That is, we should only include a variable in a model equation if we have good evidence that a relationship between the relevant variables actually exists.

(In the Moldover et al. experiment to estimate the value of $R$ it wasn't necessary to show that a relationship exists between $P$, $V$, $n$, and $T$ because many earlier experiments have provided strong evidence that this relationship exists. Therefore, the Moldover et al. experiment could take the existence of the relationship for granted.)

Sometimes a confirmation that certain variables are related can be made by merely plotting a graph of the relevant data because the relationship will be unmistakable on the graph. For example, in Figure 2 the fact that the height of the dots systematically increases with increasing pressure is unmistakable evidence that a relationship exists between speed-squared and pressure in the trials in the Moldover et al. experiment.

However, in other cases, it may be unclear whether a relationship exists between two or more variables. In such case various "statistical tests" are available to help to determine whether the data provide good evidence that a purported relationship between the variables exists. These tests work by computing a measure of the *strength of the evidence* that a relationship exists. A commonly used measure is the *p*-value for a relationship, as briefly discussed below at a few places in Appendix G. If the relevant *p*-value is less than .05, and if certain often-satisfiable assumptions are adequately satisfied, it is a widely accepted convention that we can tentatively conclude that the data provide sufficient evidence that a relationship exists between the associated variables. The lower that the *p*-value is below .05, the stronger the evidence is that a relationship between the variables actually exists in the population of entities under study.

## E.7 Estimating the Values of the Parameters of a Model Equation

After we have confirmed that a relationship exists between certain variables, and after we have determined a reasonable possible form for the model equation of the relationship, the next step is to estimate the value(s) of the parameter(s) in the model equation using one of the eight methods discussed above in sections 2 through 6.

In the case of estimating $R$, the estimation of the value of the parameter is the main goal of the research. However, estimating the value(s) of the parameter(s) in a model equation isn't always of high importance in empirical research. For example, in medical and social research the data are generally noisy so the estimated values of the parameters generally won't be very precise. Then it may be of more interest to confirm that a relationship between the variables of interest *exists* and to obtain a graph of the relationship than to obtain precise estimates of the values of the parameters of the model equation.

Although we may not be directly interested in the estimated values of the parameters, the values are always useful to assist in drawing a graph of the relationship which helps us to visualize and understand the relationship. If the response variable is continuous, a scatterplot of the relationship with the fitted line is often an effective way to illustrate a relationship between two variables. Showing both the line and the individual data points on the scatterplot enables the viewer to confirm that the line is properly fitted to the points. Also, the scatter of the data points about the line helps the viewer to estimate the likely precision of any predictions that are made on the basis of the knowledge of the relationship.

### E.8 Comparing Different Candidate Model Equations

After we have obtained a possible form of the model equation of a relationship between variables, and after we have estimated the values of the parameters in the equation, we can try out the equation on the original data or (better) on new data, observing how well the equation can predict the values of the response variable from the values of the predictor variable(s). A common case is when the response variable is continuous. In this case the prediction ability of the equation is often measured in terms of the mean-square error that the equation makes when predicting the values of the response variable from the data, as discussed below in Appendix G.14.

We can also use the square root of the mean-square error which has the advantage of being in the units of the response variable instead of being in the units of the response variable squared. This value is called the "root mean square average" of the errors in prediction, and is sensibly viewed as an estimate of the "average" error that is made if the equation is used to make predictions in circumstances that are sufficiently similar to those in which the equation was obtained.

The prediction ability of one form of a model equation can be compared with the prediction ability of other possible forms of the equation. And (assuming we take proper steps to avoid overfitting) the model equation that makes the best predictions (e.g., for continuous response variables the model equation with the lowest mean-square error) can be chosen as having the best form. This approach is an alternative to using *p*-values for choosing the best form. (Because they have the same goals, the two approaches generally deliver identical or highly similar "best" model equations.)

### E.9 Communicating the Findings

After performing the preceding steps, and assuming that the research findings are of scientific interest, a researcher communicates the findings to members of the relevant research community (and to any other people) who are interested in the findings. Communicating the findings of an empirical research project is a complicated ritual that is performed essentially the same way in all branches of empirical research. The main purpose of the ritual is to facilitate clarity of communication and ease of understanding.

As a vital part of the ritual, the main sections of the report of an empirical research project can often be named (in order):
- Abstract (or Summary)

- Purpose [or Introduction, including a review of relevant earlier research and a statement of the research hypothesis(es)]
- Methods (or Methodology)
- Results (or Observations)
- Conclusions (or Discussion).

Key considerations in writing a research report are to make the report rational, complete, succinct, and unequivocal. The discussion of a research project is rational if the ideas are presented in a logical and easy-to-understand order. The discussion is complete if it includes (a) a summary of directly relevant earlier theory and research, (b) a discussion of the motivation of the research being reported, (c) enough descriptive information to enable a careful researcher to successfully repeat the research, (d) a clear description of the results of the research, ideally with easy-to-understand graphs of the findings, (e) a discussion of the limitations of the research, and (f) a discussion of the practical and theoretical implications of the research. The discussion is succinct if it contains no unnecessary material. The discussion is unequivocal if the conclusions aren't susceptible to any reasonable alternative explanations.

Researchers often first communicate their findings to other researchers by presenting a talk about the findings at a meeting of researchers in the discipline. This enables a researcher to try out the ideas on other researchers before writing the final report. The final report of a (successful) empirical research project is usually in the form of a journal article, book chapter, or book about the research.

## APPENDIX F: THE COMPLEXITY OF DATA ANALYSIS

The mathematical steps for analyzing the data from an empirical research project (as discussed above in Appendices E.6 through E.8) are generally done by a single statistical procedure that is nowadays invariably performed by statistical software. The researcher need only make the data table available to the software, choose the correct procedure in the software, and specify a few simple instructions, perhaps by selecting items from menus. The software will then use the procedure to automatically perform all (or most) of the necessary analyses of the data. However, although this sounds easy, the analysis of research data is surprisingly complicated, with four sources of complexity.

One important source of complexity in data analysis arises because the choice of the correct statistical procedure for a research project depends on the following four aspects of the research:
1. the type of each of the variables in the model equation, where any variable can be classified as belonging to one of four possible types, as discussed above in Appendix C
2. distributional assumptions about each of the variables including whether any of the variables are manipulated or whether all are merely observed
3. the sampling design of the research
4. possibly other unique features of the research.

The distributional assumptions about a variable describe how (we believe) the values of the variable are distributed in the population of entities under study. For example, (depending on the design of the research) it might be assumed that the values of a variable have a normal distribution or it might be assumed that each value can be one of a limited number of fixed values.

The sampling design of a research project reflects how the rows in the data table are selected from the population of all possible rows that might occur. We can only draw reliable conclusions about the population if our data are from a sample that is properly representative of the population.

Because many different combinations of the above four aspects of an empirical research project are possible, it turns out that hundreds of statistical procedures or sub-procedures are available to study relationships between variables. The multiplicity of procedures makes choosing the optimal procedure for a given research project difficult for less experienced researchers.

The choice of the optimal statistical procedure for studying relationships between variables is discussed in books and courses about the statistical analysis of research data. Sheskin (2007, p. 133) gives decision tables for selecting among basic procedures. Andrews, Klem, O'Malley, Rodgers, Welch, and Davidson (1998) give a decision tree. Both Sheskin and Andrews et al. use the terms "dependent variable" and "independent variable" to refer to what this paper respectively refers to as response and predictor variables.

It would be useful if statistical software could provide an interactive application that can guide a user through a decision table or tree for selecting statistical procedures, with appropriate detailed explanations a keystroke away at each step. After helping the user to select procedures, the software could then help to specify the appropriate commands to execute the procedures to analyze the data.

(In a long jump from the physics of the ideal gas law, the Andrews et al. decision tree is aimed at social science researchers who are using SAS statistical software. However, the decision tree can be used by researchers in any area of empirical research using any standard statistical software because the statistical procedures are [except for some minor variations in terminology] highly consistent across disciplines and across instances of statistical software.)

A second source of complexity in data analysis arises because the methods for confirming that a relationship exists between or among a set of variables (as briefly discussed above in Appendix E.6) are complex and are the subject of some controversy.

A third source of complexity in data analysis arises because computer programs that perform data analyses generally report (for completeness) a profusion of different statistics that (for full understanding) the researcher must consider. (A seemingly simple analysis may generate hundreds of different statistics.) These statistics can be classified into the following nine categories:

1. measures of properties of the individual variables (e.g., mean, standard deviation, skewness, kurtosis)
2. measures of strength of the evidence that the relationship exists between two or more variables (e.g., *p*-values)
3. estimates of the values of parameters of the model equation (e.g., the estimate of *R* in the ideal gas law)
4. measures of the strength of the evidence that a parameter is different from a stated value, typically zero (e.g., *p*-values again)
5. measures of the precision of the estimates of the values of the parameters of the model equation (e.g., the estimated standard error of a parameter or an estimated confidence interval for a parameter)
6. measures of the precision of the predictions made by a model equation (e.g., the estimated root-mean-square error of the predictions)
7. measures of the strength of the relationship between or among a group of variables (e.g., the correlation coefficient for the strength of the linear relationship between two continuous variables)
8. statistics used in the computation of other statistics (e.g., sums of squares, degrees of freedom, mean squares, and *F*-ratios, which are used to compute some *p*-values)
9. other statistics (e.g., factor loadings or the Kaiser-Meyer-Olkin measure of sampling adequacy).

To reduce complexity, less experienced researchers may find it useful to print the computer output from a statistical analysis and to label each statistic (or group of statistics) on the printout with its category. The better statistical software systems have good documentation to help users perform this operation.

A fourth source of complexity in data analysis arises because data-analysis procedures are based on certain assumptions, with different procedures being based on different assumptions. Therefore, to ensure that the conclusions are trustworthy, the researcher must verify that the relevant assumptions underlying the data-analysis procedure are adequately satisfied. (The assumptions are often adequately satisfied, but not always.) This verification is difficult because the criteria of adequacy haven't yet been properly systematized, and thus less experienced researchers are unsure how to perform the verification. (Experienced researchers often perform the verification through informal judgments based on experience.) It seems likely that data-analysis procedures in statistical software will be enhanced to enable automatic testing to determine whether the assumptions underlying a statistical procedure are adequately satisfied.

## APPENDIX G: AN ANALYSIS OF THE MOLDOVER ET AL. DATA TO ESTIMATE *R*

This appendix discusses a duplication of the main analysis that Moldover et al. (1988) performed of their data in order to estimate *R*. The appendix also discusses some further analyses of the Moldover et al. data that lead to different estimates of *R*.

### G.1  The Moldover et al. Data

To set the stage, Table 3 shows four of the 70 rows of data that Moldover et al. collected in their experiment.

Table 3
Four Rows of the Moldover et al.
Data for Estimating *R*

| $P$ | $c^2$ |
|---|---|
| Pressure of Argon (Pa) | Squared Speed of Sound in Argon $(m^2/s^2)$ |
| 25,396 | 94,763.099 |
| 50,017 | 94,767.836 |
| 75,154 | 94,773.462 |
| 100,261 | 94,779.334 |
| (In the complete table 66 more rows of data appear here.) | |

The complete Moldover et al. data table, which also contains data for several more relevant variables, is given below. Corrections to typographical errors in the data listing in the Moldover et al. article are given below in Appendix I.

### G.2  The Moldover et al. Model Equation

The goal of the analysis is to fit a line to the full set of data behind Table 3 (as plotted in Figure 2) to allow us to compute the *y*-intercept of the line. This *y*-intercept will be multiplied by the conversion factor to provide the estimate of *R*.

Moldover et al. (1988, sec. 2) give a theoretical discussion that suggests that it is reasonable to model the relationship between the set of speed-squared and pressure values behind Table 3 with the following "polynomial" model equation (their equation 9.2):

$$c^2 = b_0 + b_1 P + b_2 P^2 + b_3 P^3 + b_4 P^{-1}, \qquad (3)$$

where

$c^2$ = the square of the speed of sound in argon, as given above in Table 3

$P$ = the pressure of the argon in which the speed is measured, as also given in Table 3

$b_0$, $b_1$, $b_2$, $b_3$, and $b_4$ are the five parameters of the equation and are fixed numbers.

As discussed above in section 3, as a key step in estimating *R* Moldover et al. used a computer program that used the least-squares method to fit (3) to the data behind Table 3. With an exception to be discussed below, the program provided as output an estimate of the numeric value of each of $b_0$, $b_1$, $b_2$, $b_3$, and $b_4$.

If we set the pressure variable $P$ in (3) to zero, the equation degenerates (with a qualification to be discussed below) to $c^2 = b_0$. Thus the parameter $b_0$ is the value predicted by (3) of what the speed-of-sound-squared-in-argon will be if the pressure of the argon is reduced to zero—$b_0$ is the *y*-intercept of the line.

With the exception of $b_0$, the values of the *b*'s in (3) are known or thought to be different at different tempera-tures. Thus for increased precision of $b_0$ Moldover et al. were careful to (in effect) take their speed-of-sound measurements with the argon always at the same fixed temperature (of 273.16 kelvins).

The fifth (inverse-pressure) term in (3) is unusual because polynomial equations usually don't contain inverse terms. Furthermore, this term behaves somewhat strangely because its value goes to positive infinity as its argument (i.e., pressure) goes to zero. This journey to positive infinity explains the abrupt change in the direction of the fitted line in Figure 2 (and in Figure 1) in the vicinity of zero pressure. Surprisingly, despite the trip to infinity of the fifth term, the *y*-intercept term, $b_0$, in (3) has a specific value that is "pointed to" by the line in Figure 2. This *y*-intercept can be readily estimated from the data, as discussed further below.

[As suggested in section 1, for things to work properly the pressure values must theoretically be recorded in pascals because the SI unit of pressure is the pascal. However, the procedure of fitting (3) to the Moldover et al. data is an exception to this rule because the goal in fitting (3) is to obtain the value of $b_0$, the *y*-intercept, which is the value of $c^2$ when the pressure is zero. But (assuming discussion is sensibly limited to pressure scales with a true zero) a pressure of zero is zero regardless of the unit that is used to report pressure. Therefore, the unit of pressure used in performing the fitting is irrelevant. (The pressure values can't be extremely large or extremely small numbers because that may lead to computer numerical problems.)]

### G.3  Moldover et al. May Have Overfitted their Data

I used the SAS NLIN computer procedure (program) to fit (3) to the Moldover et al. data. Statistical tests that NLIN performed suggest that (3) "overfits" the Moldover et al. data. (I discuss the details of these tests below.) Specifically, the fourth term (with the variable pressure-cubed) and the fifth term (with the variable inverse-pressure) in the equation appear to be unneeded. This raises the question of whether the fourth and fifth terms should be kept in the model equation.

Many researchers use a widely accepted rule called the principle of parsimony (also known as Occam's razor) to answer this type of question. The rule says that we should make things as simple as possible while remaining consistent with the known facts. In other words, to avoid a proliferation of unnecessary terms in model equations the principle of parsimony tells us to assume that a parameter for a multiplicative term in a model equation is zero until convincing empirical evidence to the contrary is brought forward. This rule helps to eliminate superfluous or imaginary concepts that sometimes appear in human reasoning. Since the fourth and fifth terms appear to be unneeded, the principle of parsimony implies that these terms should be omitted.

Moldover et al. discuss the possibility that the fourth term is unneeded (1988, p. 133) and note that omitting the term has only a small effect on the estimate of *R*.

Omitting the fifth term removes the puzzling rise to infinity in the line in Figure 1 in the vicinity of zero pressure. On the other hand, this rise to infinity might be theoretically correct, even though no strong evidence for the need of the term was found in the Moldover et al. data. Perhaps no strong evidence was found because the pressures in the experiment weren't in the range in which the term becomes active. It seems likely that the term would be active at lower pressures, because it zooms to infinity as the pressure approaches zero. It seems unlikely that the term would be active at higher pressures because the size of the term become progressively minutely smaller in a near-linear fashion as pressure increases, so this negligible or near-negligible effect at higher pressures can generally be (if necessary) easily handled to the required accuracy by the already-present linear term in the model equation.

Moldover et al. discuss the possibility that the parameter for the fifth term is zero, but say that "there is no a priori reason for assuming [the parameter] is exactly 0" (1988, p. 134). However, the principle of parsimony is a good reason for assuming that the parameter for the fifth term is zero. For, apart from the principle of parsimony, there is no a priori reason why *any number* of parameters for possible terms (e.g., a term in pressure to the twentieth power) are exactly zero.

Before we can conclusively omit the fourth and fifth terms from (3) we should review all the evidence why Moldover et al. and earlier researchers included these terms. This review, which is beyond the scope of this paper, may yield contravening points. If earlier data show good evidence that the terms are needed in the Moldover et al. range of pressures, we must explain why such evidence is missing in the Moldover et al. data. Or perhaps the earlier researchers also overfitted their data. (Perhaps they were unaware of the statistical tests and perhaps they overfitted for the sake of caution to ensure that all the possibly relevant terms were included in the model equation.) Or perhaps some other phenomenon has caused the discrepancy between the Moldover et al. research and the earlier research.

Since the fourth and fifth terms in (3) appear to be unnecessary, omitting them should have little effect on the *y*-intercept of the best-fitting line (provided that the data points on the line are sufficiently close to the *y*-intercept, as they are in the present case). If the fourth and fifth terms in (3) are omitted, then we are fitting a quadratic line to the data. If a quadratic line is fitted to the Moldover et al. data, the resulting estimate of *R* is 8.314 477 (instead of 8.314 471), as discussed in detail below. This slight difference is well within the error estimated by Moldover et al., so it has no practical effect on the value of the estimate. However, overfitting should generally be avoided because (a) it can give us a false sense of understanding of the nature of the relationship between the variables, (b) it generally increases the size of errors in the parameter estimates, and (c) it generally increases the size of the errors made by the equation if it is used to make predictions.

### G.4 An Experiment by He and Liu

He and Liu (2002) used a procedure that is highly similar to the Moldover et al. (1988) procedure to measure the speed of sound in argon as a function of pressure to estimate the value of *R*. However, instead of using the Moldover et al. model equation, He and Liu omitted the fourth and fifth terms and thus used a quadratic model equation (2002, eq. 19). He and Liu don't discuss why they omitted the fourth and fifth terms.

The He and Liu estimate of *R* is 8.314 39. Due to limitations of the precision of their measuring instruments their estimate is less precise than the estimates from the Moldover et al. data of 8.314 471 and 8.314 477. However, the He and Liu estimate is in good agreement with the Moldover et al. estimates in terms of He and Liu's estimated standard error, which is 0.000 30. Their article contains a copy of some of their raw data.

### G.5 Determining *R* at Lower and Lower Pressures

As noted, we are interested in determining the value of $R_P = PV/nT$ when the pressure is reduced to zero. Therefore, it is instructive to consider what happens as we try to determine the value of $R_P$ at lower and lower pressures. Figure 3 is a scatterplot showing a magnified view of the raw Moldover et al. data values behind the three points in the lower-left corner of Figure 1.

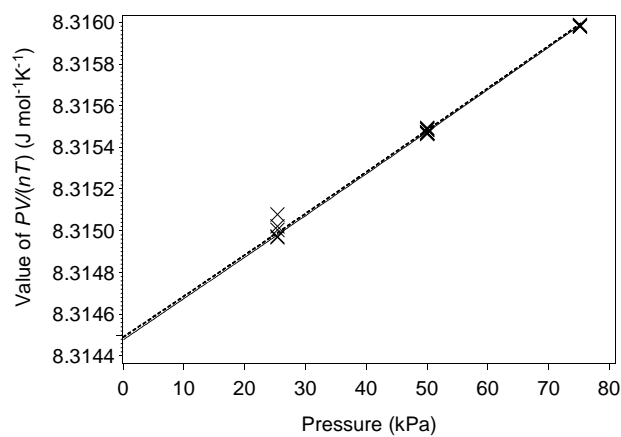

**Figure 3.** A scatterplot of the individual values of "*PV/(nT)*" versus pressure for the lowest three pressures in the Moldover et al. data showing 15 of the 70 data points as X's. The solid line is the weighted least-squares quadratic line for the 70 points. The dotted line is the unweighted least-squares quadratic line for the 70 points.

Each X on Figure 3 represents one or more data points. (At standard magnification some of the X's are invisible due to overlap.) As can be (mostly) confirmed by zooming in on the scatterplot, Moldover et al. collected five data points at 25 and 75 kilopascals and they collected ten data points at 50 kilopascals (five at 50.017 kilopascals and five at 50.041 kilopascals, as also shown on pages 40 and 42 below). Note the wide vertical spread of the points at 25 kilopascals.

The solid diagonal line on Figure 3 is the *weighted* least-squares quadratic line for the 70 data points, with a *y*-intercept of 8.314 477. This line doesn't pass through the middle of each grouping of points because it is being "steered" partly by the 50 other (more highly weighted) data points at pressures above 75 kilopascals that aren't visible on the scatterplot.

To illustrate the effect of the weights, the dotted line on Figure 3 is the *unweighted* least-square quadratic line for the 70 points, with a *y*-intercept of 8.314 490. The weights reflect the estimated relative precision of measurement of the value of "$PV/(nT)$" for each of the 70 points, and thus the weighted line is more appropriate. (The two less precise "outliers" at 25 kilopascals are pulling the dotted line "too high".) The weights are discussed in more detail below.

The cluster of five points in the upper-right corner of Figure 3 shows the tight vertical clustering of the Moldover et al. data that were collected at 75 kilopascals. Similar tight vertical clustering was obtained at each pressure above 125 kilopascals. However, as indicated by the figure, as the pressure falls below 75 kilopascals and as it approaches zero the estimates of $R$ tend to vary more widely. This suggests that in the limit of zero pressure the estimate of $R$ would have extremely high variability, which (apparently) makes the concept of $R$ at zero pressure a physical fiction—an extrapolation to an imaginary point that doesn't actually exist in the real world. However, the fictitious nature of $R$ isn't a serious practical problem because we don't need to know the (unknowable) exact "empirical" value of $R$ at zero pressure. Instead, we only need a *y*-intercept for the clearly defined lines in Figures 1 and 3 to act as a vertical "anchor" for the lines. These lines (properly anchored) are highly accurate at pressures above 25 or so kilopascals.

## G.6  Residual Scatterplots

Moldover et al. and He and Liu took account of a variable called "resonance mode", which is a measure of the type of "radial-symmetric resonance frequency" being used in the measurement of the square of the speed of sound in argon. This variable doesn't appear to be related to the speed of sound in argon, so it is less important, although it plays a role in the following discussion. At each pressure Moldover et al. used in their experiment they measured the speed of sound in argon at five different resonance mode values, which are labeled 2, 3, 4, 5, and 6.

As suggested by Figure 3, if we fit a line to the Moldover et al. data, most of the data points don't lie exactly on the line, but instead lie slight distances above or below the line. These distances between the points and the line are usually measured vertically (as opposed to either horizontally or at right angles—i.e., orthogonally) in terms of Figures 1, 2, or 3. These distances are called "residuals". They are of substantial interest because they contain information about the success or failure of the fit of the line to the data.

(It is sensible to measure residuals vertically in terms of Figures 1, 2, or 3 because the value of $R$ [or its closely related proxy] is plotted on the vertical axis of these graphs, and therefore vertical measurement assists focusing on minimizing the size of the the of the errors in the estimate of $R$. However, see Appendix G.20 below.)

Residuals are often studied graphically by plotting their values as a function of one or more other variables. Moldover et al. provide a scatterplot of their scaled (unweighted) residuals as a function of pressure and resonance mode (1988, fig. 17) and He and Liu provide a similar scatterplot (2002, fig. 3). Figure 4 is a slightly enhanced version of the Moldover et al. scatterplot.

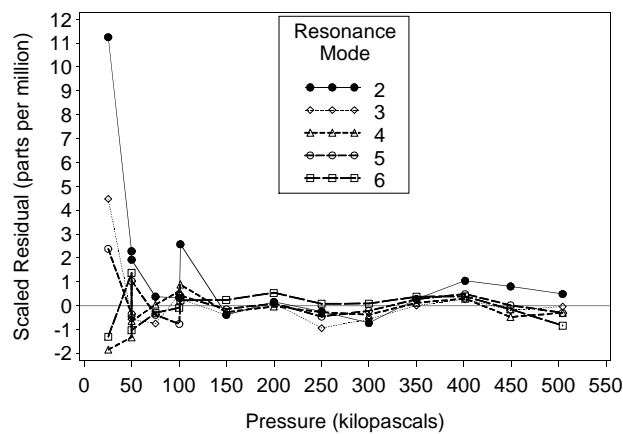

*Figure 4.* A scatterplot of the 70 (scaled) residuals obtained from the Moldover et al. data using the Moldover et al. model equation. Each residual reflects the distance between the actual speed-squared value and the value predicted by the model equation. The residuals are plotted as a function of both pressure and resonance mode.

The figure shows a greater spread of the residuals at pressures below 150 kilopascals, with especially large residuals at 25 kilopascals, as also shown on Figure 3. The figure indicates there is no good evidence that the residuals are related to resonance mode because none of the lines on the figure appears to be systematically different from any other line. That is, the changes in vertical distances between the lines at different pressures are consistent with merely being evidence of statistical noise. (However, the three high black dots at the right side of the figure are *suggestive* that a somewhat complicated relationship may exist, and are worth bearing in mind when further similar research is being designed or interpreted.)

On the other hand, the residuals in Figure 4 are clearly related to pressure, as shown by the undulating pattern that the five lines trace in unison as they move from left to right on the scatterplot. This pattern is especially noticeable at pressures above 100 kilopascals. This phenomenon (which is also visible in the He and Liu residuals) is possibly a complicated physical phenomenon related to the behavior of gases. If so, the phenomenon could be modeled by using a more complicated equation than (3) for the relationship between speed-squared and pressure. However, the pattern might also be due to extraneous factors related to either (a) the experimental apparatus (including the measuring in-struments), or (b) the mathematical manipulation of the data prior to the data analysis.

We could model the undulating pattern in the residuals by adding terms to equation (3) that could take account of the pattern. Possible approaches are given in Appendices G.17 and G.18 below. However, it would first be efficient to rule out the possibility that extraneous factors caused the pattern, although that is beyond the scope of this paper.

It is generally instructive to generate scatterplots of the residuals as a function of (in turn) *all* of the other variables available in the research. If any of the resulting residual scatterplots convincingly suggest that a relationship exists between the residuals and another variable, this suggests that the model equation might be enhanced to take account of this relationship. This thus helps to refine the equation.

(An exception is that we can omit plotting the residuals as a function of the actual [as opposed to predicted] values of the response variable because this leads to an anomaly, as discussed by Draper and Smith 1998, p. 63.)

The earlier discussion notes that in generating a graph or scatterplot researchers usually put a variable associated with a *parameter* on the vertical axis if they are using a parameter-focus approach, but they usually put the *response variable* on the vertical axis if they are using a response-variable-focus approach. When researchers are studying residuals, they usually put the *residuals* on the vertical axis of graphs or scatterplots (regardless of whether they are using a parameter-focus or response-variable-focus approach). In general, the variable that we currently view as having prime importance (there almost always is one) is given the place of honor on the vertical axis. This convention helps viewers to orient themselves to a graph or scatterplot.

It is often useful to generate residual scatterplots that show the residuals simultaneously plotted against *two* different variables, as shown in Figure 4. In this case we can plot the residuals against all possible combinations of the variables in pairs, with (when feasible) each variable in a pair plotted in turn both on the horizontal axis and as the second variable. (Resonance mode is the second variable in Figure 4.) This generates 35 different scatterplots from the Moldover et al. data, as shown below in Appendix H. Generating and studying all these scatterplots enables the researcher to thoroughly study the residuals, which may identify unexpected phenomena. (Computer programs will likely someday generate and examine these scatterplots automatically.)

For increased visibility of relationships sometimes it is useful to use lines to join similar points (as in the case of the resonance mode values in Figure 4), although sometimes the lines are non-unique, and are therefore then best omitted. Because these lines can reveal patterns in the data, they are useful for linking similar points even when the variable plotted on the horizontal axis isn't continuous.

A researcher may gain substantial insight by studying all possible residual scatterplots. However, except in the important case when relationships are present between the residuals and one or more other variables (as illustrated in Figure 4), residual scatterplots aren't normally shown in scientific research reports because they usually show little more than statistical noise, and not much information is conveyed by a figure that merely shows noise.

Figure 4 follows the Moldover et al. approach of using *scaled* residuals. Here the scaled residuals are the raw residuals divided by the associated speed-squared value and then multiplied by one million. This causes the vertical axis of the scatterplot to be in units of parts per million, which is reasonable when a parameter-focus method is used. In contrast, if a response-variable-focus method is used, the raw residuals are usually easier to understand because they are measured in the units of the response variable, which are often the most concrete and understandable units from a practical point of view.

(With the Moldover et al. data the raw residuals and the scaled residuals are virtually indistinguishable in terms of the relative positions of the points on a scatterplot. Moldover et al. scale the residuals a different way in their Figure 18, which attenuates the effects of the outliers, but doesn't remove the undulating pattern.)

### G.7 Introduction to the Computer Analyses of the Moldover et al. Data

The remaining sections of this appendix describe several computer analyses of the Moldover et al. data. This material is in two parts:

1. several sections of text discussing the computer output
2. the computer output itself, which appears below in Appendix J, beginning on page 40.

The analyses were performed with SAS statistical software (version 9.2.1), but the following discussion doesn't require familiarity with SAS. For readers who wish to see how the problem was specified to SAS or who wish to repeat or modify the analyses, the SAS program (with data) that I used to generate the output and the log of the analyses that SAS produced are available on the web (Macnaughton 2010a, 2010b).

The following discussion makes many references to the computer output. Thus for efficient reading of this material you should be able to see both the discussion and the output simultaneously without having to flip between pages. A reasonable approach is to print the discussion of the output and to view it on the printed page and to view the computer output on a computer monitor. Another approach is to print both the discussion and the output and to put them in separate three-ring binders. Alternatively, if you have a computer monitor with sufficiently high resolution, you can show both the discussion and the computer output simultaneously on the monitor by selecting Window, New Window in the reader software.

The titles on the lines at the top of each page of the output indicate what is being reported on the page.

### G.8 Data Listing

Pages 40 and 41 below give a listing of all the data used in the analysis, as copied from the Moldover et al. article (1988, app. 1 and 2) and with the corrections to ty-pographical errors in the published data that are discussed below in Appendix I. Each row in the table reflects a single "observation" of the values of all the variables. The variable names in the table should be self-explanatory, with SpeedSq reflecting the measured square of the speed of sound in argon at the indicated Pressure and at the indicated Resonance Mode. As noted above, the important variables in the table are Pressure and SpeedSq because the goal of the exercise is to fit a line to the scatterplot of SpeedSq as a function of Pressure in order to obtain the *y*-intercept of this line, which we will multiply by the conversion factor to obtain the estimate of *R*.

Some other points about the data table: The Date, Temperature, and Mean Half-Width variables were included in the table for completeness and for study in the residual scatterplots, but these variables aren't used directly in the analyses. Each of the Mean Resonance Frequency and Mean Half-Width values is the average of two values from Appendix 1 in the Moldover et al. article. Before the data listing was generated a new variable, WeightVar, was added to the data, as shown in the rightmost column of the table. This variable was computed from the other variables and is discussed below.

### G.9  Univariate Distribution of Speed-Squared at Different Pressures

Page 42 summarizes distribution of the SpeedSq values at each of the fourteen pressures used in the experiment. For example, the first row of the table indicates that at the pressure of 25.396 kilopascals the square of the speed of sound in argon was measured five times (once at each of the five resonance mode values), and the average (mean) square of the speed was 94,762.316 meters-squared per second-squared, and the standard deviation of the five values was 0.503 meters-squared per second-squared, which implies that the five values are quite close together.

For completeness, the table also reports the width of the range of the five values at each pressure. For example (as indicated on page 40), at the pressure of 25.396 kilopascals, the lowest of the five speed-squared values was 94,761.858 and the highest of the five values was 94,763.099. Thus width of the range is 94,763.099 - 94,761.858 = 1.241, as indicated in the last column of the first row of the table on page 42.

The table on page 42 is important from a practical point of view because it helps to confirm (through the *N*'s and the very small standard deviations) that no gross errors are in the pressure and speed-squared data values.

Note that the table indicates that standard deviations (and ranges) are somewhat larger at the three lowest pressures, as also shown in Figures 3, and 4.

The discussion in this paper sometimes refers to fourteen pressures in the Moldover et al. experiment and sometimes to twelve pressures. This is because, as indicated in the table on page 42, two sets of five values were collected at pressures in the vicinity of 50 kilopascals and these values tend to merge when the data are considered at a high level. Similarly, two sets of five values were collected at pressures in the vicinity of 100 kilopascals.

### G.10  The Moldover et al. Weight Variable

As noted, Moldover et al. used a "weight" variable to weight each of the 70 observations in their data (1988, sec. 9.4). The values of this variable are shown in the rightmost column of the data table on pages 40 and 41. These values are based on Moldover et al.'s knowledge of the varying precision of their measuring instruments and on the theory of propagation of errors (Taylor, 1997). The values were computed from the pressure (*P*), speed-squared ($c^2$), and mean resonance frequency (*f*) values in the table for each observation. The formula for an individual value of the weight variable is

$$\text{WeightVar} = \cfrac{1}{\left\{1.414 \times 10^{-7} c^2 \left[1 + \left(\dfrac{10^5}{P}\right)^2 \left(\dfrac{6 \times 10^3}{f}\right)^2\right]\right\}^2 + \left\{3.7 \times 10^{-7} c^2\right\}^2}.$$

The formula for the weight variable isn't intuitive. An easy way to understand the operation of the weight variable is to study a scatterplot of its values as a function of one or more of the other variables. Figure 5 shows the computed values of the weight variable as a function of pressure and resonance mode.

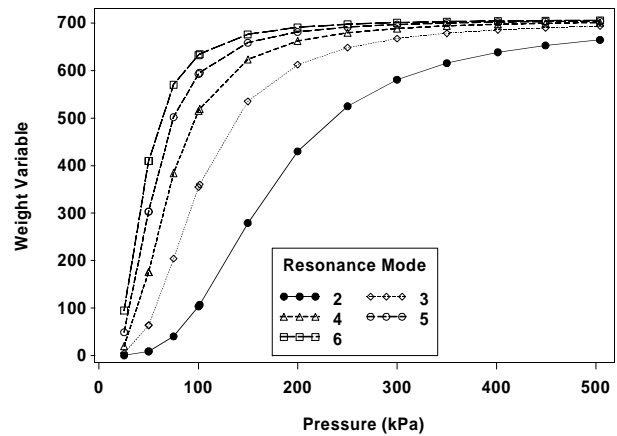

*Figure 5.* A scatterplot of the values of the Moldover et al. weight variable as a function of pressure at each of the five resonance modes used in the Moldover et al. experiment. Successive points for a given resonance mode as pressure increases are joined by straight lines.

Figure 5 implies that values of the Moldover et al. weight variable range between roughly zero and roughly 700, and the weight variable gives higher weights to observations that were obtained at higher pressures, and it gives higher weights to observations that were obtained at higher resonance modes. The increase in weight with increasing pressure is pronounced at lower pressures and tapers off at higher pressures.

Formally, the units of a weight variable are the number one, which enables weight variables to transparently participate in dimensional analyses. However, the unit one usually isn't explicitly shown (BIPM 2006, p. 120) and thus informally a weight variable is unitless.

Moldover et al. report that they derived the $3.7 \times 10^{-7}$ value of the parameter in the second term in the denominator of the formula for the weight variable empirically from the data (1998, sec. 9.4). Unfortunately, Moldover et al. appear to have omitted testing whether the value $3.7 \times 10^{-7}$ is significantly different from zero. Thus it is possible that noise in the data is a source of the value of the multiplier of $c^2$. If possible, it is useful to test whether the estimated value of each parameter used in computing a weight variable is significantly different from zero, and it is sensible to omit a term from the computation of the values of the weight variable if the estimated value of the parameter for the term isn't significantly different from zero. It is sensible to omit a term if there is insufficient evidence that it belongs because including inappropriate terms increases the error variation in the computations, thereby tending to decrease the precision. Moldover et al. appear to indirectly acknowledge this problem at the end of their paragraph about the weight variable (1988, sec. 9.4).

### G.11 Fitting a Line Using the Moldover et al. Approach

As noted, Moldover et al. used a computer program to fit the following model equation to their data:

$$c^2 = b_0 + b_1 P + b_2 P^2 + b_3 P^3 + b_4 P^{-1}, \quad (3)$$

where $c^2$ is the speed of sound in argon squared, $P$ is pressure of the argon, and $b_0$, $b_1$, $b_2$, $b_3$, and $b_4$ are the five parameters of the equation. The program finds the values of the parameters so that the equation will be the best possible fit to the data in the least-squares sense. This can be understood in terms of Figure 2 by noting that the line reflected by (3) is fitted to the set of 70 data points behind the 12 black dots on the figure. The line is the best possible fit in the sense that the sum of the squares of the (vertical) deviations of the 70 points from the line is mathematically the *smallest possible value* given that the line is constrained to have the model equation (3). This sum of squares is called the "error" sum of squares.

The weight variable participates in this operation in the sense that each squared vertical deviation is multiplied by the weight variable before being added to the other values to compute the weighted error sum of the squares—the value that is minimized.

(In the case of the ideal gas law two types of squares are being used—the *square of the speed* of sound in argon and the *square of the vertical deviation* between the predicted speed-squared and the measured speed-squared. The fact that the response variable in this analysis is already a square is unusual, but this point plays no important role in the analysis. For completeness it is useful to note that the same exercise could be performed by fitting a [different] line to the raw [un-squared] speed values.)

I fitted equation (3) to the data with the SAS NLIN procedure, which uses a highly efficient trial and error approach to perform the fitting. The procedure works by first computing (analytically) the first and second partial derivatives of the weighted error sum of squares (i.e., the sum of the weighted squared vertical deviations of the points from the line—i.e., the sum of the weighted squared residuals). The partial derivatives are computed with respect to each of the free parameters (i.e., the $b$'s) of the model equation. Next the procedure uses guesses of the values of the parameters that are provided by the user to determine where to start on the path to the best estimates. Guesses of the estimates that are near the correct values are best because they generally increase the chance that the procedure will converge on the correct values. However, I used an initial guess of 1.0 for each free parameter to illustrate how quickly and efficiently the procedure can converge on the correct values.

The procedure uses the current estimates of the values of the parameters and the partial derivatives to "aim" toward new estimates of the values of the parameters in a way that seems most likely to minimize the weighted error sum of squares. The new estimates become the current estimates and this procedure is repeated over and over until no further improvements in the estimates of the values of the parameters are possible.

The operations are analogous to walking through a valley looking for the lowest point in the valley, but the ground is non-absorbing, so all you need to do is pour some water on the ground and follow where it goes. You aren't constrained to stay on the valley floor, and can jump to new locations on the basis of where the first and second partial derivatives of all the parameters point to. The procedure doesn't always work because there may be local puddles where the procedure becomes stuck because it fails to see the bigger picture. However, the procedure works very well in the present well-defined (and puddle-free) situation.

The computer output from the NLIN procedure is shown on pages 43 through 45 below. The "Iterative Phase" table at the top of page 43 shows the two iterations that NLIN performed to estimate the values of the parameters. On the zeroth iteration (i.e., just before the first iteration) the procedure reports that the initial estimates of the values of the parameters are all the value 1.0000, which are the starting values that I specified. (It isn't possible to specify subscripts on parameter names in standard NLIN output. Therefore, the parameters that are referred to as $b_0$, $b_1$, $b_2$, ... in the present discussion are referred to as B0, B1, B2, ... in the NLIN output.) On the first iteration NLIN amazingly jumps to almost the final values and on the second iteration NLIN does some final small adjustments to the values and then (following a sensible stopping rule) stops, with the final estimates of the values of $b_0$, $b_1$, $b_2$, and $b_4$ given in the bottom row of the table.

(The value of parameter $b_3$ doesn't appear in the output because Moldover et al. fixed this value in their analysis at $1.45 \times 10^{-18}$ on the basis of work by Goodwin [1988], as discussed in section 9.4 of their article. Thus the value of $b_3$ was fixed the same way in the present analysis.)

The bottom row of the Iterative Phase table on page 43 indicates that the final estimated value of $b_0$ is 94,756.2. As discussed above, we simply multiply this value by the conversion factor to obtain the estimate of $R$. However, the *displayed* value of $b_0$ on page 43 isn't precise enough for the present exercise, although the *internal* value SAS computed is sufficiently precise. (On my Windows computer standard SAS internal numerical values are maintained with a precision of roughly fifteen decimal digits. SAS can be forgiven for the insufficient display precision because very few scientific exercises require the high precision that is required here.) Thus we ignore the parameter estimates on page 43, and we work with more precise versions of these estimates that are given on a later page of the output.

The last column of the Iterative Phase table on page 43 shows the sum of the (weighted) squared (vertical) deviations of the 70 points from the fitted model equation after each iteration. On the basis of my estimated parameter values of all 1's, the first row of the table indicates that the weighted sum of squares (SS) just before the first iteration was $5.51 \times 10^{26}$, which is a rather large number. The next two rows in the table indicate that first iteration reduced the weighted sum of squares to 85.7379, and the second iteration reduced it down to its (absolute) minimum value of 85.7373.

As noted, SAS reports that the estimated value of $b_0$ is 94,756.2. How can the statistical procedure compute the $y$-intercept of the line despite the fact that the inverse-pressure term goes to infinity as the pressure goes to zero? Shouldn't the $y$-intercept be "infinity"? It turns out that this isn't an issue because the weighted-least-squares procedure for fitting the line to the data ignores the fact that the inverse-pressure term goes to infinity. That is, the statistical fitting of the line takes place *only* at the 70 points behind the twelve black dots on Figure 2. This yields a sensible estimate for $b_0$ at the point where the smoothly extended line intercepts the vertical axis, as suggested by the figure. In other words, the $y$-intercept is computed as the vertical anchor for the line that yields the smallest (weighted) sum of squares of the residuals at the fourteen pressures.

The second table on page 43, "Estimation Summary," reports various statistics about the success of the estimation procedure. These aren't discussed here except to note that the very small value for R in this table indicates that the estimating procedure was successful. (R must be distinguished from $R$, the molar gas constant. R can be viewed as the average relative residual.) The numbers in the Estimation Summary table are defined (using the language of matrix algebra) in the manual for the NLIN procedure (SAS Institute Inc. 2009, p. 4308).

The table at the bottom of page 43 reports the results of a statistical test of whether the line that is fitted to the 70 points appears to be no different from a horizontal straight line. If the best-fitting line is a horizontal straight line, this would imply that there is no evidence of a relationship between speed-squared and pressure. We can readily see by scanning Figure 2 that the best-fitting line isn't a horizontal straight line, so this test is unnecessary in the present case, but the test is sometimes important. The key number in this table is the *p*-value, which is given in the rightmost column (labeled "Approx Pr > F"), and which is reported as being less than .0001. The fact that the *p*-value is very low implies that this test provides strong evidence that the line isn't a horizontal straight line.

The upper table on page 44 gives some information about each of the parameter estimates. However, the key values in this table also aren't precise enough for the present discussion, so we ignore it. The lower table on page 44 reports the correlations among the free parameter estimates, which aren't of direct interest, so we also ignore them.

The table on page 45 provides the key outputs from the analysis, reporting information about the parameter estimates with the appropriate precision. The number in the second column of the first row tells us that the estimate of the value of $b_0$ is $9.475\,617\,8 \times 10^4$, which is equivalent to 94,756.178.

Table 4 compares the parameter estimates on page 45 below with the parameter estimates that Moldover et al. obtained, as reported in column 1 of their Table 11 (1988, p. 133).

Table 4

Comparison Between the Values Obtained in the Present Analysis and the Values Obtained by Moldover et al.

| Parameter | Value in Present Analysis* | Moldover et al. Value | Percent Difference |
|---|---|---|---|
| $b_0$ | 94,756.178 | 94,756.178 | 0 |
| $b_1$ | $2.2503 \times 10^{-4}$ | $2.2502 \times 10^{-4}$ | 0.006 |
| $b_2$ | $5.320 \times 10^{-11}$ | $5.321 \times 10^{-11}$ | 0.01 |
| $b_4$ | $2.7 \times 10^3$ | $2.7 \times 10^3$ | 0 |

*Values are rounded to the number of significant digits reported by Moldover et el.

If the analyses are both done correctly, and if the same data are used in each analysis, the second and third columns in Table 4 should be identical. The table indicates that the values are identical in two cases and within one hundredth of one percent in the other two cases, so the values are quite close. The difference between the two $b_1$ values and the difference between the two $b_2$ values may have arisen due to differences in roundoff errors between the computer programs that Moldover et al. used and the SAS NLIN program I used for the present analysis. Or the differences may be due to small discrepancies between the data analyzed in the present analysis and the data Moldover et al. analyzed, possibly a reflection of roundoff of the published data. Or the differences may have arisen for some other reason.

As indicated in the table on page 45, NLIN computed an approximate standard error (standard uncertainty) for each of the parameter estimates. If certain often-satisfiable assumptions are adequately satisfied, we can expect the estimated value of the parameter being estimated to be within one standard error of the theoretical "true" value approximately 67% of the times that the estimate is computed if the analysis is performed over and over, each time with fresh data. Similarly, we can expect the estimated value of the parameter to be within *two* standard errors of the true value approximately 95% of the times that the estimate is computed if the analysis is performed over and over, each time with fresh data.

The lower 95% confidence limit for each parameter reported on page 45 is computed by subtracting approximately two standard errors from the associated parameter estimate. We can see that the lower 95% confidence limit for $b_4$ is less than zero, but the upper 95% confidence limit for $b_4$ is greater than zero. Thus the 95% confidence "interval" spans the value zero. This raises the possibility that the correct value for $b_4$ is actually zero. Of course, if the correct value of a multiplicative parameter is zero, this has the effect of removing the term associated with the parameter from the model equation.

The *t*-value for each parameter given in the second-last column in the table on page 45 is simply the parameter estimate divided by its estimated standard error. The *t*-value can be shown to have a central *t* "distribution" with 70 - 4 = 66 degrees of freedom if the correct value of the parameter in nature is zero and if certain often-satisfiable assumptions are adequately satisfied, as discussed by Chatterjee and Hadi (2006, ch. 4).

The *p*-value in the last column in the table on page 45 is derived from the *t*-value and is the estimated probability that the *t*-value will be equal to or exceed its estimated value *if the actual value of the parameter in the population is zero* and if the assumptions are adequately satisfied. It is a widely accepted statistical convention that we can't "reject" the "null" hypothesis that a parameter has a population value of zero unless the associated *p*-value is less than .05. This amounts to requiring that the size of the parameter must be at least approximately twice its standard error before we believe that the parameter is different from zero.

The *p*-value column on page 45 indicates that the *p*-value for $b_4$ is approximately .36, which is greater than .05. Thus the output suggests that we don't have good evidence from this test that $b_4$ is different from zero.

As noted, a *p*-value is valid only if the assumptions underlying the *p*-value are adequately satisfied. There appears to be no evidence in the Moldover et al. data that the assumptions underlying the *p*-values aren't adequately satisfied. Interestingly, however, the assumptions underlying the *p*-values aren't *completely* satisfied by the data used in the present analyses because the distribution of the 70 residuals isn't independent of the other variables, as shown by the undulating pattern in Figure 4 above and in Figures 6, and 7 below. These figures show that the residuals are somewhat related (in both spread and value) to pressure. Also, the assumption of independence of the predictor variables is violated because $P$, $P^2$, $P^3$, and $P^{-1}$ are correlated with each other through their algebraic interrelationships. However, in this example many statisticians will agree that the assumptions are *adequately* satisfied in the sense that the obtained *p*-values provide good evidence that the linear and quadratic terms are needed in the model equation (because the relevant *p*-values are extremely low—less than .0001). Also, the high *p*-value for the inverse-pressure term implies that we have no evidence that this term is needed.

(Schenker and Gentleman [2001] compare the *p*-value and confidence interval methods for determining whether the value of a parameter is significantly different from zero and conclude that the *p*-value method is preferred.)

As noted, the parameter estimate for $b_0$ on page 45 is 94,756.178. We can convert this estimate to an estimate of the molar gas constant, $R$, by multiplying it by the conversion factor, which has a value of $8.774\ 595\ 1 \times 10^{-5}$.

[The conversion factor equals $0.023\ 968\ 684$ divided by 273.16, which is $M/(T_1\gamma_0)$ in equation 1.5 on page 88 of the Moldover et al. article. The value .0239... is from the second line of Table 8 (page 124) of the article, but converted from grams per mole to kilograms per mole because the analysis is in SI units, and the kilogram is the SI unit of mass. Note that a slightly different value is given on page 129 at the end of section 7.4 and explained in section 7.5.]

Multiplying the estimated speed-squared at zero pressure (i.e., 94,756.178) by the conversion factor yields an estimated value of $R$ of 8.314 471, which is the value that is reported in the Moldover et al. article. The result of this computation is shown on page 46 below. (The result on page 46 is slightly different from the value obtained by multiplying the two values given above because the analysis on page 46 is based on the estimated value of $b_0$ with all the available significant digits instead of the only 8 significant digits for $b_0$ shown in the first sentence in this paragraph.)

### G.12 The Residuals in the Moldover et al. Analysis

Following the curve-fitting operation discussed above, NLIN generated a *predicted* speed-squared from the derived model equation for each of the 70 Moldover et al. data observations. This value was computed for each observation by substituting the measured value of pressure into the right-hand side of the derived model equation and then evaluating the expression. (The predictions are different at different pressures, but at any of the fourteen pressures all of the five predicted speed-squared values made by the equation are obviously the same because the pressure is the same.) NLIN also computed a residual for each observation, which (as noted above) is the actual measured speed-squared value for the observation *minus* the predicted speed-squared value for the observation. (These residuals are generally different for each of the 70 observations.)

Pages 47 through 51 show a copy of the data table on pages 40 and 41, but with four new columns of numbers on the right side of the table. The predicted speed-squared values and the residuals are shown as the third and second columns from the right of the lower section of this table on each page.

The residual for the first observation on page 47 is computed as follows: The SpeedSq (third) column of the table indicates that the measured speed-squared value for the first observation is 94,763.099. In contrast, the speed-squared value predicted for this observation by the fitted polynomial line is 94,762.032, as shown in the PredSpeedSq column of the table. Therefore, the residual is 94,763.099 - 94,762.032 = 1.067, as shown in the PredSpeedSqResid column of the table.

The last column in the lower section on pages 47 through 51 shows the scaled residuals, which are described above on page 20. Figure 4 above is a scatterplot of the scaled residuals as a function of the values shown in the Mode and Pressure (converted to kilopascals) columns of the same data table.

### G.13 Fitting the Line With an Unconstrained Pressure-Cubed Parameter

The analysis above fitted (3) to the Moldover et al. data with $b_3$ having its value fixed (constrained) at $1.45 \times 10^{-18}$ which, as noted, is an estimate for $b_3$ that was obtained by Goodwin (1988). It is of interest to fit the same line, but with $b_3$ unconstrained and estimated from the data, just like the other four parameters of the model equation. Pages 52 through 55 below show the results of this analysis.

The Iterative Phase table on page 52 shows that convergence was achieved in two iterations. The Estimation Summary table on page 52 implies that the estimation was successful. Page 54 reports the estimates of the obtained parameters to the required number of significant digits. In this case the estimate for $b_0$ is 94,756.406. If we multiply this number by the conversion factor, we obtain an estimated value of $R$ of 8.314 491, as shown on page 55. This is somewhat higher than the Moldover et al. estimate of 8.314 471.

Note the interesting fact on page 54 that the parameter estimate for $b_3$, the parameter for the cubic term, is a negative number ($-6.88 \times 10^{-18}$), while the parameter estimate that Moldover et al. used in their model equation was a positive number ($1.45 \times 10^{-18}$). This change of sign of the estimate for $b_3$ is unexpected, and if things were working properly, we would expect the estimated value of the cubic parameter to have the same sign as the value Moldover used from the earlier research, even though the two values would almost certainly be somewhat different due to differing measurement errors in the two research projects. This change in sign suggests that the cubic parameter is waggling around zero and thus the cubic term may be unnecessary. Similarly, the estimate for $b_4$, the parameter for the inverse-pressure term, is now negative ($-3.54 \times 10^{3}$) although the Moldover et al. estimate was positive ($2.7 \times 10^{3}$), which suggests that the inverse pressure term may be also be waggling around zero and thus may also be unnecessary.

The ideas in the preceding paragraph are reinforced by the fact that the table on page 54 indicates that the *p*-value for $b_3$ is approximately .10 and the *p*-value for $b_4$ is approximately .40. Since both of these *p*-values are greater than .05, this suggests that we have no evidence that either the cubic or inverse-pressure term is needed in the model equation.

### G.14 Fitting the Quadratic Line

The preceding discussion concludes that there is no evidence in the analysis of the Moldover et al. data with the Moldover et al. model equation that the cubic and inverse-pressure terms are needed in the equation. Thus the analysis was repeated with these two terms omitted, and thus a (weighted) quadratic line (model equation) was fitted to the data. The results of this analysis are reported on pages 56 through 59.

The Iterative Phase and Estimation Summary tables on page 56 imply that the line was successfully fitted. The *y*-intercept of this line is reported with the necessary precision in the table on page 58 as 94,756.251. If we multiply the *y*-intercept by the conversion factor we obtain an estimate of $R$ of 8.314 477, as shown on page 59, which is quite close to the Moldover et al. value of 8.314 471. Thus including or omitting the cubic and inverse terms in the model equation has only a small effect on the estimate of $R$.

Figure 6 shows the scaled residuals from the quadratic line as a function of pressure and resonance mode. These residuals are almost the same as the residuals from the Moldover et al. line, which are shown above in Figure 4. (The residuals in Figure 6 are sometimes larger and sometimes smaller than the corresponding residuals in Figure 4.)

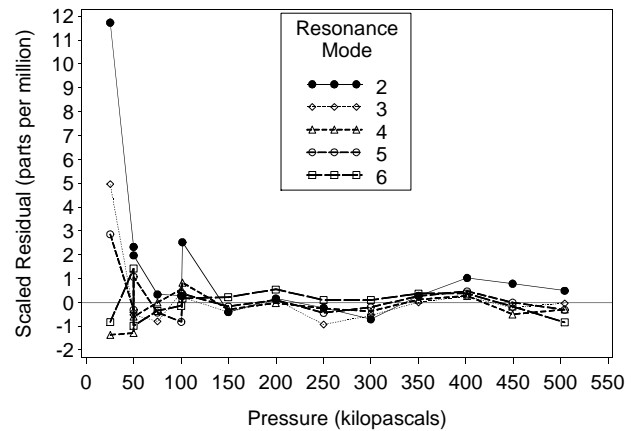

*Figure 6.* A scatterplot of the scaled residuals from the quadratic model equation as a function of pressure and resonance mode.

The goodness of the fit of the Moldover et al. model equation and the quadratic model equation can be compared by considering the "average" length of the residuals in Figures 6 and 4. A standard measure of the average length of the residuals is the weighted mean-square error, which is the sum of weighted squared residuals divided by their "degrees of freedom". (The degrees of freedom is the number of observations in the data minus the number of parameters estimated = 70 - 3 = 67 for the quadratic model. This number is used as the divisor because it can be shown to give an unbiased estimate of the true mean-square error.) The weighted mean-square error of the Moldover et al.

model equation is computed (on page 43) as 1.2990, but the weighted mean-square error of the quadratic equation is computed (on page 56) as 1.2609, which is slightly lower than the Moldover et al. value. The lower weighted mean-square error value for the quadratic model occurs because the forced positive value for the pressure-cubed parameter in the Moldover et al. model (although the data reveal that a negative value or zero is more appropriate) worsens the fit. Thus from the point of view of minimizing the weighted mean-square error we find that the quadratic model equation fits the data slightly better than the Moldover et al. model equation.

(In saying that the quadratic model fits better we needn't attempt to demonstrate that the quadratic model is *significantly* better than the Moldover et al. model because the simpler quadratic model isn't required to perform this test to support its worthiness. Instead, the principle of parsimony implies that the onus lies on the more complicated Moldover et al. model (3) to establish through appropriate statistical tests *its* worthiness over the simpler quadratic model. As discussed above in Appendix G.13, the high *p*-values on the two relevant statistical tests imply that the Moldover et al. model was unsuccessful at demonstrating its worthiness.)

Figure 7 shows the raw residuals multiplied by the square roots of the values of the weight variable. (The square roots of the values are used because the weight variable is applied to the *squared* residuals in fitting the line to the data.)

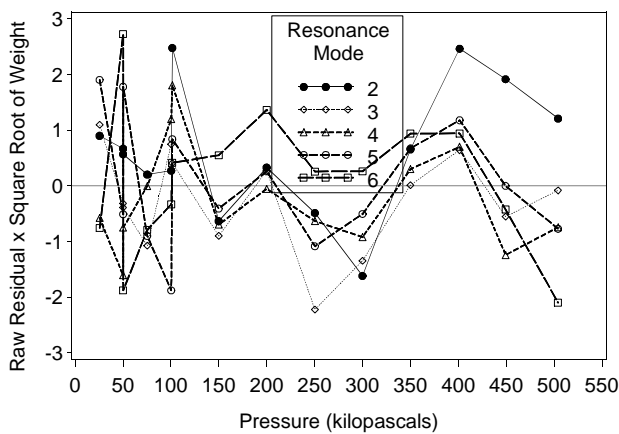

*Figure 7.* A scatterplot of the weighted raw residuals from the quadratic model equation as a function of pressure and resonance mode.

The figure shows that the weight variable successfully makes the spread of the residuals somewhat more consistent across the different pressures. However, the residuals still show more variation at lower pressure and the undulating pattern in the residuals above 100 kilopascals is still present.

### G.15 Fitting the Quadratic Line with a Linear Regression Program

For completeness, pages 60 through 62 show fitting the quadratic line with the SAS linear regression procedure (called REG), which is the more standard regression program approach in situations like the situation studied here. Because the same line is fitted, this analysis gives exactly the same estimates for the values of the parameters as can be seen by comparing the parameter estimates in the tables on pages 58 and 62.

(The estimates aren't *precisely* the same as can be seen by comparing the *t*-values in the two analyses, which are all slightly different. The differences arise because the two procedures are using different algorithms to estimate the values of the parameters and their standard errors. That is, REG uses a closed-form algorithm, and NLIN uses an iterative algorithm. Although these algorithms theoretically converge on the same solution, their results can differ slightly due to the effects of computer roundoff errors.)

These analyses were begun with NLIN as opposed to REG because (as is typical for a regression program) REG can't directly handle the Moldover et al. approach of setting the parameter for the cubic term in the model equation to a fixed value. However, NLIN can directly handle this situation.

For the analysis on pages 60 through 62 SAS gives a warning in its log that the range of SpeedSq is small relative to its mean, and thus computing precision may be lost. Therefore, pages 63 through 65 repeat the analysis with a rescaled version of the variable SpeedSq, called SpeedSqResc, which is simply SpeedSq values minus an approximate average of their values, which was taken to be 94,808.35. The values of SpeedSqResc are shown in the fourth-last column in the data table on pages 47 through 51. After the rescaling is reversed this analysis gives exactly the same *y*-intercept (to the allowable number of significant digits) and thus exactly the same estimate for *R*, as documented in the text at the bottom of page 65.

The SAS REG procedure can perform an analysis of the "collinearity" of the predictor variables to determine whether the predictor variables are highly correlated with each other. This is of interest because high correlations among the predictor variables can make the parameter estimates and their standard errors highly variable. The "variance inflation" factor is a measure of collinearity and is given on page 63 as 21.2. This value is greater than 10, which Chatterjee and Hadi (2006) suggest is a threshold above which the data may have collinearity problems.

A second form of collinearity analysis is performed on page 64. Here the largest condition index for the more reasonable "intercept adjusted" analysis is 9.1, which is less than 30, which Schabenberger and Pierce (2002) suggest is a sensible maximum value for this number before collinearity should be considered. Chatterjee and Hadi (2006) suggest a maximum of 15. Since the variance inflation factor limit is exceeded, but neither of the condition index limits is exceeded, the data are on the borderline of having worri-some collinearity. Friendly and Kwan (2009) discuss effective ways of visualizing collinearity diagnostic statistics.

### G.16   A Simulation of the Moldover et al. Analysis to Examine the Effects of Collinearity in the Quadratic Model

The analysis in the preceding subsection suggests that collinearity between the two predictor variables might lead to problems with the results. Pages 67 through 79 show the results of a simulation to check whether the collinearity might lead to a biased estimate of $R$ or might lead to a biased estimate of the error in $R$. As we shall see, no evidence of bias is found.

The simulation uses a random number generator to generate 100,000 different sets of 70 rows of data that are highly similar to the Moldover et al. data. The data are generated from a quadratic model equation with parameters equal to the parameters given above except that the intercept is set to zero to avoid the warning in the SAS log. (This change of intercept doesn't affect the correctness of the following results.) This simulation then performs a separate quadratic regression analysis on each of the 100,000 datasets and then examines the distribution of the 100,000 estimated intercept parameters to see if the expected (i.e., average) intercept value is actually equal to the intercept value used to generate the data (i.e., zero).

Pages 67 through 70 show the data for the first two of the 100,000 datasets. The two datasets are indexed by the variable in the column labeled "Iteration Number".

After the 100,000 datasets are generated, a separate quadratic regression analysis is performed on each dataset—i.e., 100,000 independent regression analyses are performed. Pages 71 through 75 show the output from the first 5 of the 100,000 regression analyses. These analyses are indexed by the variable "Iteration Number" given near the top of each page. As can be seen on page 71, in the first regression analysis the intercept is estimated as -0.005 46 and the standard error of the intercept is estimated as 0.298 97. Note how the estimates of the intercept and its standard error are slightly different on pages 72 through 75 because the datasets are all slightly different from each other due to the use of the random number generator to generate the data.

The estimated intercept and the estimated standard error of the intercept from each of the 100,000 regressions are collected in a new dataset. The square of the estimated standard error of each intercept (i.e., the estimated variance) is added as a new variable to the new dataset for later use. The first 30 of the 100,000 rows in the new dataset are shown on page 76. As before, these 100,000 rows are indexed by the variable in the column labeled "Iteration Number". Note how the values in the second and third columns in the first five rows of the table are identical to the corresponding values on pages 71 through 75.

Page 77 shows the mean, standard deviation, skewness, and kurtosis of the 100,000 values of the three variables in the dataset. Page 78 shows another version of the information on page 77 to assist with checking the computations behind page 79.

Page 79 summarizes key computations based on the data on pages 77 and 78. In this analysis we see that the $p$-value for the one-sample $t$-test of the null hypothesis that the mean estimated intercept equals zero is 0.712. Since this $p$-value isn't less than 0.05, there is no evidence that the estimated intercept is different from zero. Thus we have no evidence from the simulation of any difference between the average value of the estimated intercept and the actual correct value used to generate the data.

Similarly, we see the $p$-value for the variance ratio test of the null hypothesis that the mean of the variances estimated by the regression program is equal to the actual variance of the estimates. This $p$-value is 0.509. Since this $p$-value isn't less than 0.05, there is no evidence of any difference between the estimated and actual variances.

Thus the simulation implies that we have no evidence that the moderate collinearity in the data creates a bias in the estimate of $R$. Similarly, we have no evidence that the moderate collinearity creates a bias in the estimated standard error of the estimate of $R$.

The simulation analysis can be rerun with different values of the "seed" for the random number generator, which gives slightly different results. Also, the preceding analysis was run with only 100,000 regressions to keep execution time in the example to a reasonable length. However, it is easy to change the number of regressions to, say, 20 million, which can be performed and the results analyzed with a desktop computer in around 6 hours. This brings the mean of the estimated intercepts still closer to zero. Even in such large runs there is no significant evidence of bias in the estimate of the intercept or in the estimate of the standard error of the intercept.

### G.17   Fitting a Ninth-Degree Polynomial Line

Pages 80 through 88 report the successful fitting of a ninth-degree polynomial line to the data. The results on pages 80 through 85 show the progress through the 18 necessary iterations. The results on page 88 show the surprising result that all of the terms above the linear term have $p$-values that are greater than .05. The reason why the quadratic term (which has a very low $p$-value in the earlier analyses) doesn't have a $p$-value less than .05 in the present analysis is that the operation of this term has been spread (in a mathematically efficient fashion) over all the higher-degree terms, which drains significance from the quadratic term. Thus we have no evidence from this analysis that any of the second-degree through ninth-degree terms are needed in the model equation.

As shown on page 86, the weighted mean-square error from the ninth-degree equation is 1.1262 which is somewhat less than the value of 1.2609 that was obtained using the quadratic equation, as shown on page 56. Figure 8 shows the distribution of the scaled residuals from the ninth-degree equation as a function of pressure and resonance mode.

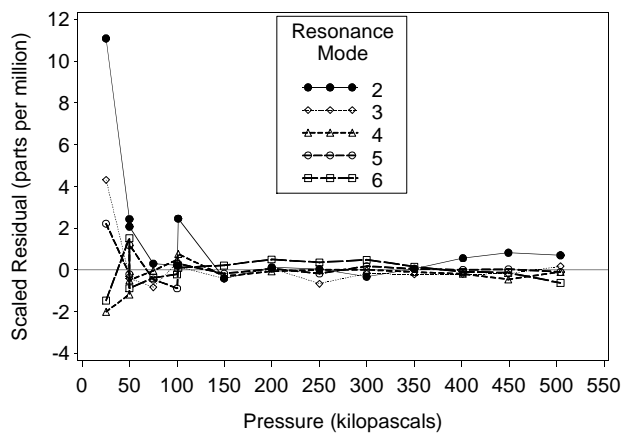

*Figure 8.* A scatterplot of the scaled residuals from the ninth-degree polynomial model equation as a function of pressure and resonance mode.

The pattern of the residuals in Figure 8 is similar to the pattern in Figures 4 and 6, although the undulating pattern has vanished. Thus the ninth-degree polynomial equation succeeds in modeling the undulating pattern in the residuals. But this isn't surprising in view of all the terms included and in view of the number of permissible undulations in a ninth-degree polynomial line. Thus we lack good evidence that the ninth-degree polynomial line is the correct line, even though it fits the data well.

Pages 89 through 91 show the fitting of a ninth degree polynomial line with SAS PROC REG, which is a commonly used SAS program for fitting linear model equations. The table at the bottom of page 89 indicates that the $p$-values (labeled Pr > |t|) for all the terms above the linear term are again greater than .05. Note how the procedure refers on page 90 to "singularities" in the data due to the high correlations between the nine predictor variables. Note also the condition indices greater than 30 in the table on page 91 indicating the presence of high collinearity. The estimated values of the parameters on page 89 are close to the estimates on page 88, but not identical.

Page 92 show fitting of a ninth degree polynomial with SAS PROC ORTHOREG, which is a procedure designed to deal with high correlations among the predictor variables. Note how the parameter estimates are similar to but different from the estimates in the two earlier runs and again the $p$-values for terms above the linear term are all greater than .05. Note also that ORTHOREG decided to omit the last term (pressure to the ninth power).

Each of the three ninth-degree analyses yields $p$-values for the terms above the linear term that are all greater than .05. However, the earlier analyses imply that a quadratic term is needed. Therefore, fitting a ninth-degree line seems inappropriate.

### G.18  Fitting a Line with Stepwise Regression

In the ninth-degree polynomial fitting in the preceding subsection it seems likely that too many terms were included in the model equation. This suggests that we could enter terms into the model equation one at a time in a rea-sonable order, keeping terms in the equation only if they are statistically significant. This is an example of the "stepwise" approach to determining the best model equation and is illustrated on pages 93 to 103 of the output.

The procedure starts with a model equation that contains only the intercept term. Then the $p$-value is computed for the parameter for each candidate term if the term were "entered" by itself as a term in the equation. Then the term with the parameter with the lowest $p$-value is entered into the equation (assuming that the $p$-value is less than the value 0.15 that I specified in the instructions to the program). The process is repeated with the remaining candidate terms to enter a second term (assuming that the $p$-value is again less 0.15) and possibly to enter further terms, one at a time. At each step, the parameters of the terms that are in the equation are examined, and if any parameter has a $p$-value greater than 0.05, the associated term is removed from the equation. (Using different $p$-values for entry and removal of terms facilitates identification of "near-significant" terms.) This process is repeated until no more new candidate terms for entry into the model equation have parameters with $p$-values that are less than 0.15.

In the present analysis the following 17 functions of pressure ($P$) were provided to the program as candidate variables for entry in separate terms in the model equation: $P^{1/3}$, $P^{1/2}$, $P$, $P^2$, $P^3$, $P^4$, $P^5$, $P^6$, $P^7$, $P^8$, $P^9$, $P^{10}$, $P^{11}$, $P^{12}$, $P^{13}$, $P^{14}$, and $1/P$. Page 93 shows the list of the candidate variables and the $p$-values of their parameters (in the "Pr > F" column) at the beginning of the first step of the procedure. PCbRt stands for $P^{1/3}$ and P2 stands for $P^2$ etc. We see that the $p$-values are all so low that they aren't reported exactly, but are only reported as being less than 0.0001. However, we can infer which $p$-value is lowest from the $F$-values because the $p$-values are (for any fixed degrees of freedom, DF) monotonically inversely related to the $F$-values. That is, the term with the highest $F$-value has the lowest $p$-value. Thus we infer that the term with pressure to the first power ($P$) has the lowest $p$-value because this term has the highest $F$-value, which is 96,845.9. Thus the program enters this term into the model equation, as noted in the second line from the bottom of page 93. Some statistics for the model equation with only the term for the $P$ variable (and the intercept) in it are given in the top half of page 94.

SAS reports that the *variable P* is entered into the equation, but this paper says that the *term* with the variable $P$ is entered. The term is the product of the variable (or variables or a function of the variables) and the parameter associated with the term. Some other sources also refer to entering variables. However, referring to terms instead of variables is slightly more self-explanatory and better handles the case when multiple variables appear within a single term.

The bottom half of page 94 reports the beginning of the second step, which examines the 16 remaining candidate terms to determine which term to enter next. We see that once again all the terms that aren't in the equation have very low $p$-values, but the $P^2$ term has the lowest $p$-value (as indicated by the fact that it has the highest $F$-value).

Thus the program enters the $P^2$ term into the equation in this step, as reported near the middle of page 95. (Note the $F$-value on page 94 for the parameter for the $P^3$ term is close to the $F$-value for the parameter for the $P^2$ term, so the program came close to entering the $P^3$ term instead of the $P^2$ term. If new similar data were analyzed, the program might enter the $P^3$ term in this step due to random variation in the data.)

Page 97 reports that the $P^9$ term is entered next, with a $p$-value for the parameter of .035, which is only slightly less than the critical value of .05. The program reports an examination of the remaining 14 candidate variables on page 99 and concludes that no more terms can be entered or removed according to the $p$-value rules I specified for entry and removal of terms. Therefore, the program stops entering terms and prints a summary of the steps that were executed and prints some details about the chosen model equation on pages 100 through 102. Page 101 indicates that the weighted mean-square error for the chosen model equation is 1.1963, which is lower than the 1.2609 obtained for the quadratic equation, but higher than the 1.1262 obtained for the 9th degree polynomial equation (pages 86 and 89) with all the standard polynomial terms.

The stepwise analysis has arrived at the quadratic model discussed above, but with the addition of one more term—a term for the variable $P^9$. This additional term is included to take account of the undulating pattern in the residuals shown in Figure 6. The stepwise procedure has resulted in a more parsimonious and better-behaved model equation than the equation discussed in the preceding subsection in which *all* the terms with powers of pressure between the 1st and 9th power are included in the equation. Page 103 reports that the estimated value of $R$ with the chosen model equation is 8.314 480.

We can see that the chosen model equation has collinearity problems because the largest variance inflation factor on pages 101 and 102 is 74.9, which is greater than the recommended maximum value of 10 proposed by Chatterjee and Hadi (2006). Also, the largest condition index for the intercept-adjusted approach on page 102 is 18.5, which is greater than the value of 15 recommended by Chatterjee and Hadi (2006), although it is less than the value of 30 recommended by Schabenberger and Pierce (2002).

As noted above in Appendix G.6, it would be premature to accept the addition of the 9th-degree pressure term to the quadratic model equation until the possibility that extraneous factors caused the undulating pattern in the residuals is ruled out.

The preceding discussion used a $p$-value criterion for choosing terms to enter into the model equation. Other sensible criteria for choosing terms in stepwise regression are also available in statistical software packages. It is useful to try different criteria to see how well they agree among themselves as to the best form of the model equation. Also, polynomial equations aren't the only option for modeling the undulating pattern in the residuals, and other model equations (e.g., with trigonometric terms) might also be explored to model the pattern.

In making terms available to a stepwise variable selection procedure it is generally sensible to ensure that all the simplest possible terms are included in the set of candidate terms for inclusion in the equation. This ensures that the simplest terms will be included in the equation if they belong, which satisfies the principle of parsimony. Also, the intercept term is almost always included in the equation because omitting it implies that we are studying the situation in which the line for the relationship passes through the origin of the coordinate system, which is a situation that occurs only rarely.

The stepwise approach is practical, and the $p$-values it provides for the parameters in the final model equation can be viewed as sensible measures of the weight of evidence for the inclusion of the associated terms in the model equation. However, the iterative operation of the approach implies that the $p$-values are technically no longer correct estimates of the probabilities they purport to estimate. (The $p$-values tend to be somewhat lower than they should be.) Thus any terms in the model equation whose parameters have borderline statistical significance are questionable. A sensible resolution of this problem is to perform appropriate new research to study whether further evidence can be found to support the inclusion of any questionable terms in the model equation.

Including terms with variables with large powers in a model equation will lead to computer numerical problems if the computed values approach or exceed the computer's absolute numerical magnitude limit. If the value of the variable associated with a term is greater than one, then high powers of the variable will tend to become very large, perhaps larger than the computer can handle. In contrast, if the value of the variable associated with a term is less than one, then high powers of the variable will tend to become very small, perhaps smaller than the computer can handle. For example, the largest pressure value in the Moldover et al. data is roughly 504,000 pascals. If we raise this number to the 14th power, we get a value of roughly $6.8 \times 10^{79}$. On IBM mainframe computers the largest number that can be handled in the number format used by SAS is roughly $7.2 \times 10^{75}$, which is thus less than $P^{14}$ for a pressure value of 504,000 pascals. Thus the analysis would be unreliable on an IBM mainframe computer. In contrast, on personal computers such as mine that use the IEEE 754 number formats (which are used by both Windows and Unix) the largest number that can be handled in the number format used by SAS is roughly $10^{308}$ so (assuming no other problems) the analysis should be reliable on such a computer.

### G.19  An Analysis in Units of $R$

For completeness, pages 104 and 105 report a quadratic regression analysis with the response variable in units of $R$ (as opposed to the units of speed-squared used above), and pages 106 and 107 repeat the analysis with the response variable in rescaled units of $R$ (to eliminate the warning in the SAS log). Simple mathematical statistical considerations imply that exactly the same estimated value for *R* should be obtained from these analyses (except for possible insubstantial differences in roundoff errors) as was obtained in the quadratic regression with the response variable in units of speed-squared. This equivalence of the approaches is shown on page 108.

## G.20 Fitting the Quadratic Line with Scaled Orthogonal Regression

The preceding analyses are all *standard* least-squares analyses. For comparison, this section discusses the output from a scaled orthogonal least-squares analysis (Method 3 in section 2 above) using the quadratic model equation for the relationship between speed-squared and pressure. The analysis uses the approach given by Björck (1996, p. 354). The analysis was done with the SAS NLP (nonlinear programming) procedure, which is part of the mathematical programming subsystem of the optional OR (operations research) component of SAS. NLP is a general program that can estimate the values of the parameters of an equation in such a way that the sum of the values of a user-specified function of the parameters across all the observations in the research will have the lowest possible value.

Moldover et al. indicate in section 8.1 of their article that the pressure values have an "imprecision" of no greater than 100 pascals. This was interpreted to mean that the standard error of the pressure values is roughly 100 pascals. Similarly, Moldover et al. indicate in sections 4.2 and 4.3 that the errors in the resonance frequencies (from which the speed-squared measurements are derived) are one part per million when the pressure is 25 kilopascals. Converting one part per million in frequency units to speed-squared units implies that the speed-squared values have a standard error of roughly 0.2 meters-squared per second-squared at a pressure of 25 kilopascals. The speed-squared standard error becomes lower at higher pressures as indicated by Moldover et al.'s equation 4.3.

The scaling aspect of the orthogonal method requires that the two variables have approximately the same standard error. Satisfying this requirement causes each variable to play an equal role in the analysis. We can cause the two variables to have approximately the same standard error by scaling one of the variables—we multiply each value of the variable by the same appropriate scaling factor. In the present case we can cause the pressure and speed-squared values to have approximately the same standard error if we multiply the pressure values by .001, which (by coincidence) implies that we use pressures measured in kilopascals. This yields a standard error of the pressure values of roughly 0.1 kilopascals, which is reasonably close to the maximum standard error of the speed-squared values of 0.2 meters-squared per second-squared. Thus unlike the earlier analyses, the present analysis was done using the pressure values measured in kilopascals.

(The scaling operation described in the preceding paragraph is approximate. The operation could be done more accurately if more accurate information about the standard errors were available. More accurate scaling might slightly change the estimate of *R*.

(In orthogonal regression a change in the measurement units of a variable generally causes a change in the "orthogonal direction" between each point and the fitted surface, which generally causes changes in the estimated value of the parameter[s]. As suggested above, if two continuous variables don't have the same standard errors, we can multiply all the values of one of the variables by the appropriate scaling factor to scale up or scale down the standard errors to equal the standard errors of the other variable. Although this scaling approach generally slightly changes [in a non-proportional sense] the values of parameter estimates in *orthogonal* regression, it has [if the scaling is properly taken into account] theoretically no equivalent effect on *standard* least-squares regression, although it can lead to computer roundoff problems if not done with care.)

Because an analytic solution usually isn't available in the problems that NLP is designed to solve, the procedure works iteratively in the same manner as discussed above for the SAS NLIN procedure. Page 109 reports (using the technical names for two matrices) that NLP was able to compute analytical first and second partial derivatives (with respect to each of the parameters) of the function that is minimized in the orthogonal regression (using the quadratic model equation). If the function had been more complicated, NLP might have been unable to compute these analytical partial derivatives, and thus would have attempted to compute them numerically, which must be done repeatedly as the iterations progress, and which is therefore more computer intensive and somewhat less reliable.

The scaled orthogonal method with the quadratic model equation has the same three parameters of the quadratic line ($b_0$, $b_1$, and $b_2$) as the standard least-squares approach with the quadratic line, and the main parameter of interest is still $b_0$ because its final estimated value is rescaled and multiplied by the conversion factor to obtain the estimate of *R*. However, the scaled orthogonal method has 70 additional "parameters"—one for each of the 70 observations in the data. In specifying the analysis to NLIN I called these 70 additional parameters $PErr_1$ though $PErr_{70}$. Each $PErr$ estimates the size of the measurement error in the pressure value in one of the 70 observations in the data. Pages 110 through 112 report the initial estimates that I specified for the 73 parameters and indicate that I specified an initial estimate of 1.0 for the three parameters of the quadratic equation (i.e., $b_0$, $b_1$, and $b_2$) and I specified a reasonable initial estimate of 0 for each of the $PErr$s.

The equation is fitted to the data in a way that minimizes

$$\sum_{i=1}^{70} w_i(e_i^2 + PErr_i^2), \qquad (5)$$

where

$w_i=$ the value of the weight variable for the *i*th row of data in the 70 rows of data, as given in the rightmost column of the table on pages 40 and 41 $e_i^2$ = the squared error in the prediction made by the (quadratic) model equation of the value of speed-squared for the $i$th row in the data (or, omitting the idea of 'prediction', the estimated error in the $i$th speed-squared value)

$PErr_i^2$ = the estimated squared error in the $i$th pressure value.

Some of the details of the minimization are illustrated in the SAS program code used to perform the orthogonal fitting (Macnaughton 2010a).

Note that if the $PErr^2$ term is omitted from (5), the expression degenerates to the expression that is minimized in a standard weighted least-squares regression analysis. Thus the smaller the estimated $PErr^2$'s are relative to the estimated $e^2$'s, the closer the solution for the estimated values of the $b$'s will be to the standard weighted least-squares estimates.

Pages 113 and 114 summarize the 25 iterations that NLP performed to find the optimal fit of the quadratic line to the data while taking account of the 70 pressure errors. Note how the objective function (the function that is minimized) becomes smaller after each successive iteration, being reduced from approximately $5.53 \times 10^{14}$ prior to the first iteration to approximately 79.6 after the 25th iteration.

The table of "Optimization Results" on page 114 reports key properties of the final fit of the equation to the data. The small maximum absolute gradient element value and the small slope of the search direction imply that the orthogonal fit of the quadratic line to the data is stable.

Pages 115 through 117 report the final estimates of the values of the 73 parameters, with the important centered (for numerical precision) estimate of $b_0$ being approximately -52.1. The estimated values of the $PErr$s are all reasonable small numbers that are centered around zero.

Finally, page 118 reports the computed value of $R$ from the estimated value of $b_0$ with all the available digits. This value is slightly larger than the value reported on page 108 from the standard least-squares regression analysis using the same quadratic model equation. However, if the two estimated values are rounded to the appropriate precision of seven significant digits, they are identical.

The two values are identical because, as shown using the "Sum Observations" from pages 119 and 123, at the solution of optimum fit the weighted sum of the squares of the $e$'s in (5) is approximately 16 times larger than the weighted sum of squares of the $PErr$s, even though the pressure values were scaled to show approximately the same *theoretical* standard error as the speed-squared values. Because the sum of squares of the $e$'s is larger, the absolute values of the $e$'s themselves are generally substantially larger than the absolute values of the $PErr$s, and thus the optimization routine gives more emphasis to minimizing the $e$'s, and thus the solution is close to the standard solution in which *only* the $e$'s are minimized.

If the original pressure values in pascals are scaled by a factor of .0001 (instead of the .001 that is used in the present analysis), the last significant digit of the estimate of $R$ increases from 7 to 8, which illustrates how the choice of the scaling factor can affect the value of a parameter estimate.

Although the scaled orthogonal regression procedure uses 73 parameters, the 70 $PErr$ parameters aren't formal parameters of the relationship between speed-squared and pressure. Rather, the $PErr$s estimate errors, and (unlike $b_0$, $b_1$, and $b_2$) they would be excepted to have completely different values if new data were gathered using the Moldover et al. research design.

Generally, if we fit a model equation to a set of data, we use only a few parameters, typically six or fewer. Thus the use of 73 parameters in the present example is unusual, especially because we have only 70 observations (rows) in the data table. Using such a large number of parameters can give rise to numerical instability. However, instability is less likely to occur if most of the parameters are estimating measurement errors, as in the present example. Stability obviously isn't a problem in this example because the approach leads to the same estimate of $R$ as the standard approach. However, numerical stability can be a serious problem in scaled orthogonal regression if the data aren't well behaved, as discussed by Van Huffel and Vandewalle (1991, pp. 248-250).

Because the scaled orthogonal method yields (after appropriate rounding) the same estimate of $R$ as the standard least-squares method, the complicated scaled orthogonal method and the equivalent maximum-likelihood method aren't useful in a practical sense in the analysis of the Moldover et al. data with the quadratic model equation and with the scaling that I used above. However, the method is of significant value in some situations. Van Huffel and Vandewalle (1991) discuss the scaled orthogonal method from an engineering perspective under the name of "total least squares". They say that in some cases the approach can lead to "gains of 10-15 percent in accuracy" (p. xi) and refer to simulations comparing the approach with the standard least-square approach (p. 5). They also list applications of the approach in their section 1.2. Fuller (1987) and Carroll, Ruppert, and Stefanski (1995) discuss orthogonal regression from a statistical perspective. Gustafson (2004) discusses dealing with errors in predictor variables from a Bayesian statistical perspective. Orthogonal regression is sometimes called an "errors-in-variables" approach because it takes account of the errors in the measured values of the predictor variables, which are ignored in standard approaches. (Errors in the values of the *response* variable are taken account of in both the standard and scaled orthogonal approaches.)

Orthogonal regression is appropriate when the goal is to estimate the values of one or more of the parameters of the model equation, as in the case of estimating the $y$-intercept parameter of the relationship between speed-squared and pressure as a means to estimating $R$. However, in some cases we are less interested in estimating the values of the parameters in a model equation, and we are more interested in using the equation to make predictions of new values of the response variable from the values of the predictor variables in new similar situations. In these cases standard regression is generally more appropriate than orthogonal regression, as discussed by Van Huffel and Vandewalle (1991, p. 5).

Pages 126 through 138 repeat the preceding analysis except that the mathematically asymmetrical roles of pressure and speed-squared are reversed. That is, the quadratic equation

$$c^2 = b_0 + b_1 P + b_2 P^2 \tag{6}$$

is solved for $P$, which gives

$$P = \frac{-b_1 \pm \sqrt{b_1^2 - 4b_2(b_0 - c^2)}}{2b_2} \tag{7}$$

Since the relationship between $c^2$ and $P$ is an increasing relationship (as shown in Figure 2), therefore the positive square root in (7) provides the solution that is consistent with the data, and therefore the program uses equation (7) with the positive square root.

The analysis based on (7) isn't as robust as the analysis based on (6) because it fails unless good initial estimates are given for the values of the parameters and it requires 67 iterations as opposed to the 25 iterations that were required for the analysis based on (6). Also, the program reports at the bottom of page 132 that the solution has failed to satisfy one of the criteria for a good solution. The estimated value of $R$ from this analysis is theoretically identical to the value from the preceding analysis because the same two errors terms are minimized, albeit through different means. The estimated value of $R$ is given on page 138. This value differs from the value given for the preceding analysis (on page 118) by 1 unit in the last (fifteenth) available digit.

### G.21  An Analysis Reversing the Roles of $P$ and $c^2$

Pages 139 through 142 discuss an analysis (not an *orthogonal* analysis) that uses the quadratic model equation, but instead of using $c^2$ in the role of the response variable, the model equation places $P$ in the role of the response variable, using equation (7).

Page 142 gives the estimated value of $R$ under this approach. When the value is rounded to the appropriate seven significant digits it is very slightly larger than the earlier estimates with the quadratic model: 8.314 478 versus 8.314 477. This difference is of no practical significance, but it illustrates that the estimated value of $R$ depends on the point of view we take.

### G.22  Repeated Measurements Analysis

Pages 143 through 149 report a repeated measurements quadratic analysis with SAS PROC MIXED. This analysis takes account of the statistically relevant fact that all the speed-squared values at each pressure value were measured consecutively before the pressure was changed to the next selected value. The need for this enhanced model is tested by the "Null Model Likelihood Ratio Test" on page 146, which yields a $p$-value of 0.0263. This $p$-value is only slightly below 0.05, so the need for the enhanced model is somewhat questionable. The computed value of $R$ from this analysis is shown on page 149 to be (after appropriate rounding) 8.314 477, which is the same as the earlier result obtained with the quadratic model that ignores the repeated measurements of the speed-squared values at each pressure value.

### G.23  Other Analyses

Pages 150 through the end of the output show various other analyses of the data that may be of interest to some readers. These analyses are documented in the output.

The following analyses are not reported in the output, but are reported near the end of the log for the program run (Macnaughton 2010b):

- the computation of the standard errors of the four estimates of $R$ derived from the Moldover et al. data
- the computation of the weighted averages of the estimates of $R$ obtained by combining (a) an estimate based on the Moldover et al. data and (b) the estimate obtained by Colclough, Quinn, and Chandler (1979)
- the computation of the standard errors of the weighted average estimates of $R$.

Other measurement models for the data are possible, including modeling the differing spread of the residuals at different pressures and taking direct account of the weight variable in computing covariance matrices.

### G.24  Is It Appropriate to Use Multiple Approaches to Analyze a Set of Data?

The preceding subsections use various approaches to analyze the Moldover et al. data. Some authors suggest that it is less appropriate to use multiple approaches, and the researcher should choose the analysis approach before collecting the data, and that approach alone should be used.

This paper takes the view that it is always instructive to contrast the results obtained using multiple analysis approaches provided that this is done responsibly. That is, all (sensible) analyses of the data that are performed should be reported, especially if they disagree among themselves. In other words, a researcher shouldn't analyze the data several ways and then select the analysis that gives the "best" results and then report only it—such a report would be biased and might be reporting incorrect results. Instead, the researcher should report the range of results that were obtained. If the results all suggest the same conclusion, this somewhat strengthens the conclusion. However, if some of the results contradict each other, this makes the conclusion questionable and thus the finding should be treated with skepticism until it is independently confirmed.

### G.25  Shrinkage

"Shrinkage" is a statistical method that is sometimes used to improve predictions made by a linear model equation. The method works by slightly "shrinking" or reducing the absolute values of the parameters in the equation, which (counter-intuitively) can be shown to be sometimes effective. Thus the question arises whether shrinkage might be used in the present analyses to improve the estimate of $R$.

However, shrinkage is designed to improve *predictions* made by an equation, which it does by slightly worsening (in a least-squares sense) the estimated values of the pa-rameters of the equation. In addition, shrinkage isn't generally applied to an intercept in an equation. But, as noted, the value of *R* is estimated through the estimate of the value of the $b_0$ (intercept) parameter of equation (3) or equation (6). Therefore, shrinkage can't be used to improve the estimate of *R*.

Shrinkage works by reducing the effects that squaring of the errors has on the estimates of the values of the parameters—shrinkage deemphasizes the undue effects of the larger errors. This is related to the choice of the least-squares method, as discussed in Method 4 in section 2, and related to the emphasis on minimizing variance, as discussed above in Appendix D. Shrinkage has the problem that it is somewhat ad hoc in the sense that the amount of shrinkage to perform must be chosen, typically by choosing a value for a "shrinkage parameter" that indicates the desired extent of the shrinkage. The ad hoc aspect can be removed through the use of cross-validation to estimate the optimal value of the shrinkage parameter (Tibshirani, 2009).

### G.26  Summing Up

This appendix introduced the Moldover et al. data and the Moldover et al. model equation and discussed several analyses of the data. The analyses suggest that there is no evidence in the data that cubic and inverse-pressure terms are needed in the equation to model the relationship between speed-of-sound-squared and pressure. If the cubic and inverse-pressure terms are omitted from the model equation used to estimate *R,* this yields a change in the last significant digit of the estimate, giving 8.314 477 instead of 8.314 471.

### APPENDIX H:  RESIDUAL SCATTERPLOTS FROM THE MOLDOVER ET AL. DATA

Following are 35 residual scatterplots derived from the Moldover et al. data. These scatterplots provide 35 different views of the (same) 70 residuals from the least-squares best-fitting quadratic model equation for the data. Plotting symbols aren't used (to prevent obscuring points) and the different values of the "auxiliary" variable on each scatterplot are shown by different colors of lines joining the points. Thus if printed versions of the scatterplots are to be studied, they should be printed on a color printer. The scatterplots are self-explanatory so are presented with figure numbers for ease of reference, but without descriptive captions.

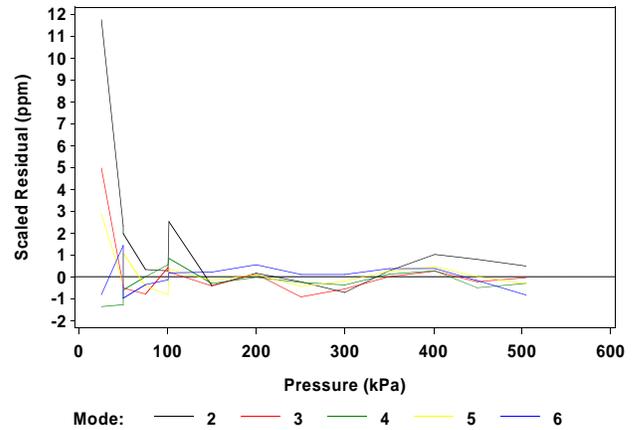

*Figure 9.*

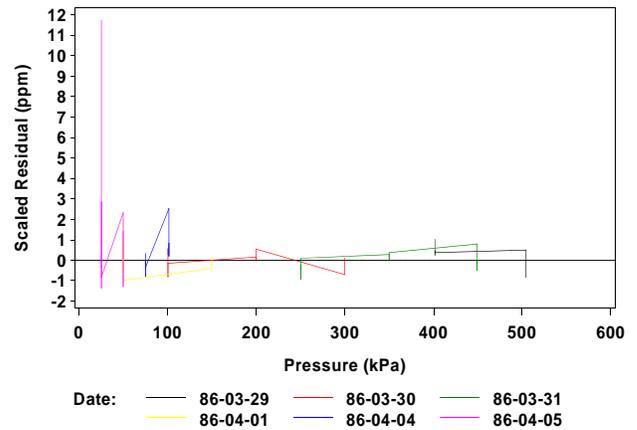

*Figure 10.*

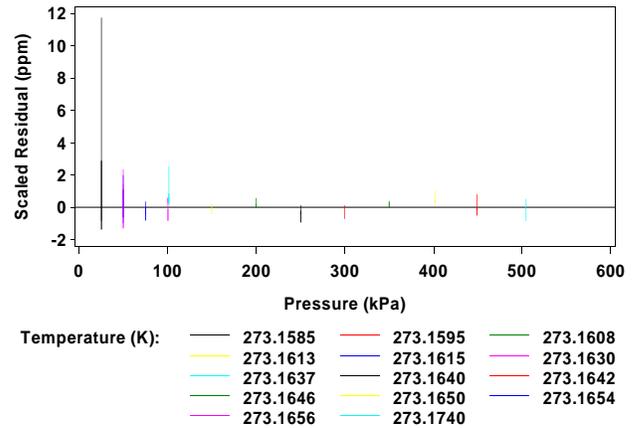

*Figure 11.*

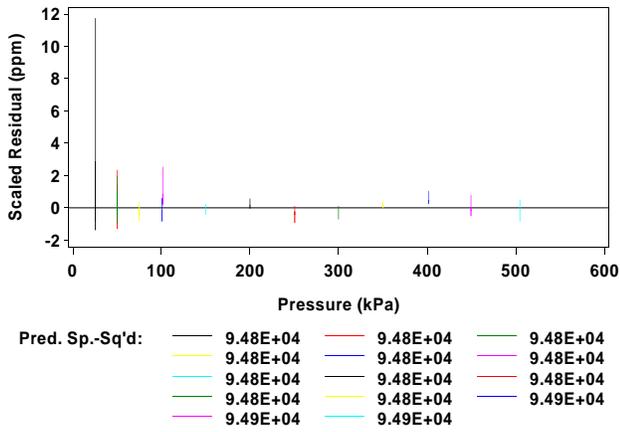

*Figure 12.*

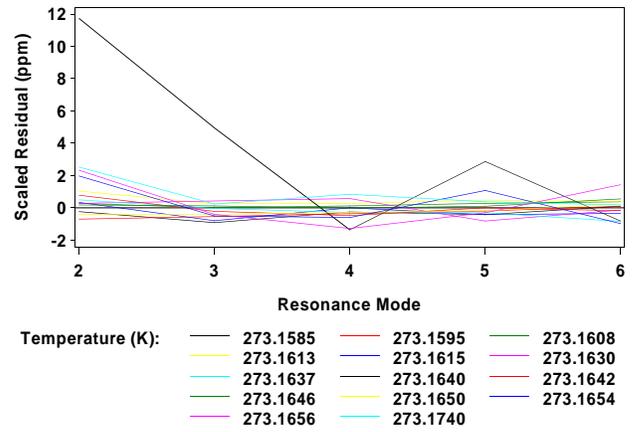

*Figure 15.*

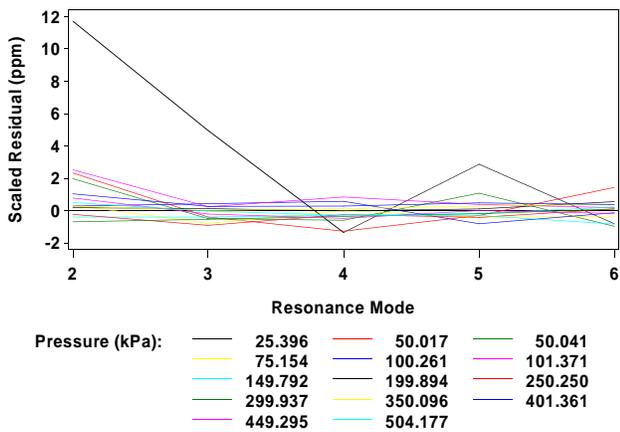

*Figure 13.*

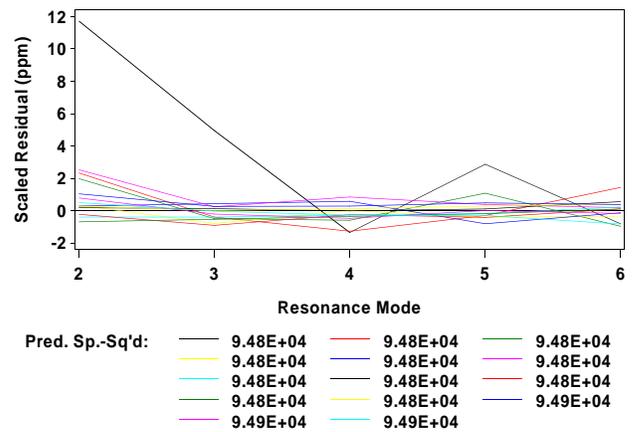

*Figure 16.*

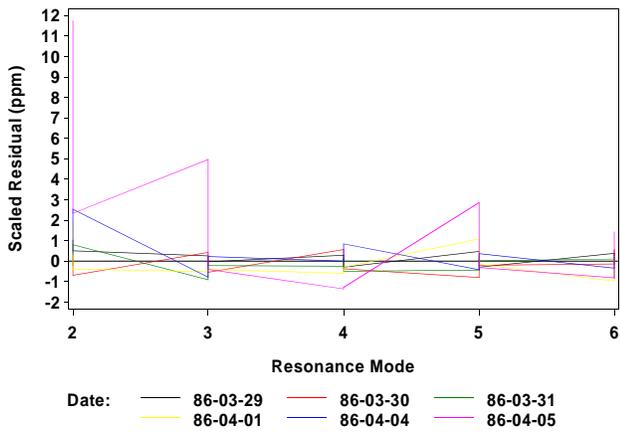

*Figure 14.*

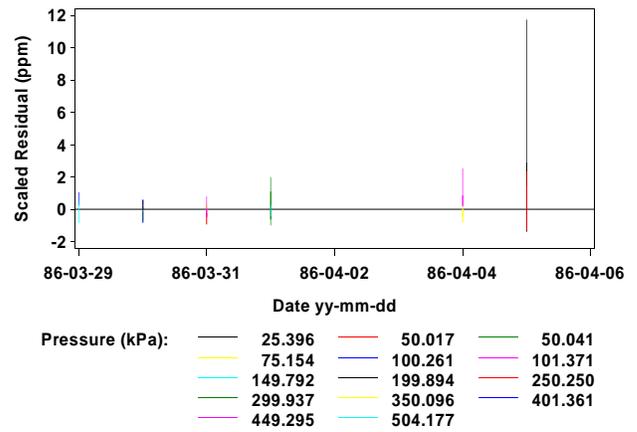

*Figure 17.*

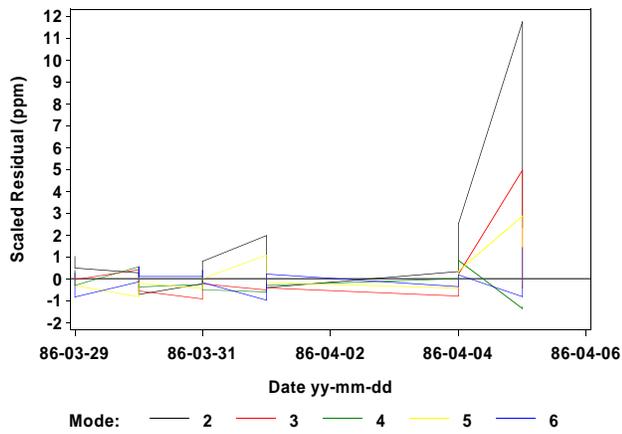

*Figure 18.*

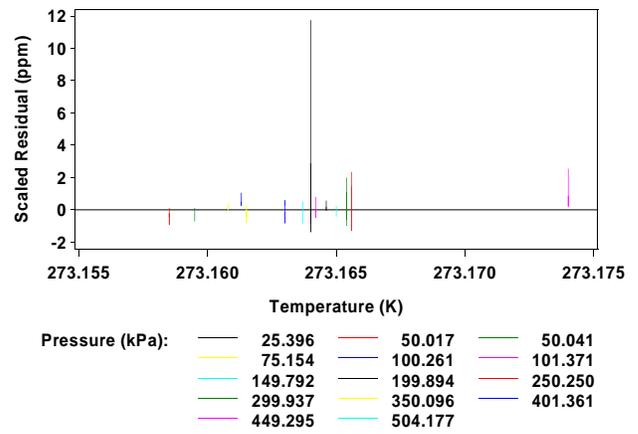

*Figure 21.*

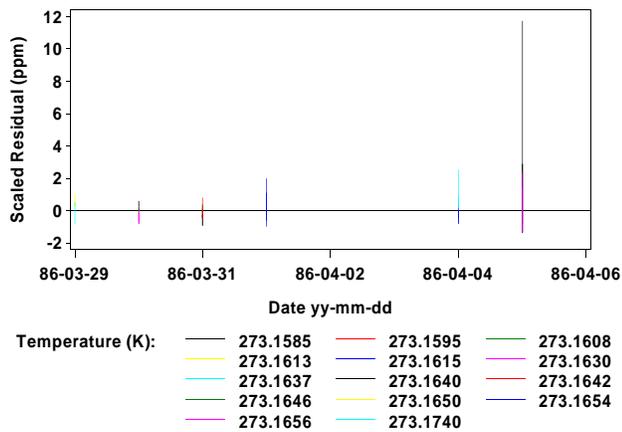

*Figure 19.*

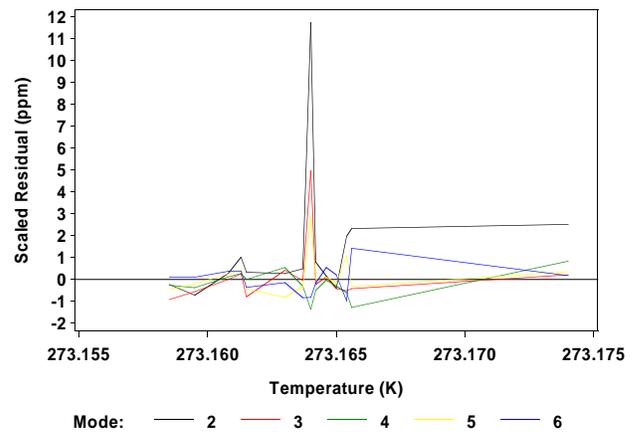

*Figure 22.*

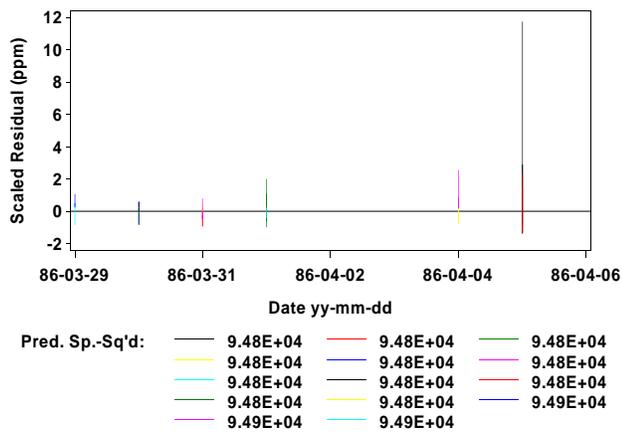

*Figure 20.*

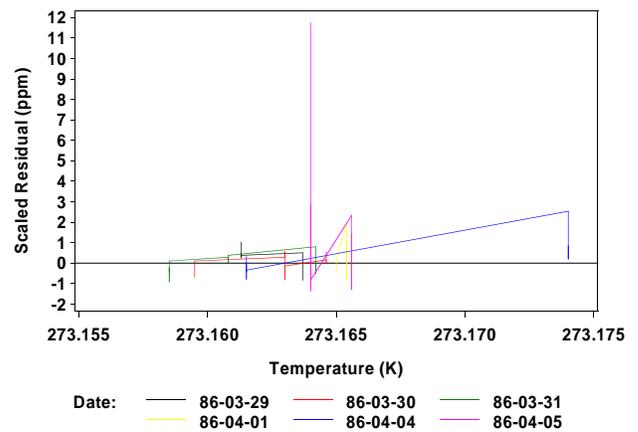

*Figure 23.*

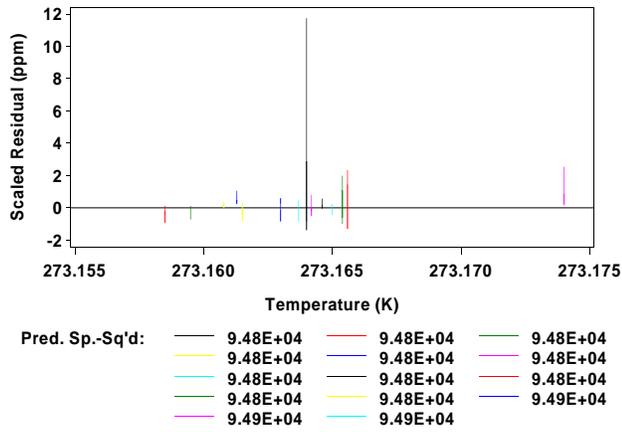

*Figure 24.*

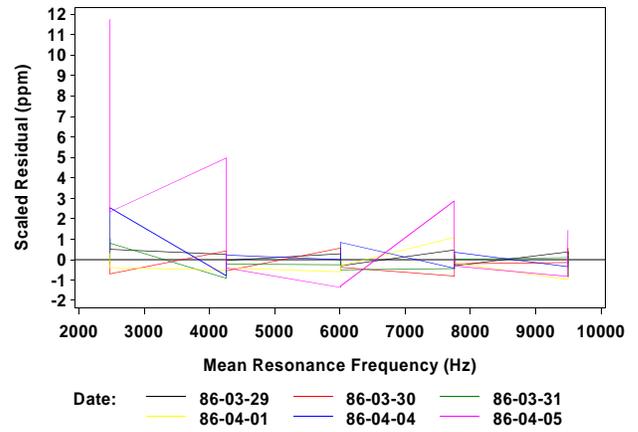

*Figure 27.*

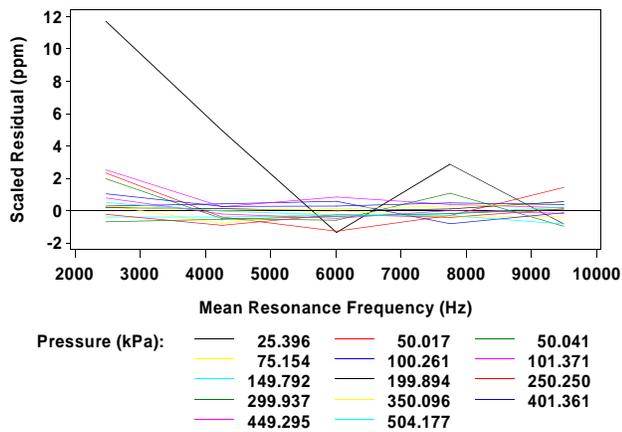

*Figure 25.*

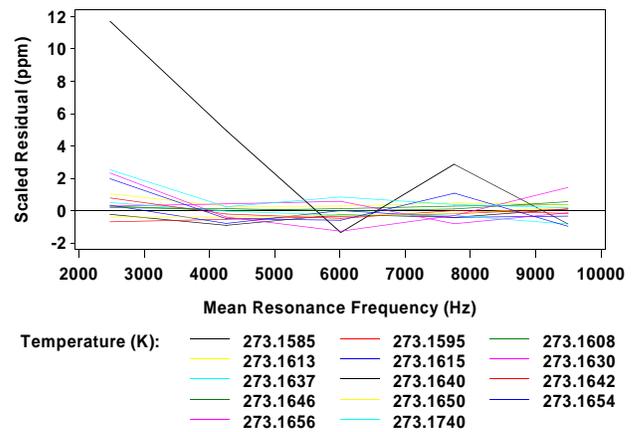

*Figure 28.*

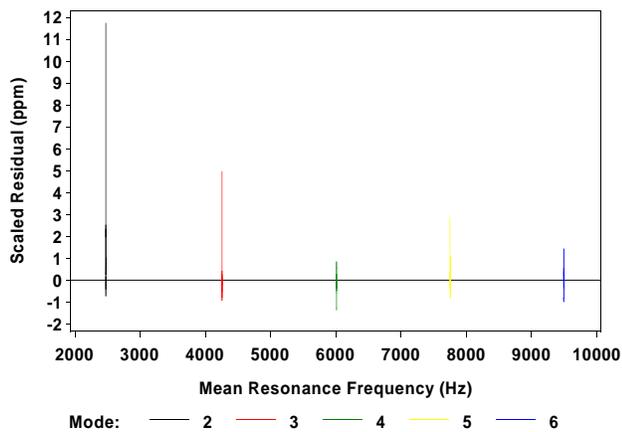

*Figure 26.*

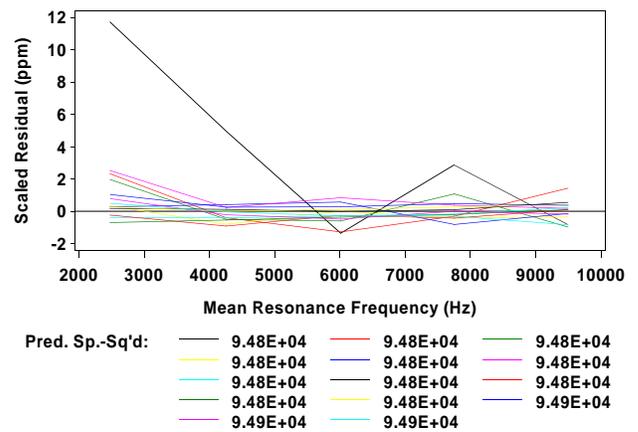

*Figure 29.*

Figure 30.

Figure 33.

Figure 31.

Figure 34.

Figure 32.

Figure 35.

Figure 36.

Figure 39.

Figure 37.

Figure 40.

Figure 38.

Figure 41.

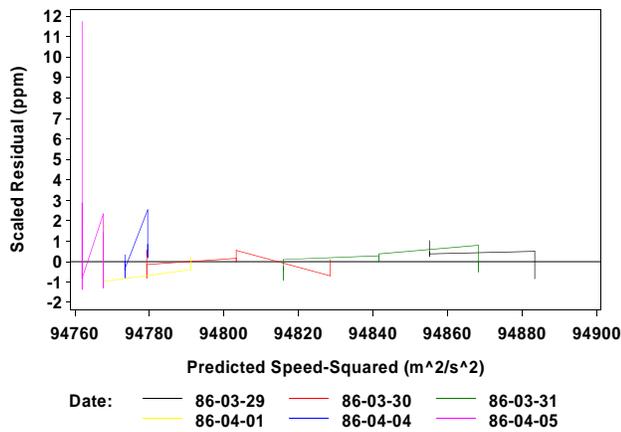

*Figure 42.*

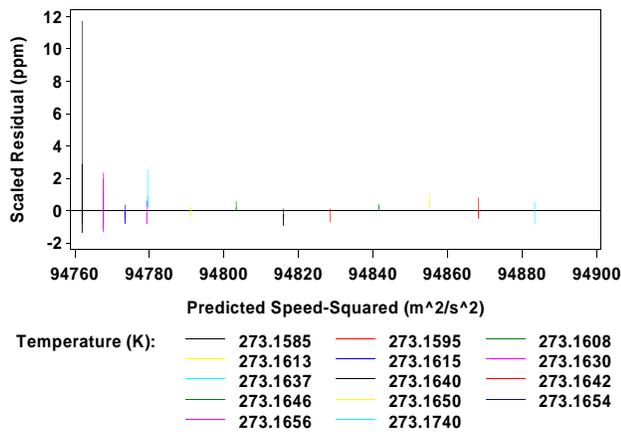

*Figure 43.*

The similarity between Figures 9 and 41 is expected because predicted speed-squared is strongly related to pressure.

### APPENDIX I:  CORRECTIONS TO THE MOLDOVER ET AL. DATA

Appendices 1 and 2 of the Moldover et al. article (1988) give the main data that Moldover et al. used to estimate *R*, and these data can be easily directly pasted (in steps) from the article into statistical analysis software for analysis. A copy of the data also appears in the output below and in the SAS program I used to generate the output (2010a).

Ideally, any regression program used to analyze the Moldover et al. data should allow user-specified weighting of the individual observations (rows) in the data table and should also allow (to deal with the fixed cubic parameter) user-specified fixed parameter values.

(Alternatively, the fixed cubic parameter can be dealt with in a standard regression program if the values of the cubic term are computed and subtracted from the values of speed-squared and then the remainders used as the response variable in a regression analysis with a *y*-intercept and terms reflecting linear, quadratic, and inverse functions of pressure.)

If you wish to analyze the Moldover et al. data, Dr. Moldover has advised me (July 15, 2005) that the following errors are in the data given in the 1988 article:

- In Appendix 2 (page 143) in the table for the pressure of 504.177 kPa the speed-squared value 94,833.320 should be 94,883.320.

- In Appendix 2 (page 143) in the table for the pressure of 199.894 kPa the speed-squared value 95,803.362 should be 94,803.362.

- The half-width data in Appendix 1 (page 142) aren't used in the analyses discussed in this paper. However, in case these data might be used in other analyses, the half-width data, *g*, for March 30, 1986 at the pressure of 299.937 kilopascals should be corrected as follows:

| Existing | Corrected |
|----------|-----------|
| 0.6392   | 0.6388    |
| 0.5796   | 0.5815    |
| 0.4974   | 0.4974    |
| 0.4309   | 0.4304    |
| 0.3567   | 0.3564    |
| 0.2681   | 0.2683    |

### APPENDIX J:  COMPUTER LISTING OUTPUT

The computer listing output begins on the following page.

Listing of the Merged Data Table for Proofreading
The corrections in M. Moldover's email of 2005/7/15 were made.

| Obs | Pressure (Pa) | SpeedSq (m^2/s^2) | Resonance Mode | Date yy-mm-dd | Temperature (K) | Mean Resonance Frequency (Hz) | Mean Half-Width (Hz) | Weight Var |
|---|---|---|---|---|---|---|---|---|
| 1 | 25,396 | 94,763.099 | 2 | 86-04-05 | 273.1640 | 2475.4602 | 0.89690 | 0.7 |
| 2 | 25,396 | 94,762.457 | 3 | 86-04-05 | 273.1640 | 4256.2611 | 1.22805 | 5.5 |
| 3 | 25,396 | 94,761.858 | 4 | 86-04-05 | 273.1640 | 6007.9130 | 1.51695 | 20.0 |
| 4 | 25,396 | 94,762.258 | 5 | 86-04-05 | 273.1640 | 7750.3681 | 1.80160 | 49.4 |
| 5 | 25,396 | 94,761.909 | 6 | 86-04-05 | 273.1640 | 9488.6729 | 2.06520 | 94.9 |
| 6 | 50,017 | 94,767.836 | 2 | 86-04-05 | 273.1656 | 2475.7775 | 0.64685 | 9.2 |
| 7 | 50,017 | 94,767.576 | 3 | 86-04-05 | 273.1656 | 4256.7116 | 0.86790 | 64.2 |
| 8 | 50,017 | 94,767.494 | 4 | 86-04-05 | 273.1656 | 6008.4908 | 1.06175 | 175.6 |
| 9 | 50,017 | 94,767.586 | 5 | 86-04-05 | 273.1656 | 7751.0391 | 1.23335 | 303.1 |
| 10 | 50,017 | 94,767.750 | 6 | 86-04-05 | 273.1656 | 9489.4653 | 1.42080 | 409.6 |
| 11 | 50,041 | 94,767.808 | 2 | 86-04-01 | 273.1654 | 2475.7765 | 0.65060 | 9.2 |
| 12 | 50,041 | 94,767.572 | 3 | 86-04-01 | 273.1654 | 4256.7103 | 0.86755 | 64.3 |
| 13 | 50,041 | 94,767.564 | 4 | 86-04-01 | 273.1654 | 6008.4913 | 1.05675 | 175.9 |
| 14 | 50,041 | 94,767.723 | 5 | 86-04-01 | 273.1654 | 7751.0425 | 1.23640 | 303.3 |
| 15 | 50,041 | 94,767.528 | 6 | 86-04-01 | 273.1654 | 9489.4515 | 1.41800 | 409.8 |
| 16 | 75,154 | 94,773.462 | 2 | 86-04-04 | 273.1615 | 2475.9457 | 0.52825 | 40.7 |
| 17 | 75,154 | 94,773.355 | 3 | 86-04-04 | 273.1615 | 4256.9568 | 0.70765 | 204.3 |
| 18 | 75,154 | 94,773.430 | 4 | 86-04-04 | 273.1615 | 6008.8076 | 0.85615 | 384.2 |
| 19 | 75,154 | 94,773.390 | 5 | 86-04-04 | 273.1615 | 7751.4138 | 0.99800 | 501.9 |
| 20 | 75,154 | 94,773.397 | 6 | 86-04-04 | 273.1615 | 9489.8903 | 1.13795 | 570.3 |
| 21 | 100,261 | 94,779.334 | 2 | 86-03-30 | 273.1630 | 2476.0962 | 0.45795 | 103.8 |
| 22 | 100,261 | 94,779.347 | 3 | 86-03-30 | 273.1630 | 4257.1901 | 0.60950 | 354.5 |
| 23 | 100,261 | 94,779.360 | 4 | 86-03-30 | 273.1630 | 6009.1145 | 0.73845 | 514.8 |
| 24 | 100,261 | 94,779.230 | 5 | 86-03-30 | 273.1630 | 7751.7887 | 0.85825 | 592.7 |
| 25 | 100,261 | 94,779.294 | 6 | 86-03-30 | 273.1630 | 9490.3359 | 0.98100 | 632.7 |
| 26 | 101,371 | 94,779.808 | 2 | 86-04-04 | 273.1740 | 2476.1542 | 0.45620 | 107.2 |
| 27 | 101,371 | 94,779.590 | 3 | 86-04-04 | 273.1740 | 4257.2837 | 0.60640 | 360.4 |
| 28 | 101,371 | 94,779.648 | 4 | 86-04-04 | 273.1740 | 6009.2472 | 0.73410 | 519.0 |
| 29 | 101,371 | 94,779.603 | 5 | 86-04-04 | 273.1740 | 7751.9628 | 0.85440 | 595.3 |
| 30 | 101,371 | 94,779.585 | 6 | 86-04-04 | 273.1740 | 9490.5444 | 0.96920 | 634.4 |
| 31 | 149,792 | 94,791.066 | 2 | 86-04-01 | 273.1650 | 2476.3363 | 0.37660 | 279.4 |
| 32 | 149,792 | 94,791.065 | 3 | 86-04-01 | 273.1650 | 4257.5695 | 0.49955 | 535.2 |
| 33 | 149,792 | 94,791.076 | 4 | 86-04-01 | 273.1650 | 6009.6257 | 0.60480 | 623.1 |
| 34 | 149,792 | 94,791.088 | 5 | 86-04-01 | 273.1650 | 7752.4331 | 0.69950 | 658.6 |
| 35 | 149,792 | 94,791.125 | 6 | 86-04-01 | 273.1650 | 9491.1030 | 0.79785 | 675.9 |
| 36 | 199,894 | 94,803.326 | 2 | 86-03-30 | 273.1646 | 2476.5411 | 0.32545 | 430.0 |
| 37 | 199,894 | 94,803.321 | 3 | 86-03-30 | 273.1646 | 4257.9014 | 0.43395 | 612.3 |
| 38 | 199,894 | 94,803.308 | 4 | 86-03-30 | 273.1646 | 6010.0786 | 0.52385 | 661.8 |
| 39 | 199,894 | 94,803.320 | 5 | 86-03-30 | 273.1646 | 7753.0042 | 0.60495 | 681.2 |
| 40 | 199,894 | 94,803.362 | 6 | 86-03-30 | 273.1646 | 9491.7884 | 0.69620 | 690.7 |

| Obs | Pressure (Pa) | SpeedSq (m^2/s^2) | Resonance Mode | Date yy-mm-dd | Temperature (K) | Mean Resonance Frequency (Hz) | Mean Half-Width (Hz) | Weight Var |
|---|---|---|---|---|---|---|---|---|
| 41 | 250,250 | 94,815.834 | 2 | 86-03-31 | 273.1585 | 2476.7078 | 0.29185 | 524.9 |
| 42 | 250,250 | 94,815.768 | 3 | 86-03-31 | 273.1585 | 4258.1729 | 0.38895 | 648.3 |
| 43 | 250,250 | 94,815.831 | 4 | 86-03-31 | 273.1585 | 6010.4538 | 0.46975 | 679.2 |
| 44 | 250,250 | 94,815.814 | 5 | 86-03-31 | 273.1585 | 7753.4774 | 0.54240 | 691.3 |
| 45 | 250,250 | 94,815.865 | 6 | 86-03-31 | 273.1585 | 9492.3575 | 0.62435 | 697.3 |
| 46 | 299,937 | 94,828.439 | 2 | 86-03-30 | 273.1595 | 2476.8992 | 0.26820 | 580.8 |
| 47 | 299,937 | 94,828.454 | 3 | 86-03-30 | 273.1595 | 4258.4937 | 0.35655 | 667.1 |
| 48 | 299,937 | 94,828.471 | 4 | 86-03-30 | 273.1595 | 6010.8974 | 0.43065 | 688.3 |
| 49 | 299,937 | 94,828.487 | 5 | 86-03-30 | 273.1595 | 7754.0436 | 0.49740 | 696.6 |
| 50 | 299,937 | 94,828.516 | 6 | 86-03-30 | 273.1595 | 9493.0408 | 0.58055 | 700.7 |
| 51 | 350,096 | 94,841.578 | 2 | 86-03-31 | 273.1608 | 2477.0935 | 0.24930 | 615.5 |
| 52 | 350,096 | 94,841.552 | 3 | 86-03-31 | 273.1608 | 4258.8189 | 0.33065 | 678.3 |
| 53 | 350,096 | 94,841.563 | 4 | 86-03-31 | 273.1608 | 6011.3499 | 0.40035 | 693.6 |
| 54 | 350,096 | 94,841.577 | 5 | 86-03-31 | 273.1608 | 7754.6212 | 0.46330 | 699.7 |
| 55 | 350,096 | 94,841.587 | 6 | 86-03-31 | 273.1608 | 9493.7392 | 0.54275 | 702.7 |
| 56 | 401,361 | 94,855.267 | 2 | 86-03-29 | 273.1613 | 2477.2876 | 0.23390 | 638.4 |
| 57 | 401,361 | 94,855.194 | 3 | 86-03-29 | 273.1613 | 4259.1450 | 0.30960 | 685.3 |
| 58 | 401,361 | 94,855.196 | 4 | 86-03-29 | 273.1613 | 6011.8048 | 0.37495 | 697.1 |
| 59 | 401,361 | 94,855.214 | 5 | 86-03-29 | 273.1613 | 7755.2028 | 0.43450 | 701.6 |
| 60 | 401,361 | 94,855.205 | 6 | 86-03-29 | 273.1613 | 9494.4431 | 0.51675 | 703.9 |
| 61 | 449,295 | 94,868.238 | 2 | 86-03-31 | 273.1642 | 2477.4800 | 0.22150 | 652.8 |
| 62 | 449,295 | 94,868.142 | 3 | 86-03-31 | 273.1642 | 4259.4702 | 0.29375 | 690.1 |
| 63 | 449,295 | 94,868.116 | 4 | 86-03-31 | 273.1642 | 6012.2587 | 0.35600 | 699.2 |
| 64 | 449,295 | 94,868.163 | 5 | 86-03-31 | 273.1642 | 7755.7851 | 0.41290 | 702.8 |
| 65 | 449,295 | 94,868.147 | 6 | 86-03-31 | 273.1642 | 9495.1495 | 0.49295 | 704.6 |
| 66 | 504,177 | 94,883.396 | 2 | 86-03-29 | 273.1637 | 2477.6853 | 0.21005 | 664.4 |
| 67 | 504,177 | 94,883.346 | 3 | 86-03-29 | 273.1637 | 4259.8192 | 0.27835 | 693.6 |
| 68 | 504,177 | 94,883.321 | 4 | 86-03-29 | 273.1637 | 6012.7472 | 0.33690 | 700.8 |
| 69 | 504,177 | 94,883.320 | 5 | 86-03-29 | 273.1637 | 7756.4088 | 0.39230 | 703.7 |
| 70 | 504,177 | 94,883.270 | 6 | 86-03-29 | 273.1637 | 9495.9048 | 0.47570 | 705.1 |

Analysis of the Moldover et al. Data to Estimate R                 42
Univariate Distribution of Speed-Squared at Each Pressure

N is the number of values used to compute the mean and standard deviation.

The MEANS Procedure

Analysis Variable : SpeedSq (m^2/s^2)

| Pressure (kPa) | N | Mean | Std Dev | Range |
|---|---|---|---|---|
| 25.396 | 5 | 94762.316 | 0.503 | 1.241 |
| 50.017 | 5 | 94767.648 | 0.140 | 0.342 |
| 50.041 | 5 | 94767.639 | 0.120 | 0.280 |
| 75.154 | 5 | 94773.407 | 0.041 | 0.107 |
| 100.261 | 5 | 94779.313 | 0.053 | 0.130 |
| 101.371 | 5 | 94779.647 | 0.093 | 0.223 |
| 149.792 | 5 | 94791.084 | 0.025 | 0.060 |
| 199.894 | 5 | 94803.327 | 0.020 | 0.054 |
| 250.250 | 5 | 94815.822 | 0.036 | 0.097 |
| 299.937 | 5 | 94828.473 | 0.030 | 0.077 |
| 350.096 | 5 | 94841.571 | 0.014 | 0.035 |
| 401.361 | 5 | 94855.215 | 0.030 | 0.073 |
| 449.295 | 5 | 94868.161 | 0.046 | 0.122 |
| 504.177 | 5 | 94883.331 | 0.046 | 0.126 |

Analysis of the Moldover et al. Data to Estimate R                                           43
Fit the Line with the Moldover et al. Equation Using the Weights

The NLIN Procedure
Dependent Variable SpeedSq
Method: Gauss-Newton

Iterative Phase

| Iter | B0 | B1 | B2 | B4 | Weighted SS |
|------|------|------|------|------|------|
| 0 | 1.0000 | 1.0000 | 1.0000 | 1.0000 | 5.51E26 |
| 1 | 94756.2 | 0.000225 | 5.32E-11 | 2616.0 | 85.7379 |
| 2 | 94756.2 | 0.000225 | 5.32E-11 | 2679.8 | 85.7373 |

NOTE: Convergence criterion met.

Estimation Summary

| Method | Gauss-Newton |
|------|------|
| Iterations | 2 |
| R | 6.33E-11 |
| PPC | 9.33E-11 |
| RPC(B4) | 0.024373 |
| Object | 7.781E-6 |
| Objective | 85.73727 |
| Observations Read | 70 |
| Observations Used | 70 |
| Observations Missing | 0 |

| Source | DF | Sum of Squares | Mean Square | F Value | Approx Pr > F |
|------|------|------|------|------|------|
| Model | 3 | 47749337 | 15916446 | 1.225E7 | <.0001 |
| Error | 66 | 85.7373 | 1.2990 | | |
| Corrected Total | 69 | 47749423 | | | |

Analysis of the Moldover et al. Data to Estimate R                          44
Fit the Line with the Moldover et al. Equation Using the Weights

The NLIN Procedure

| Parameter | Estimate | Approx Std Error | Approximate 95% Confidence Limits | | Skewness |
|-----------|----------|------------------|-----------|-----------|----------|
| B0 | 94756.2 | 0.0647 | 94756.0 | 94756.3 | 0 |
| B1 | 0.000225 | 3.533E-7 | 0.000224 | 0.000226 | 0 |
| B2 | 5.32E-11 | 5.16E-13 | 5.22E-11 | 5.42E-11 | 0 |
| B4 | 2679.8 | 2886.5 | -3083.4 | 8442.9 | 0 |

Approximate Correlation Matrix

| | B0 | B1 | B2 | B4 |
|-----|-----------|-----------|-----------|-----------|
| B0 | 1.0000000 | -0.9590698 | 0.8934886 | -0.9357800 |
| B1 | -0.9590698 | 1.0000000 | -0.9810378 | 0.8370753 |
| B2 | 0.8934886 | -0.9810378 | 1.0000000 | -0.7501044 |
| B4 | -0.9357800 | 0.8370753 | -0.7501044 | 1.0000000 |

Analysis of the Moldover et al. Data to Estimate R                                           45
Fit the Line with the Moldover et al. Equation Using the Weights

Here is a more accurate table of the values of the parameters of the above model.
This table was obtained via the SAS output delivery system (ODS).

| Parameter | Parameter Estimate | Approximate Standard Error | Approximate Lower 95% Confidence Limit | Approximate Upper 95% Confidence Limit | t | p |
|---|---|---|---|---|---|---|
| B0 | 9.4756178E+04 | 6.47E-02 | 9.4756048E+04 | 9.4756307E+04 | 1464230.49 | <.0001 |
| B1 | 2.2503316E-04 | 3.53E-07 | 2.2432771E-04 | 2.2573862E-04 | 636.88 | <.0001 |
| B2 | 5.3204471E-11 | 5.16E-13 | 5.2173275E-11 | 5.4235667E-11 | 103.01 | <.0001 |
| B4 | 2.6797889E+03 | 2.89E+03 | -3.0833583E+03 | 8.4429361E+03 | 0.93 | 0.3566 |

Analysis of the Moldover et al. Data to Estimate R
Fit the Line with the Moldover et al. Equation Using the Weights

Here is the estimate of R from the Moldover et al. analysis with full
significant digits:
This was computed by proc SQL.

| Estimate of Intercept from Moldover et al. Model | Estimate of Standard Error of Intercept from Moldover et al. Model | Estimate of R from Moldover et al. Model |
|---|---|---|
| 9.4756177677121000E+04 | 6.4713976431392000E-02 | 8.3144709320207000E+00 |

Analysis of the Moldover et al. Data to Estimate R                                    47
Fit the Line with the Moldover et al. Equation Using the Weights

Here is a listing of the output data table (dataset) from the analysis.

| Obs | Pressure (Pa) | SpeedSq (m^2/s^2) | Resonance Mode | Date yy-mm-dd | Temperature (K) | Resonance Frequency (Hz) |
|---|---|---|---|---|---|---|
| 1 | 25,396 | 94,763.099 | 2 | 86-04-05 | 273.1640 | 2,475.46 |
| 2 | 25,396 | 94,762.457 | 3 | 86-04-05 | 273.1640 | 4,256.26 |
| 3 | 25,396 | 94,761.858 | 4 | 86-04-05 | 273.1640 | 6,007.91 |
| 4 | 25,396 | 94,762.258 | 5 | 86-04-05 | 273.1640 | 7,750.37 |
| 5 | 25,396 | 94,761.909 | 6 | 86-04-05 | 273.1640 | 9,488.67 |
| 6 | 50,017 | 94,767.836 | 2 | 86-04-05 | 273.1656 | 2,475.78 |
| 7 | 50,017 | 94,767.576 | 3 | 86-04-05 | 273.1656 | 4,256.71 |
| 8 | 50,017 | 94,767.494 | 4 | 86-04-05 | 273.1656 | 6,008.49 |
| 9 | 50,017 | 94,767.586 | 5 | 86-04-05 | 273.1656 | 7,751.04 |
| 10 | 50,017 | 94,767.750 | 6 | 86-04-05 | 273.1656 | 9,489.47 |
| 11 | 50,041 | 94,767.808 | 2 | 86-04-01 | 273.1654 | 2,475.78 |
| 12 | 50,041 | 94,767.572 | 3 | 86-04-01 | 273.1654 | 4,256.71 |
| 13 | 50,041 | 94,767.564 | 4 | 86-04-01 | 273.1654 | 6,008.49 |
| 14 | 50,041 | 94,767.723 | 5 | 86-04-01 | 273.1654 | 7,751.04 |
| 15 | 50,041 | 94,767.528 | 6 | 86-04-01 | 273.1654 | 9,489.45 |
| 16 | 75,154 | 94,773.462 | 2 | 86-04-04 | 273.1615 | 2,475.95 |

| Obs | Half Width (Hz) | Weight Var | SpeedSq Resc (m^2/s^2) | PredSpeedSq (m^2/s^2) | Pred SpeedSq Resid (m^2/s^2) | Scaled Residual (parts per million) |
|---|---|---|---|---|---|---|
| 1 | 0.89690 | 0.7 | -45.251 | 94,762.032 | 1.067 | 11.2546 |
| 2 | 1.22805 | 5.5 | -45.893 | 94,762.032 | 0.425 | 4.4799 |
| 3 | 1.51695 | 20.0 | -46.492 | 94,762.032 | -0.174 | -1.8412 |
| 4 | 1.80160 | 49.4 | -46.092 | 94,762.032 | 0.226 | 2.3799 |
| 5 | 2.06520 | 94.9 | -46.441 | 94,762.032 | -0.123 | -1.3030 |
| 6 | 0.64685 | 9.2 | -40.514 | 94,767.620 | 0.216 | 2.2790 |
| 7 | 0.86790 | 64.2 | -40.774 | 94,767.620 | -0.044 | -0.4645 |
| 8 | 1.06175 | 175.6 | -40.856 | 94,767.620 | -0.126 | -1.3298 |
| 9 | 1.23335 | 303.1 | -40.764 | 94,767.620 | -0.034 | -0.3590 |
| 10 | 1.42080 | 409.6 | -40.600 | 94,767.620 | 0.130 | 1.3715 |
| 11 | 0.65060 | 9.2 | -40.542 | 94,767.626 | 0.182 | 1.9255 |
| 12 | 0.86755 | 64.3 | -40.778 | 94,767.626 | -0.054 | -0.5648 |
| 13 | 1.05675 | 175.9 | -40.786 | 94,767.626 | -0.062 | -0.6492 |
| 14 | 1.23640 | 303.3 | -40.627 | 94,767.626 | 0.097 | 1.0286 |
| 15 | 1.41800 | 409.8 | -40.822 | 94,767.626 | -0.098 | -1.0291 |
| 16 | 0.52825 | 40.7 | -34.888 | 94,773.427 | 0.035 | 0.3735 |

Analysis of the Moldover et al. Data to Estimate R                                    48
Fit the Line with the Moldover et al. Equation Using the Weights

Here is a listing of the output data table (dataset) from the analysis.

| Obs | Pressure (Pa) | SpeedSq (m^2/s^2) | Resonance Mode | Date yy-mm-dd | Temperature (K) | Resonance Frequency (Hz) |
|---|---|---|---|---|---|---|
| 17 | 75,154 | 94,773.355 | 3 | 86-04-04 | 273.1615 | 4,256.96 |
| 18 | 75,154 | 94,773.430 | 4 | 86-04-04 | 273.1615 | 6,008.81 |
| 19 | 75,154 | 94,773.390 | 5 | 86-04-04 | 273.1615 | 7,751.41 |
| 20 | 75,154 | 94,773.397 | 6 | 86-04-04 | 273.1615 | 9,489.89 |
| 21 | 100,261 | 94,779.334 | 2 | 86-03-30 | 273.1630 | 2,476.10 |
| 22 | 100,261 | 94,779.347 | 3 | 86-03-30 | 273.1630 | 4,257.19 |
| 23 | 100,261 | 94,779.360 | 4 | 86-03-30 | 273.1630 | 6,009.11 |
| 24 | 100,261 | 94,779.230 | 5 | 86-03-30 | 273.1630 | 7,751.79 |
| 25 | 100,261 | 94,779.294 | 6 | 86-03-30 | 273.1630 | 9,490.34 |
| 26 | 101,371 | 94,779.808 | 2 | 86-04-04 | 273.1740 | 2,476.15 |
| 27 | 101,371 | 94,779.590 | 3 | 86-04-04 | 273.1740 | 4,257.28 |
| 28 | 101,371 | 94,779.648 | 4 | 86-04-04 | 273.1740 | 6,009.25 |
| 29 | 101,371 | 94,779.603 | 5 | 86-04-04 | 273.1740 | 7,751.96 |
| 30 | 101,371 | 94,779.585 | 6 | 86-04-04 | 273.1740 | 9,490.54 |
| 31 | 149,792 | 94,791.066 | 2 | 86-04-01 | 273.1650 | 2,476.34 |
| 32 | 149,792 | 94,791.065 | 3 | 86-04-01 | 273.1650 | 4,257.57 |

| Obs | Half Width (Hz) | Weight Var | SpeedSq Resc (m^2/s^2) | PredSpeedSq (m^2/s^2) | Pred SpeedSq Resid (m^2/s^2) | Scaled Residual (parts per million) |
|---|---|---|---|---|---|---|
| 17 | 0.70765 | 204.3 | -34.995 | 94,773.427 | -0.072 | -0.75546 |
| 18 | 0.85615 | 384.2 | -34.920 | 94,773.427 | 0.003 | 0.03590 |
| 19 | 0.99800 | 501.9 | -34.960 | 94,773.427 | -0.037 | -0.38616 |
| 20 | 1.13795 | 570.3 | -34.953 | 94,773.427 | -0.030 | -0.31230 |
| 21 | 0.45795 | 103.8 | -29.016 | 94,779.303 | 0.031 | 0.32979 |
| 22 | 0.60950 | 354.5 | -29.003 | 94,779.303 | 0.044 | 0.46696 |
| 23 | 0.73845 | 514.8 | -28.990 | 94,779.303 | 0.057 | 0.60412 |
| 24 | 0.85825 | 592.7 | -29.120 | 94,779.303 | -0.073 | -0.76749 |
| 25 | 0.98100 | 632.7 | -29.056 | 94,779.303 | -0.009 | -0.09224 |
| 26 | 0.45620 | 107.2 | -28.542 | 94,779.564 | 0.244 | 2.57235 |
| 27 | 0.60640 | 360.4 | -28.760 | 94,779.564 | 0.026 | 0.27228 |
| 28 | 0.73410 | 519.0 | -28.702 | 94,779.564 | 0.084 | 0.88423 |
| 29 | 0.85440 | 595.3 | -28.747 | 94,779.564 | 0.039 | 0.40944 |
| 30 | 0.96920 | 634.4 | -28.765 | 94,779.564 | 0.021 | 0.21953 |
| 31 | 0.37660 | 279.4 | -17.284 | 94,791.102 | -0.036 | -0.38391 |
| 32 | 0.49955 | 535.2 | -17.285 | 94,791.102 | -0.037 | -0.39446 |

Analysis of the Moldover et al. Data to Estimate R                                              49
Fit the Line with the Moldover et al. Equation Using the Weights

Here is a listing of the output data table (dataset) from the analysis.

| Obs | Pressure (Pa) | SpeedSq (m^2/s^2) | Resonance Mode | Date yy-mm-dd | Temperature (K) | Resonance Frequency (Hz) |
|-----|-----|-----|-----|-----|-----|-----|
| 33 | 149,792 | 94,791.076 | 4 | 86-04-01 | 273.1650 | 6,009.63 |
| 34 | 149,792 | 94,791.088 | 5 | 86-04-01 | 273.1650 | 7,752.43 |
| 35 | 149,792 | 94,791.125 | 6 | 86-04-01 | 273.1650 | 9,491.10 |
| 36 | 199,894 | 94,803.326 | 2 | 86-03-30 | 273.1646 | 2,476.54 |
| 37 | 199,894 | 94,803.321 | 3 | 86-03-30 | 273.1646 | 4,257.90 |
| 38 | 199,894 | 94,803.308 | 4 | 86-03-30 | 273.1646 | 6,010.08 |
| 39 | 199,894 | 94,803.320 | 5 | 86-03-30 | 273.1646 | 7,753.00 |
| 40 | 199,894 | 94,803.362 | 6 | 86-03-30 | 273.1646 | 9,491.79 |
| 41 | 250,250 | 94,815.834 | 2 | 86-03-31 | 273.1585 | 2,476.71 |
| 42 | 250,250 | 94,815.768 | 3 | 86-03-31 | 273.1585 | 4,258.17 |
| 43 | 250,250 | 94,815.831 | 4 | 86-03-31 | 273.1585 | 6,010.45 |
| 44 | 250,250 | 94,815.814 | 5 | 86-03-31 | 273.1585 | 7,753.48 |
| 45 | 250,250 | 94,815.865 | 6 | 86-03-31 | 273.1585 | 9,492.36 |
| 46 | 299,937 | 94,828.439 | 2 | 86-03-30 | 273.1595 | 2,476.90 |
| 47 | 299,937 | 94,828.454 | 3 | 86-03-30 | 273.1595 | 4,258.49 |
| 48 | 299,937 | 94,828.471 | 4 | 86-03-30 | 273.1595 | 6,010.90 |

| Obs | Half Width (Hz) | Weight Var | SpeedSq Resc (m^2/s^2) | PredSpeedSq (m^2/s^2) | Pred SpeedSq Resid (m^2/s^2) | Scaled Residual (parts per million) |
|-----|-----|-----|-----|-----|-----|-----|
| 33 | 0.60480 | 623.1 | -17.274 | 94,791.102 | -0.026 | -0.27842 |
| 34 | 0.69950 | 658.6 | -17.262 | 94,791.102 | -0.014 | -0.15182 |
| 35 | 0.79785 | 675.9 | -17.225 | 94,791.102 | 0.023 | 0.23851 |
| 36 | 0.32545 | 430.0 | -5.024 | 94,803.311 | 0.015 | 0.15434 |
| 37 | 0.43395 | 612.3 | -5.029 | 94,803.311 | 0.010 | 0.10160 |
| 38 | 0.52385 | 661.8 | -5.042 | 94,803.311 | -0.003 | -0.03552 |
| 39 | 0.60495 | 681.2 | -5.030 | 94,803.311 | 0.009 | 0.09106 |
| 40 | 0.69620 | 690.7 | -4.988 | 94,803.311 | 0.051 | 0.53408 |
| 41 | 0.29185 | 524.9 | 7.484 | 94,815.858 | -0.024 | -0.24882 |
| 42 | 0.38895 | 648.3 | 7.418 | 94,815.858 | -0.090 | -0.94491 |
| 43 | 0.46975 | 679.2 | 7.481 | 94,815.858 | -0.027 | -0.28046 |
| 44 | 0.54240 | 691.3 | 7.464 | 94,815.858 | -0.044 | -0.45976 |
| 45 | 0.62435 | 697.3 | 7.515 | 94,815.858 | 0.007 | 0.07813 |
| 46 | 0.26820 | 580.8 | 20.089 | 94,828.508 | -0.069 | -0.72658 |
| 47 | 0.35655 | 667.1 | 20.104 | 94,828.508 | -0.054 | -0.56840 |
| 48 | 0.43065 | 688.3 | 20.121 | 94,828.508 | -0.037 | -0.38913 |

Fit the Line with the Moldover et al. Equation Using the Weights

Here is a listing of the output data table (dataset) from the analysis.

| Obs | Pressure (Pa) | SpeedSq (m^2/s^2) | Resonance Mode | Date yy-mm-dd | Temperature (K) | Resonance Frequency (Hz) |
|---|---|---|---|---|---|---|
| 49 | 299,937 | 94,828.487 | 5 | 86-03-30 | 273.1595 | 7,754.04 |
| 50 | 299,937 | 94,828.516 | 6 | 86-03-30 | 273.1595 | 9,493.04 |
| 51 | 350,096 | 94,841.578 | 2 | 86-03-31 | 273.1608 | 2,477.09 |
| 52 | 350,096 | 94,841.552 | 3 | 86-03-31 | 273.1608 | 4,258.82 |
| 53 | 350,096 | 94,841.563 | 4 | 86-03-31 | 273.1608 | 6,011.35 |
| 54 | 350,096 | 94,841.577 | 5 | 86-03-31 | 273.1608 | 7,754.62 |
| 55 | 350,096 | 94,841.587 | 6 | 86-03-31 | 273.1608 | 9,493.74 |
| 56 | 401,361 | 94,855.267 | 2 | 86-03-29 | 273.1613 | 2,477.29 |
| 57 | 401,361 | 94,855.194 | 3 | 86-03-29 | 273.1613 | 4,259.15 |
| 58 | 401,361 | 94,855.196 | 4 | 86-03-29 | 273.1613 | 6,011.80 |
| 59 | 401,361 | 94,855.214 | 5 | 86-03-29 | 273.1613 | 7,755.20 |
| 60 | 401,361 | 94,855.205 | 6 | 86-03-29 | 273.1613 | 9,494.44 |
| 61 | 449,295 | 94,868.238 | 2 | 86-03-31 | 273.1642 | 2,477.48 |
| 62 | 449,295 | 94,868.142 | 3 | 86-03-31 | 273.1642 | 4,259.47 |
| 63 | 449,295 | 94,868.116 | 4 | 86-03-31 | 273.1642 | 6,012.26 |
| 64 | 449,295 | 94,868.163 | 5 | 86-03-31 | 273.1642 | 7,755.79 |

| Obs | Half Width (Hz) | Weight Var | SpeedSq Resc (m^2/s^2) | PredSpeedSq (m^2/s^2) | Pred SpeedSq Resid (m^2/s^2) | Scaled Residual (parts per million) |
|---|---|---|---|---|---|---|
| 49 | 0.49740 | 696.6 | 20.137 | 94,828.508 | -0.021 | -0.22040 |
| 50 | 0.58055 | 700.7 | 20.166 | 94,828.508 | 0.008 | 0.08541 |
| 51 | 0.24930 | 615.5 | 33.228 | 94,841.552 | 0.026 | 0.27535 |
| 52 | 0.33065 | 678.3 | 33.202 | 94,841.552 | 0.000 | 0.00121 |
| 53 | 0.40035 | 693.6 | 33.213 | 94,841.552 | 0.011 | 0.11719 |
| 54 | 0.46330 | 699.7 | 33.227 | 94,841.552 | 0.025 | 0.26480 |
| 55 | 0.54275 | 702.7 | 33.237 | 94,841.552 | 0.035 | 0.37024 |
| 56 | 0.23390 | 638.4 | 46.917 | 94,855.168 | 0.099 | 1.03966 |
| 57 | 0.30960 | 685.5 | 46.844 | 94,855.168 | 0.026 | 0.27006 |
| 58 | 0.37495 | 697.1 | 46.846 | 94,855.168 | 0.028 | 0.29115 |
| 59 | 0.43450 | 701.6 | 46.864 | 94,855.168 | 0.046 | 0.48091 |
| 60 | 0.51675 | 703.9 | 46.855 | 94,855.168 | 0.037 | 0.38603 |
| 61 | 0.22150 | 652.8 | 59.888 | 94,868.162 | 0.076 | 0.80531 |
| 62 | 0.29375 | 690.1 | 59.792 | 94,868.162 | -0.020 | -0.20662 |
| 63 | 0.35600 | 699.2 | 59.766 | 94,868.162 | -0.046 | -0.48069 |
| 64 | 0.41290 | 702.8 | 59.813 | 94,868.162 | 0.001 | 0.01474 |

Fit the Line with the Moldover et al. Equation Using the Weights

Here is a listing of the output data table (dataset) from the analysis.

| Obs | Pressure (Pa) | SpeedSq (m^2/s^2) | Resonance Mode | Date yy-mm-dd | Temperature (K) | Resonance Frequency (Hz) |
|---|---|---|---|---|---|---|
| 65 | 449,295 | 94,868.147 | 6 | 86-03-31 | 273.1642 | 9,495.15 |
| 66 | 504,177 | 94,883.396 | 2 | 86-03-29 | 273.1637 | 2,477.69 |
| 67 | 504,177 | 94,883.346 | 3 | 86-03-29 | 273.1637 | 4,259.82 |
| 68 | 504,177 | 94,883.321 | 4 | 86-03-29 | 273.1637 | 6,012.75 |
| 69 | 504,177 | 94,883.320 | 5 | 86-03-29 | 273.1637 | 7,756.41 |
| 70 | 504,177 | 94,883.270 | 6 | 86-03-29 | 273.1637 | 9,495.90 |

| Obs | Half Width (Hz) | Weight Var | SpeedSq Resc (m^2/s^2) | PredSpeedSq (m^2/s^2) | Pred SpeedSq Resid (m^2/s^2) | Scaled Residual (parts per million) |
|---|---|---|---|---|---|---|
| 65 | 0.49295 | 704.6 | 59.797 | 94,868.162 | -0.015 | -0.15392 |
| 66 | 0.21005 | 664.4 | 75.046 | 94,883.350 | 0.046 | 0.48850 |
| 67 | 0.27835 | 693.6 | 74.996 | 94,883.350 | -0.004 | -0.03846 |
| 68 | 0.33690 | 700.8 | 74.971 | 94,883.350 | -0.029 | -0.30194 |
| 69 | 0.39230 | 703.7 | 74.970 | 94,883.350 | -0.030 | -0.31248 |
| 70 | 0.47570 | 705.1 | 74.920 | 94,883.350 | -0.080 | -0.83945 |

Fit the Moldover et al. equation, but with the pressure-cubed term unconstrained.
This is Analysis 2 in supporting online material in the short article.

The NLIN Procedure
Dependent Variable SpeedSq
Method: Gauss-Newton

Iterative Phase

| Iter | B0 | B1 | B2 | B3 | B4 | Weighted SS |
|------|------|------|------|------|------|------|
| 0 | 1.0000 | 1.0000 | 1.0000 | 1.0000 | 1.0000 | 1.094E38 |
| 1 | 96692.5 | -0.0211 | 7.468E-8 | -797E-16 | -4.568E7 | 5.4297E8 |
| 2 | 94756.4 | 0.000223 | 6.1E-11 | -688E-20 | -3536.1 | 80.5307 |

NOTE: Convergence criterion met.

Estimation Summary

| Method | Gauss-Newton |
|--------|-------------|
| Iterations | 2 |
| R | 2.292E-9 |
| PPC(B4) | 1.491E-8 |
| RPC(B1) | 1.010522 |
| Object | 1 |
| Objective | 80.53072 |
| Observations Read | 70 |
| Observations Used | 70 |
| Observations Missing | 0 |

| Source | DF | Sum of Squares | Mean Square | F Value | Approx Pr > F |
|--------|-----|-----|-----|-----|-----|
| Model | 4 | 47749343 | 11937336 | 9635165 | <.0001 |
| Error | 65 | 80.5307 | 1.2389 | | |
| Corrected Total | 69 | 47749423 | | | |

Analysis of the Moldover et al. Data to Estimate R                      53
Fit the Moldover et al. equation, but with the pressure-cubed term unconstrained.
This is Analysis 2 in supporting online material in the short article.

The NLIN Procedure

| Parameter | Estimate | Approx Std Error | Approximate 95% Confidence Limits | | Skewness |
|-----------|----------|------------------|-----------------------------------|---|----------|
| B0 | 94756.4 | 0.1280 | 94756.2 | 94756.7 | 0 |
| B1 | 0.000223 | 1.173E-6 | 0.000220 | 0.000225 | 0 |
| B2 | 6.1E-11 | 3.86E-12 | 5.33E-11 | 6.88E-11 | 0 |
| B3 | -688E-20 | 4.07E-18 | -15E-18 | 1.24E-18 | 0 |
| B4 | -3536.1 | 4140.1 | -11804.4 | 4732.3 | 0 |

Approximate Correlation Matrix

|    | B0 | B1 | B2 | B3 | B4 |
|----|-----------|-----------|-----------|-----------|-----------|
| B0 | 1.0000000 | -0.9703996 | 0.9197266 | -0.8695157 | -0.9515187 |
| B1 | -0.9703996 | 1.0000000 | -0.9852593 | 0.9557353 | 0.8676630 |
| B2 | 0.9197266 | -0.9852593 | 1.0000000 | -0.9914252 | -0.7928455 |
| B3 | -0.8695157 | 0.9557353 | -0.9914252 | 1.0000000 | 0.7323845 |
| B4 | -0.9515187 | 0.8676630 | -0.7928455 | 0.7323845 | 1.0000000 |

Analysis of the Moldover et al. Data to Estimate R                                      54
Fit the Moldover et al. equation, but with the pressure-cubed term unconstrained.
This is Analysis 2 in supporting online material in the short article.

Here is a more accurate table of the values of the parameters:

| Parameter | Parameter Estimate | Approximate Standard Error | Approximate Lower 95% Confidence Limit | Approximate Upper 95% Confidence Limit | t | p |
|---|---|---|---|---|---|---|
| B0 | 9.4756406E+04 | 1.28E-01 | 9.4756150E+04 | 9.4756661E+04 | 740530.36 | <.0001 |
| B1 | 2.2273541E-04 | 1.17E-06 | 2.2039321E-04 | 2.2507761E-04 | 189.92 | <.0001 |
| B2 | 6.1049355E-11 | 3.86E-12 | 5.3340608E-11 | 6.8758102E-11 | 15.82 | <.0001 |
| B3 | -6.8835721E-18 | 4.07E-18 | -1.5002308E-17 | 1.2351640E-18 | -1.69 | 0.0952 |
| B4 | -3.5360509E+03 | 4.14E+03 | -1.1804383E+04 | 4.7322811E+03 | -0.85 | 0.3962 |

Analysis of the Moldover et al. Data to Estimate R
Fit the Moldover et al. equation, but with the pressure-cubed term unconstrained.
This is Analysis 2 in supporting online material in the short article.

Here are the estimates from the full model with the estimated cubic term with more significant digits:

| Estimate of Intercept from Full Model with Cubic Term Estimated | Estimate of Standard Error of Intercept from Full Model with Cubic Term Estimated | Estimate of R from Full Model with Cubic Term Estimated |
|---|---|---|
| 9.4756405760680000E+04 | 1.2795748871819000E-01 | 8.3144909454294000E+00 |

Analysis of the Moldover et al. Data to Estimate R                    56
Fit the weighted line with only the intercept, linear, and quadratic terms.
This is Analysis 3 in supporting online material in the short article.

The NLIN Procedure
Dependent Variable SpeedSq
Method: Gauss-Newton

Iterative Phase

| Iter | B0 | B1 | B2 | Weighted SS |
|------|------|------|------|------|
| 0 | 1.0000 | 1.0000 | 1.0000 | 5.51E26 |
| 1 | 94756.3 | 0.000224 | 5.48E-11 | 84.4841 |
| 2 | 94756.3 | 0.000224 | 5.48E-11 | 84.4813 |

NOTE: Convergence criterion met.

Estimation Summary

| Method | Gauss-Newton |
|---|---|
| Iterations | 2 |
| R | 1.73E-10 |
| PPC | 0 |
| RPC(B1) | 0.000037 |
| Object | 0.000033 |
| Objective | 84.48133 |
| Observations Read | 70 |
| Observations Used | 70 |
| Observations Missing | 0 |

| Source | DF | Sum of Squares | Mean Square | F Value | Approx Pr > F |
|---|---|---|---|---|---|
| Model | 2 | 47749339 | 23874669 | 1.893E7 | <.0001 |
| Error | 67 | 84.4813 | 1.2609 | | |
| Corrected Total | 69 | 47749423 | | | |

Analysis of the Moldover et al. Data to Estimate R                    57
Fit the weighted line with only the intercept, linear, and quadratic terms.
This is Analysis 3 in supporting online material in the short article.

The NLIN Procedure

| Parameter | Estimate | Approx Std Error | Approximate 95% Confidence Limits | | Skewness |
|-----------|----------|------------------|---------------|-------------|----------|
| B0 | 94756.3 | 0.0225 | 94756.2 | 94756.3 | 0 |
| B1 | 0.000224 | 1.904E-7 | 0.000224 | 0.000225 | 0 |
| B2 | 5.48E-11 | 3.37E-13 | 5.41E-11 | 5.55E-11 | 0 |

Approximate Correlation Matrix

| | B0 | B1 | B2 |
|-----|-----------|-----------|-----------|
| B0 | 1.0000000 | -0.9111268 | 0.8215256 |
| B1 | -0.9111268 | 1.0000000 | -0.9760758 |
| B2 | 0.8215256 | -0.9760758 | 1.0000000 |

Fit the weighted line with only the intercept, linear, and quadratic terms.
This is Analysis 3 in supporting online material in the short article.

Here is a more accurate table of the values of the parameters:

| Parameter | Parameter Estimate | Approximate Standard Error | Approximate Lower 95% Confidence Limit | Approximate Upper 95% Confidence Limit | t | p |
|-----------|-----------|-----------|-----------|-----------|-----------|-----------|
| B0 | 9.4756251E+04 | 2.25E-02 | 9.4756206E+04 | 9.4756296E+04 | 4215184.2 | <.0001 |
| B1 | 2.2446960E-04 | 1.90E-07 | 2.2408946E-04 | 2.2484973E-04 | 1178.64 | <.0001 |
| B2 | 5.4783954E-11 | 3.37E-13 | 5.4112275E-11 | 5.5455632E-11 | 162.80 | <.0001 |

Analysis of the Moldover et al. Data to Estimate R
Fit the weighted line with only the intercept, linear, and quadratic terms.
This is Analysis 3 in supporting online material in the short article.

Here are the estimates from the quadratic analysis with more
significant digits:

| Estimate of Intercept from Quadratic Model from Proc NLIN | Estimate of Standard Error of Intercept from Quadratic Model from Proc NLIN | Estimate of R from Quadratic Model from Proc NLIN |
|---|---|---|
| 9.4756250893324000E+04 | 2.2479741205803000E-02 | 8.3144773564460000E+00 |

Analysis of the Moldover et al. Data to Estimate R 60
Fit the weighted quadratic line with PROC REG with weights to confirm the
parallelism between PROC NLIN and PROC REG.
This analysis generates a warning in the SAS log.

The REG Procedure
Model: MODEL1
Dependent Variable: SpeedSq (m^2/s^2)

Number of Observations Read          70
Number of Observations Used          70

Weight: WeightVar Weight Variable

### Analysis of Variance

| Source | DF | Sum of Squares | Mean Square | F Value | Pr > F |
|--------|-----|---------------|-------------|---------|--------|
| Model | 2 | 47749339 | 23874669 | 1.893E7 | <.0001 |
| Error | 67 | 84.51549 | 1.26143 | | |
| Corrected Total | 69 | 47749423 | | | |

| | | | |
|--------|--------|----------|--------|
| Root MSE | 1.12313 | R-Square | 1.0000 |
| Dependent Mean | 94823 | Adj R-Sq | 1.0000 |
| Coeff Var | 0.00118 | | |

### Parameter Estimates

| Variable | Label | DF | Parameter Estimate | Standard Error | t Value | Pr > |t| |
|----------|-------|-----|--------------------|-----------------|---------|---------|
| Intercept | Intercept | 1 | 94756 | 0.02248 | 4214332 | <.0001 |
| Pressure | Pressure (Pa) | 1 | 0.00022447 | 1.904861E-7 | 1178.40 | <.0001 |
| PressSq | Pressure Squared | 1 | 5.4784E-11 | 3.36579E-13 | 162.77 | <.0001 |

### Parameter Estimates

| Variable | Label | DF | Variance Inflation |
|----------|-------|-----|--------------------|
| Intercept | Intercept | 1 | 0 |
| Pressure | Pressure (Pa) | 1 | 21.15237 |
| PressSq | Pressure Squared | 1 | 21.15237 |

Analysis of the Moldover et al. Data to Estimate R                    61
Fit the weighted quadratic line with PROC REG with weights to confirm the
parallelism between PROC NLIN and PROC REG.
This analysis generates a warning in the SAS log.

The REG Procedure
Model: MODEL1
Dependent Variable: SpeedSq (m^2/s^2)

### Collinearity Diagnostics

| Number | Eigenvalue | Condition Index | Intercept | Pressure | PressSq |
|--------|-----------|-----------------|-----------|----------|---------|
| 1 | 2.74044 | 1.00000 | 0.00815 | 0.00139 | 0.00244 |
| 2 | 0.25305 | 3.29081 | 0.17774 | 0.00099369 | 0.02849 |
| 3 | 0.00651 | 20.52008 | 0.81411 | 0.99761 | 0.96907 |

Collinearity Diagnostics header: ---------Proportion of Variation---------

### Collinearity Diagnostics (intercept adjusted)

| Number | Eigenvalue | Condition Index | Pressure | PressSq |
|--------|-----------|-----------------|----------|---------|
| 1 | 1.97608 | 1.00000 | 0.01196 | 0.01196 |
| 2 | 0.02392 | 9.08831 | 0.98804 | 0.98804 |

Proportion of Variation header: --Proportion of Variation-

Analysis of the Moldover et al. Data to Estimate R                       62
Fit the weighted quadratic line with PROC REG with weights to confirm the
parallelism between PROC NLIN and PROC REG.
This analysis generates a warning in the SAS log.

Here is a more accurate table of the values of the parameters:

| Label | Parameter Estimate | Standard Error | t | p |
|---|---|---|---|---|
| Intercept | 9.4756251E+04 | 2.25E-02 | 4214332.3 | <.0001 |
| Pressure (Pa) | 2.2446960E-04 | 1.90E-07 | 1178.40 | <.0001 |
| Pressure Squared | 5.4783954E-11 | 3.37E-13 | 162.77 | <.0001 |

Analysis of the Moldover et al. Data to Estimate R                    63
Repeat the preceding quadratic fit, but with a rescaled version of SpeedSq in which
SpeedSqResc = SpeedSq - 94,808.35, where 94,808.35 is the rough average
of the SpeedSq values. This eliminates the warning in the log.

The REG Procedure
Model: MODEL1
Dependent Variable: SpeedSqResc Rescaled Speed-Squared = SpeedSq - 94,808.35

Number of Observations Read          70
Number of Observations Used          70

Weight: WeightVar Weight Variable

### Analysis of Variance

| Source | DF | Sum of Squares | Mean Square | F Value | Pr > F |
|---|---|---|---|---|---|
| Model | 2 | 47749339 | 23874669 | 1.893E7 | <.0001 |
| Error | 67 | 84.48133 | 1.26092 | | |
| Corrected Total | 69 | 47749423 | | | |

| | | | |
|---|---|---|---|
| Root MSE | 1.12290 | R-Square | 1.0000 |
| Dependent Mean | 14.38129 | Adj R-Sq | 1.0000 |
| Coeff Var | 7.80809 | | |

### Parameter Estimates

| Variable | Label | DF | Parameter Estimate | Standard Error | t Value |
|---|---|---|---|---|---|
| Intercept | Intercept | 1 | -52.09911 | 0.02248 | -2317.6 |
| Pressure | Pressure (Pa) | 1 | 0.00022447 | 1.904476E-7 | 1178.64 |
| PressSq | Pressure Squared | 1 | 5.4784E-11 | 3.36511E-13 | 162.80 |

### Parameter Estimates

| Variable | Label | DF | Pr > \|t\| | Variance Inflation |
|---|---|---|---|---|
| Intercept | Intercept | 1 | <.0001 | 0 |
| Pressure | Pressure (Pa) | 1 | <.0001 | 21.15237 |
| PressSq | Pressure Squared | 1 | <.0001 | 21.15237 |

Analysis of the Moldover et al. Data to Estimate R                                    64
Repeat the preceding quadratic fit, but with a rescaled version of SpeedSq in which
SpeedSqResc = SpeedSq - 94,808.35, where 94,808.35 is the rough average
of the SpeedSq values.  This eliminates the warning in the log.

The REG Procedure
Model: MODEL1
Dependent Variable: SpeedSqResc Rescaled Speed-Squared = SpeedSq - 94,808.35

### Collinearity Diagnostics

| Number | Eigenvalue | Condition Index | Intercept | Pressure | PressSq |
|--------|-----------|-----------------|-----------|----------|---------|
| 1 | 2.74044 | 1.00000 | 0.00815 | 0.00139 | 0.00244 |
| 2 | 0.25305 | 3.29081 | 0.17774 | 0.00099369 | 0.02849 |
| 3 | 0.00651 | 20.52008 | 0.81411 | 0.99761 | 0.96907 |

### Collinearity Diagnostics (intercept adjusted)

| Number | Eigenvalue | Condition Index | Pressure | PressSq |
|--------|-----------|-----------------|----------|---------|
| 1 | 1.97608 | 1.00000 | 0.01196 | 0.01196 |
| 2 | 0.02392 | 9.08831 | 0.98804 | 0.98804 |

Analysis of the Moldover et al. Data to Estimate R                                65
Repeat the preceding quadratic fit, but with a rescaled version of SpeedSq in which
SpeedSqResc = SpeedSq - 94,808.35, where 94,808.35 is the rough average
of the SpeedSq values.  This eliminates the warning in the log.

Here is a more accurate table of the values of the parameters:

| Variable | DF | Parameter Estimate | Standard Error | t | p | Variance Inflation |
|----------|----|-----|-----|-----|-----|-----|
| Intercept | 1 | -5.2099107E+01 | 2.25E-02 | -2317.6 | <.0001 | 0 |
| Pressure | 1 | 2.2446960E-04 | 1.90E-07 | 1178.6 | <.0001 | 21.15237 |
| PressSq | 1 | 5.4783954E-11 | 3.37E-13 | 162.8 | <.0001 | 21.15237 |

Note that (-52.099107 + 94,808.35) = 94,756.251, as above.  Thus despite
the warning from SAS, the earlier analysis provided a correct estimate of
the intercept of the best-fitting line to the required number of
significant digits.

Note that the two analyses are not identical because the error mean square
from two pages above is 1.26092, but the error mean square from the earlier
analysis with PROC REG without the rescaling is 1.26143, although the two
values should theoretically be identical.  The difference is due to roundoff
errors accumulated in the unscaled analysis.

Analysis of the Moldover et al. Data to Estimate R
Repeat the preceding quadratic fit, but with a rescaled version of SpeedSq in which
SpeedSqResc = SpeedSq - 94,808.35, where 94,808.35 is the rough average
of the SpeedSq values.  This eliminates the warning in the log.
Here are the estimates from the quadratic analysis with proc REG with
rescaled speed-squared and with more significant digits:

| Estimate of Intercept from Rescaled Quadratic Model from Proc REG | Estimate of Standard Error of Intercept from Rescaled Quadratic Model from Proc REG | Estimate of R from Rescaled Quadratic Model from Proc REG |
|---|---|---|
| 9.4756250893324000E+04 | 2.2479741202120000E-02 | 8.3144773564460000E+00 |

Analysis of the Moldover et al. Data to Estimate R                                        67
Simulation of Quadratic Regression: Seed = 88218421  PSD = 50  SSD = 1.1
Listing of the 70 Data Values in the First Two of the 100,000 Analyzed Datasets

The speed-squared values are rescaled so that the correct intercept is exactly zero.

| Obs | Iteration Number | Simulated Measured Value of Pressure | Square of Simulated Measured Value of Pressure | Simulated Measured Speed-Squared (rescaled) from Quadratic Model |
|-----|-----|-----|-----|-----|
| 1  | 1 | 24,995.54  | 624,777,167.45    | 5.9035  |
| 2  | 1 | 24,995.54  | 624,777,167.45    | 4.9729  |
| 3  | 1 | 24,995.54  | 624,777,167.45    | 5.6905  |
| 4  | 1 | 24,995.54  | 624,777,167.45    | 7.0391  |
| 5  | 1 | 24,995.54  | 624,777,167.45    | 5.9092  |
| 6  | 1 | 49,990.28  | 2,499,027,996.10  | 11.3744 |
| 7  | 1 | 49,990.28  | 2,499,027,996.10  | 10.6348 |
| 8  | 1 | 49,990.28  | 2,499,027,996.10  | 10.9341 |
| 9  | 1 | 49,990.28  | 2,499,027,996.10  | 10.5257 |
| 10 | 1 | 49,990.28  | 2,499,027,996.10  | 11.7002 |
| 11 | 1 | 50,061.57  | 2,506,160,943.19  | 12.0061 |
| 12 | 1 | 50,061.57  | 2,506,160,943.19  | 12.4718 |
| 13 | 1 | 50,061.57  | 2,506,160,943.19  | 10.5508 |
| 14 | 1 | 50,061.57  | 2,506,160,943.19  | 11.2237 |
| 15 | 1 | 50,061.57  | 2,506,160,943.19  | 11.5087 |
| 16 | 1 | 75,118.15  | 5,642,737,005.80  | 18.1917 |
| 17 | 1 | 75,118.15  | 5,642,737,005.80  | 16.7325 |
| 18 | 1 | 75,118.15  | 5,642,737,005.80  | 16.6000 |
| 19 | 1 | 75,118.15  | 5,642,737,005.80  | 16.5988 |
| 20 | 1 | 75,118.15  | 5,642,737,005.80  | 16.7315 |
| 21 | 1 | 99,924.79  | 9,984,963,716.25  | 23.6524 |
| 22 | 1 | 99,924.79  | 9,984,963,716.25  | 21.6485 |
| 23 | 1 | 99,924.79  | 9,984,963,716.25  | 24.5494 |
| 24 | 1 | 99,924.79  | 9,984,963,716.25  | 22.9884 |
| 25 | 1 | 99,924.79  | 9,984,963,716.25  | 25.0259 |
| 26 | 1 | 99,980.49  | 9,996,098,208.79  | 23.4343 |
| 27 | 1 | 99,980.49  | 9,996,098,208.79  | 21.9606 |
| 28 | 1 | 99,980.49  | 9,996,098,208.79  | 22.7679 |
| 29 | 1 | 99,980.49  | 9,996,098,208.79  | 21.7449 |
| 30 | 1 | 99,980.49  | 9,996,098,208.79  | 23.2329 |
| 31 | 1 | 149,897.96 | 22,469,399,601.90 | 36.1273 |
| 32 | 1 | 149,897.96 | 22,469,399,601.90 | 35.8105 |
| 33 | 1 | 149,897.96 | 22,469,399,601.90 | 34.9473 |
| 34 | 1 | 149,897.96 | 22,469,399,601.90 | 33.4092 |
| 35 | 1 | 149,897.96 | 22,469,399,601.90 | 35.7095 |

The speed-squared values are rescaled so that the correct intercept is exactly zero.

| Obs | Iteration Number | Simulated Measured Value of Pressure | Square of Simulated Measured Value of Pressure | Simulated Measured Speed-Squared (rescaled) from Quadratic Model |
|---|---|---|---|---|
| 36 | 1 | 199,933.01 | 39,973,208,975.56 | 47.150 |
| 37 | 1 | 199,933.01 | 39,973,208,975.56 | 47.784 |
| 38 | 1 | 199,933.01 | 39,973,208,975.56 | 45.259 |
| 39 | 1 | 199,933.01 | 39,973,208,975.56 | 46.595 |
| 40 | 1 | 199,933.01 | 39,973,208,975.56 | 46.933 |
| 41 | 1 | 250,060.00 | 62,530,002,732.57 | 59.948 |
| 42 | 1 | 250,060.00 | 62,530,002,732.57 | 59.005 |
| 43 | 1 | 250,060.00 | 62,530,002,732.57 | 60.140 |
| 44 | 1 | 250,060.00 | 62,530,002,732.57 | 61.511 |
| 45 | 1 | 250,060.00 | 62,530,002,732.57 | 59.256 |
| 46 | 1 | 299,969.47 | 89,981,682,929.55 | 70.948 |
| 47 | 1 | 299,969.47 | 89,981,682,929.55 | 72.541 |
| 48 | 1 | 299,969.47 | 89,981,682,929.55 | 74.512 |
| 49 | 1 | 299,969.47 | 89,981,682,929.55 | 74.348 |
| 50 | 1 | 299,969.47 | 89,981,682,929.55 | 73.143 |
| 51 | 1 | 349,945.61 | 122,461,931,475.90 | 86.198 |
| 52 | 1 | 349,945.61 | 122,461,931,475.90 | 85.544 |
| 53 | 1 | 349,945.61 | 122,461,931,475.90 | 84.459 |
| 54 | 1 | 349,945.61 | 122,461,931,475.90 | 86.448 |
| 55 | 1 | 349,945.61 | 122,461,931,475.90 | 84.729 |
| 56 | 1 | 400,016.37 | 160,013,100,038.49 | 98.013 |
| 57 | 1 | 400,016.37 | 160,013,100,038.49 | 97.897 |
| 58 | 1 | 400,016.37 | 160,013,100,038.49 | 97.497 |
| 59 | 1 | 400,016.37 | 160,013,100,038.49 | 98.840 |
| 60 | 1 | 400,016.37 | 160,013,100,038.49 | 100.069 |
| 61 | 1 | 450,017.04 | 202,515,337,723.16 | 114.145 |
| 62 | 1 | 450,017.04 | 202,515,337,723.16 | 112.103 |
| 63 | 1 | 450,017.04 | 202,515,337,723.16 | 111.387 |
| 64 | 1 | 450,017.04 | 202,515,337,723.16 | 111.880 |
| 65 | 1 | 450,017.04 | 202,515,337,723.16 | 113.167 |
| 66 | 1 | 500,012.45 | 250,012,448,529.98 | 127.171 |
| 67 | 1 | 500,012.45 | 250,012,448,529.98 | 127.905 |
| 68 | 1 | 500,012.45 | 250,012,448,529.98 | 126.465 |
| 69 | 1 | 500,012.45 | 250,012,448,529.98 | 125.664 |
| 70 | 1 | 500,012.45 | 250,012,448,529.98 | 125.984 |

Analysis of the Moldover et al. Data to Estimate R                                    69
Simulation of Quadratic Regression: Seed = 88218421  PSD = 50  SSD = 1.1
Listing of the 70 Data Values in the First Two of the 100,000 Analyzed Datasets

The speed-squared values are rescaled so that the correct intercept is exactly zero.

| Obs | Iteration Number | Simulated Measured Value of Pressure | Square of Simulated Measured Value of Pressure | Simulated Measured Speed-Squared (rescaled) from Quadratic Model |
|---|---|---|---|---|
| 71 | 2 | 24,990.03 | 624,501,552.94 | 5.2922 |
| 72 | 2 | 24,990.03 | 624,501,552.94 | 7.0448 |
| 73 | 2 | 24,990.03 | 624,501,552.94 | 4.6014 |
| 74 | 2 | 24,990.03 | 624,501,552.94 | 4.0089 |
| 75 | 2 | 24,990.03 | 624,501,552.94 | 7.0285 |
| 76 | 2 | 50,011.92 | 2,501,192,022.07 | 8.7626 |
| 77 | 2 | 50,011.92 | 2,501,192,022.07 | 11.5355 |
| 78 | 2 | 50,011.92 | 2,501,192,022.07 | 11.7434 |
| 79 | 2 | 50,011.92 | 2,501,192,022.07 | 9.9424 |
| 80 | 2 | 50,011.92 | 2,501,192,022.07 | 10.0660 |
| 81 | 2 | 50,010.48 | 2,501,048,085.42 | 12.9148 |
| 82 | 2 | 50,010.48 | 2,501,048,085.42 | 11.7456 |
| 83 | 2 | 50,010.48 | 2,501,048,085.42 | 11.6477 |
| 84 | 2 | 50,010.48 | 2,501,048,085.42 | 9.7784 |
| 85 | 2 | 50,010.48 | 2,501,048,085.42 | 10.1573 |
| 86 | 2 | 75,002.15 | 5,625,322,002.58 | 18.1187 |
| 87 | 2 | 75,002.15 | 5,625,322,002.58 | 14.7542 |
| 88 | 2 | 75,002.15 | 5,625,322,002.58 | 16.1778 |
| 89 | 2 | 75,002.15 | 5,625,322,002.58 | 17.5150 |
| 90 | 2 | 75,002.15 | 5,625,322,002.58 | 15.7944 |
| 91 | 2 | 99,968.32 | 9,993,665,110.98 | 22.2935 |
| 92 | 2 | 99,968.32 | 9,993,665,110.98 | 25.5248 |
| 93 | 2 | 99,968.32 | 9,993,665,110.98 | 22.8044 |
| 94 | 2 | 99,968.32 | 9,993,665,110.98 | 21.5424 |
| 95 | 2 | 99,968.32 | 9,993,665,110.98 | 25.3722 |
| 96 | 2 | 99,940.58 | 9,988,119,965.01 | 24.4714 |
| 97 | 2 | 99,940.58 | 9,988,119,965.01 | 21.5509 |
| 98 | 2 | 99,940.58 | 9,988,119,965.01 | 21.8420 |
| 99 | 2 | 99,940.58 | 9,988,119,965.01 | 24.4124 |
| 100 | 2 | 99,940.58 | 9,988,119,965.01 | 19.7638 |
| 101 | 2 | 150,044.07 | 22,513,223,921.73 | 34.5760 |
| 102 | 2 | 150,044.07 | 22,513,223,921.73 | 35.2747 |
| 103 | 2 | 150,044.07 | 22,513,223,921.73 | 34.8083 |
| 104 | 2 | 150,044.07 | 22,513,223,921.73 | 34.4136 |
| 105 | 2 | 150,044.07 | 22,513,223,921.73 | 34.0312 |

The speed-squared values are rescaled so that the correct intercept is exactly zero.

| Obs | Iteration Number | Simulated Measured Value of Pressure | Square of Simulated Measured Value of Pressure | Simulated Measured Speed-Squared (rescaled) from Quadratic Model |
|-----|------|-----------|--------------------|---------|
| 106 | 2 | 200,012.00 | 40,004,799,826.30 | 47.207 |
| 107 | 2 | 200,012.00 | 40,004,799,826.30 | 47.034 |
| 108 | 2 | 200,012.00 | 40,004,799,826.30 | 47.256 |
| 109 | 2 | 200,012.00 | 40,004,799,826.30 | 45.812 |
| 110 | 2 | 200,012.00 | 40,004,799,826.30 | 47.178 |
| 111 | 2 | 250,055.42 | 62,527,711,334.49 | 58.814 |
| 112 | 2 | 250,055.42 | 62,527,711,334.49 | 59.959 |
| 113 | 2 | 250,055.42 | 62,527,711,334.49 | 59.168 |
| 114 | 2 | 250,055.42 | 62,527,711,334.49 | 58.928 |
| 115 | 2 | 250,055.42 | 62,527,711,334.49 | 58.879 |
| 116 | 2 | 299,995.90 | 89,997,539,429.68 | 72.760 |
| 117 | 2 | 299,995.90 | 89,997,539,429.68 | 73.236 |
| 118 | 2 | 299,995.90 | 89,997,539,429.68 | 71.134 |
| 119 | 2 | 299,995.90 | 89,997,539,429.68 | 74.526 |
| 120 | 2 | 299,995.90 | 89,997,539,429.68 | 71.917 |
| 121 | 2 | 349,982.62 | 122,487,837,272.37 | 84.997 |
| 122 | 2 | 349,982.62 | 122,487,837,272.37 | 84.624 |
| 123 | 2 | 349,982.62 | 122,487,837,272.37 | 84.627 |
| 124 | 2 | 349,982.62 | 122,487,837,272.37 | 84.972 |
| 125 | 2 | 349,982.62 | 122,487,837,272.37 | 85.757 |
| 126 | 2 | 399,931.01 | 159,944,813,541.78 | 97.082 |
| 127 | 2 | 399,931.01 | 159,944,813,541.78 | 99.193 |
| 128 | 2 | 399,931.01 | 159,944,813,541.78 | 98.567 |
| 129 | 2 | 399,931.01 | 159,944,813,541.78 | 99.470 |
| 130 | 2 | 399,931.01 | 159,944,813,541.78 | 97.348 |
| 131 | 2 | 449,996.49 | 202,496,840,160.35 | 111.706 |
| 132 | 2 | 449,996.49 | 202,496,840,160.35 | 112.116 |
| 133 | 2 | 449,996.49 | 202,496,840,160.35 | 112.229 |
| 134 | 2 | 449,996.49 | 202,496,840,160.35 | 110.413 |
| 135 | 2 | 449,996.49 | 202,496,840,160.35 | 112.004 |
| 136 | 2 | 500,006.17 | 250,006,174,125.99 | 125.979 |
| 137 | 2 | 500,006.17 | 250,006,174,125.99 | 125.223 |
| 138 | 2 | 500,006.17 | 250,006,174,125.99 | 125.016 |
| 139 | 2 | 500,006.17 | 250,006,174,125.99 | 126.099 |
| 140 | 2 | 500,006.17 | 250,006,174,125.99 | 126.426 |

Analysis of the Moldover et al. Data to Estimate R                              71
Simulation of Quadratic Regression: Seed = 88218421  PSD = 50  SSD = 1.1
Analyze the multiple sets of data with PROC REG using BY group processing on Iter.
Listing of the Output from the First Five of the 100,000 Independent Regressions

The true intercept is exactly zero for convenience and greater accuracy.
Thus the theoretical expected value of the intercept is zero, which
is why the p-values for the intercept (as given in the Pr > |T| column)
are generally greater than .05.

Iteration Number=1

The REG Procedure
Model: MODEL1
Dependent Variable: MeasSSqResc

Number of Observations Read          70
Number of Observations Used          70

### Analysis of Variance

| Source | DF | Sum of Squares | Mean Square | F Value | Pr > F |
|---|---|---|---|---|---|
| Model | 2 | 108343 | 54171 | 59539.4 | <.0001 |
| Error | 67 | 60.95942 | 0.90984 | | |
| Corrected Total | 69 | 108404 | | | |

| | | | |
|---|---|---|---|
| Root MSE | 0.95386 | R-Square | 0.9994 |
| Dependent Mean | 52.12780 | Adj R-Sq | 0.9994 |
| Coeff Var | 1.82984 | | |

### Parameter Estimates

| Variable | DF | Parameter Estimate | Standard Error | t Value | Pr > \|t\| |
|---|---|---|---|---|---|
| Intercept | 1 | -0.00546 | 0.29897 | -0.02 | 0.9855 |
| MeasuredP | 1 | 0.00022508 | 0.0000315 | 71.53 | <.0001 |
| MeasuredPSq | 1 | 5.56875E-11 | 6.11151E-12 | 9.11 | <.0001 |

Analysis of the Moldover et al. Data to Estimate R                              72
Simulation of Quadratic Regression: Seed = 88218421  PSD = 50  SSD = 1.1
Analyze the multiple sets of data with PROC REG using BY group processing on Iter.
Listing of the Output from the First Five of the 100,000 Independent Regressions

The true intercept is exactly zero for convenience and greater accuracy.
Thus the theoretical expected value of the intercept is zero, which
is why the p-values for the intercept (as given in the Pr > |T| column)
are generally greater than .05.

Iteration Number=2

The REG Procedure
Model: MODEL1
Dependent Variable: MeasSSqResc

Number of Observations Read            70
Number of Observations Used            70

### Analysis of Variance

| Source | DF | Sum of Squares | Mean Square | F Value | Pr > F |
|--------|-----|----------|----------|---------|--------|
| Model | 2 | 107681 | 53841 | 42060.1 | <.0001 |
| Error | 67 | 85.76594 | 1.28009 | | |
| Corrected Total | 69 | 107767 | | | |

| | | | |
|--------|--------|--------|--------|
| Root MSE | 1.13141 | R-Square | 0.9992 |
| Dependent Mean | 51.69956 | Adj R-Sq | 0.9992 |
| Coeff Var | 2.18843 | | |

### Parameter Estimates

| Variable | DF | Parameter Estimate | Standard Error | t Value | Pr > |t| |
|----------|-----|----------|----------|---------|--------|
| Intercept | 1 | -0.45165 | 0.35453 | -1.27 | 0.2071 |
| MeasuredP | 1 | 0.00022674 | 0.0000373 | 60.76 | <.0001 |
| MeasuredPSq | 1 | 5.08757E-11 | 7.248E-12 | 7.02 | <.0001 |

Analysis of the Moldover et al. Data to Estimate R                          73
Simulation of Quadratic Regression: Seed = 88218421  PSD = 50  SSD = 1.1
Analyze the multiple sets of data with PROC REG using BY group processing on Iter.
Listing of the Output from the First Five of the 100,000 Independent Regressions

The true intercept is exactly zero for convenience and greater accuracy.
Thus the theoretical expected value of the intercept is zero, which
is why the p-values for the intercept (as given in the Pr > |T| column)
are generally greater than .05.

Iteration Number=3

The REG Procedure
Model: MODEL1
Dependent Variable: MeasSSqResc

Number of Observations Read          70
Number of Observations Used          70

## Analysis of Variance

| Source | DF | Sum of Squares | Mean Square | F Value | Pr > F |
|--------|----|----|----|----|----|
| Model | 2 | 107882 | 53941 | 36734.4 | <.0001 |
| Error | 67 | 98.38285 | 1.46840 | | |
| Corrected Total | 69 | 107980 | | | |

| | | | |
|--------|----|----|----|
| Root MSE | 1.21178 | R-Square | 0.9991 |
| Dependent Mean | 51.80683 | Adj R-Sq | 0.9991 |
| Coeff Var | 2.33903 | | |

## Parameter Estimates

| Variable | DF | Parameter Estimate | Standard Error | t Value | Pr > |t| |
|--------|----|----|----|----|----|
| Intercept | 1 | -0.11120 | 0.37973 | -0.29 | 0.7705 |
| MeasuredP | 1 | 0.00022323 | 0.00000400 | 55.85 | <.0001 |
| MeasuredPSq | 1 | 5.82751E-11 | 7.76408E-12 | 7.51 | <.0001 |

Analysis of the Moldover et al. Data to Estimate R
Simulation of Quadratic Regression: Seed = 88218421  PSD = 50  SSD = 1.1
Analyze the multiple sets of data with PROC REG using BY group processing on Iter.
Listing of the Output from the First Five of the 100,000 Independent Regressions

The true intercept is exactly zero for convenience and greater accuracy.
Thus the theoretical expected value of the intercept is zero, which
is why the p-values for the intercept (as given in the Pr > |T| column)
are generally greater than .05.

Iteration Number=4

The REG Procedure
Model: MODEL1
Dependent Variable: MeasSSqResc

Number of Observations Read          70
Number of Observations Used          70

### Analysis of Variance

| Source | DF | Sum of Squares | Mean Square | F Value | Pr > F |
|--------|----|----|----|----|----|
| Model | 2 | 106573 | 53286 | 58075.5 | <.0001 |
| Error | 67 | 61.47483 | 0.91753 | | |
| Corrected Total | 69 | 106634 | | | |

| | | | |
|---|---|---|---|
| Root MSE | 0.95788 | R-Square | 0.9994 |
| Dependent Mean | 52.06006 | Adj R-Sq | 0.9994 |
| Coeff Var | 1.83995 | | |

### Parameter Estimates

| Variable | DF | Parameter Estimate | Standard Error | t Value | Pr > |t| |
|----------|----|----|----|----|----|
| Intercept | 1 | -0.03343 | 0.30001 | -0.11 | 0.9116 |
| MeasuredP | 1 | 0.00022851 | 0.0000316 | 72.36 | <.0001 |
| MeasuredPSq | 1 | 4.47013E-11 | 6.13408E-12 | 7.29 | <.0001 |

Analysis of the Moldover et al. Data to Estimate R                          75
Simulation of Quadratic Regression: Seed = 88218421  PSD = 50  SSD = 1.1
Analyze the multiple sets of data with PROC REG using BY group processing on Iter.
Listing of the Output from the First Five of the 100,000 Independent Regressions

The true intercept is exactly zero for convenience and greater accuracy.
Thus the theoretical expected value of the intercept is zero, which
is why the p-values for the intercept (as given in the Pr > |T| column)
are generally greater than .05.

Iteration Number=5

The REG Procedure
Model: MODEL1
Dependent Variable: MeasSSqResc

Number of Observations Read          70
Number of Observations Used          70

### Analysis of Variance

| Source | DF | Sum of Squares | Mean Square | F Value | Pr > F |
|--------|----|----|----|----|----|
| Model | 2 | 107373 | 53687 | 55543.6 | <.0001 |
| Error | 67 | 64.76010 | 0.96657 | | |
| Corrected Total | 69 | 107438 | | | |

| | | | |
|---|---|---|---|
| Root MSE | 0.98314 | R-Square | 0.9994 |
| Dependent Mean | 51.91925 | Adj R-Sq | 0.9994 |
| Coeff Var | 1.89360 | | |

### Parameter Estimates

| Variable | DF | Parameter Estimate | Standard Error | t Value | Pr > |t| |
|----------|----|----|----|----|----|
| Intercept | 1 | -0.18805 | 0.30812 | -0.61 | 0.5437 |
| MeasuredP | 1 | 0.00022676 | 0.0000324 | 69.93 | <.0001 |
| MeasuredPSq | 1 | 5.01209E-11 | 6.29878E-12 | 7.96 | <.0001 |

Analysis of the Moldover et al. Data to Estimate R                                76
Simulation of Quadratic Regression: Seed = 88218421  PSD = 50  SSD = 1.1

Listing of the first 30 estimates of the intercept b0, its estimated standard
error, and the square of its estimated standard error.
If the entire output were given here, it would contain 100,000 rows of data.
Assuming 30 rows of data per page, this output would span 3,334 pages.

Note how the second and third columns of the first five rows of the table give
the same information as is given on the preceding five pages.

| Iteration Number | Estimated Centered Intercept from Regression Program | Estimated Standard Error of Intercept from Regression Program | Square of Estimated Standard Error from Regression Program = Variance |
|---|---|---|---|
| 1 | -0.00546 | 0.29897 | 0.08938 |
| 2 | -0.45165 | 0.35453 | 0.12569 |
| 3 | -0.11120 | 0.37973 | 0.14419 |
| 4 | -0.03343 | 0.30001 | 0.09000 |
| 5 | -0.18805 | 0.30812 | 0.09494 |
| 6 | -0.48600 | 0.36881 | 0.13602 |
| 7 | -0.38210 | 0.32424 | 0.10513 |
| 8 | -0.02117 | 0.35103 | 0.12322 |
| 9 | 0.33448 | 0.37552 | 0.14102 |
| 10 | 0.23473 | 0.36097 | 0.13030 |
| 11 | 0.00212 | 0.37059 | 0.13734 |
| 12 | -0.76544 | 0.33390 | 0.11149 |
| 13 | 0.02022 | 0.27106 | 0.07347 |
| 14 | -0.35951 | 0.33501 | 0.11223 |
| 15 | 0.22661 | 0.37117 | 0.13777 |
| 16 | 0.40781 | 0.29867 | 0.08921 |
| 17 | -1.07484 | 0.31535 | 0.09945 |
| 18 | 0.95609 | 0.33732 | 0.11379 |
| 19 | 0.34538 | 0.32250 | 0.10401 |
| 20 | 0.10853 | 0.29919 | 0.08952 |
| 21 | 0.02602 | 0.30564 | 0.09341 |
| 22 | 0.34225 | 0.37490 | 0.14055 |
| 23 | -0.04790 | 0.28029 | 0.07856 |
| 24 | -0.71476 | 0.40046 | 0.16037 |
| 25 | 0.18732 | 0.29004 | 0.08412 |
| 26 | -0.34077 | 0.34182 | 0.11684 |
| 27 | -0.14516 | 0.34107 | 0.11633 |
| 28 | 0.49366 | 0.31564 | 0.09963 |
| 29 | 0.53524 | 0.34354 | 0.11802 |
| 30 | -0.22100 | 0.37993 | 0.14435 |

Analysis of the Moldover et al. Data to Estimate R                      77
Simulation of Quadratic Regression: Seed = 88218421  PSD = 50  SSD = 1.1
Distribution of the intercept parameter and related statistics from the
simulation with 100,000 independent regressions

Statistics Computed Across the 100,000 Rows in the Dataset

The MEANS Procedure

| Variable | Label | N | Mean | Std Dev | Skewness |
|----------|-------|---|------|---------|----------|
| Estimate | Estimated Centered Intercept | 100000 | -0.000402 | 0.345184 | 0.008718 |
| StdErr | Estimated Std. Err. of Int. | 100000 | 0.343398 | 0.029615 | 0.096127 |
| StdErrSq | Est. Std. Err. of Int. Sq'd | 100000 | 0.118799 | 0.020461 | 0.355557 |

| Variable | Label | Kurtosis |
|----------|-------|----------|
| Estimate | Estimated Centered Intercept | -0.002703 |
| StdErr | Estimated Std. Err. of Int. | 0.015407 |
| StdErrSq | Est. Std. Err. of Int. Sq'd | 0.212514 |

Simulation of Quadratic Regression: Seed = 88218421  PSD = 50  SSD = 1.1
Distribution of the intercept parameter and related statistics from the
simulation with 100,000 independent regressions

This is a listing of the output datatset from the analysis on the preceding page.

By naming all the variables in the output dataset this listing helps in writing
and checking the SQL statement to generate the table on the following page.

| Obs | VName_Estimate | Label_Estimate | Estimate_N | Estimate_Mean | Estimate_StdDev |
|---|---|---|---|---|---|
| 1 | Estimate | Estimated Centered Intercept | 100000 | -0.000402 | 0.345184 |

| Obs | Estimate_Skew | Estimate_Kurt | VName_StdErr | Label_StdErr | Std Err_N | StdErr_Mean |
|---|---|---|---|---|---|---|
| 1 | 0.008718 | -0.002703 | StdErr | Estimated Std. Err. of Int. | 100000 | 0.343398 |

| Obs | StdErr_StdDev | StdErr_Skew | StdErr_Kurt | VName_StdErrSq | Label_StdErrSq |
|---|---|---|---|---|---|
| 1 | 0.029615 | 0.096127 | 0.015407 | StdErrSq | Est. Std. Err. of Int. Sq'd |

| Obs | Std ErrSq_N | StdErrSq_Mean | StdErrSq_StdDev | StdErrSq_Skew | StdErrSq_Kurt |
|---|---|---|---|---|---|
| 1 | 100000 | 0.118799 | 0.020461 | 0.355557 | 0.212514 |

Analysis of the Moldover et al. Data to Estimate R
Simulation of Quadratic Regression: Seed = 88218421  PSD = 50  SSD = 1.1
Distribution of the intercept parameter and related statistics from the
simulation with 100,000 independent regressions

Key statistics computed from the output dataset.  Note that if n is large, then
- the mean estimate of the intercept is generally not significantly different
  from zero
- the estimated variance and the actual variance are generally not significantly
  different from each other.

| Mean of 100,000 Estimated Intercepts (Correct Value = 0) | Actual Standard Deviation of the 100,000 Estimated Intercepts | p-value for Null Hypothesis that Expected Intercept = 0 (Using One-Sample t-Test) | Square Root of Mean ESTIMATED Variance of the 100,000 Estimated Intercepts | Relative Difference Between Second and Fourth Columns | p-value for Null Hypothesis that Mean Estimated Variance = Actual Variance (Using F-Test) |
|---|---|---|---|---|---|
| -.0004022931 | 0.34518386 | 0.712 | 0.34467248 | 0.001481 | 0.509 |

Weighted Ninth Degree Polynomial with Proc NLIN and Centered SpeedSq

The NLIN Procedure
Dependent Variable SpeedSqResc
Method: Gauss-Newton

Iterative Phase

| Iter | B0 | B1 | B2 | B3 | B4 | B5 | B6 | B7 | B8 |
|---|---|---|---|---|---|---|---|---|---|
| 0 | -52.0000 | 0.000225 | 5.36E-11 | 1.0000 | 1.0000 | 1.0000 | 1.0000 | 1.0000 | 1.0000 |

Iterative Phase

| Iter | B9 | Weighted SS |
|---|---|---|
| 0 | 1.0000 | 1.76E106 |

Iterative Phase

| Iter | B0 | B1 | B2 | B3 | B4 | B5 | B6 | B7 | B8 |
|---|---|---|---|---|---|---|---|---|---|
| 1 | -3.61E43 | 2.704E39 | -7.72E34 | 1.123E30 | -9.36E24 | 4.721E19 | -1.46E14 | 2.7192E8 | -277.9 |

Iterative Phase

| Iter | B9 | Weighted SS |
|---|---|---|
| 1 | 0.000120 | 8.149E87 |

Iterative Phase

| Iter | B0 | B1 | B2 | B3 | B4 | B5 | B6 | B7 | B8 |
|---|---|---|---|---|---|---|---|---|---|
| 2 | -4.21E40 | 3.174E36 | -9.11E31 | 1.332E27 | -1.11E22 | 5.64E16 | -1.75E11 | 326507 | -0.3343 |

Iterative Phase

| Iter | B9 | Weighted SS |
|---|---|---|
| 2 | 1.445E-7 | 1.171E82 |

Iterative Phase

| Iter | B0 | B1 | B2 | B3 | B4 | B5 | B6 | B7 | B8 |
|---|---|---|---|---|---|---|---|---|---|
| 3 | -5.05E37 | 3.809E33 | -1.09E29 | 1.599E24 | -1.34E19 | 6.769E13 | -2.103E8 | 391.8 | -0.00040 |

Weighted Ninth Degree Polynomial with Proc NLIN and Centered SpeedSq

The NLIN Procedure
Dependent Variable SpeedSqResc
Method: Gauss-Newton

Iterative Phase

| Iter | B9 | Weighted SS |
|------|------|------|
| 3 | 1.73E-10 | 1.687E76 |

Iterative Phase

| Iter | B0 | B1 | B2 | B3 | B4 | B5 | B6 | B7 | B8 |
|------|------|------|------|------|------|------|------|------|------|
| 4 | -6.06E34 | 4.571E30 | -1.31E26 | 1.919E21 | -1.61E16 | 8.122E10 | -252367 | 0.4702 | -4.81E-7 |

Iterative Phase

| Iter | B9 | Weighted SS |
|------|------|------|
| 4 | 2.08E-13 | 2.429E70 |

Iterative Phase

| Iter | B0 | B1 | B2 | B3 | B4 | B5 | B6 | B7 | B8 |
|------|------|------|------|------|------|------|------|------|------|
| 5 | -7.27E31 | 5.486E27 | -1.58E23 | 2.303E18 | -1.93E13 | 97472396 | -302.8 | 0.000564 | -578E-12 |

Iterative Phase

| Iter | B9 | Weighted SS |
|------|------|------|
| 5 | 2.5E-16 | 3.499E64 |

Iterative Phase

| Iter | B0 | B1 | B2 | B3 | B4 | B5 | B6 | B7 | B8 |
|------|------|------|------|------|------|------|------|------|------|
| 6 | -8.72E28 | 6.583E24 | -1.89E20 | 2.763E15 | -2.31E10 | 116970 | -0.3634 | 6.771E-7 | -693E-15 |

Iterative Phase

| Iter | B9 | Weighted SS |
|------|------|------|
| 6 | 3E-19 | 5.038E58 |

Weighted Ninth Degree Polynomial with Proc NLIN and Centered SpeedSq

The NLIN Procedure
Dependent Variable SpeedSqResc
Method: Gauss-Newton

### Iterative Phase

| Iter | B0 | B1 | B2 | B3 | B4 | B5 | B6 | B7 | B8 |
|------|------|------|--------|--------|--------|-------|----------|---------|---------|
| 7 | -1.05E26 | 7.9E21 | -2.27E17 | 3.316E12 | -2.774E7 | 140.4 | -0.00044 | 8.13E-10 | -832E-18 |

### Iterative Phase

| Iter | B9 | Weighted SS |
|------|--------|-----------|
| 7 | 3.6E-22 | 7.255E52 |

### Iterative Phase

| Iter | B0 | B1 | B2 | B3 | B4 | B5 | B6 | B7 | B8 |
|------|----------|---------|---------|---------|---------|--------|---------|----------|----------|
| 8 | -1.26E23 | 9.48E18 | -2.72E14 | 3.9792E9 | -33294.5 | 0.1684 | -5.23E-7 | 9.75E-13 | -999E-21 |

### Iterative Phase

| Iter | B9 | Weighted SS |
|------|---------|-----------|
| 8 | 4.32E-25 | 1.045E47 |

### Iterative Phase

| Iter | B0 | B1 | B2 | B3 | B4 | B5 | B6 | B7 | B8 |
|------|----------|----------|---------|---------|---------|----------|----------|---------|---------|
| 9 | -1.51E20 | 1.138E16 | -3.27E11 | 4775184 | -39.9545 | 0.000202 | -628E-12 | 1.17E-15 | -12E-22 |

### Iterative Phase

| Iter | B9 | Weighted SS |
|------|---------|-----------|
| 9 | 5.18E-28 | 1.505E41 |

### Iterative Phase

| Iter | B0 | B1 | B2 | B3 | B4 | B5 | B6 | B7 | B8 |
|------|----------|----------|--------|--------|---------|----------|----------|--------|----------|
| 10 | -1.81E17 | 1.365E13 | -3.92E8 | 5730.4 | -0.0479 | 2.426E-7 | -754E-15 | 1.4E-18 | -144E-26 |

The NLIN Procedure
Dependent Variable SpeedSqResc
Method: Gauss-Newton

Iterative Phase

| Iter | B9 | Weighted SS |
|------|------|------|
| 10 | 6.22E-31 | 2.167E35 |

Iterative Phase

| Iter | B0 | B1 | B2 | B3 | B4 | B5 | B6 | B7 | B8 |
|------|------|------|------|------|------|------|------|------|------|
| 11 | -2.17E14 | 1.638E10 | -470410 | 6.8766 | -0.00006 | 2.91E-10 | -904E-18 | 1.69E-21 | -173E-29 |

Iterative Phase

| Iter | B9 | Weighted SS |
|------|------|------|
| 11 | 7.46E-34 | 3.12E29 |

Iterative Phase

| Iter | B0 | B1 | B2 | B3 | B4 | B5 | B6 | B7 | B8 |
|------|------|------|------|------|------|------|------|------|------|
| 12 | -2.61E11 | 19659560 | -564.5 | 0.00825 | -6.9E-8 | 3.49E-13 | -109E-20 | 2.02E-24 | -207E-32 |

Iterative Phase

| Iter | B9 | Weighted SS |
|------|------|------|
| 12 | 8.95E-37 | 4.494E23 |

Iterative Phase

| Iter | B0 | B1 | B2 | B3 | B4 | B5 | B6 | B7 | B8 |
|------|------|------|------|------|------|------|------|------|------|
| 13 | -3.126E8 | 23592.1 | -0.6774 | 9.903E-6 | -829E-13 | 4.19E-16 | -13E-22 | 2.43E-27 | -248E-35 |

Iterative Phase

| Iter | B9 | Weighted SS |
|------|------|------|
| 13 | 1.07E-39 | 6.471E17 |

Analysis of the Moldover et al. Data to Estimate R                84
Weighted Ninth Degree Polynomial with Proc NLIN and Centered SpeedSq

The NLIN Procedure
Dependent Variable SpeedSqResc
Method: Gauss-Newton

### Iterative Phase

| Iter | B0 | B1 | B2 | B3 | B4 | B5 | B6 | B7 | B8 |
|------|------|------|------|------|------|------|------|------|------|
| 14 | -375241 | 28.3115 | -0.00081 | 1.188E-8 | -994E-16 | 5.03E-19 | -156E-26 | 2.91E-30 | -298E-38 |

### Iterative Phase

| Iter | B9 | Weighted SS |
|------|------|------|
| 14 | 1.29E-42 | 9.319E11 |

### Iterative Phase

| Iter | B0 | B1 | B2 | B3 | B4 | B5 | B6 | B7 | B8 |
|------|------|------|------|------|------|------|------|------|------|
| 15 | -501.7 | 0.0342 | -9.74E-7 | 1.42E-11 | -119E-18 | 6.03E-22 | -187E-29 | 3.49E-33 | -357E-41 |

### Iterative Phase

| Iter | B9 | Weighted SS |
|------|------|------|
| 15 | 1.55E-45 | 1342075 |

### Iterative Phase

| Iter | B0 | B1 | B2 | B3 | B4 | B5 | B6 | B7 | B8 |
|------|------|------|------|------|------|------|------|------|------|
| 16 | -52.0224 | 0.000223 | 2.7E-11 | 1.07E-15 | -121E-22 | 6.96E-26 | -235E-33 | 4.73E-37 | -527E-45 |

### Iterative Phase

| Iter | B9 | Weighted SS |
|------|------|------|
| 16 | 2.49E-49 | 69.5043 |

### Iterative Phase

| Iter | B0 | B1 | B2 | B3 | B4 | B5 | B6 | B7 | B8 |
|------|------|------|------|------|------|------|------|------|------|
| 17 | -51.4828 | 0.000182 | 1.196E-9 | -16E-15 | 1.31E-19 | -654E-27 | 2.01E-30 | -372E-38 | 3.76E-42 |

Analysis of the Moldover et al. Data to Estimate R                    85
Weighted Ninth Degree Polynomial with Proc NLIN and Centered SpeedSq

The NLIN Procedure
Dependent Variable SpeedSqResc
Method: Gauss-Newton

        Iterative Phase

                    Weighted
  Iter      B9          SS

    17  -16E-49     67.5717

                                        Iterative Phase

  Iter      B0        B1        B2        B3        B4        B5        B6        B7        B8

    18  -51.4821  0.000182  1.198E-9  -16E-15  1.31E-19  -655E-27  2.02E-30  -372E-38  3.77E-42

        Iterative Phase

                    Weighted
  Iter      B9          SS

    18  -161E-50    67.5717

NOTE: Convergence criterion met.

        Estimation Summary

Method                  Gauss-Newton
Iterations                        18
R                             2.434E-7
PPC(B1)                       3.205E-7
RPC(B1)                       0.000267
Object                        4.119E-8
Objective                     67.57169
Observations Read                   70
Observations Used                   70
Observations Missing                 0

Analysis of the Moldover et al. Data to Estimate R                                86
Weighted Ninth Degree Polynomial with Proc NLIN and Centered SpeedSq

The NLIN Procedure

| Source | DF | Sum of Squares | Mean Square | F Value | Approx Pr > F |
|---|---|---|---|---|---|
| Model | 9 | 47749356 | 5305484 | 4710982 | <.0001 |
| Error | 60 | 67.5717 | 1.1262 | | |
| Corrected Total | 69 | 47749423 | | | |

| Parameter | Estimate | Approx Std Error | Approximate 95% Confidence Limits | | Skewness |
|---|---|---|---|---|---|
| B0 | -51.4821 | 0.5419 | -52.5661 | -50.3982 | 0 |
| B1 | 0.000182 | 0.000036 | 0.000110 | 0.000254 | 0 |
| B2 | 1.198E-9 | 9.66E-10 | -735E-12 | 3.13E-9 | 0 |
| B3 | -16E-15 | 1.36E-14 | -432E-16 | 1.11E-14 | 0 |
| B4 | 1.31E-19 | 1.11E-19 | -91E-21 | 3.53E-19 | 0 |
| B5 | -655E-27 | 5.56E-25 | -177E-26 | 4.57E-25 | 0 |
| B6 | 2.02E-30 | 1.72E-30 | -142E-32 | 5.45E-30 | 0 |
| B7 | -372E-38 | 3.2E-36 | -101E-37 | 2.68E-36 | 0 |
| B8 | 3.77E-42 | 3.28E-42 | -28E-43 | 1.03E-41 | 0 |
| B9 | -161E-50 | 1.42E-48 | -445E-50 | 1.24E-48 | 0 |

Approximate Correlation Matrix

| | B0 | B1 | B2 | B3 | B4 |
|---|---|---|---|---|---|
| B0 | 1.0000000 | -0.9798872 | 0.9381773 | -0.8922410 | 0.8497007 |
| B1 | -0.9798872 | 1.0000000 | -0.9876563 | 0.9616049 | -0.9321421 |
| B2 | 0.9381773 | -0.9876563 | 1.0000000 | -0.9924809 | 0.9763143 |
| B3 | -0.8922410 | 0.9616049 | -0.9924809 | 1.0000000 | -0.9953466 |
| B4 | 0.8497007 | -0.9321421 | 0.9763143 | -0.9953466 | 1.0000000 |
| B5 | -0.8126348 | 0.9036466 | -0.9572728 | 0.9849765 | -0.9969791 |
| B6 | 0.7809304 | -0.8775821 | 0.9379666 | -0.9722093 | 0.9899925 |
| B7 | -0.7538673 | 0.8542403 | -0.9195024 | 0.9587137 | -0.9810613 |
| B8 | 0.7306752 | -0.8334957 | 0.9023127 | -0.9453347 | 0.9713101 |
| B9 | -0.7106883 | 0.8150986 | -0.8865286 | 0.9325008 | -0.9613810 |

Approximate Correlation Matrix

| | B5 | B6 | B7 | B8 | B9 |
|---|---|---|---|---|---|
| B0 | -0.8126348 | 0.7809304 | -0.7538673 | 0.7306752 | -0.7106883 |
| B1 | 0.9036466 | -0.8775821 | 0.8542403 | -0.8334957 | 0.8150986 |
| B2 | -0.9572728 | 0.9379666 | -0.9195024 | 0.9023127 | -0.8865286 |
| B3 | 0.9849765 | -0.9722093 | 0.9587137 | -0.9453347 | 0.9325008 |

Analysis of the Moldover et al. Data to Estimate R                              87
Weighted Ninth Degree Polynomial with Proc NLIN and Centered SpeedSq

The NLIN Procedure

Approximate Correlation Matrix

|    | B5 | B6 | B7 | B8 | B9 |
|----|----|----|----|----|----|
| B4 | -0.9969791 | 0.9899925 | -0.9810613 | 0.9713101 | -0.9613810 |
| B5 | 1.0000000 | -0.9979351 | 0.9930116 | -0.9865351 | 0.9792944 |
| B6 | -0.9979351 | 1.0000000 | -0.9985269 | 0.9949336 | -0.9901075 |
| B7 | 0.9930116 | -0.9985269 | 1.0000000 | -0.9989150 | 0.9962242 |
| B8 | -0.9865351 | 0.9949336 | -0.9989150 | 1.0000000 | -0.9991821 |
| B9 | 0.9792944 | -0.9901075 | 0.9962242 | -0.9991821 | 1.0000000 |

Here is a more accurate table of the values of the parameters with the p-value
for each parameter in the last column:

| Parameter | Parameter Estimate | Approximate Standard Error | Approximate Lower 95% Confidence Limit | Approximate Upper 95% Confidence Limit | t | p |
|---|---|---|---|---|---|---|
| B0 | -5.1482137E+01 | 5.42E-01 | -5.2566071E+01 | -5.0398202E+01 | -95.01 | <.0001 |
| B1 | 1.8202043E-04 | 3.62E-05 | 1.0969433E-04 | 2.5434652E-04 | 5.03 | <.0001 |
| B2 | 1.1977044E-09 | 9.66E-10 | -7.3501207E-10 | 3.1304208E-09 | 1.24 | 0.2200 |
| B3 | -1.6039161E-14 | 1.36E-14 | -4.3176459E-14 | 1.1098138E-14 | -1.18 | 0.2418 |
| B4 | 1.3113898E-19 | 1.11E-19 | -9.1023284E-20 | 3.5330125E-19 | 1.18 | 0.2424 |
| B5 | -6.5486618E-25 | 5.56E-25 | -1.7664742E-24 | 4.5674184E-25 | -1.18 | 0.2433 |
| B6 | 2.0157821E-30 | 1.72E-30 | -1.4219578E-30 | 5.4535221E-30 | 1.17 | 0.2455 |
| B7 | -3.7205725E-36 | 3.20E-36 | -1.0120966E-35 | 2.6798209E-36 | -1.16 | 0.2495 |
| B8 | 3.7675250E-42 | 3.28E-42 | -2.7978670E-42 | 1.0332917E-41 | 1.15 | 0.2556 |
| B9 | -1.6068415E-48 | 1.42E-48 | -4.4540411E-48 | 1.2403581E-48 | -1.13 | 0.2634 |

Analysis of the Moldover et al. Data to Estimate R                                        89
Weighted Ninth-Degree Polynomial with Proc REG and Centered SpeedSq

The REG Procedure
Model: MODEL1
Dependent Variable: SpeedSqResc

Number of Observations Read          70
Number of Observations Used          70

Weight: WeightVar

### Analysis of Variance

| Source | DF | Sum of Squares | Mean Square | F Value | Pr > F |
|--------|----|----|----|----|----|
| Model | 9 | 47749356 | 5305484 | 4711030 | <.0001 |
| Error | 60 | 67.57101 | 1.12618 | | |
| Corrected Total | 69 | 47749423 | | | |

| | | | |
|---|---|---|---|
| Root MSE | 1.06122 | R-Square | 1.0000 |
| Dependent Mean | 14.38129 | Adj R-Sq | 1.0000 |
| Coeff Var | 7.37916 | | |

### Parameter Estimates

| Variable | DF | Parameter Estimate | Standard Error | t Value | Pr > \|t\| | Tolerance | Variance Inflation |
|----------|----|----|----|----|----|----|----|
| Intercept | 1 | -51.48190 | 0.54216 | -94.96 | <.0001 | . | 0 |
| Pressure | 1 | 0.00018200 | 0.00003618 | 5.03 | <.0001 | 0.00000117 | 854735 |
| P2 | 1 | 1.198192E-9 | 9.66907E-10 | 1.24 | 0.2201 | 5.11438E-9 | 195527114 |
| P3 | 1 | -1.6046E-14 | 1.35771E-14 | -1.18 | 0.2419 | 1.02088E-10 | 9795498574 |
| P4 | 1 | 1.31196E-19 | 1.11153E-19 | 1.18 | 0.2425 | 6.19185E-12 | 1.615027E11 |
| P5 | 1 | -6.5515E-25 | 5.5617E-25 | -1.18 | 0.2435 | 1.00607E-12 | 9.939626E11 |
| P6 | 1 | 2.01666E-30 | 1.72E-30 | 1.17 | 0.2456 | 4.25665E-13 | 2.349267E12 |
| P7 | 1 | -3.7222E-36 | 3.2023E-36 | -1.16 | 0.2497 | 4.93992E-13 | 2.024324E12 |
| P8 | 1 | 3.76917E-42 | 3.28483E-42 | 1.15 | 0.2558 | 1.87872E-12 | 5.322766E11 |
| P9 | 1 | -1.6075E-48 | 1.42451E-48 | -1.13 | 0.2636 | 3.98043E-11 | 25122929309 |

Weighted Ninth-Degree Polynomial with Proc REG and Centered SpeedSq

The REG Procedure
Model: MODEL1
Dependent Variable: SpeedSqResc

### Collinearity Diagnostics

| Number | Eigenvalue | Condition Index | Intercept | Pressure | P2 | P3 |
|---|---|---|---|---|---|---|
| 1 | 8.82773 | 1.00000 | 0.00000137 | 8.140248E-9 | 1.13297E-10 | 4.09901E-12 |
| 2 | 1.00964 | 2.95693 | 0.00011594 | 1.516392E-7 | 4.29672E-10 | 1.15236E-12 |
| 3 | 0.15094 | 7.64749 | 0.00058357 | 3.135928E-7 | 1.088889E-8 | 2.54932E-10 |
| 4 | 0.01118 | 28.09518 | 0.00160 | 0.00002579 | 1.778842E-8 | 1.612195E-9 |
| 5 | 0.00048364 | 135.10249 | 0.00581 | 0.00037950 | 0.00000355 | 9.266971E-8 |
| 6 | 0.00001599 | 742.93972 | 0.01888 | 0.00302 | 0.00018958 | 7.843123E-7 |
| 7 | 3.341064E-7 | 5140.22534 | 0.05527 | 0.01881 | 0.00372 | 0.00029337 |
| 8 | 4.156979E-9 | 46082 | 0.20445 | 0.12633 | 0.05570 | 0.01547 |
| 9 | 3.78039E-11 | 483233 | 0.55945 | 0.57279 | 0.48663 | 0.32851 |
| 10 | 1E-12 | 2971150 | 0.15386 | 0.27864 | 0.45376 | 0.65573 |

### Collinearity Diagnostics

| Number | P4 | P5 | P6 | P7 | P8 | P9 |
|---|---|---|---|---|---|---|
| 1 | 3.53203E-13 | 6.9422E-14 | 3.1753E-14 | 3.70196E-14 | 1.35437E-13 | 2.69811E-12 |
| 2 | 6.70224E-14 | 9.08219E-14 | 9.03318E-14 | 1.65667E-13 | 8.2064E-13 | 2.03384E-11 |
| 3 | 7.98382E-12 | 1.65028E-13 | 6.53012E-14 | 6.20974E-13 | 5.62185E-12 | 1.94961E-10 |
| 4 | 3.34257E-10 | 5.12889E-11 | 7.46969E-12 | 1.86998E-14 | 4.6453E-11 | 3.625667E-9 |
| 5 | 3.755E-12 | 9.52742E-10 | 6.77021E-10 | 3.26257E-10 | 5.86612E-11 | 8.586366E-8 |
| 6 | 3.191592E-7 | 5.487121E-9 | 9.675485E-9 | 2.748126E-8 | 1.924407E-8 | 0.00000224 |
| 7 | 6.664483E-8 | 0.00000307 | 2.058557E-7 | 8.663522E-7 | 0.00000497 | 0.00007706 |
| 8 | 0.00184 | 6.158358E-7 | 0.00012554 | 0.00000126 | 0.00067357 | 0.00313 |
| 9 | 0.16090 | 0.04435 | 0.00085743 | 0.01598 | 0.06346 | 0.12230 |
| 10 | 0.83726 | 0.95565 | 0.99902 | 0.98402 | 0.93586 | 0.87449 |

NOTE: Singularities or near singularities caused grossly large variance calculations.
      To provide diagnostics, eigenvalues are inflated to a minimum of 1e-12.

Analysis of the Moldover et al. Data to Estimate R                              91
Weighted Ninth-Degree Polynomial with Proc REG and Centered SpeedSq

The REG Procedure
Model: MODEL1
Dependent Variable: SpeedSqResc

### Collinearity Diagnostics (intercept adjusted)

| Number | Eigenvalue | Condition Index | Pressure | P2 | P3 | P4 |
|---|---|---|---|---|---|---|
| 1 | 8.43302 | 1.00000 | 3.473364E-8 | 2.40684E-10 | 6.4978E-12 | 4.70407E-13 |
| 2 | 0.53886 | 3.95599 | 0.00000248 | 6.065046E-9 | 3.90861E-11 | 1.25748E-13 |
| 3 | 0.02712 | 17.63228 | 0.00004002 | 8.977498E-9 | 2.555372E-9 | 1.91376E-10 |
| 4 | 0.00096604 | 93.43143 | 0.00042509 | 0.00000582 | 4.034851E-8 | 5.09017E-10 |
| 5 | 0.00002798 | 549.03227 | 0.00298 | 0.00019848 | 0.00000155 | 2.106803E-7 |
| 6 | 5.316141E-7 | 3982.84504 | 0.01778 | 0.00346 | 0.00028293 | 6.33368E-7 |
| 7 | 6.245931E-9 | 36745 | 0.11624 | 0.04910 | 0.01318 | 0.00160 |
| 8 | 5.42286E-11 | 394346 | 0.51724 | 0.41679 | 0.26645 | 0.12566 |
| 9 | 1E-12 | 2903967 | 0.34529 | 0.53045 | 0.72009 | 0.87274 |

### Collinearity Diagnostics (intercept adjusted)

| Number | P5 | P6 | P7 | P8 | P9 |
|---|---|---|---|---|---|
| 1 | 8.29752E-14 | 3.56506E-14 | 4.02503E-14 | 1.45254E-13 | 2.88341E-12 |
| 2 | 7.94026E-14 | 1.66579E-13 | 3.91104E-13 | 2.21074E-12 | 5.97696E-11 |
| 3 | 1.66887E-11 | 1.06076E-12 | 7.11E-13 | 2.79815E-11 | 1.541792E-9 |
| 4 | 7.37354E-10 | 3.32033E-10 | 1.06855E-10 | 9.83378E-11 | 4.411969E-8 |
| 5 | 7.36886E-10 | 7.849238E-9 | 1.513732E-8 | 7.012055E-9 | 0.00000132 |
| 6 | 0.00000217 | 7.076841E-8 | 6.265033E-7 | 0.00000278 | 0.00004988 |
| 7 | 6.661608E-7 | 0.00008563 | 4.765637E-8 | 0.00043164 | 0.00216 |
| 8 | 0.03481 | 0.00099346 | 0.01028 | 0.04424 | 0.08852 |
| 9 | 0.96519 | 0.99892 | 0.98972 | 0.95532 | 0.90927 |

NOTE: Singularities or near singularities caused grossly large variance calculations.
      To provide diagnostics, eigenvalues are inflated to a minimum of 1e-12.

Analysis of the Moldover et al. Data to Estimate R                        92
Weighted Ninth-Degree Polynomial with Proc ORTHOREG and Centered SpeedSq

The ORTHOREG Procedure

Dependent Variable: SpeedSqResc   Rescaled Speed-Squared = SpeedSq - 94,808.35

Weight: WeightVar  Weight Variable

| Source | DF | Sum of Squares | Mean Square | F Value | Pr > F |
|--------|-----|----------------|-------------|---------|--------|
| Model | 8 | 47749354.122 | 5968669.2652 | 5276253 | <.0001 |
| Error | 61 | 69.005190433 | 1.13123263 | | |
| Corrected Total | 69 | 47749423.127 | | | |

Root MSE     1.0635942037
R-Square     0.9999985548

| Parameter | DF | Parameter Estimate | Standard Error | t Value | Pr > \|t\| |
|-----------|-----|---------------------|----------------|---------|-----------|
| Intercept | 1 | -51.9169237547118 | 0.382073706 | -135.88 | <.0001 |
| P | 1 | 0.00021529340263 | 0.0000209938 | 10.26 | <.0001 |
| P**2 | 1 | 2.306725025984E-10 | 4.480425E-10 | 0.51 | 0.6085 |
| P**3 | 1 | -1.75712181673E-15 | 4.910797E-15 | -0.36 | 0.7217 |
| P**4 | 1 | 1.059776244314E-20 | 3.063572E-20 | 0.35 | 0.7306 |
| P**5 | 1 | -4.04950601749E-26 | 1.127527E-25 | -0.36 | 0.7207 |
| P**6 | 1 | 9.48241018079E-32 | 2.416802E-31 | 0.39 | 0.6962 |
| P**7 | 1 | -1.22056797247E-37 | 2.784166E-37 | -0.44 | 0.6626 |
| P**8 | 1 | 6.530236196212E-44 | 1.330194E-43 | 0.49 | 0.6252 |
| P**9 | 0 | 0 | . | . | . |

NOTE: Model is not full rank. Least-squares solutions for the parameters are not
      unique. Some statistics will be misleading. A reported DF of 0 means that the
      estimate is biased.

Repeat the regression using a stepwise procedure with the following candidate
terms:  Cube Root of Pressure, Square Root of Pressure, Pressure, Pressure^2,
Pressure^3, ..., Pressure^14, and Inverse Pressure.

The REG Procedure
Model: MODEL1
Dependent Variable: SpeedSqResc Rescaled Speed-Squared = SpeedSq - 94,808.35

Number of Observations Read        70
Number of Observations Used        70

Stepwise Selection: Step 1

Statistics for Entry
DF = 1,68

| Variable | Tolerance | Model R-Square | F Value | Pr > F |
|----------|-----------|----------------|---------|--------|
| PCbRt | 1.000000 | 0.9544 | 1422.05 | <.0001 |
| PSqRt | 1.000000 | 0.9731 | 2462.83 | <.0001 |
| P | 1.000000 | 0.9993 | 96845.9 | <.0001 |
| P2 | 1.000000 | 0.9633 | 1785.56 | <.0001 |
| P3 | 1.000000 | 0.8829 | 512.88 | <.0001 |
| P4 | 1.000000 | 0.8001 | 272.25 | <.0001 |
| P5 | 1.000000 | 0.7268 | 180.89 | <.0001 |
| P6 | 1.000000 | 0.6645 | 134.67 | <.0001 |
| P7 | 1.000000 | 0.6121 | 107.31 | <.0001 |
| P8 | 1.000000 | 0.5681 | 89.46 | <.0001 |
| P9 | 1.000000 | 0.5311 | 77.02 | <.0001 |
| P10 | 1.000000 | 0.4998 | 67.94 | <.0001 |
| P11 | 1.000000 | 0.4731 | 61.07 | <.0001 |
| P12 | 1.000000 | 0.4731 | 61.07 | <.0001 |
| P13 | 1.000000 | 0.4309 | 51.50 | <.0001 |
| P14 | 1.000000 | 0.4142 | 48.08 | <.0001 |
| PInv | 1.000000 | 0.6037 | 103.61 | <.0001 |

Variable P Entered: R-Square = 0.9993 and C(p) = 31083.05

Weight: WeightVar Weight Variable

Analysis of the Moldover et al. Data to Estimate R                                  94

Repeat the regression using a stepwise procedure with the following candidate
terms:  Cube Root of Pressure, Square Root of Pressure, Pressure, Pressure^2,
Pressure^3, ..., Pressure^14, and Inverse Pressure.

The REG Procedure
Model: MODEL1
Dependent Variable: SpeedSqResc Rescaled Speed-Squared = SpeedSq - 94,808.35

Stepwise Selection: Step 1

### Analysis of Variance

| Source | DF | Sum of Squares | Mean Square | F Value | Pr > F |
|--------|----|---------------:|------------:|--------:|-------:|
| Model | 1 | 47715920 | 47715920 | 96845.9 | <.0001 |
| Error | 68 | 33504 | 492.69945 | | |
| Corrected Total | 69 | 47749423 | | | |

| Variable | Parameter Estimate | Standard Error | Type II SS | F Value | Pr > F |
|----------|-------------------:|---------------:|-----------:|--------:|-------:|
| Intercept | -55.10564 | 0.25336 | 23306885 | 47304.5 | <.0001 |
| P | 0.00025473 | 8.185479E-7 | 47715920 | 96845.9 | <.0001 |

Bounds on condition number: 1, 1
--------------------------------------------------------------------------------

Stepwise Selection: Step 2

### Statistics for Entry
#### DF = 1,67

| Variable | Tolerance | Model R-Square | F Value | Pr > F |
|----------|----------:|---------------:|--------:|-------:|
| PCbRt | 0.035723 | 0.9999 | 527.94 | <.0001 |
| PSqRt | 0.019244 | 0.9999 | 695.80 | <.0001 |
| P2 | 0.047276 | 1.0000 | 26503.8 | <.0001 |
| P3 | 0.134380 | 1.0000 | 2586.36 | <.0001 |
| P4 | 0.220630 | 1.0000 | 888.88 | <.0001 |
| P5 | 0.295552 | 0.9999 | 486.23 | <.0001 |
| P6 | 0.358397 | 0.9999 | 324.32 | <.0001 |
| P7 | 0.410737 | 0.9998 | 241.01 | <.0001 |

Analysis of the Moldover et al. Data to Estimate R                                    95

Repeat the regression using a stepwise procedure with the following candidate
terms:  Cube Root of Pressure, Square Root of Pressure, Pressure, Pressure^2,
Pressure^3, ..., Pressure^14, and Inverse Pressure.

The REG Procedure
Model: MODEL1
Dependent Variable: SpeedSqResc Rescaled Speed-Squared = SpeedSq - 94,808.35

Stepwise Selection: Step 2

                        Statistics for Entry
                           DF = 1,67

                                    Model
Variable          Tolerance      R-Square      F Value      Pr > F

P8                0.454401        0.9998        191.60      <.0001
P9                0.490994        0.9998        159.47      <.0001
P10               0.521821        0.9998        137.20      <.0001
P11               0.547929        0.9997        121.02      <.0001
P12               0.547929        0.9997        121.02      <.0001
P13               0.589150        0.9997         99.45      <.0001
P14               0.605464        0.9997         92.02      <.0001
PInv              0.376681        0.9997         86.90      <.0001

Variable P2 Entered: R-Square = 1.0000 and C(p) = 14.5443

Weight: WeightVar Weight Variable

                        Analysis of Variance

                              Sum of          Mean
Source            DF        Squares        Square     F Value      Pr > F

Model              2       47749339      23874669     1.893E7      <.0001
Error             67       84.48133       1.26092
Corrected Total   69       47749423

Analysis of the Moldover et al. Data to Estimate R                                      96

Repeat the regression using a stepwise procedure with the following candidate
terms:  Cube Root of Pressure, Square Root of Pressure, Pressure, Pressure^2,
Pressure^3, ..., Pressure^14, and Inverse Pressure.

The REG Procedure
Model: MODEL1
Dependent Variable: SpeedSqResc Rescaled Speed-Squared = SpeedSq - 94,808.35

Stepwise Selection: Step 2

| Variable | Parameter Estimate | Standard Error | Type II SS | F Value | Pr > F |
|---|---|---|---|---|---|
| Intercept | -52.09911 | 0.02248 | 6772732 | 5371282 | <.0001 |
| P | 0.00022447 | 1.904476E-7 | 1751661 | 1389198 | <.0001 |
| P2 | 5.4784E-11 | 3.36511E-13 | 33419 | 26503.8 | <.0001 |

Bounds on condition number: 21.152, 84.609
-------------------------------------------------------------------------------

Stepwise Selection: Step 3

Statistics for Removal
DF = 1,67

| Variable | Partial R-Square | Model R-Square | F Value | Pr > F |
|---|---|---|---|---|
| P | 0.0367 | 0.9633 | 1389198 | <.0001 |
| P2 | 0.0007 | 0.9993 | 26503.8 | <.0001 |

Statistics for Entry
DF = 1,66

| Variable | Tolerance | Model R-Square | F Value | Pr > F |
|---|---|---|---|---|
| PCbRt | 0.004056 | 1.0000 | 0.65 | 0.4217 |
| PSqRt | 0.001703 | 1.0000 | 0.73 | 0.3968 |
| P3 | 0.002702 | 1.0000 | 2.47 | 0.1209 |
| P4 | 0.013794 | 1.0000 | 3.19 | 0.0786 |
| P5 | 0.032967 | 1.0000 | 3.77 | 0.0565 |
| P6 | 0.057394 | 1.0000 | 4.18 | 0.0449 |
| P7 | 0.084357 | 1.0000 | 4.44 | 0.0389 |

Analysis of the Moldover et al. Data to Estimate R                        97

Repeat the regression using a stepwise procedure with the following candidate
terms:  Cube Root of Pressure, Square Root of Pressure, Pressure, Pressure^2,
Pressure^3, ..., Pressure^14, and Inverse Pressure.

The REG Procedure
Model: MODEL1
Dependent Variable: SpeedSqResc Rescaled Speed-Squared = SpeedSq - 94,808.35

Stepwise Selection: Step 3

                        Statistics for Entry
                           DF = 1,66

                                   Model
Variable            Tolerance    R-Square     F Value     Pr > F

P8                  0.111886      1.0000        4.58      0.0361
P9                  0.138724      1.0000        4.62      0.0353
P10                 0.164140      1.0000        4.60      0.0357
P11                 0.187755      1.0000        4.54      0.0369
P12                 0.187755      1.0000        4.54      0.0369
P13                 0.229113      1.0000        4.34      0.0410
P14                 0.246912      1.0000        4.24      0.0435
PInv                0.164739      1.0000        0.31      0.5780

Variable P9 Entered: R-Square = 1.0000 and C(p) = 11.4059

Weight: WeightVar Weight Variable

                        Analysis of Variance

                              Sum of         Mean
Source               DF      Squares        Square     F Value     Pr > F

Model                 3     47749344      15916448      1.33E7     <.0001
Error                66     78.95452       1.19628
Corrected Total      69     47749423

Analysis of the Moldover et al. Data to Estimate R                                        98

Repeat the regression using a stepwise procedure with the following candidate
terms:  Cube Root of Pressure, Square Root of Pressure, Pressure, Pressure^2,
Pressure^3, ..., Pressure^14, and Inverse Pressure.

The REG Procedure
Model: MODEL1
Dependent Variable: SpeedSqResc Rescaled Speed-Squared = SpeedSq - 94,808.35

Stepwise Selection: Step 3

| Variable | Parameter Estimate | Standard Error | Type II SS | F Value | Pr > F |
|----------|-------------------|----------------|-----------|---------|--------|
| Intercept | -52.06546 | 0.02692 | 4475839 | 3741463 | <.0001 |
| P | 0.00022402 | 2.811528E-7 | 759459 | 634850 | <.0001 |
| P2 | 5.59066E-11 | 6.16644E-13 | 9833.10996 | 8219.74 | <.0001 |
| P9 | -5.3455E-53 | 2.48695E-53 | 5.52681 | 4.62 | 0.0353 |

Bounds on condition number: 74.866, 391.99
-------------------------------------------------------------------------------

Stepwise Selection: Step 4

Statistics for Removal
DF = 1,66

| Variable | Partial R-Square | Model R-Square | F Value | Pr > F |
|----------|------------------|----------------|---------|--------|
| P | 0.0159 | 0.9841 | 634850 | <.0001 |
| P2 | 0.0002 | 0.9998 | 8219.74 | <.0001 |
| P9 | 0.0000 | 1.0000 | 4.62 | 0.0353 |

Analysis of the Moldover et al. Data to Estimate R                                99

Repeat the regression using a stepwise procedure with the following candidate
terms:  Cube Root of Pressure, Square Root of Pressure, Pressure, Pressure^2,
Pressure^3, ..., Pressure^14, and Inverse Pressure.

The REG Procedure
Model: MODEL1
Dependent Variable: SpeedSqResc Rescaled Speed-Squared = SpeedSq - 94,808.35

Stepwise Selection: Step 4

Statistics for Entry
DF = 1,65

| Variable | Tolerance | Model R-Square | F Value | Pr > F |
|----------|-----------|----------------|---------|--------|
| PCbRt | 0.002124 | 1.0000 | 0.81 | 0.3722 |
| PSqRt | 0.000844 | 1.0000 | 0.85 | 0.3601 |
| P3 | 0.000458 | 1.0000 | 0.77 | 0.3826 |
| P4 | 0.001370 | 1.0000 | 0.55 | 0.4597 |
| P5 | 0.001771 | 1.0000 | 0.35 | 0.5567 |
| P6 | 0.001475 | 1.0000 | 0.19 | 0.6633 |
| P7 | 0.000824 | 1.0000 | 0.09 | 0.7715 |
| P8 | 0.000235 | 1.0000 | 0.02 | 0.8750 |
| P10 | 0.000260 | 1.0000 | 0.00 | 0.9446 |
| P11 | 0.001041 | 1.0000 | 0.03 | 0.8698 |
| P12 | 0.001041 | 1.0000 | 0.03 | 0.8698 |
| P13 | 0.003945 | 1.0000 | 0.10 | 0.7492 |
| P14 | 0.005891 | 1.0000 | 0.15 | 0.7014 |
| PInv | 0.122863 | 1.0000 | 0.34 | 0.5614 |

All variables left in the model are significant at the 0.0500 level.

No other variable met the 0.1500 significance level for entry into the model.

Weight: WeightVar Weight Variable

Analysis of the Moldover et al. Data to Estimate R                                100

Repeat the regression using a stepwise procedure with the following candidate
terms:  Cube Root of Pressure, Square Root of Pressure, Pressure, Pressure^2,
Pressure^3, ..., Pressure^14, and Inverse Pressure.

The REG Procedure
Model: MODEL1
Dependent Variable: SpeedSqResc Rescaled Speed-Squared = SpeedSq - 94,808.35

### Summary of Stepwise Selection

| Step | Variable Entered | Variable Removed | Label | Number Vars In |
|------|------------------|------------------|-----------|----------------|
| 1 | P | | Pressure | 1 |
| 2 | P2 | | Pressure^2 | 2 |
| 3 | P9 | | Pressure^9 | 3 |

### Summary of Stepwise Selection

| Step | Partial R-Square | Model R-Square | C(p) | F Value | Pr > F |
|------|------------------|----------------|---------|---------|--------|
| 1 | 0.9993 | 0.9993 | 31083.0 | 96845.9 | <.0001 |
| 2 | 0.0007 | 1.0000 | 14.5443 | 26503.8 | <.0001 |
| 3 | 0.0000 | 1.0000 | 11.4059 | 4.62 | 0.0353 |

Analysis of the Moldover et al. Data to Estimate R                                   101

Repeat the regression using a stepwise procedure with the following candidate terms:  Cube Root of Pressure, Square Root of Pressure, Pressure, Pressure^2, Pressure^3, ..., Pressure^14, and Inverse Pressure.

The REG Procedure
Model: MODEL1
Dependent Variable: SpeedSqResc Rescaled Speed-Squared = SpeedSq - 94,808.35

| | |
|---|---|
| Number of Observations Read | 70 |
| Number of Observations Used | 70 |

Weight: WeightVar Weight Variable

### Analysis of Variance

| Source | DF | Sum of Squares | Mean Square | F Value | Pr > F |
|---|---|---|---|---|---|
| Model | 3 | 47749344 | 15916448 | 1.33E7 | <.0001 |
| Error | 66 | 78.95452 | 1.19628 | | |
| Corrected Total | 69 | 47749423 | | | |

| | | | |
|---|---|---|---|
| Root MSE | 1.09375 | R-Square | 1.0000 |
| Dependent Mean | 14.38129 | Adj R-Sq | 1.0000 |
| Coeff Var | 7.60534 | | |

### Parameter Estimates

| Variable | Label | DF | Parameter Estimate | Standard Error | t Value |
|---|---|---|---|---|---|
| Intercept | Intercept | 1 | -52.06546 | 0.02692 | -1934.3 |
| P | Pressure | 1 | 0.00022402 | 2.811528E-7 | 796.77 |
| P2 | Pressure^2 | 1 | 5.59066E-11 | 6.16644E-13 | 90.66 |
| P9 | Pressure^9 | 1 | -5.3455E-53 | 2.48695E-53 | -2.15 |

### Parameter Estimates

| Variable | Label | DF | Pr > |t| | Variance Inflation |
|---|---|---|---|---|
| Intercept | Intercept | 1 | <.0001 | 0 |
| P | Pressure | 1 | <.0001 | 48.58986 |

Repeat the regression using a stepwise procedure with the following candidate terms:  Cube Root of Pressure, Square Root of Pressure, Pressure, Pressure^2, Pressure^3, ..., Pressure^14, and Inverse Pressure.

The REG Procedure
Model: MODEL1
Dependent Variable: SpeedSqResc Rescaled Speed-Squared = SpeedSq - 94,808.35

### Parameter Estimates

| Variable | Label | DF | Pr > \|t\| | Variance Inflation |
|---|---|---|---|---|
| P2 | Pressure^2 | 1 | <.0001 | 74.86563 |
| P9 | Pressure^9 | 1 | 0.0353 | 7.20854 |

### Collinearity Diagnostics

| Number | Eigenvalue | Condition Index | Proportion of Variation Intercept | P | P2 | P9 |
|---|---|---|---|---|---|---|
| 1 | 3.31385 | 1.00000 | 0.00319 | 0.00040317 | 0.00049910 | 0.00664 |
| 2 | 0.57675 | 2.39703 | 0.03404 | 0.00026766 | 0.00016738 | 0.10330 |
| 3 | 0.10697 | 5.56585 | 0.14827 | 0.00646 | 0.01463 | 0.25272 |
| 4 | 0.00243 | 36.91474 | 0.81449 | 0.99287 | 0.98471 | 0.63735 |

### Collinearity Diagnostics (intercept adjusted)

| Number | Eigenvalue | Condition Index | Proportion of Variation P | P2 | P9 |
|---|---|---|---|---|---|
| 1 | 2.68132 | 1.00000 | 0.00260 | 0.00182 | 0.01530 |
| 2 | 0.31081 | 2.93713 | 0.01903 | 0.00204 | 0.29692 |
| 3 | 0.00787 | 18.45873 | 0.97837 | 0.99614 | 0.68778 |

Analysis of the Moldover et al. Data to Estimate R     103

Repeat the regression using a stepwise procedure with the following candidate terms:  Cube Root of Pressure, Square Root of Pressure, Pressure, Pressure^2, Pressure^3, ..., Pressure^14, and Inverse Pressure.

Here is the estimate of R from the stepwise regression.

```
    Estimate of R from
   Quadratic Model Plus
     Ninth-Degree Term
 ─────────────────────────
 8.3144803091335000E+00
```

Analysis of the Moldover et al. Data to Estimate R                                      104
Perform a quadratic regression using R as the response variable instead of Speed-Squared

This run uses the unscaled estimates of R and generates a warning in the SAS log.
Note that pressure here is in kilopascals instead of pascals.

The REG Procedure
Model: MODEL1
Dependent Variable: EstR Value of "PV/(nT)"

Number of Observations Read          70
Number of Observations Used          70

Weight: WeightVar Weight Variable

### Analysis of Variance

| Source | DF | Sum of Squares | Mean Square | F Value | Pr > F |
|--------|-----|--------------|-------------|---------|--------|
| Model | 2 | 0.36764 | 0.18382 | 1.892E7 | <.0001 |
| Error | 67 | 6.509222E-7 | 9.715256E-9 | | |
| Corrected Total | 69 | 0.36764 | | | |

| | | | |
|---|---|---|---|
| Root MSE | 0.00009857 | R-Square | 1.0000 |
| Dependent Mean | 8.32031 | Adj R-Sq | 1.0000 |
| Coeff Var | 0.00118 | | |

### Parameter Estimates

| Variable | Label | DF | Parameter Estimate | Standard Error | t Value | Pr > \|t\| |
|----------|-------|-----|-------------------|----------------|---------|-----------|
| Intercept | Intercept | 1 | 8.31448 | 0.00000197 | 4213660 | <.0001 |
| PressureK | Pressure (kPa) | 1 | 0.00001970 | 1.671705E-8 | 1178.22 | <.0001 |
| PressKSq | Pressure Squared (kPa**2) | 1 | 4.80707E-9 | 2.95381E-11 | 162.74 | <.0001 |

Perform a quadratic regression using R as the response variable instead of Speed-Squared

This run uses the unscaled estimates of R and generates a warning in the SAS log.
Here is a more accurate table of the values of the parameters:

| Variable | Estimate | StdErr | tValue | Probt | Label |
|----------|----------|--------|--------|-------|-------|
| Intercept | 8.314477E+00 | 1.97E-06 | 4,213,660.0 | <.0001 | Intercept |
| PressureK | 1.969630E-05 | 1.67E-08 | 1,178.2 | <.0001 | Pressure (kPa) |
| PressKSq | 4.807070E-09 | 2.95E-11 | 162.7 | <.0001 | Pressure Squared (kPa**2) |

Analysis of the Moldover et al. Data to Estimate R    106
Perform a quadratic regression using R as the response variable instead of Speed-Squared

This run uses the rescaled estimates of R to avoid the warning in the log.

The rescaled value is the value of "PV/(nT)" minus 8.31447.

The REG Procedure
Model: MODEL1
Dependent Variable: RescaledEstR Value of "PV/(nT)" - 8.31447

```
Number of Observations Read          70
Number of Observations Used          70
```

Weight: WeightVar Weight Variable

### Analysis of Variance

| Source | DF | Sum of Squares | Mean Square | F Value | Pr > F |
|---|---|---|---|---|---|
| Model | 2 | 0.36764 | 0.18382 | 1.893E7 | <.0001 |
| Error | 67 | 6.504515E-7 | 9.708231E-9 | | |
| Corrected Total | 69 | 0.36764 | | | |

| | | | |
|---|---|---|---|
| Root MSE | 0.00009853 | R-Square | 1.0000 |
| Dependent Mean | 0.00584 | Adj R-Sq | 1.0000 |
| Coeff Var | 1.68695 | | |

### Parameter Estimates

| Variable | Label | DF | Parameter Estimate | Standard Error | t Value |
|---|---|---|---|---|---|
| Intercept | Intercept | 1 | 0.00000736 | 0.00000197 | 3.73 |
| PressureK | Pressure (kPa) | 1 | 0.00001970 | 1.671101E-8 | 1178.64 |
| PressKSq | Pressure Squared (kPa**2) | 1 | 4.80707E-9 | 2.95275E-11 | 162.80 |

### Parameter Estimates

| Variable | Label | DF | Pr > \|t\| |
|---|---|---|---|
| Intercept | Intercept | 1 | 0.0004 |
| PressureK | Pressure (kPa) | 1 | <.0001 |
| PressKSq | Pressure Squared (kPa**2) | 1 | <.0001 |

Analysis of the Moldover et al. Data to Estimate R                    107
Perform a quadratic regression using R as the response variable instead of Speed-Squared

This run uses the rescaled estimates of R to avoid the warning in the log.

The rescaled value is the value of "PV/(nT)" minus 8.31447.

Here is a more accurate table of the values of the parameters.

| Variable | Estimate | StdErr | tValue | Probt |
|---|---|---|---|---|
| Intercept | 7.356446028947900E-06 | 1.97E-06 | 3.73 | 0.0004 |
| PressureK | 1.969629845431000E-05 | 1.67E-08 | 1178.64 | <.0001 |
| PressKSq | 4.807070123094100E-09 | 2.95E-11 | 162.80 | <.0001 |

Because the value 8.31447 was subtracted from each estimated value of R
(to deal with the warning produced in the preceding analysis), this value must
be added back to the intercept to get the final estimate of R, as follows:
 7.356 446 028 947 900E-06 + 8.31447 = 8.314477, as obtained above.

With more digits, this is 8.314 477 356 446 028 947 900, which can
be rounded to                8.314 477 356 446 03 and which can be compared
with the value on the next page.

Analysis of the Moldover et al. Data to Estimate R          108

Here (from proc SQL) is the value of the correctly adjusted intercept from the proc REG run fitting rescaled speed-squared versus pressure with a quadratic equation. The value has been multiplied by the conversion factor and all the available digits are shown for comparison with the value in the second line from the bottom of the preceding page.  The values agree to 14 significant digits, and differ by 1 in the 15th significant digit.

   Computed Value of R
   ___________________

 8.314477356446040000

The different ways of computing the value help to ensure that things are working properly and illustrate the very small difference in the estimated values due to differing computer roundoff errors between the two approaches.

Analysis of the Moldover et al. Data to Estimate R     109

Weighted Scaled Orthogonal Regression With Pressure in kPa and Centered SpeedSq
using the quadratic model equation.

PROC NLP: Nonlinear Minimization

Gradient is computed using analytic formulas.
Hessian is computed using analytic formulas.

Analysis of the Moldover et al. Data to Estimate R     110

Weighted Scaled Orthogonal Regression With Pressure in kPa and Centered SpeedSq
using the quadratic model equation.

PROC NLP: Nonlinear Minimization

Optimization Start
Parameter Estimates

| N | Parameter | Estimate | Gradient Objective Function |
|---|---|---|---|
| 1 | B0 | 1.000000 | 6603073314 |
| 2 | B1 | 1.000000 | 2.6147332E12 |
| 3 | B2 | 1.000000 | 1.1042922E15 |
| 4 | PErrs1 | 0 | 48712 |
| 5 | PErrs2 | 0 | 406151 |
| 6 | PErrs3 | 0 | 1490210 |
| 7 | PErrs4 | 0 | 3670105 |
| 8 | PErrs5 | 0 | 7057167 |
| 9 | PErrs6 | 0 | 4815749 |
| 10 | PErrs7 | 0 | 33622520 |
| 11 | PErrs8 | 0 | 92049361 |
| 12 | PErrs9 | 0 | 158833787 |
| 13 | PErrs10 | 0 | 214632939 |
| 14 | PErrs11 | 0 | 4831374 |
| 15 | PErrs12 | 0 | 33722866 |
| 16 | PErrs13 | 0 | 92287517 |
| 17 | PErrs14 | 0 | 159184206 |
| 18 | PErrs15 | 0 | 215080668 |
| 19 | PErrs16 | 0 | 70969843 |
| 20 | PErrs17 | 0 | 356109867 |
| 21 | PErrs18 | 0 | 669565239 |
| 22 | PErrs19 | 0 | 874776909 |
| 23 | PErrs20 | 0 | 993987866 |
| 24 | PErrs21 | 0 | 425910499 |
| 25 | PErrs22 | 0 | 1455048662 |
| 26 | PErrs23 | 0 | 2112924258 |
| 27 | PErrs24 | 0 | 2432353855 |
| 28 | PErrs25 | 0 | 2596456086 |
| 29 | PErrs26 | 0 | 454737053 |
| 30 | PErrs27 | 0 | 1528327920 |
| 31 | PErrs28 | 0 | 2200764222 |
| 32 | PErrs29 | 0 | 2524469771 |
| 33 | PErrs30 | 0 | 2690344088 |
| 34 | PErrs31 | 0 | 3796394550 |
| 35 | PErrs32 | 0 | 7273044503 |
| 36 | PErrs33 | 0 | 8468102567 |

Analysis of the Moldover et al. Data to Estimate R 111

Weighted Scaled Orthogonal Regression With Pressure in kPa and Centered SpeedSq using the quadratic model equation.

PROC NLP: Nonlinear Minimization

Optimization Start
Parameter Estimates

| N | Parameter | Estimate | Gradient Objective Function |
|---|---|---|---|
| 37 | PErrs34 | 0 | 8949671025 |
| 38 | PErrs35 | 0 | 9185747676 |
| 39 | PErrs36 | 0 | 13842043906 |
| 40 | PErrs37 | 0 | 19713603707 |
| 41 | PErrs38 | 0 | 21307441996 |
| 42 | PErrs39 | 0 | 21930346972 |
| 43 | PErrs40 | 0 | 22235795959 |
| 44 | PErrs41 | 0 | 33096876314 |
| 45 | PErrs42 | 0 | 40879540333 |
| 46 | PErrs43 | 0 | 42830975494 |
| 47 | PErrs44 | 0 | 43593064171 |
| 48 | PErrs45 | 0 | 43968461926 |
| 49 | PErrs46 | 0 | 62982390609 |
| 50 | PErrs47 | 0 | 72349679808 |
| 51 | PErrs48 | 0 | 74643836133 |
| 52 | PErrs49 | 0 | 75543452720 |
| 53 | PErrs50 | 0 | 75988385863 |
| 54 | PErrs51 | 0 | 106074496333 |
| 55 | PErrs52 | 0 | 116897760566 |
| 56 | PErrs53 | 0 | 119536218578 |
| 57 | PErrs54 | 0 | 120575438961 |
| 58 | PErrs55 | 0 | 121090955367 |
| 59 | PErrs56 | 0 | 165662694005 |
| 60 | PErrs57 | 0 | 177907804218 |
| 61 | PErrs58 | 0 | 180898894289 |
| 62 | PErrs59 | 0 | 182081415999 |
| 63 | PErrs60 | 0 | 182669374277 |
| 64 | PErrs61 | 0 | 237560922607 |
| 65 | PErrs62 | 0 | 251109208796 |
| 66 | PErrs63 | 0 | 254431236312 |
| 67 | PErrs64 | 0 | 255747996044 |
| 68 | PErrs65 | 0 | 256403824205 |
| 69 | PErrs66 | 0 | 341507875055 |
| 70 | PErrs67 | 0 | 356535791521 |
| 71 | PErrs68 | 0 | 360237828341 |
| 72 | PErrs69 | 0 | 361709238918 |

Analysis of the Moldover et al. Data to Estimate R                    112

Weighted Scaled Orthogonal Regression With Pressure in kPa and Centered SpeedSq
using the quadratic model equation.

PROC NLP: Nonlinear Minimization

Optimization Start
Parameter Estimates

| N Parameter | Estimate | Gradient Objective Function |
|---|---|---|
| 73 PErrs70 | 0 | 362443168962 |

Value of Objective Function = 5.5330506E14

Analysis of the Moldover et al. Data to Estimate R                                113

Weighted Scaled Orthogonal Regression With Pressure in kPa and Centered SpeedSq
using the quadratic model equation.

PROC NLP: Nonlinear Minimization

Newton-Raphson Optimization with Line Search

Without Parameter Scaling

Parameter Estimates                 73
Functions (Observations)            70

Optimization Start

Active Constraints              0  Objective Function           5.5330506E14
Max Abs Gradient Element    1.1042922E15

| Iter | Restarts | Function Calls | Active Constraints | Objective Function | Objective Function Change | Max Abs Gradient Element | Step Size | Slope of Search Direction |
|---|---|---|---|---|---|---|---|---|
| 1* | 0 | 3 | 0 | 2.33427E14 | 3.199E14 | 7.887E14 | 10.000 | -161E13 |
| 2* | 0 | 4 | 0 | 8.0493E12 | 2.254E14 | 1.559E14 | 1.000 | -505E12 |
| 3* | 0 | 5 | 0 | 2.62443E11 | 7.787E12 | 6.905E11 | 1.000 | -156E11 |
| 4* | 0 | 6 | 0 | 1.93545E11 | 6.89E10 | 1.416E10 | 1.000 | -743E8 |
| 5* | 0 | 7 | 0 | 1.04129E10 | 1.831E11 | 4.69E10 | 1.000 | -297E9 |
| 6* | 0 | 8 | 0 | 212172956 | 1.02E10 | 1.7286E9 | 1.000 | -18E9 |
| 7* | 0 | 9 | 0 | 33432648 | 1.7874E8 | 30499167 | 1.000 | -3.34E8 |
| 8* | 0 | 10 | 0 | 32530617 | 902031 | 1445919 | 1.000 | -1.73E6 |
| 9* | 0 | 11 | 0 | 32512437 | 18180.0 | 311317 | 1.000 | -19338 |
| 10* | 0 | 14 | 0 | 12618530 | 19893906 | 1.278E10 | 100.0 | -336705 |
| 11* | 0 | 15 | 0 | 11165699 | 1452831 | 2.8756E8 | 1.000 | -1.61E6 |
| 12* | 0 | 16 | 0 | 2473631 | 8692068 | 2.8039E9 | 1.000 | -1.19E7 |
| 13* | 0 | 17 | 0 | 85365 | 2388266 | 1.4917E9 | 1.000 | -4.04E6 |
| 14* | 0 | 18 | 0 | 963.96683 | 84401.2 | 98766360 | 1.000 | -153368 |
| 15* | 0 | 19 | 0 | 82.31979 | 881.6 | 1004077 | 1.000 | -1674.2 |
| 16* | 0 | 20 | 0 | 79.66203 | 2.6578 | 1985.9 | 1.000 | -5.115 |
| 17* | 0 | 21 | 0 | 79.63389 | 0.0281 | 1.4715 | 1.000 | -0.0397 |
| 18* | 0 | 22 | 0 | 79.61539 | 0.0185 | 0.2231 | 1.000 | -0.0239 |
| 19* | 0 | 23 | 0 | 79.60073 | 0.0147 | 0.1130 | 1.000 | -0.0202 |
| 20* | 0 | 24 | 0 | 79.59694 | 0.00380 | 0.0533 | 1.000 | -0.0064 |
| 21* | 0 | 25 | 0 | 79.59681 | 0.000125 | 0.00204 | 1.000 | -0.0002 |
| 22* | 0 | 26 | 0 | 79.59681 | 1.214E-6 | 0.000095 | 1.000 | -231E-8 |
| 23* | 0 | 27 | 0 | 79.59681 | 3.258E-9 | 0.000030 | 1.000 | -63E-10 |
| 24* | 0 | 28 | 0 | 79.59681 | 6.11E-13 | 0.000016 | 1.000 | -45E-13 |

Weighted Scaled Orthogonal Regression With Pressure in kPa and Centered SpeedSq
using the quadratic model equation.

PROC NLP: Nonlinear Minimization

| Iter | Restarts | Function Calls | Active Constraints | Objective Function | Objective Function Change | Max Abs Gradient Element | Step Size | Slope of Search Direction |
|------|----------|----------------|--------------------|--------------------|---------------------------|--------------------------|-----------|---------------------------|
| 25*  | 0        | 29             | 0                  | 79.59681           | 2.81E-12                  | 3.342E-6                 | 1.000     | -82E-17                   |

Optimization Results

| | | | |
|---|---|---|---|
| Iterations | 25 | Function Calls | 30 |
| Hessian Calls | 26 | Active Constraints | 0 |
| Objective Function | 79.596810187 | Max Abs Gradient Element | 3.3415854E-6 |
| Slope of Search Direction | -8.21426E-16 | Ridge | 0.0188262286 |

ABSGCONV convergence criterion satisfied.

Analysis of the Moldover et al. Data to Estimate R                     115

Weighted Scaled Orthogonal Regression With Pressure in kPa and Centered SpeedSq
using the quadratic model equation.

PROC NLP: Nonlinear Minimization

Optimization Results
Parameter Estimates

| N | Parameter | Estimate | Gradient Objective Function |
|---|-----------|----------|-----------------------------|
| 1 | B0 | -52.098942 | -4.16378E-11 |
| 2 | B1 | 0.224468 | -1.092849E-8 |
| 3 | B2 | 0.000054787 | -0.000003342 |
| 4 | PErrs1 | 0.240324 | -4.5619E-10 |
| 5 | PErrs2 | 0.101572 | -5.00716E-15 |
| 6 | PErrs3 | -0.027871 | -4.17192E-15 |
| 7 | PErrs4 | 0.058567 | 4.112148E-15 |
| 8 | PErrs5 | -0.016851 | 1.712386E-14 |
| 9 | PErrs6 | 0.048196 | -5.48421E-15 |
| 10 | PErrs7 | -0.008589 | -1.40233E-14 |
| 11 | PErrs8 | -0.026498 | 0 |
| 12 | PErrs9 | -0.006405 | -1.7876E-14 |
| 13 | PErrs10 | 0.029413 | -5.68412E-14 |
| 14 | PErrs11 | 0.040875 | -5.23865E-15 |
| 15 | PErrs12 | -0.010668 | -1.53829E-14 |
| 16 | PErrs13 | -0.012416 | -1.09819E-14 |
| 17 | PErrs14 | 0.022311 | -3.57824E-14 |
| 18 | PErrs15 | -0.020278 | 2.274987E-14 |
| 19 | PErrs16 | 0.007028 | -2.25343E-14 |
| 20 | PErrs17 | -0.016592 | -5.24586E-14 |
| 21 | PErrs18 | -0.000035596 | -1.20597E-13 |
| 22 | PErrs19 | -0.008866 | 6.182E-13 |
| 23 | PErrs20 | -0.007320 | -1.35543E-13 |
| 24 | PErrs21 | 0.005986 | 6.138951E-14 |
| 25 | PErrs22 | 0.008886 | 2.361733E-13 |
| 26 | PErrs23 | 0.011786 | -4.69778E-13 |
| 27 | PErrs24 | -0.017215 | -4.9761E-13 |
| 28 | PErrs25 | -0.002937 | -5.44904E-13 |
| 29 | PErrs26 | 0.053435 | -9.97056E-14 |
| 30 | PErrs27 | 0.004779 | 3.025849E-13 |
| 31 | PErrs28 | 0.017724 | -4.14113E-13 |
| 32 | PErrs29 | 0.007680 | -4.35781E-13 |
| 33 | PErrs30 | 0.003663 | -4.28097E-13 |
| 34 | PErrs31 | -0.008616 | -5.2725E-13 |
| 35 | PErrs32 | -0.008844 | 7.687102E-13 |
| 36 | PErrs33 | -0.006339 | 8.874533E-13 |

Weighted Scaled Orthogonal Regression With Pressure in kPa and Centered SpeedSq
using the quadratic model equation.

PROC NLP: Nonlinear Minimization

Optimization Results
Parameter Estimates

| N | Parameter | Estimate | Gradient Objective Function |
|---|-----------|----------|------------------------------|
| 37 | PErrs34 | -0.003607 | -1.13842E-12 |
| 38 | PErrs35 | 0.004817 | 8.782406E-13 |
| 39 | PErrs36 | 0.003716 | 2.237562E-13 |
| 40 | PErrs37 | 0.002554 | 3.165453E-13 |
| 41 | PErrs38 | -0.000465 | -1.79508E-12 |
| 42 | PErrs39 | 0.002322 | 3.568667E-13 |
| 43 | PErrs40 | 0.012078 | 3.067233E-13 |
| 44 | PErrs41 | -0.005020 | 6.519195E-13 |
| 45 | PErrs42 | -0.020652 | -1.48448E-12 |
| 46 | PErrs43 | -0.005730 | -1.60601E-12 |
| 47 | PErrs44 | -0.009757 | -1.62619E-12 |
| 48 | PErrs45 | 0.002323 | 8.297735E-13 |
| 49 | PErrs46 | -0.016183 | -3.38502E-13 |
| 50 | PErrs47 | -0.012563 | -9.48972E-14 |
| 51 | PErrs48 | -0.008460 | -5.27739E-12 |
| 52 | PErrs49 | -0.004598 | -4.16888E-13 |
| 53 | PErrs50 | 0.002401 | -1.10609E-13 |
| 54 | PErrs51 | 0.006522 | -3.08584E-13 |
| 55 | PErrs52 | 0.000130 | -3.575E-13 |
| 56 | PErrs53 | 0.002834 | -3.53762E-13 |
| 57 | PErrs54 | 0.006276 | -3.59265E-13 |
| 58 | PErrs55 | 0.008735 | -3.48612E-13 |
| 59 | PErrs56 | 0.024438 | 2.989891E-12 |
| 60 | PErrs57 | 0.006159 | 3.112185E-12 |
| 61 | PErrs58 | 0.006659 | -2.54056E-12 |
| 62 | PErrs59 | 0.011167 | -2.54377E-12 |
| 63 | PErrs60 | 0.008913 | 3.208912E-12 |
| 64 | PErrs61 | 0.019094 | 4.516308E-12 |
| 65 | PErrs62 | -0.005350 | -4.9678E-13 |
| 66 | PErrs63 | -0.011971 | -9.41207E-13 |
| 67 | PErrs64 | -0.000003327 | -8.49644E-13 |
| 68 | PErrs65 | -0.004077 | -5.03587E-13 |
| 69 | PErrs66 | 0.012145 | -2.02848E-12 |
| 70 | PErrs67 | -0.000826 | -1.84385E-12 |
| 71 | PErrs68 | -0.007312 | -1.91724E-12 |
| 72 | PErrs69 | -0.007571 | -1.93362E-12 |

Weighted Scaled Orthogonal Regression With Pressure in kPa and Centered SpeedSq
using the quadratic model equation.

PROC NLP: Nonlinear Minimization

Optimization Results
Parameter Estimates

| N Parameter | Estimate | Gradient Objective Function |
|---|---|---|
| 73 PErrs70 | -0.020542 | 4.149082E-12 |

Value of Objective Function = 79.596810187

Analysis of the Moldover et al. Data to Estimate R          118

Weighted Scaled Orthogonal Regression With Pressure in kPa and Centered SpeedSq
using the quadratic model equation.

Here (from proc SQL), and using the computed value of B0 from NLP
is the estimated value of R with all the available digits:

Computed Value of R
______________________

8.314477370879840000

Analysis of the Moldover et al. Data to Estimate R                               119

Weighted Scaled Orthogonal Regression With Pressure in kPa and Centered SpeedSq
using the quadratic model equation.

Distribution of the Weighted Squared Speed-Squared and Pressure Errors

The UNIVARIATE Procedure
Variable:  SpeedSqErrSq  (Squared Estimated Error in Speed-Squared Values (m^4/s^4))

Weight:  WeightVar  (Weight Variable)

### Weighted Moments

| | | | |
|---|---|---|---|
| N | 70 | Sum Weights | 34366.8707 |
| Mean | 0.00218234 | Sum Observations | 75.0000911 |
| Std Deviation | 0.16113699 | Variance | 0.02596513 |
| Skewness | 3.91962724 | Kurtosis | 16.368106 |
| Uncorrected SS | 1.95526941 | Corrected SS | 1.79159395 |
| Coeff Variation | 7383.68984 | Std Error Mean | 0.00086921 |

### Weighted Basic Statistical Measures

| Location | | Variability | |
|---|---|---|---|
| Mean | 0.002182 | Std Deviation | 0.16114 |
| Median | 0.000693 | Variance | 0.02597 |
| Mode | . | Range | 1.11810 |
| | | Interquartile Range | 0.00169 |

### Weighted Tests for Location: Mu0=0

| Test | | -Statistic- | -----p Value------ | |
|---|---|---|---|---|
| Student's t | t | 2.51071 | Pr > \|t\| | 0.0144 |

### Weighted Quantiles

| Quantile | Estimate |
|---|---|
| 100% Max | 1.11810E+00 |
| 99% | 1.63604E-02 |
| 95% | 8.28723E-03 |
| 90% | 5.39334E-03 |
| 75% Q3 | 1.91290E-03 |
| 50% Median | 6.92606E-04 |

Analysis of the Moldover et al. Data to Estimate R                              120

Weighted Scaled Orthogonal Regression With Pressure in kPa and Centered SpeedSq
using the quadratic model equation.

Distribution of the Weighted Squared Speed-Squared and Pressure Errors

The UNIVARIATE Procedure
Variable:  SpeedSqErrSq  (Squared Estimated Error in Speed-Squared Values (m^4/s^4))

Weight:  WeightVar  (Weight Variable)

      Weighted Quantiles

Quantile          Estimate

25% Q1            2.27455E-04
10%               8.50578E-05
5%                2.45308E-07
1%                1.47754E-10
0% Min            1.47754E-10

            Extreme Observations

--------Lowest-------     ------Highest------

      Value      Obs          Value      Obs

   1.47754E-10     64       0.0439274      6
   2.33991E-08     18       0.0514490     26
   2.45308E-07     52       0.0664150      4
   3.56576E-06     38       0.1997534      2
   8.72608E-06     67       1.1181028      1

Analysis of the Moldover et al. Data to Estimate R 121

Weighted Scaled Orthogonal Regression With Pressure in kPa and Centered SpeedSq
using the quadratic model equation.

Distribution of the Weighted Squared Speed-Squared and Pressure Errors

The UNIVARIATE Procedure
Variable: SpeedSqErrSq (Squared Estimated Error in Speed-Squared Values (m^4/s^4))

Weight: WeightVar (Weight Variable)

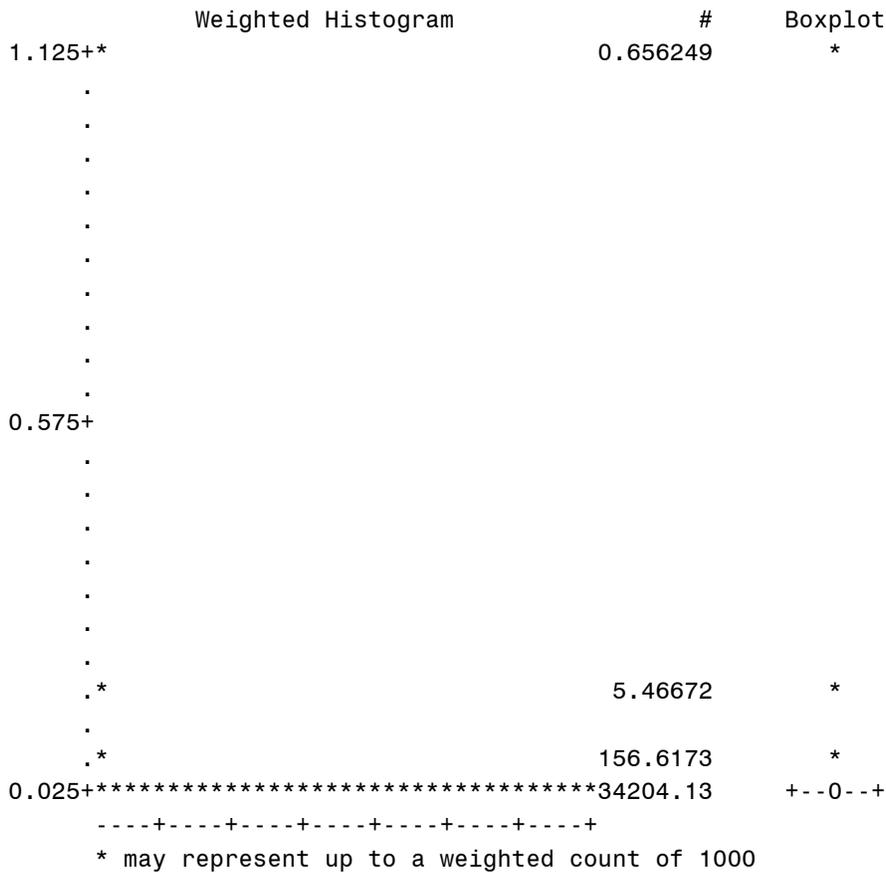

```
            Weighted Histogram            #   Boxplot
    1.125+*                          0.656249      *
         .
         .
         .
         .
         .
         .
         .
         .
         .
    0.575+
         .
         .
         .
         .
         .
         .
         .*                          5.46672       *
         .
         .*                          156.6173      *
    0.025+*******************************34204.13   +--0--+
         ----+----+----+----+----+----+----+
          * may represent up to a weighted count of 1000
```

Weighted Scaled Orthogonal Regression With Pressure in kPa and Centered SpeedSq
using the quadratic model equation.

Distribution of the Weighted Squared Speed-Squared and Pressure Errors

The UNIVARIATE Procedure
Variable:  SpeedSqErrSq  (Squared Estimated Error in Speed-Squared Values (m^4/s^4))

Weight:  WeightVar  (Weight Variable)

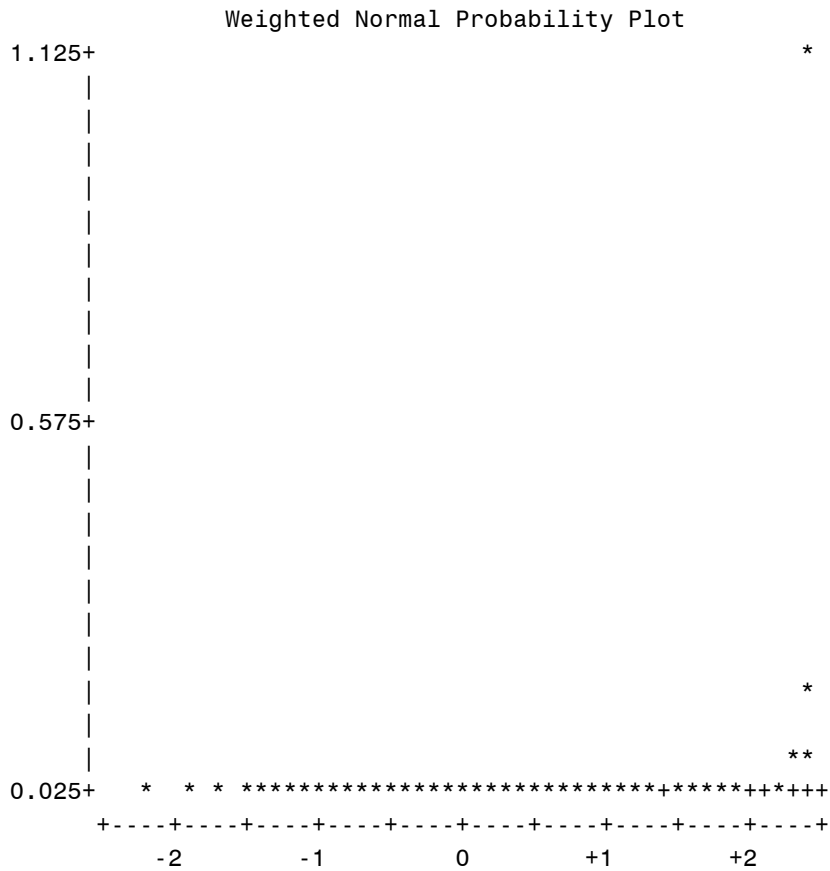

Weighted Scaled Orthogonal Regression With Pressure in kPa and Centered SpeedSq
using the quadratic model equation.

Distribution of the Weighted Squared Speed-Squared and Pressure Errors

The UNIVARIATE Procedure
Variable:  PErrSq  (Squared Estimated Error in Pressure Values (kPa^2))

Weight:  WeightVar  (Weight Variable)

### Weighted Moments

| | | | |
|---|---|---|---|
| N | 70 | Sum Weights | 34366.8707 |
| Mean | 0.00013375 | Sum Observations | 4.59671907 |
| Std Deviation | 0.00859117 | Variance | 0.00007381 |
| Skewness | 3.685488 | Kurtosis | 14.4910626 |
| Uncorrected SS | 0.0057076 | Corrected SS | 0.00509276 |
| Coeff Variation | 6423.09375 | Std Error Mean | 0.00004634 |

### Weighted Basic Statistical Measures

| Location | | Variability | |
|---|---|---|---|
| Mean | 0.000134 | Std Deviation | 0.00859 |
| Median | 0.000053 | Variance | 0.0000738 |
| Mode | . | Range | 0.05776 |
| | | Interquartile Range | 0.0001295 |

### Weighted Tests for Location: Mu0=0

| Test | -Statistic- | | -----p Value------ | |
|---|---|---|---|---|
| Student's t | t | 2.886195 | Pr > \|t\| | 0.0052 |

### Weighted Quantiles

| Quantile | Estimate |
|---|---|
| 100% Max | 5.77554E-02 |
| 99% | 8.65103E-04 |
| 95% | 4.97762E-04 |
| 90% | 4.11195E-04 |
| 75% Q3 | 1.43296E-04 |
| 50% Median | 5.34603E-05 |

Analysis of the Moldover et al. Data to Estimate R                              124

Weighted Scaled Orthogonal Regression With Pressure in kPa and Centered SpeedSq
using the quadratic model equation.

Distribution of the Weighted Squared Speed-Squared and Pressure Errors

The UNIVARIATE Procedure
Variable:  PErrSq  (Squared Estimated Error in Pressure Values (kPa^2))

Weight:  WeightVar  (Weight Variable)

     Weighted Quantiles

Quantile          Estimate

25% Q1            1.38063E-05
10%               5.39189E-06
5%                1.69457E-08
1%                1.10684E-11
0% Min            1.10684E-11

          Extreme Observations

--------Lowest-------      -------Highest------

     Value       Obs          Value      Obs

   1.10684E-11    64       0.00232283      6
   1.26708E-09    18       0.00285535     26
   1.69457E-08    52       0.00343005      4
   2.16437E-07    38       0.01031684      2
   6.82721E-07    67       0.05775539      1

Weighted Scaled Orthogonal Regression With Pressure in kPa and Centered SpeedSq
using the quadratic model equation.

Distribution of the Weighted Squared Speed-Squared and Pressure Errors

The UNIVARIATE Procedure
Variable:  PErrSq  (Squared Estimated Error in Pressure Values (kPa^2))

Weight:  WeightVar  (Weight Variable)

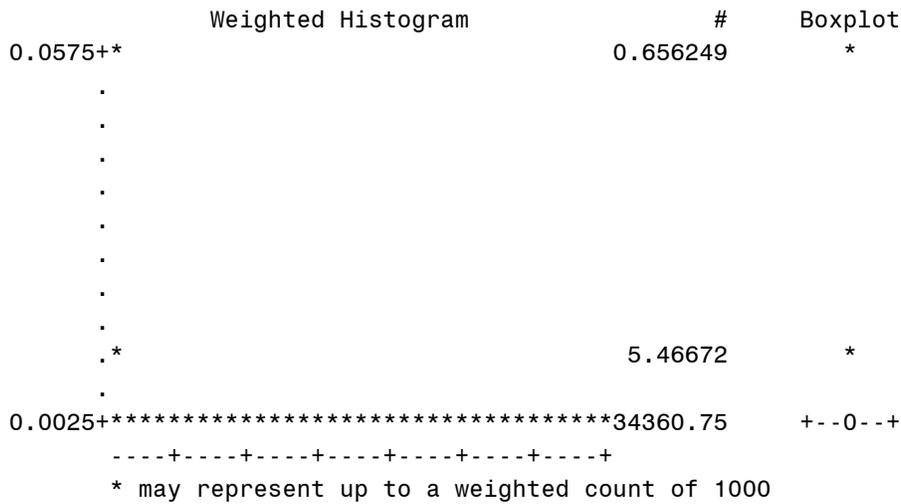

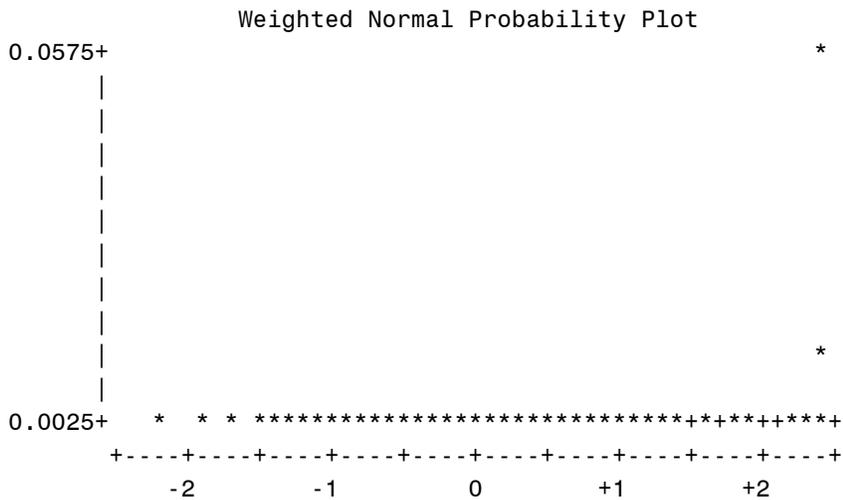

Analysis of the Moldover et al. Data to Estimate R                126

Weighted Scaled Orthogonal Regression With Pressure in kPa
using the quadratic model equation.

Repeat the Preceding Analysis Reversing the Roles of Pressure and Speed-Squared.

PROC NLP: Nonlinear Minimization

Gradient is computed using analytic formulas.
Hessian is computed using analytic formulas.

Weighted Scaled Orthogonal Regression With Pressure in kPa
using the quadratic model equation.

Repeat the Preceding Analysis Reversing the Roles of Pressure and Speed-Squared.

PROC NLP: Nonlinear Minimization

Optimization Start
Parameter Estimates

| N | Parameter | Estimate | Gradient Objective Function |
|---|-----------|----------|------------------------------|
| 1 | B0 | 94760 | -6.75623E13 |
| 2 | B1 | 0.000225 | -2.262061E19 |
| 3 | B2 | 5.47E-11 | -8.629696E24 |
| 4 | CErrs1 | 0 | 79398940 |
| 5 | CErrs2 | 0 | 525219656 |
| 6 | CErrs3 | 0 | 1457768361 |
| 7 | CErrs4 | 0 | 4362329686 |
| 8 | CErrs5 | 0 | 7092951439 |
| 9 | CErrs6 | 0 | 2770712496 |
| 10 | CErrs7 | 0 | 18715163199 |
| 11 | CErrs8 | 0 | 50693135643 |
| 12 | CErrs9 | 0 | 88525418108 |
| 13 | CErrs10 | 0 | 122159536898 |
| 14 | CErrs11 | 0 | 2766065007 |
| 15 | CErrs12 | 0 | 18734857190 |
| 16 | CErrs13 | 0 | 51217538314 |
| 17 | CErrs14 | 0 | 90163943960 |
| 18 | CErrs15 | 0 | 118807894300 |
| 19 | CErrs16 | 0 | 20729050898 |
| 20 | CErrs17 | 0 | 103217774530 |
| 21 | CErrs18 | 0 | 195120976779 |
| 22 | CErrs19 | 0 | 254191931015 |
| 23 | CErrs20 | 0 | 288974459642 |
| 24 | CErrs21 | 0 | 74533472551 |
| 25 | CErrs22 | 0 | 254792323158 |
| 26 | CErrs23 | 0 | 370227236070 |
| 27 | CErrs24 | 0 | 423492965252 |
| 28 | CErrs25 | 0 | 453486238262 |
| 29 | CErrs26 | 0 | 78790237466 |
| 30 | CErrs27 | 0 | 262055023608 |
| 31 | CErrs28 | 0 | 378409046294 |
| 32 | CErrs29 | 0 | 433129848227 |
| 33 | CErrs30 | 0 | 461189176482 |
| 34 | CErrs31 | 0 | 311436130981 |

Analysis of the Moldover et al. Data to Estimate R                       128

Weighted Scaled Orthogonal Regression With Pressure in kPa
using the quadratic model equation.

Repeat the Preceding Analysis Reversing the Roles of Pressure and Speed-Squared.

PROC NLP: Nonlinear Minimization

Optimization Start
Parameter Estimates

| N | Parameter | Estimate | Gradient Objective Function |
|---|-----------|----------|------------------------------|
| 35 | CErrs32 | 0 | 596624576207 |
| 36 | CErrs33 | 0 | 694882447126 |
| 37 | CErrs34 | 0 | 734658212559 |
| 38 | CErrs35 | 0 | 754856205439 |
| 39 | CErrs36 | 0 | 645741191669 |
| 40 | CErrs37 | 0 | 919560007573 |
| 41 | CErrs38 | 0 | 993643011062 |
| 42 | CErrs39 | 0 | 1.0229413E12 |
| 43 | CErrs40 | 0 | 1.0380759E12 |
| 44 | CErrs41 | 0 | 981988239468 |
| 45 | CErrs42 | 0 | 1.2116769E12 |
| 46 | CErrs43 | 0 | 1.2707418E12 |
| 47 | CErrs44 | 0 | 1.2930159E12 |
| 48 | CErrs45 | 0 | 1.3051676E12 |
| 49 | CErrs46 | 0 | 1.2888146E12 |
| 50 | CErrs47 | 0 | 1.4807666E12 |
| 51 | CErrs48 | 0 | 1.5280344E12 |
| 52 | CErrs49 | 0 | 1.5467492E12 |
| 53 | CErrs50 | 0 | 1.5564038E12 |
| 54 | CErrs51 | 0 | 1.5755081E12 |
| 55 | CErrs52 | 0 | 1.735821E12 |
| 56 | CErrs53 | 0 | 1.7751914E12 |
| 57 | CErrs54 | 0 | 1.7908708E12 |
| 58 | CErrs55 | 0 | 1.7987042E12 |
| 59 | CErrs56 | 0 | 1.8462477E12 |
| 60 | CErrs57 | 0 | 1.9815349E12 |
| 61 | CErrs58 | 0 | 2.0148825E12 |
| 62 | CErrs59 | 0 | 2.0283514E12 |
| 63 | CErrs60 | 0 | 2.0347518E12 |
| 64 | CErrs61 | 0 | 2.0816022E12 |
| 65 | CErrs62 | 0 | 2.1988443E12 |
| 66 | CErrs63 | 0 | 2.2275292E12 |
| 67 | CErrs64 | 0 | 2.2397922E12 |
| 68 | CErrs65 | 0 | 2.2452851E12 |

Analysis of the Moldover et al. Data to Estimate R                    129

Weighted Scaled Orthogonal Regression With Pressure in kPa
using the quadratic model equation.

Repeat the Preceding Analysis Reversing the Roles of Pressure and Speed-Squared.

PROC NLP: Nonlinear Minimization

                    Optimization Start
                    Parameter Estimates
                                        Gradient
                                        Objective
    N Parameter            Estimate      Function

   69 CErrs66                    0    2.334789E12
   70 CErrs67                    0    2.4368081E12
   71 CErrs68                    0    2.4617454E12
   72 CErrs69                    0    2.4717858E12
   73 CErrs70                    0    2.4760667E12

Value of Objective Function = 3.0137143E15

Weighted Scaled Orthogonal Regression With Pressure in kPa
using the quadratic model equation.

Repeat the Preceding Analysis Reversing the Roles of Pressure and Speed-Squared.

PROC NLP: Nonlinear Minimization

Newton-Raphson Optimization with Line Search

Without Parameter Scaling

Parameter Estimates                73
Functions (Observations)           70

Optimization Start

Active Constraints                 0  Objective Function            3.0137143E15
Max Abs Gradient Element    8.6296962E24

| Iter | Restarts | Function Calls | Active Constraints | Objective Function | Objective Function Change | Max Abs Gradient Element | Step Size | Slope of Search Direction |
|------|----------|----------------|--------------------|--------------------|---------------------------|--------------------------|-----------|---------------------------|
| 1*   | 0        | 4              | 0                  | 4.6894E14          | 2.545E15                  | 1.284E23                 | 20.015    | -119E13                   |
| 2*   | 0        | 5              | 0                  | 1.31975E14         | 3.37E14                   | 9.168E21                 | 5.807     | -221E12                   |
| 3*   | 0        | 6              | 0                  | 7.82868E13         | 5.369E13                  | 3.155E21                 | 1.426     | -643E11                   |
| 4*   | 0        | 7              | 0                  | 5.26139E13         | 2.567E13                  | 1.409E21                 | 1.000     | -385E11                   |
| 5*   | 0        | 8              | 0                  | 3.52806E13         | 1.733E13                  | 6.285E20                 | 1.000     | -26E12                    |
| 6*   | 0        | 9              | 0                  | 2.36131E13         | 1.167E13                  | 2.804E20                 | 1.000     | -175E11                   |
| 7*   | 0        | 10             | 0                  | 1.57798E13         | 7.833E12                  | 1.251E20                 | 1.000     | -117E11                   |
| 8*   | 0        | 11             | 0                  | 1.0533E13          | 5.247E12                  | 5.589E19                 | 1.000     | -7852E9                   |
| 9*   | 0        | 12             | 0                  | 7.02634E12         | 3.507E12                  | 2.502E19                 | 1.000     | -5242E9                   |
| 10*  | 0        | 13             | 0                  | 4.68783E12         | 2.339E12                  | 1.124E19                 | 1.000     | -3488E9                   |
| 11*  | 0        | 14             | 0                  | 3.1319E12          | 1.556E12                  | 5.085E18                 | 1.000     | -2313E9                   |
| 12*  | 0        | 15             | 0                  | 2.09919E12         | 1.033E12                  | 2.325E18                 | 1.000     | -1527E9                   |
| 13*  | 0        | 16             | 0                  | 1.41538E12         | 6.838E11                  | 1.08E18                  | 1.000     | -1003E9                   |
| 14*  | 0        | 17             | 0                  | 9.63344E11         | 4.52E11                   | 5.141E17                 | 1.000     | -655E9                    |
| 15*  | 0        | 18             | 0                  | 6.64349E11         | 2.99E11                   | 2.523E17                 | 1.000     | -427E9                    |
| 16*  | 0        | 19             | 0                  | 4.50235E11         | 2.141E11                  | 1.206E17                 | 1.000     | -309E9                    |
| 17*  | 0        | 20             | 0                  | 3.09068E11         | 1.412E11                  | 5.956E16                 | 1.000     | -201E9                    |
| 18*  | 0        | 21             | 0                  | 2.07787E11         | 1.013E11                  | 2.856E16                 | 1.000     | -146E9                    |
| 19*  | 0        | 22             | 0                  | 1.41339E11         | 6.645E10                  | 1.415E16                 | 1.000     | -943E8                    |
| 20*  | 0        | 23             | 0                  | 9.37496E10         | 4.759E10                  | 6.791E15                 | 1.000     | -686E8                    |
| 21*  | 0        | 24             | 0                  | 6.27707E10         | 3.098E10                  | 3.368E15                 | 1.000     | -439E8                    |
| 22*  | 0        | 25             | 0                  | 4.07356E10         | 2.204E10                  | 1.613E15                 | 1.000     | -318E8                    |

Weighted Scaled Orthogonal Regression With Pressure in kPa
using the quadratic model equation.

Repeat the Preceding Analysis Reversing the Roles of Pressure and Speed-Squared.

PROC NLP: Nonlinear Minimization

| Iter | Restarts | Function Calls | Active Constraints | Objective Function | Objective Function Change | Max Abs Gradient Element | Step Size | Slope of Search Direction |
|------|----------|----------------|---------------------|--------------------|---------------------------|--------------------------|-----------|---------------------------|
| 23* | 0 | 26 | 0 | 2.65728E10 | 1.416E10 | 7.978E14 | 1.000 | -201E8 |
| 24* | 0 | 27 | 0 | 1.66465E10 | 9.9263E9 | 3.8E14 | 1.000 | -143E8 |
| 25* | 0 | 28 | 0 | 1.04021E10 | 6.2444E9 | 1.865E14 | 1.000 | -8.88E9 |
| 26* | 0 | 29 | 0 | 6148049533 | 4.254E9 | 8.775E13 | 1.000 | -6.17E9 |
| 27* | 0 | 30 | 0 | 3573892702 | 2.5742E9 | 4.232E13 | 1.000 | -3.68E9 |
| 28* | 0 | 31 | 0 | 2038135220 | 1.5358E9 | 2.092E13 | 1.000 | -2.17E9 |
| 29* | 0 | 32 | 0 | 1048740995 | 9.8939E8 | 9.628E12 | 1.000 | -1.44E9 |
| 30* | 0 | 33 | 0 | 521621263 | 5.2712E8 | 4.45E12 | 1.000 | -7.62E8 |
| 31* | 0 | 34 | 0 | 260273771 | 2.6135E8 | 2.03E12 | 1.000 | -3.77E8 |
| 32* | 0 | 35 | 0 | 144892540 | 1.1538E8 | 8.843E11 | 1.000 | -1.68E8 |
| 33* | 0 | 36 | 0 | 103303097 | 41589443 | 3.447E11 | 1.000 | -6.22E7 |
| 34* | 0 | 37 | 0 | 92914397 | 10388701 | 1.064E11 | 1.000 | -1.63E7 |
| 35* | 0 | 38 | 0 | 91560513 | 1353883 | 2.114E10 | 1.000 | -2.3E6 |
| 36* | 0 | 39 | 0 | 91497675 | 62838.5 | 2.1381E9 | 1.000 | -113150 |
| 37* | 0 | 40 | 0 | 91493341 | 4333.7 | 7.9531E8 | 1.000 | -4920.8 |
| 38* | 0 | 44 | 0 | 78446371 | 13046970 | 9.963E10 | 1000.0 | -14756 |
| 39* | 0 | 45 | 0 | 76756419 | 1689952 | 1.191E10 | 1.000 | -3.34E6 |
| 40* | 0 | 47 | 0 | 72731955 | 4024464 | 9.942E10 | 10.000 | -577303 |
| 41* | 0 | 48 | 0 | 61467704 | 11264250 | 2.53E10 | 1.000 | -1.32E7 |
| 42* | 0 | 49 | 0 | 33488333 | 27979372 | 2.318E10 | 1.000 | -3.12E7 |
| 43* | 0 | 50 | 0 | 5291471 | 28196862 | 1.389E11 | 1.000 | -4.17E7 |
| 44* | 0 | 51 | 0 | 922151 | 4369320 | 2.1025E9 | 1.000 | -7.39E6 |
| 45* | 0 | 52 | 0 | 717857 | 204293 | 1.4881E9 | 1.000 | -365689 |
| 46* | 0 | 53 | 0 | 714702 | 3155.7 | 13176712 | 1.000 | -5526.4 |
| 47* | 0 | 55 | 0 | 706100 | 8601.3 | 1.0714E8 | 10.000 | -962.8 |
| 48* | 0 | 57 | 0 | 599946 | 106155 | 1.3705E9 | 10.000 | -19887 |
| 49* | 0 | 58 | 0 | 346644 | 253302 | 1.189E9 | 1.000 | -342133 |
| 50* | 0 | 59 | 0 | 103799 | 242845 | 3.3141E8 | 1.000 | -322286 |
| 51* | 0 | 60 | 0 | 9425 | 94373.4 | 2.8142E8 | 1.000 | -151671 |
| 52* | 0 | 61 | 0 | 1511 | 7914.3 | 24079231 | 1.000 | -13331 |
| 53* | 0 | 62 | 0 | 731.41022 | 779.8 | 1147977 | 1.000 | -1099.0 |
| 54* | 0 | 63 | 0 | 333.46458 | 397.9 | 37625.4 | 1.000 | -557.3 |
| 55* | 0 | 64 | 0 | 149.07657 | 184.4 | 23212.4 | 1.000 | -264.9 |
| 56* | 0 | 65 | 0 | 101.03613 | 48.0404 | 23408.2 | 1.000 | -68.530 |
| 57* | 0 | 66 | 0 | 85.50240 | 15.5337 | 8466.5 | 1.000 | -23.305 |

Weighted Scaled Orthogonal Regression With Pressure in kPa
using the quadratic model equation.

Repeat the Preceding Analysis Reversing the Roles of Pressure and Speed-Squared.

PROC NLP: Nonlinear Minimization

| Iter | Restarts | Function Calls | Active Constraints | Objective Function | Objective Function Change | Max Abs Gradient Element | Step Size | Slope of Search Direction |
|------|----------|----------------|--------------------|--------------------|---------------------------|--------------------------|-----------|---------------------------|
| 58* | 0 | 67 | 0 | 81.23562 | 4.2668 | 923.8 | 1.000 | -5.723 |
| 59* | 0 | 68 | 0 | 79.68341 | 1.5522 | 459.8 | 1.000 | -2.525 |
| 60* | 0 | 69 | 0 | 79.59828 | 0.0851 | 28.6048 | 1.000 | -0.151 |
| 61* | 0 | 70 | 0 | 79.59682 | 0.00146 | 0.8550 | 1.000 | -0.0027 |
| 62* | 0 | 71 | 0 | 79.59681 | 7.122E-6 | 0.1198 | 1.000 | -137E-7 |
| 63* | 0 | 72 | 0 | 79.59681 | 3.197E-9 | 0.1141 | 1.000 | -18E-9 |
| 64* | 0 | 73 | 0 | 79.59681 | 6.062E-9 | 0.1908 | 1.000 | -62E-13 |
| 65* | 0 | 75 | 0 | 79.59681 | 1.801E-9 | 0.6087 | 0.119 | -22E-16 |
| 66* | 0 | 110 | 0 | 79.59681 | 1.52E-10 | 0.7629 | 0.0287 | -17E-16 |
| 67 | 0 | 121 | 0 | 79.59681 | 0 | 0.7629 | 2.16E-8 | -17E-16 |

Optimization Results

| | | | |
|----|----|----|----|
| Iterations | 67 | Function Calls | 122 |
| Hessian Calls | 68 | Active Constraints | 0 |
| Objective Function | 79.596810184 | Max Abs Gradient Element | 0.7628505677 |
| Slope of Search Direction | -1.65861E-15 | Ridge | 0 |

FCONV convergence criterion satisfied.

NOTE: At least one element of the (projected) gradient is greater than 1e-3.

Weighted Scaled Orthogonal Regression With Pressure in kPa
using the quadratic model equation.

Repeat the Preceding Analysis Reversing the Roles of Pressure and Speed-Squared.

PROC NLP: Nonlinear Minimization

Optimization Results
Parameter Estimates

| N | Parameter | Estimate | Gradient Objective Function |
|---|-----------|----------|------------------------------|
| 1 | B0 | 94756 | 0.000008157 |
| 2 | B1 | 0.224468 | 0.002210 |
| 3 | B2 | 0.000054787 | 0.762851 |
| 4 | CErrs1 | -1.057404 | 0.000000203 |
| 5 | CErrs2 | -0.446938 | 1.0179297E-9 |
| 6 | CErrs3 | 0.122646 | -6.227895E-9 |
| 7 | CErrs4 | -0.257711 | -1.274252E-8 |
| 8 | CErrs5 | 0.074150 | -3.414107E-9 |
| 9 | CErrs6 | -0.209589 | -9.58551E-10 |
| 10 | CErrs7 | 0.037354 | 7.1702807E-9 |
| 11 | CErrs8 | 0.115236 | 2.0496378E-8 |
| 12 | CErrs9 | 0.027856 | 3.9732211E-8 |
| 13 | CErrs10 | -0.127908 | -0.000000190 |
| 14 | CErrs11 | -0.177753 | -1.08134E-9 |
| 15 | CErrs12 | 0.046394 | -2.480492E-8 |
| 16 | CErrs13 | 0.053993 | 2.6905699E-8 |
| 17 | CErrs14 | -0.097022 | -4.328149E-8 |
| 18 | CErrs15 | 0.088185 | 1.2522532E-8 |
| 19 | CErrs16 | -0.030203 | 2.0789646E-9 |
| 20 | CErrs17 | 0.071300 | -0.000000101 |
| 21 | CErrs18 | 0.000153 | -2.955124E-8 |
| 22 | CErrs19 | 0.038098 | 2.3320532E-8 |
| 23 | CErrs20 | 0.031458 | 5.7602403E-8 |
| 24 | CErrs21 | -0.025424 | -4.586683E-8 |
| 25 | CErrs22 | -0.037741 | -0.000000220 |
| 26 | CErrs23 | -0.050058 | -0.000000162 |
| 27 | CErrs24 | 0.073113 | -0.000000116 |
| 28 | CErrs25 | 0.012475 | -0.000000179 |
| 29 | CErrs26 | -0.226824 | -6.438802E-8 |
| 30 | CErrs27 | -0.020286 | -0.000000133 |
| 31 | CErrs28 | -0.075237 | -0.000000344 |
| 32 | CErrs29 | -0.032603 | -0.000000233 |
| 33 | CErrs30 | -0.015549 | -0.000000345 |
| 34 | CErrs31 | 0.035769 | -1.034172E-8 |

Analysis of the Moldover et al. Data to Estimate R 134

Weighted Scaled Orthogonal Regression With Pressure in kPa
using the quadratic model equation.

Repeat the Preceding Analysis Reversing the Roles of Pressure and Speed-Squared.

PROC NLP: Nonlinear Minimization

Optimization Results
Parameter Estimates

| N | Parameter | Estimate | Gradient Objective Function |
|---|---|---|---|
| 35 | CErrs32 | 0.036714 | -0.000000329 |
| 36 | CErrs33 | 0.026317 | -0.000000111 |
| 37 | CErrs34 | 0.014976 | 0.000000123 |
| 38 | CErrs35 | -0.019995 | -0.000000261 |
| 39 | CErrs36 | -0.015082 | -0.000000210 |
| 40 | CErrs37 | -0.010368 | -0.000000126 |
| 41 | CErrs38 | 0.001888 | -0.000000283 |
| 42 | CErrs39 | -0.009425 | -0.000000198 |
| 43 | CErrs40 | -0.049021 | -1.589126E-8 |
| 44 | CErrs41 | 0.019928 | -5.101827E-8 |
| 45 | CErrs42 | 0.081990 | -0.000000187 |
| 46 | CErrs43 | 0.022749 | -0.000000218 |
| 47 | CErrs44 | 0.038735 | -0.000000131 |
| 48 | CErrs45 | -0.009223 | -0.000000161 |
| 49 | CErrs46 | 0.062887 | -0.000000154 |
| 50 | CErrs47 | 0.048819 | 1.6473532E-8 |
| 51 | CErrs48 | 0.032874 | -0.000000153 |
| 52 | CErrs49 | 0.017868 | 5.6794905E-8 |
| 53 | CErrs50 | -0.009331 | -0.000000259 |
| 54 | CErrs51 | -0.024815 | -0.000000322 |
| 55 | CErrs52 | -0.000495 | -0.000000228 |
| 56 | CErrs53 | -0.010785 | -0.000000162 |
| 57 | CErrs54 | -0.023880 | -0.000000304 |
| 58 | CErrs55 | -0.033234 | -0.000000291 |
| 59 | CErrs56 | -0.091034 | -0.000000174 |
| 60 | CErrs57 | -0.022941 | -2.021358E-9 |
| 61 | CErrs58 | -0.024807 | -0.000000244 |
| 62 | CErrs59 | -0.041597 | -0.000000160 |
| 63 | CErrs60 | -0.033202 | -5.908617E-8 |
| 64 | CErrs61 | -0.069761 | -0.000000105 |
| 65 | CErrs62 | 0.019549 | -0.000000134 |
| 66 | CErrs63 | 0.043737 | -0.000000208 |
| 67 | CErrs64 | 0.000012155 | -4.553695E-8 |
| 68 | CErrs65 | 0.014897 | -9.899059E-8 |

Analysis of the Moldover et al. Data to Estimate R                135

Weighted Scaled Orthogonal Regression With Pressure in kPa
using the quadratic model equation.

Repeat the Preceding Analysis Reversing the Roles of Pressure and Speed-Squared.

PROC NLP: Nonlinear Minimization

Optimization Results
Parameter Estimates

| N | Parameter | Estimate | Gradient Objective Function |
|---|-----------|----------|-----------------------------|
| 69 | CErrs66 | -0.043418 | -0.000000273 |
| 70 | CErrs67 | 0.002954 | -0.000000321 |
| 71 | CErrs68 | 0.026140 | -0.000000129 |
| 72 | CErrs69 | 0.027067 | -0.000000260 |
| 73 | CErrs70 | 0.073439 | 0.000000104 |

Value of Objective Function = 79.596810184

Weighted Scaled Orthogonal Regression With Pressure in kPa
using the quadratic model equation.

Repeat the Preceding Analysis Reversing the Roles of Pressure and Speed-Squared.

Here is a listing of the parameter estimates after the analysis
with sufficient significant digits.

| Obs | Number | Parameter | Estimate | GradObj |
|---|---|---|---|---|
| 74 | 1 | B0 | 9.4756251E+04 | 0.000008157 |
| 75 | 2 | B1 | 2.2446788E-01 | 0.002210 |
| 76 | 3 | B2 | 5.4787137E-05 | 0.762851 |
| 77 | 4 | CErrs1 | -1.0574038E+00 | 0.000000203 |
| 78 | 5 | CErrs2 | -4.4693778E-01 | 1.0179297E-9 |
| 79 | 6 | CErrs3 | 1.2264574E-01 | -6.227895E-9 |
| 80 | 7 | CErrs4 | -2.5771111E-01 | -1.274252E-8 |
| 81 | 8 | CErrs5 | 7.4150111E-02 | -3.414107E-9 |
| 82 | 9 | CErrs6 | -2.0958868E-01 | -9.58551E-10 |
| 83 | 10 | CErrs7 | 3.7353565E-02 | 7.1702807E-9 |
| 84 | 11 | CErrs8 | 1.1523556E-01 | 2.0496378E-8 |
| 85 | 12 | CErrs9 | 2.7855767E-02 | 3.9732211E-8 |
| 86 | 13 | CErrs10 | -1.2790789E-01 | -0.000000190 |
| 87 | 14 | CErrs11 | -1.7775312E-01 | -1.08134E-9 |
| 88 | 15 | CErrs12 | 4.6394270E-02 | -2.480492E-8 |
| 89 | 16 | CErrs13 | 5.3992502E-02 | 2.6905699E-8 |
| 90 | 17 | CErrs14 | -9.7022164E-02 | -4.328149E-8 |
| 91 | 18 | CErrs15 | 8.8184554E-02 | 1.2522532E-8 |
| 92 | 19 | CErrs16 | -3.0203218E-02 | 2.0789646E-9 |
| 93 | 20 | CErrs17 | 7.1300340E-02 | -0.000000101 |
| 94 | 21 | CErrs18 | 1.5296769E-04 | -2.955124E-8 |
| 95 | 22 | CErrs19 | 3.8098222E-02 | 2.3320532E-8 |
| 96 | 23 | CErrs20 | 3.1457801E-02 | 5.7602403E-8 |
| 97 | 24 | CErrs21 | -2.5423809E-02 | -4.586683E-8 |
| 98 | 25 | CErrs22 | -3.7740957E-02 | -0.000000220 |
| 99 | 26 | CErrs23 | -5.0058103E-02 | -0.000000162 |
| 100 | 27 | CErrs24 | 7.3113471E-02 | -0.000000116 |
| 101 | 28 | CErrs25 | 1.2475125E-02 | -0.000000179 |
| 102 | 29 | CErrs26 | -2.2682370E-01 | -6.438802E-8 |
| 103 | 30 | CErrs27 | -2.0286137E-02 | -0.000000133 |
| 104 | 31 | CErrs28 | -7.5236569E-02 | -0.000000344 |
| 105 | 32 | CErrs29 | -3.2602618E-02 | -0.000000233 |
| 106 | 33 | CErrs30 | -1.5549028E-02 | -0.000000345 |
| 107 | 34 | CErrs31 | 3.5768995E-02 | -1.034172E-8 |
| 108 | 35 | CErrs32 | 3.6714154E-02 | -0.000000329 |
| 109 | 36 | CErrs33 | 2.6317406E-02 | -0.000000111 |
| 110 | 37 | CErrs34 | 1.4975502E-02 | 0.000000123 |

Weighted Scaled Orthogonal Regression With Pressure in kPa
using the quadratic model equation.

Repeat the Preceding Analysis Reversing the Roles of Pressure and Speed-Squared.

Here is a listing of the parameter estimates after the analysis
with sufficient significant digits.

| Obs | Number | Parameter | Estimate | GradObj |
|-----|--------|-----------|----------|---------|
| 111 | 38 | CErrs35 | -1.9995355E-02 | -0.000000261 |
| 112 | 39 | CErrs36 | -1.5081621E-02 | -0.000000210 |
| 113 | 40 | CErrs37 | -1.0367749E-02 | -0.000000126 |
| 114 | 41 | CErrs38 | 1.8883225E-03 | -0.000000283 |
| 115 | 42 | CErrs39 | -9.4249739E-03 | -0.000000198 |
| 116 | 43 | CErrs40 | -4.9021493E-02 | -1.589126E-8 |
| 117 | 44 | CErrs41 | 1.9927786E-02 | -5.101827E-8 |
| 118 | 45 | CErrs42 | 8.1990122E-02 | -0.000000187 |
| 119 | 46 | CErrs43 | 2.2748800E-02 | -0.000000218 |
| 120 | 47 | CErrs44 | 3.8734546E-02 | -0.000000131 |
| 121 | 48 | CErrs45 | -9.2226789E-03 | -0.000000161 |
| 122 | 49 | CErrs46 | 6.2886967E-02 | -0.000000154 |
| 123 | 50 | CErrs47 | 4.8818565E-02 | 1.6473532E-8 |
| 124 | 51 | CErrs48 | 3.2874381E-02 | -0.000000153 |
| 125 | 52 | CErrs49 | 1.7868095E-02 | 5.6794905E-8 |
| 126 | 53 | CErrs50 | -9.3307866E-03 | -0.000000259 |
| 127 | 54 | CErrs51 | -2.4815272E-02 | -0.000000322 |
| 128 | 55 | CErrs52 | -4.9528607E-04 | -0.000000228 |
| 129 | 56 | CErrs53 | -1.0784512E-02 | -0.000000162 |
| 130 | 57 | CErrs54 | -2.3879888E-02 | -0.000000304 |
| 131 | 58 | CErrs55 | -3.3233725E-02 | -0.000000291 |
| 132 | 59 | CErrs56 | -9.1034230E-02 | -0.000000174 |
| 133 | 60 | CErrs57 | -2.2941344E-02 | -2.021358E-9 |
| 134 | 61 | CErrs58 | -2.4806904E-02 | -0.000000244 |
| 135 | 62 | CErrs59 | -4.1596939E-02 | -0.000000160 |
| 136 | 63 | CErrs60 | -3.3201922E-02 | -5.908617E-8 |
| 137 | 64 | CErrs61 | -6.9760996E-02 | -0.000000105 |
| 138 | 65 | CErrs62 | 1.9548658E-02 | -0.000000134 |
| 139 | 66 | CErrs63 | 4.3736721E-02 | -0.000000208 |
| 140 | 67 | CErrs64 | 1.2155419E-05 | -4.553695E-8 |
| 141 | 68 | CErrs65 | 1.4897109E-02 | -9.899059E-8 |
| 142 | 69 | CErrs66 | -4.3417884E-02 | -0.000000273 |
| 143 | 70 | CErrs67 | 2.9539938E-03 | -0.000000321 |
| 144 | 71 | CErrs68 | 2.6139952E-02 | -0.000000129 |
| 145 | 72 | CErrs69 | 2.7067390E-02 | -0.000000260 |
| 146 | 73 | CErrs70 | 7.3439346E-02 | 0.000000104 |

Analysis of the Moldover et al. Data to Estimate R                    138

Weighted Scaled Orthogonal Regression With Pressure in kPa
using the quadratic model equation.

Repeat the Preceding Analysis Reversing the Roles of Pressure and Speed-Squared.

Here (from proc SQL), and using the computed value of BO from NLP
is the estimated value of R with all the available digits:

    Computed Value of R
_______________________________

    8.314477370879850000

Analysis of the Moldover et al. Data to Estimate R                              139

Analysis with P as Response Variable Instead of c^2 With Pressure in kPa
using the same quadratic model equation, but as SOLVED for P instead of as
solved for c^2.

The NLIN Procedure
Dependent Variable PressureK
Method: Gauss-Newton

Iterative Phase

| Iter | B1 | B2 | B0 | Weighted SS |
|------|------|------|------|------|
| 0 | 1.0000 | 1.0000 | 1.0000 | 7.76E8 |
| 1 | 364.1 | -0.1811 | 1.0000 | 7.7508E8 |
| 2 | 669.9 | -1.1757 | 1.0000 | 7.3219E8 |
| 3 | -379.9 | 0.7921 | 139978 | 7.2869E8 |
| 4 | -18.4800 | 0.0652 | 95184.1 | 7.1548E8 |
| 5 | -33.0515 | 0.0746 | 98356.0 | 6.7882E8 |
| 6 | -5.3813 | 0.0184 | 94962.6 | 6.5453E8 |
| 7 | 2.8648 | -0.00136 | 94161.3 | 5.7361E8 |
| 8 | 0.9361 | 0.00112 | 94496.9 | 5.1304E8 |
| 9 | 0.5248 | 0.000560 | 94642.8 | 3.5258E8 |
| 10 | 0.4106 | 0.000038 | 94708.7 | 1.2342E8 |
| 11 | 0.2595 | 0.000097 | 94744.2 | 25211004 |
| 12 | 0.2239 | 0.000039 | 94757.5 | 1114107 |
| 13 | 0.2249 | 0.000053 | 94756.2 | 3861.9 |
| 14 | 0.2244 | 0.000055 | 94756.3 | 1403.4 |
| 15 | 0.2244 | 0.000055 | 94756.3 | 1403.3 |

NOTE: Convergence criterion met.

Estimation Summary

| Method | Gauss-Newton |
|--------|--------------|
| Iterations | 15 |
| Subiterations | 24 |
| Average Subiterations | 1.6 |
| R | 1.806E-6 |
| PPC(B2) | 9.276E-8 |
| RPC(B2) | 0.00013 |
| Object | 0.000022 |
| Objective | 1403.346 |
| Observations Read | 70 |
| Observations Used | 70 |
| Observations Missing | 0 |

Analysis of the Moldover et al. Data to Estimate R 140

Analysis with P as Response Variable Instead of c^2 With Pressure in kPa
using the same quadratic model equation, but as SOLVED for P instead of as
solved for c^2.

The NLIN Procedure

NOTE: An intercept was not specified for this model.

| Source | DF | Sum of Squares | Mean Square | F Value | Approx Pr > F |
|--------|-----|-----|-----|-----|-----|
| Model | 3 | 3.2926E9 | 1.0975E9 | 5.24E7 | <.0001 |
| Error | 67 | 1403.3 | 20.9455 | | |
| Uncorrected Total | 70 | 3.2926E9 | | | |

| Parameter | Estimate | Approx Std Error | Approximate 95% Confidence Limits | | Skewness |
|--------|-----|-----|-----|-----|-----|
| B1 | 0.2244 | 0.000194 | 0.2241 | 0.2248 | 0.000998 |
| B2 | 0.000055 | 3.505E-7 | 0.000054 | 0.000056 | 0.000092 |
| B0 | 94756.3 | 0.0220 | 94756.2 | 94756.3 | -0.00251 |

Approximate Correlation Matrix

| | B1 | B2 | B0 |
|--------|-----|-----|-----|
| B1 | 1.0000000 | -0.9751416 | -0.9095811 |
| B2 | -0.9751416 | 1.0000000 | 0.8197672 |
| B0 | -0.9095811 | 0.8197672 | 1.0000000 |

Analysis of the Moldover et al. Data to Estimate R                        141

Analysis with P as Response Variable Instead of c^2 With Pressure in kPa
using the same quadratic model equation, but as SOLVED for P instead of as
solved for c^2.

Here is a listing of the parameter estimates after the analysis
with sufficient significant digits.

| Obs | Parameter | Estimate | StdErr | Alpha | LowerCL | UpperCL | Skewness | tValue | Probt |
|-----|-----------|----------|--------|-------|---------|---------|----------|--------|-------|
| 1 | B1 | 2.2444672E-01 | 0.000194 | 0.05 | 0.2241 | 0.2248 | 0.000998 | 1157.67 | <.0001 |
| 2 | B2 | 5.4826470E-05 | 3.505E-7 | 0.05 | 0.000054 | 0.000056 | 0.000092 | 156.44 | <.0001 |
| 3 | B0 | 9.4756253E+04 | 0.0220 | 0.05 | 94756.2 | 94756.3 | -0.00251 | 4311829 | <.0001 |

Analysis of the Moldover et al. Data to Estimate R                                 142

Analysis with P as Response Variable Instead of c^2 With Pressure in kPa
using the same quadratic model equation, but as SOLVED for P instead of as
solved for c^2.

Here (from proc SQL) is the estimated value of R with all the available digits:

 Computed Value of R with
Roles of P & c^2 Reversed
────────────────────────────
      8.314477546870290000

Analysis of the Moldover et al. Data to Estimate R                                143
Perform a Repeated Measurements Quadratic Analysis with Proc Mixed
The SINGRES option of the model statement is needed to yield non-zero
estimates of the standard errors of the Pressure and Pressure-squared parameters.
This is Analysis 4 in supporting online material in the short article.

The Mixed Procedure

Model Information

Data Set                      WORK.MERGED
Dependent Variable            SpeedSq
Weight Variable               WeightVar
Covariance Structure          Compound Symmetry
Subject Effect                Pressure
Estimation Method             REML
Residual Variance Method      Profile
Fixed Effects SE Method       Model-Based
Degrees of Freedom Method     Between-Within

Dimensions

Covariance Parameters         2
Columns in X                  3
Columns in Z                  0
Subjects                      14
Max Obs Per Subject           5

Number of Observations

Number of Observations Read              70
Number of Observations Used              70
Number of Observations Not Used           0

Iteration History

| Iteration | Evaluations | -2 Res Log Like | Criterion |
|-----------|-------------|-----------------|-----------|
| 0 | 1 | -96.87722971 | |
| 1 | 2 | -101.81022097 | 0.00000000 |

Convergence criteria met.

Analysis of the Moldover et al. Data to Estimate R     144
Perform a Repeated Measurements Quadratic Analysis with Proc Mixed
The SINGRES option of the model statement is needed to yield non-zero
estimates of the standard errors of the Pressure and Pressure-squared parameters.
This is Analysis 4 in supporting online material in the short article.

The Mixed Procedure

   Estimated R Matrix for Subject 1/Weighted by WeightVar

| Row | Col1 | Col2 | Col3 | Col4 | Col5 |
|-----|------|------|------|------|------|
| 1 | 2.0005 | 0.1639 | 0.08558 | 0.05452 | 0.03933 |
| 2 | 0.1639 | 0.2401 | 0.02965 | 0.01889 | 0.01363 |
| 3 | 0.08558 | 0.02965 | 0.06551 | 0.009866 | 0.007116 |
| 4 | 0.05452 | 0.01889 | 0.009866 | 0.02658 | 0.004533 |
| 5 | 0.03933 | 0.01363 | 0.007116 | 0.004533 | 0.01383 |

   Estimated R Matrix for Subject 2/Weighted by WeightVar

| Row | Col1 | Col2 | Col3 | Col4 | Col5 |
|-----|------|------|------|------|------|
| 1 | 0.1429 | 0.01278 | 0.007725 | 0.005881 | 0.005059 |
| 2 | 0.01278 | 0.02046 | 0.002924 | 0.002226 | 0.001915 |
| 3 | 0.007725 | 0.002924 | 0.007474 | 0.001345 | 0.001157 |
| 4 | 0.005881 | 0.002226 | 0.001345 | 0.004332 | 0.000881 |
| 5 | 0.005059 | 0.001915 | 0.001157 | 0.000881 | 0.003205 |

   Estimated R Matrix for Subject 3/Weighted by WeightVar

| Row | Col1 | Col2 | Col3 | Col4 | Col5 |
|-----|------|------|------|------|------|
| 1 | 0.1426 | 0.01276 | 0.007714 | 0.005873 | 0.005053 |
| 2 | 0.01276 | 0.02043 | 0.002920 | 0.002223 | 0.001913 |
| 3 | 0.007714 | 0.002920 | 0.007466 | 0.001344 | 0.001156 |
| 4 | 0.005873 | 0.002223 | 0.001344 | 0.004328 | 0.000880 |
| 5 | 0.005053 | 0.001913 | 0.001156 | 0.000880 | 0.003203 |

   Estimated R Matrix for Subject 4/Weighted by WeightVar

| Row | Col1 | Col2 | Col3 | Col4 | Col5 |
|-----|------|------|------|------|------|
| 1 | 0.03224 | 0.003403 | 0.002481 | 0.002171 | 0.002037 |
| 2 | 0.003403 | 0.006425 | 0.001108 | 0.000969 | 0.000909 |
| 3 | 0.002481 | 0.001108 | 0.003417 | 0.000707 | 0.000663 |
| 4 | 0.002171 | 0.000969 | 0.000707 | 0.002616 | 0.000580 |

Analysis of the Moldover et al. Data to Estimate R 145
Perform a Repeated Measurements Quadratic Analysis with Proc Mixed
The SINGRES option of the model statement is needed to yield non-zero
estimates of the standard errors of the Pressure and Pressure-squared parameters.
This is Analysis 4 in supporting online material in the short article.

The Mixed Procedure

Estimated R Matrix for Subject 4/Weighted by WeightVar

| Row | Col1 | Col2 | Col3 | Col4 | Col5 |
|-----|------|------|------|------|------|
| 5 | 0.002037 | 0.000909 | 0.000663 | 0.000580 | 0.002302 |

Estimated R Matrix for Subject 5/Weighted by WeightVar

| Row | Col1 | Col2 | Col3 | Col4 | Col5 |
|-----|------|------|------|------|------|
| 1 | 0.01265 | 0.001618 | 0.001343 | 0.001251 | 0.001211 |
| 2 | 0.001618 | 0.003703 | 0.000726 | 0.000677 | 0.000655 |
| 3 | 0.001343 | 0.000726 | 0.002550 | 0.000562 | 0.000544 |
| 4 | 0.001251 | 0.000677 | 0.000562 | 0.002215 | 0.000507 |
| 5 | 0.001211 | 0.000655 | 0.000544 | 0.000507 | 0.002075 |

Estimated R Matrix for Subject 14/Weighted by WeightVar

| Row | Col1 | Col2 | Col3 | Col4 | Col5 |
|-----|------|------|------|------|------|
| 1 | 0.001976 | 0.000457 | 0.000455 | 0.000454 | 0.000453 |
| 2 | 0.000457 | 0.001893 | 0.000445 | 0.000444 | 0.000444 |
| 3 | 0.000455 | 0.000445 | 0.001873 | 0.000442 | 0.000442 |
| 4 | 0.000454 | 0.000444 | 0.000442 | 0.001866 | 0.000441 |
| 5 | 0.000453 | 0.000444 | 0.000442 | 0.000441 | 0.001862 |

Covariance Parameter Estimates

| Cov Parm | Subject | Estimate | Standard Error | Z Value | Pr Z | Alpha | Lower | Upper |
|----------|---------|----------|----------------|---------|------|-------|-------|-------|
| CS | Pressure | 0.3104 | 0.2201 | 1.41 | 0.1585 | 0.05 | -0.1210 | 0.7418 |
| Residual | | 1.0024 | 0.1902 | 5.27 | <.0001 | 0.05 | 0.7136 | 1.5113 |

Analysis of the Moldover et al. Data to Estimate R                              146
Perform a Repeated Measurements Quadratic Analysis with Proc Mixed
The SINGRES option of the model statement is needed to yield non-zero
estimates of the standard errors of the Pressure and Pressure-squared parameters.
This is Analysis 4 in supporting online material in the short article.

The Mixed Procedure

### Fit Statistics

| | |
|---|---|
| -2 Res Log Likelihood | -101.8 |
| AIC (smaller is better) | -97.8 |
| AICC (smaller is better) | -97.6 |
| BIC (smaller is better) | -96.5 |

### Null Model Likelihood Ratio Test

| DF | Chi-Square | Pr > ChiSq |
|---|---|---|
| 1 | 4.93 | 0.0263 |

### Solution for Fixed Effects

| Effect | Estimate | Standard Error | DF | t Value | Pr > \|t\| | Alpha |
|---|---|---|---|---|---|---|
| Intercept | 94756 | 0.02956 | 11 | 3205167 | <.0001 | 0.05 |
| Pressure | 0.000225 | 2.585E-7 | 11 | 868.62 | <.0001 | 0.05 |
| Pressure*Pressure | 5.47E-11 | 4.63E-13 | 11 | 118.08 | <.0001 | 0.05 |

### Solution for Fixed Effects

| Effect | Lower | Upper |
|---|---|---|
| Intercept | 94756 | 94756 |
| Pressure | 0.000224 | 0.000225 |
| Pressure*Pressure | 5.37E-11 | 5.57E-11 |

### Covariance Matrix for Fixed Effects

| Row | Effect | Col1 | Col2 | Col3 |
|---|---|---|---|---|
| 1 | Intercept | 0.000874 | -6.92E-9 | 1.11E-14 |
| 2 | Pressure | -6.92E-9 | 6.68E-14 | -117E-21 |
| 3 | Pressure*Pressure | 1.11E-14 | -117E-21 | 2.15E-25 |

Analysis of the Moldover et al. Data to Estimate R                       147
Perform a Repeated Measurements Quadratic Analysis with Proc Mixed
The SINGRES option of the model statement is needed to yield non-zero
estimates of the standard errors of the Pressure and Pressure-squared parameters.
This is Analysis 4 in supporting online material in the short article.

The Mixed Procedure

Type 3 Tests of Fixed Effects

| Effect | Num DF | Den DF | F Value | Pr > F |
|---|---|---|---|---|
| Pressure | 1 | 11 | 754498 | <.0001 |
| Pressure*Pressure | 1 | 11 | 13943.3 | <.0001 |

Analysis of the Moldover et al. Data to Estimate R                 148
Perform a Repeated Measurements Quadratic Analysis with Proc Mixed
The SINGRES option of the model statement is needed to yield non-zero
estimates of the standard errors of the Pressure and Pressure-squared parameters.
This is Analysis 4 in supporting online material in the short article.

| Obs | Effect | Estimate | StdErr | DF | tValue | Probt |
|-----|--------|----------|--------|-----|--------|-------|
| 1 | Intercept | 9.4756242E+04 | 2.96E-02 | 11 | 3205167.13 | <.0001 |
| 2 | Pressure | 2.2452508E-04 | 2.58E-07 | 11 | 868.62 | <.0001 |
| 3 | Pressure*Pressure | 5.4701530E-11 | 4.63E-13 | 11 | 118.08 | <.0001 |

Analysis of the Moldover et al. Data to Estimate R
Perform a Repeated Measurements Quadratic Analysis with Proc Mixed
The SINGRES option of the model statement is needed to yield non-zero
estimates of the standard errors of the Pressure and Pressure-squared parameters.
This is Analysis 4 in supporting online material in the short article.

| Estimate of Intercept from Rep. Meas. Quadratic Model | Estimate of Standard Error of Intercept from Rep. Meas. Quadratic Model | Estimate of R from Rep. Meas. Quadratic Model |
|---|---|---|
| 9.4756242297604000E+04 | 2.9563588519671000E-02 | 8.3144766022064000E+00 |

Analysis of the Moldover et al. Data to Estimate R                    150
Linear Regression of Raw Frequency on Rescaled Speed and Mode
This enables determination of the corrected frequency values, which
are used to compute the corrected values of the weight variable.
If the rescaling is omitted, SAS gives the rescaling warning.  This model
fits better (with higher F- and t-values) than using the squared values.
D3, D4, D5, and D6 are dummy (indicator) variables used to specify the
five resonance modes. (Only four dummy variables are necessary to specify
the five modes.)  The Ds are computed near the beginning of the program.

The REG Procedure
Model: MODEL1
Dependent Variable: Freq Mean Resonance Frequency (Hz)

Number of Observations Read          70
Number of Observations Used          70

### Analysis of Variance

| Source | DF | Sum of Squares | Mean Square | F Value | Pr > F |
|--------|----|----|----|----|----|
| Model | 5 | 430025731 | 86005146 | 2.396E8 | <.0001 |
| Error | 64 | 22.96984 | 0.35890 | | |
| Corrected Total | 69 | 430025754 | | | |

| | | | | |
|--|--|--|--|--|
| Root MSE | 0.59909 | R-Square | 1.0000 | |
| Dependent Mean | 5997.87860 | Adj R-Sq | 1.0000 | |
| Coeff Var | 0.00999 | | | |

### Parameter Estimates

| Variable | Label | DF | Parameter Estimate | Standard Error | t Value |
|----------|-------|----|----|----|----|
| Intercept | Intercept | 1 | 2476.19345 | 0.16091 | 15389.0 |
| SpeedAtTempResc | SpeedAtTemp - 307.9 (m/s) | 1 | 23.22761 | 1.14581 | 20.27 |
| D3 | Indicator Variable (0/1) for Resonance Mode 3 | 1 | 1781.38069 | 0.22643 | 7867.13 |
| D4 | Indicator Variable (0/1) for Resonance Mode 4 | 1 | 3533.58032 | 0.22643 | 15605.4 |
| D5 | Indicator Variable (0/1) for Resonance Mode 5 | 1 | 5276.52963 | 0.22643 | 23302.8 |
| D6 | Indicator Variable (0/1) for Resonance Mode 6 | 1 | 7015.33708 | 0.22643 | 30981.9 |

Analysis of the Moldover et al. Data to Estimate R                    151
Linear Regression of Raw Frequency on Rescaled Speed and Mode
This enables determination of the corrected frequency values, which
are used to compute the corrected values of the weight variable.
If the rescaling is omitted, SAS gives the rescaling warning.  This model
fits better (with higher F- and t-values) than using the squared values.
D3, D4, D5, and D6 are dummy (indicator) variables used to specify the
five resonance modes. (Only four dummy variables are necessary to specify
the five modes.)  The Ds are computed near the beginning of the program.

The REG Procedure
Model: MODEL1
Dependent Variable: Freq Mean Resonance Frequency (Hz)

                        Parameter Estimates

| Variable | Label | DF | Pr > \|t\| |
|---|---|---|---|
| Intercept | Intercept | 1 | <.0001 |
| SpeedAtTempResc | SpeedAtTemp - 307.9 (m/s) | 1 | <.0001 |
| D3 | Indicator Variable (0/1) for Resonance Mode 3 | 1 | <.0001 |
| D4 | Indicator Variable (0/1) for Resonance Mode 4 | 1 | <.0001 |
| D5 | Indicator Variable (0/1) for Resonance Mode 5 | 1 | <.0001 |
| D6 | Indicator Variable (0/1) for Resonance Mode 6 | 1 | <.0001 |

Analysis of the Moldover et al. Data to Estimate R                          152
Univariate distribution of the temperatures.
Note in the stem and leaf plot on the third page of this output
the five observations at the substantially higher temperature
of 273.1740 kelvins, which are shown as 273174 00000
These five observations account for the dip at the pressure of
307.8787 on the plot of residuals versus SpeedAtTemp generated above.

The UNIVARIATE Procedure
Variable:  Temp  (Temperature (K))

### Moments

| | | | |
|---|---|---|---|
| N | 70 | Sum Weights | 70 |
| Mean | 273.16365 | Sum Observations | 19121.4555 |
| Std Deviation | 0.00360731 | Variance | 0.00001301 |
| Skewness | 1.38352179 | Kurtosis | 2.75707224 |
| Uncorrected SS | 5223286.58 | Corrected SS | 0.00089788 |
| Coeff Variation | 0.00132057 | Std Error Mean | 0.00043116 |

### Basic Statistical Measures

| Location | | Variability | |
|---|---|---|---|
| Mean | 273.1637 | Std Deviation | 0.00361 |
| Median | 273.1639 | Variance | 0.0000130 |
| Mode | 273.1585 | Range | 0.01550 |
| | | Interquartile Range | 0.00370 |

NOTE: The mode displayed is the smallest of 14 modes with a count of 5.

### Tests for Location: Mu0=0

| Test | | -Statistic- | | -----p Value------ | | |
|---|---|---|---|---|---|---|
| Student's t | t | 633561.1 | Pr > \|t\| | <.0001 |
| Sign | M | 35 | Pr >= \|M\| | <.0001 |
| Signed Rank | S | 1242.5 | Pr >= \|S\| | <.0001 |

### Quantiles (Definition 5)

| Quantile | Estimate |
|---|---|
| 100% Max | 273.174 |
| 99% | 273.174 |
| 95% | 273.174 |

Analysis of the Moldover et al. Data to Estimate R                          153
Univariate distribution of the temperatures.
Note in the stem and leaf plot on the third page of this output
the five observations at the substantially higher temperature
of 273.1740 kelvins, which are shown as 273174 00000
These five observations account for the dip at the pressure of
307.8787 on the plot of residuals versus SpeedAtTemp generated above.

The UNIVARIATE Procedure
Variable:  Temp  (Temperature (K))

Quantiles (Definition 5)

| Quantile | Estimate |
|----------|----------|
| 90% | 273.166 |
| 75% Q3 | 273.165 |
| 50% Median | 273.164 |
| 25% Q1 | 273.161 |
| 10% | 273.160 |
| 5% | 273.159 |
| 1% | 273.159 |
| 0% Min | 273.159 |

Extreme Observations

| ------Lowest----- | | -----Highest----- | |
|------|------|------|------|
| Value | Obs | Value | Obs |
| 273.159 | 45 | 273.174 | 26 |
| 273.159 | 44 | 273.174 | 27 |
| 273.159 | 43 | 273.174 | 28 |
| 273.159 | 42 | 273.174 | 29 |
| 273.159 | 41 | 273.174 | 30 |

Analysis of the Moldover et al. Data to Estimate R                                   154
Univariate distribution of the temperatures.
Note in the stem and leaf plot on the third page of this output
the five observations at the substantially higher temperature
of 273.1740 kelvins, which are shown as 273174 00000
These five observations account for the dip at the pressure of
307.8787 on the plot of residuals versus SpeedAtTemp generated above.

The UNIVARIATE Procedure
Variable:  Temp  (Temperature (K))

```
  Stem Leaf                        #  Boxplot
273174 00000                       5     0
273173
273172
273171
273170
273169
273168
273167
273166
273165 000004444466666           15  +-----+
273164 000002222266666           15  |     |
273163 0000077777              10  *--+--*
273162                            |     |
273161 3333355555              10  +-----+
273160 88888                     5     |
273159 55555                     5     |
273158 55555                     5     |
       ----+----+----+----+
    Multiply Stem.Leaf by 10**-3
```

Univariate distribution of the temperatures.
Note in the stem and leaf plot on the third page of this output
the five observations at the substantially higher temperature
of 273.1740 kelvins, which are shown as 273174 00000
These five observations account for the dip at the pressure of
307.8787 on the plot of residuals versus SpeedAtTemp generated above.

The UNIVARIATE Procedure
Variable:  Temp  (Temperature (K))

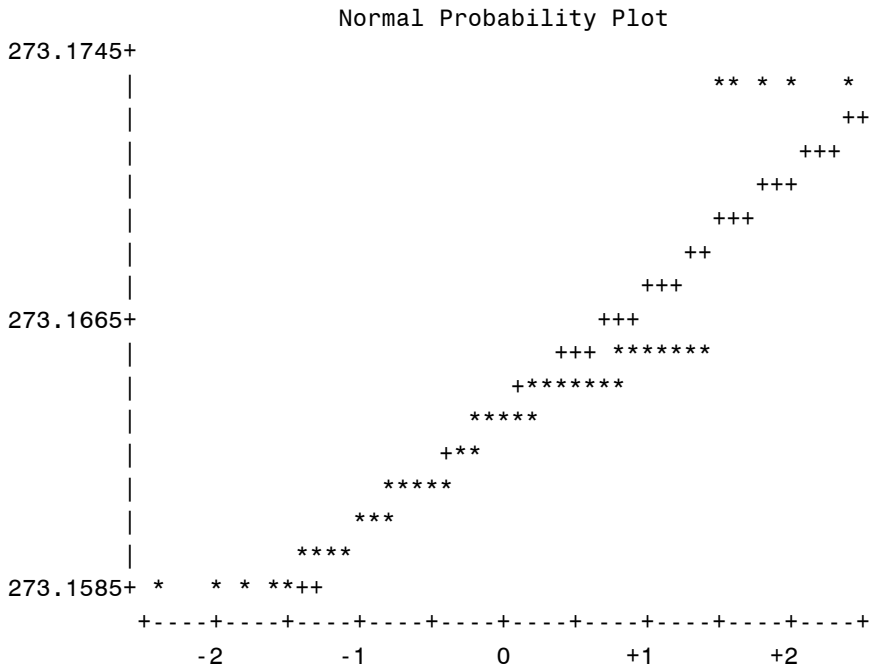

Analysis of the Moldover et al. Data to Estimate R                                              156
Rerun the preceding regression with the corrected frequencies to confirm that the
fit with the corrected frequencies is essentially perfect.  Note the F and t test
statistics are now all "Infty" (i.e., Infinity), which implies a rather good fit.
This confirms that the corrected frequency values were correctly computed.
(The corrected frequency values were computed near the beginning of the program
using the results of an unprinted run of the preceding regression.)

The REG Procedure
Model: MODEL1
Dependent Variable: FreqCor Corrected Resonance Frequency (Hz)

| | |
|---|---|
| Number of Observations Read | 70 |
| Number of Observations Used | 70 |

### Analysis of Variance

| Source | DF | Sum of Squares | Mean Square | F Value | Pr > F |
|---|---|---|---|---|---|
| Model | 5 | 430025731 | 86005146 | Infty | <.0001 |
| Error | 64 | 0 | 0 | | |
| Corrected Total | 69 | 430025731 | | | |

| | | | | |
|---|---|---|---|---|
| Root MSE | 0 | R-Square | 1.0000 | |
| Dependent Mean | 5997.87860 | Adj R-Sq | 1.0000 | |
| Coeff Var | 0 | | | |

### Parameter Estimates

| Variable | Label | DF | Parameter Estimate | Standard Error | t Value |
|---|---|---|---|---|---|
| Intercept | Intercept | 1 | 2476.19345 | 0 | Infty |
| SpeedAtTempResc | SpeedAtTemp - 307.9 (m/s) | 1 | 23.22761 | 0 | Infty |
| D3 | Indicator Variable (0/1) for Resonance Mode 3 | 1 | 1781.38069 | 0 | Infty |
| D4 | Indicator Variable (0/1) for Resonance Mode 4 | 1 | 3533.58032 | 0 | Infty |
| D5 | Indicator Variable (0/1) for Resonance Mode 5 | 1 | 5276.52963 | 0 | Infty |
| D6 | Indicator Variable (0/1) for Resonance Mode 6 | 1 | 7015.33708 | 0 | Infty |

Analysis of the Moldover et al. Data to Estimate R                              157
Rerun the preceding regression with the corrected frequencies to confirm that the
fit with the corrected frequencies is essentially perfect.  Note the F and t test
statistics are now all "Infty" (i.e., Infinity), which implies a rather good fit.
This confirms that the corrected frequency values were correctly computed.
(The corrected frequency values were computed near the beginning of the program
using the results of an unprinted run of the preceding regression.)

The REG Procedure
Model: MODEL1
Dependent Variable: FreqCor Corrected Resonance Frequency (Hz)

                        Parameter Estimates

| Variable | Label | DF | Pr > \|t\| |
|---|---|---|---|
| Intercept | Intercept | 1 | <.0001 |
| SpeedAtTempResc | SpeedAtTemp - 307.9 (m/s) | 1 | <.0001 |
| D3 | Indicator Variable (0/1) for Resonance Mode 3 | 1 | <.0001 |
| D4 | Indicator Variable (0/1) for Resonance Mode 4 | 1 | <.0001 |
| D5 | Indicator Variable (0/1) for Resonance Mode 5 | 1 | <.0001 |
| D6 | Indicator Variable (0/1) for Resonance Mode 6 | 1 | <.0001 |

Analysis of the Moldover et al. Data to Estimate R          158
Regenerate the Moldover et al. Parameter Estimates with Proc REG
using the response variable minus the value of the cubic term as the
value of the response variable.  These values are almost identical to the
values from the NLIN run, although the t-values are slightly different.

The REG Procedure
Model: MODEL1
Dependent Variable: SpSqMPCub Speed-Squared Minus Cubic Term

Number of Observations Read          70
Number of Observations Used          70

Weight: WeightVar Weight Variable

### Analysis of Variance

| Source | DF | Sum of Squares | Mean Square | F Value | Pr > F |
|--------|----|----|----|----|----|
| Model | 3 | 47603797 | 15867932 | 1.223E7 | <.0001 |
| Error | 66 | 85.62003 | 1.29727 | | |
| Corrected Total | 69 | 47603883 | | | |

| | | | | |
|---|---|---|---|---|
| Root MSE | 1.13898 | R-Square | 1.0000 | |
| Dependent Mean | 94823 | Adj R-Sq | 1.0000 | |
| Coeff Var | 0.00120 | | | |

### Parameter Estimates

| Variable | Label | DF | Parameter Estimate | Standard Error | t Value |
|----------|-------|----|----|----|----|
| Intercept | Intercept | 1 | 94756 | 0.06467 | 1465233 |
| Pressure | Pressure (Pa) | 1 | 0.00022503 | 3.530948E-7 | 637.32 |
| PressSq | Pressure Squared | 1 | 5.32045E-11 | 5.16132E-13 | 103.08 |
| PressInv | Inverse Pressure | 1 | 2679.78879 | 2884.55774 | 0.93 |

### Parameter Estimates

| Variable | Label | DF | Pr > |t| |
|----------|-------|----|----|
| Intercept | Intercept | 1 | <.0001 |
| Pressure | Pressure (Pa) | 1 | <.0001 |
| PressSq | Pressure Squared | 1 | <.0001 |
| PressInv | Inverse Pressure | 1 | 0.3563 |

Analysis of the Moldover et al. Data to Estimate R                    159
Regenerate the Moldover et al. Parameter Estimates with Proc REG
using the response variable minus the value of the cubic term as the
value of the response variable.  These values are almost identical to the
values from the NLIN run, although the t-values are slightly different.
Here is a more accurate table of the values of the parameters:

| Label | Parameter Estimate | Standard Error | t | p |
|---|---|---|---|---|
| Intercept | 9.4756178E+04 | 6.47E-02 | 1465232.7 | <.0001 |
| Pressure (Pa) | 2.2503316E-04 | 3.53E-07 | 637.32 | <.0001 |
| Pressure Squared | 5.3204471E-11 | 5.16E-13 | 103.08 | <.0001 |
| Inverse Pressure | 2.6797888E+03 | 2.88E+03 | 0.93 | 0.3563 |

Analysis of the Moldover et al. Data to Estimate R                    160
Compare Sizes of Moldover et al. and Quadratic Residuals
Observations are in order of increasing pressure and within pressure in
order of increasing resonance mode.  Thus observations 1 through 5 are
at a pressure of 25.396 kPa and observations 21 through 25 are at
at pressure of 100.261 kPa.

The UNIVARIATE Procedure
Variable:  QuadMMold  (Quadratic Residual - Moldover et al. Residual)

### Moments

| | | | |
|---|---|---|---|
| N | 70 | Sum Weights | 70 |
| Mean | 0.00324002 | Sum Observations | 0.22680118 |
| Std Deviation | 0.01217809 | Variance | 0.00014831 |
| Skewness | 3.08119 | Kurtosis | 8.49696128 |
| Uncorrected SS | 0.01096795 | Corrected SS | 0.01023311 |
| Coeff Variation | 375.865053 | Std Error Mean | 0.00145556 |

### Basic Statistical Measures

| Location | | Variability | |
|---|---|---|---|
| Mean | 0.00324 | Std Deviation | 0.01218 |
| Median | 0.00047 | Variance | 0.0001483 |
| Mode | -0.00440 | Range | 0.05002 |
| | | Interquartile Range | 0.00381 |

NOTE: The mode displayed is the smallest of 14 modes with a count of 5.

### Tests for Location: Mu0=0

| Test | -Statistic- | | -----p Value------ | |
|---|---|---|---|---|
| Student's t | t | 2.225959 | Pr > \|t\| | 0.0293 |
| Sign | M | 5 | Pr >= \|M\| | 0.2820 |
| Signed Rank | S | 202.5 | Pr >= \|S\| | 0.2384 |

### Tests for Normality

| Test | --Statistic--- | | -----p Value------ | |
|---|---|---|---|---|
| Shapiro-Wilk | W | 0.498085 | Pr < W | <0.0001 |
| Kolmogorov-Smirnov | D | 0.378282 | Pr > D | <0.0100 |
| Cramer-von Mises | W-Sq | 2.384472 | Pr > W-Sq | <0.0050 |
| Anderson-Darling | A-Sq | 13.04146 | Pr > A-Sq | <0.0050 |

Analysis of the Moldover et al. Data to Estimate R                      161
Compare Sizes of Moldover et al. and Quadratic Residuals
Observations are in order of increasing pressure and within pressure in
order of increasing resonance mode.  Thus observations 1 through 5 are
at a pressure of 25.396 kPa and observations 21 through 25 are at
at pressure of 100.261 kPa.

The UNIVARIATE Procedure
Variable:  QuadMMold  (Quadratic Residual - Moldover et al. Residual)

 Quantiles (Definition 5)

| Quantile | Estimate |
|---|---|
| 100% Max | 0.045621266 |
| 99% | 0.045621266 |
| 95% | 0.045621266 |
| 90% | 0.004779237 |
| 75% Q3 | 0.002333492 |
| 50% Median | 0.000468796 |
| 25% Q1 | -0.001475043 |
| 10% | -0.004372017 |
| 5% | -0.004400475 |
| 1% | -0.004400475 |
| 0% Min | -0.004400475 |

                    Extreme Observations

| -------Lowest------- | | -------Highest------ | |
|---|---|---|---|
| Value | Obs | Value | Obs |
| -0.00440048 | 25 | 0.00477924 | 6 |
| -0.00440048 | 24 | 0.00477924 | 7 |
| -0.00440048 | 23 | 0.00477924 | 8 |
| -0.00440048 | 22 | 0.00477924 | 9 |
| -0.00440048 | 21 | 0.00477924 | 10 |
| -0.00437202 | 30 | 0.04562127 | 1 |
| -0.00437202 | 29 | 0.04562127 | 2 |
| -0.00437202 | 28 | 0.04562127 | 3 |
| -0.00437202 | 27 | 0.04562127 | 4 |
| -0.00437202 | 26 | 0.04562127 | 5 |

Compare Sizes of Moldover et al. and Quadratic Residuals
Observations are in order of increasing pressure and within pressure in
order of increasing resonance mode.  Thus observations 1 through 5 are
at a pressure of 25.396 kPa and observations 21 through 25 are at
at pressure of 100.261 kPa.

The UNIVARIATE Procedure
Variable:  QuadMMold  (Quadratic Residual - Moldover et al. Residual)

```
  Stem Leaf                      #  Boxplot
    44 66666                     5     *
    42
    40
    38
    36
    34
    32
    30
    28
    26
    24
    22
    20
    18
    16
    14
    12
    10
     8
     6
     4 8888888888             10     |
     2 33333                   5   +--+--+
     0 44444666663333388888   20   *-----*
    -0 555554444400000        15   +-----+
    -2 55555                   5     |
    -4 4444444444             10     |
       ----+----+----+----+
  Multiply Stem.Leaf by 10**-3
```

Analysis of the Moldover et al. Data to Estimate R                    163
Compare Sizes of Moldover et al. and Quadratic Residuals
Observations are in order of increasing pressure and within pressure in
order of increasing resonance mode.  Thus observations 1 through 5 are
at a pressure of 25.396 kPa and observations 21 through 25 are at
at pressure of 100.261 kPa.

The UNIVARIATE Procedure
Variable:  QuadMMold  (Quadratic Residual - Moldover et al. Residual)

```
                         Normal Probability Plot
      0.045+                                   ** *  *    *
           |
           |
           |
      0.035+
           |                                           ++
           |                                         ++
           |                                        +
           |                                      ++
      0.025+                                    +
           |                                  ++
           |                                 ++
           |                               +
           |                             ++
      0.015+                           ++
           |                          +
           |                        ++
           |                      ++
           |                     +
      0.005+                   ++    *******
           |                  ++    ***
           |                      ********
           |              *******
           |            ***    +
     -0.005+  *    *  * ******   ++
           +----+----+----+----+----+----+----+----+----+----+
               -2        -1         0        +1        +2
```

Analysis of the Moldover et al. Data to Estimate R     164
Fit the Line with the Cubic Equation Using the Weights.

The NLIN Procedure
Dependent Variable SpeedSq
Method: Gauss-Newton

Iterative Phase

| Iter | B0 | B1 | B2 | B3 | Weighted SS |
|---|---|---|---|---|---|
| 0 | 1.0000 | 1.0000 | 1.0000 | 1.0000 | 1.094E38 |
| 1 | 95349.2 | -0.00985 | 4.092E-8 | -469E-16 | 3.9205E8 |
| 2 | 94756.3 | 0.000224 | 5.84E-11 | -434E-20 | 81.4345 |

NOTE: Convergence criterion met.

Estimation Summary

| Method | Gauss-Newton |
|---|---|
| Iterations | 2 |
| R | 1.066E-9 |
| PPC | 2.07E-11 |
| RPC(B1) | 1.022605 |
| Object | 1 |
| Objective | 81.43451 |
| Observations Read | 70 |
| Observations Used | 70 |
| Observations Missing | 0 |

| Source | DF | Sum of Squares | Mean Square | F Value | Approx Pr > F |
|---|---|---|---|---|---|
| Model | 3 | 47749342 | 15916447 | 1.29E7 | <.0001 |
| Error | 66 | 81.4345 | 1.2339 | | |
| Corrected Total | 69 | 47749423 | | | |

Fit the Line with the Cubic Equation Using the Weights.

The NLIN Procedure

| Parameter | Estimate | Approx Std Error | Approximate 95% Confidence Limits | | Skewness |
|-----------|----------|------------------|------------------|------------------|----------|
| B0 | 94756.3 | 0.0393 | 94756.2 | 94756.4 | 0 |
| B1 | 0.000224 | 5.819E-7 | 0.000222 | 0.000225 | 0 |
| B2 | 5.84E-11 | 2.35E-12 | 5.37E-11 | 6.31E-11 | 0 |
| B3 | -434E-20 | 2.76E-18 | -986E-20 | 1.17E-18 | 0 |

Approximate Correlation Matrix

| | B0 | B1 | B2 | B3 |
|----|------------|------------|------------|------------|
| B0 | 1.0000000 | -0.9469152 | 0.8819249 | -0.8243007 |
| B1 | -0.9469152 | 1.0000000 | -0.9813861 | 0.9461322 |
| B2 | 0.8819249 | -0.9813861 | 1.0000000 | -0.9898949 |
| B3 | -0.8243007 | 0.9461322 | -0.9898949 | 1.0000000 |

Fit the Line with the Cubic Equation Using the Weights.

Here is a more accurate table of the values of the parameters of the above model.

| Parameter | Parameter Estimate | Approximate Standard Error | Approximate Lower 95% Confidence Limit | Approximate Upper 95% Confidence Limit | t | p |
|---|---|---|---|---|---|---|
| B0 | 9.4756302E+04 | 3.93E-02 | 9.4756223E+04 | 9.4756380E+04 | 2412463.63 | <.0001 |
| B1 | 2.2360452E-04 | 5.82E-07 | 2.2244281E-04 | 2.2476623E-04 | 384.30 | <.0001 |
| B2 | 5.8435552E-11 | 2.35E-12 | 5.3748650E-11 | 6.3122454E-11 | 24.89 | <.0001 |
| B3 | -4.3406787E-18 | 2.76E-18 | -9.8557311E-18 | 1.1743737E-18 | -1.57 | 0.1209 |

Analysis of the Moldover et al. Data to Estimate R                              167
Quadratic Analysis with Proc GLM and Repeated Measurements
Note that WeightVar cannot be used in the standard way because it applies
at the "subject" level as opposed to the "observation" level.

The GLM Procedure

```
Number of Observations Read        14
Number of Observations Used        14
```

Analysis of the Moldover et al. Data to Estimate R                    168
Quadratic Analysis with Proc GLM and Repeated Measurements
Note that WeightVar cannot be used in the standard way because it applies
at the "subject" level as opposed to the "observation" level.

The GLM Procedure

Dependent Variable: SpeedSq2   Speed of Sound in Argon (mode 0,2) Squared (m^2/s^2)

| Source | DF | Sum of Squares | Mean Square | F Value | Pr > F |
|---|---|---|---|---|---|
| Model | 2 | 21512.95558 | 10756.47779 | 186015 | <.0001 |
| Error | 11 | 0.63609 | 0.05783 | | |
| Corrected Total | 13 | 21513.59166 | | | |

| R-Square | Coeff Var | Root MSE | SpeedSq2 Mean |
|---|---|---|---|
| 0.999970 | 0.000254 | 0.240470 | 94808.46 |

| Source | DF | Type I SS | Mean Square | F Value | Pr > F |
|---|---|---|---|---|---|
| Pressure | 1 | 21493.44232 | 21493.44232 | 371692 | <.0001 |
| PressSq | 1 | 19.51325 | 19.51325 | 337.45 | <.0001 |

| Source | DF | Type III SS | Mean Square | F Value | Pr > F |
|---|---|---|---|---|---|
| Pressure | 1 | 903.9258337 | 903.9258337 | 15631.8 | <.0001 |
| PressSq | 1 | 19.5132538 | 19.5132538 | 337.45 | <.0001 |

| Parameter | Estimate | Standard Error | t Value | Pr > |t| | 95% Confidence Limits | |
|---|---|---|---|---|---|---|
| Intercept | 94756.84377 | 0.16825275 | 563182 | <.0001 | 94756.47345 | 94757.21409 |
| Pressure | 0.00022 | 0.00000176 | 125.03 | <.0001 | 0.00022 | 0.00022 |
| PressSq | 0.00000 | 0.00000000 | 18.37 | <.0001 | 0.00000 | 0.00000 |

Analysis of the Moldover et al. Data to Estimate R                    169
Quadratic Analysis with Proc GLM and Repeated Measurements
Note that WeightVar cannot be used in the standard way because it applies
at the "subject" level as opposed to the "observation" level.

The GLM Procedure

Dependent Variable: SpeedSq3   Speed of Sound in Argon (mode O,3) Squared (m^2/s^2)

| Source | DF | Sum of Squares | Mean Square | F Value | Pr > F |
|---|---|---|---|---|---|
| Model | 2 | 21603.77828 | 10801.88914 | 627570 | <.0001 |
| Error | 11 | 0.18933 | 0.01721 | | |
| Corrected Total | 13 | 21603.96762 | | | |

| R-Square | Coeff Var | Root MSE | SpeedSq3 Mean |
|---|---|---|---|
| 0.999991 | 0.000138 | 0.131195 | 94808.34 |

| Source | DF | Type I SS | Mean Square | F Value | Pr > F |
|---|---|---|---|---|---|
| Pressure | 1 | 21587.23740 | 21587.23740 | 1254180 | <.0001 |
| PressSq | 1 | 16.54088 | 16.54088 | 961.00 | <.0001 |

| Source | DF | Type III SS | Mean Square | F Value | Pr > F |
|---|---|---|---|---|---|
| Pressure | 1 | 929.0806191 | 929.0806191 | 53977.9 | <.0001 |
| PressSq | 1 | 16.5408834 | 16.5408834 | 961.00 | <.0001 |

| Parameter | Estimate | Standard Error | t Value | Pr > \|t\| | 95% Confidence Limits | |
|---|---|---|---|---|---|---|
| Intercept | 94756.41476 | 0.09179513 | 1032260 | <.0001 | 94756.21272 | 94756.61680 |
| Pressure | 0.00022 | 0.00000096 | 232.33 | <.0001 | 0.00022 | 0.00023 |
| PressSq | 0.00000 | 0.00000000 | 31.00 | <.0001 | 0.00000 | 0.00000 |

Quadratic Analysis with Proc GLM and Repeated Measurements
Note that WeightVar cannot be used in the standard way because it applies
at the "subject" level as opposed to the "observation" level.

The GLM Procedure

Dependent Variable: SpeedSq4   Speed of Sound in Argon (mode O,4) Squared (m^2/s^2)

| Source | DF | Sum of Squares | Mean Square | F Value | Pr > F |
|--------|----|----|----|----|----|
| Model | 2 | 21652.69220 | 10826.34610 | 3530381 | <.0001 |
| Error | 11 | 0.03373 | 0.00307 | | |
| Corrected Total | 13 | 21652.72593 | | | |

| R-Square | Coeff Var | Root MSE | SpeedSq4 Mean |
|----|----|----|----|
| 0.999998 | 0.000058 | 0.055377 | 94808.30 |

| Source | DF | Type I SS | Mean Square | F Value | Pr > F |
|--------|----|----|----|----|----|
| Pressure | 1 | 21638.34755 | 21638.34755 | 7056084 | <.0001 |
| PressSq | 1 | 14.34465 | 14.34465 | 4677.67 | <.0001 |

| Source | DF | Type III SS | Mean Square | F Value | Pr > F |
|--------|----|----|----|----|----|
| Pressure | 1 | 948.2329354 | 948.2329354 | 309211 | <.0001 |
| PressSq | 1 | 14.3446497 | 14.3446497 | 4677.67 | <.0001 |

| Parameter | Estimate | Standard Error | t Value | Pr > |t| | 95% Confidence Limits | |
|--------|----|----|----|----|----|----|
| Intercept | 94756.16619 | 0.03874638 | 2445549 | <.0001 | 94756.08091 | 94756.25147 |
| Pressure | 0.00023 | 0.00000040 | 556.07 | <.0001 | 0.00022 | 0.00023 |
| PressSq | 0.00000 | 0.00000000 | 68.39 | <.0001 | 0.00000 | 0.00000 |

Quadratic Analysis with Proc GLM and Repeated Measurements
Note that WeightVar cannot be used in the standard way because it applies
at the "subject" level as opposed to the "observation" level.

The GLM Procedure

Dependent Variable: SpeedSq5   Speed of Sound in Argon (mode O,5) Squared (m^2/s^2)

| Source | DF | Sum of Squares | Mean Square | F Value | Pr > F |
|---|---|---|---|---|---|
| Model | 2 | 21616.07000 | 10808.03500 | 1571557 | <.0001 |
| Error | 11 | 0.07565 | 0.00688 | | |
| Corrected Total | 13 | 21616.14565 | | | |

| R-Square | Coeff Var | Root MSE | SpeedSq5 Mean |
|---|---|---|---|
| 0.999997 | 0.000087 | 0.082929 | 94808.34 |

| Source | DF | Type I SS | Mean Square | F Value | Pr > F |
|---|---|---|---|---|---|
| Pressure | 1 | 21600.17146 | 21600.17146 | 3140801 | <.0001 |
| PressSq | 1 | 15.89854 | 15.89854 | 2311.75 | <.0001 |

| Source | DF | Type III SS | Mean Square | F Value | Pr > F |
|---|---|---|---|---|---|
| Pressure | 1 | 934.4434506 | 934.4434506 | 135874 | <.0001 |
| PressSq | 1 | 15.8985416 | 15.8985416 | 2311.75 | <.0001 |

| Parameter | Estimate | Standard Error | t Value | Pr > \|t\| | 95% Confidence Limits | |
|---|---|---|---|---|---|---|
| Intercept | 94756.35859 | 0.05802422 | 1633048 | <.0001 | 94756.23088 | 94756.48630 |
| Pressure | 0.00022 | 0.00000061 | 368.61 | <.0001 | 0.00022 | 0.00022 |
| PressSq | 0.00000 | 0.00000000 | 48.08 | <.0001 | 0.00000 | 0.00000 |

Analysis of the Moldover et al. Data to Estimate R                              172
Quadratic Analysis with Proc GLM and Repeated Measurements
Note that WeightVar cannot be used in the standard way because it applies
at the "subject" level as opposed to the "observation" level.

The GLM Procedure

Dependent Variable: SpeedSq6   Speed of Sound in Argon (mode 0,6) Squared (m^2/s^2)

| Source | DF | Sum of Squares | Mean Square | F Value | Pr > F |
|--------|----|----------------|-------------|---------|--------|
| Model | 2 | 21638.40625 | 10819.20313 | 3345253 | <.0001 |
| Error | 11 | 0.03558 | 0.00323 | | |
| Corrected Total | 13 | 21638.44183 | | | |

| R-Square | Coeff Var | Root MSE | SpeedSq6 Mean |
|----------|-----------|----------|---------------|
| 0.999998 | 0.000060 | 0.056870 | 94808.32 |

| Source | DF | Type I SS | Mean Square | F Value | Pr > F |
|--------|----|-----------|-------------|---------|--------|
| Pressure | 1 | 21624.17540 | 21624.17540 | 6686106 | <.0001 |
| PressSq | 1 | 14.23086 | 14.23086 | 4400.12 | <.0001 |

| Source | DF | Type III SS | Mean Square | F Value | Pr > F |
|--------|----|-------------|-------------|---------|--------|
| Pressure | 1 | 948.4387372 | 948.4387372 | 293253 | <.0001 |
| PressSq | 1 | 14.2308559 | 14.2308559 | 4400.12 | <.0001 |

| Parameter | Estimate | Standard Error | t Value | Pr > |t| | 95% Confidence Limits | |
|-----------|----------|----------------|---------|---------|----------------------|--|
| Intercept | 94756.19764 | 0.03979093 | 2381352 | <.0001 | 94756.11006 | 94756.28522 |
| Pressure | 0.00023 | 0.00000042 | 541.53 | <.0001 | 0.00022 | 0.00023 |
| PressSq | 0.00000 | 0.00000000 | 66.33 | <.0001 | 0.00000 | 0.00000 |

Analysis of the Moldover et al. Data to Estimate R                    173
Quadratic Analysis with Proc GLM and Repeated Measurements
Note that WeightVar cannot be used in the standard way because it applies
at the "subject" level as opposed to the "observation" level.

The GLM Procedure
Repeated Measures Analysis of Variance

Repeated Measures Level Information

Dependent Variable     SpeedSq2 SpeedSq3 SpeedSq4 SpeedSq5 SpeedSq6

    Level of Mode          1         2         3         4         5

MANOVA Test Criteria and Exact F Statistics for the Hypothesis of no Mode Effect
                    H = Type III SSCP Matrix for Mode
                         E = Error SSCP Matrix

                         S=1    M=1    N=3

Statistic                      Value    F Value    Num DF    Den DF    Pr > F

Wilks' Lambda              0.23315921      6.58         4         8    0.0120
Pillai's Trace             0.76684079      6.58         4         8    0.0120
Hotelling-Lawley Trace     3.28891482      6.58         4         8    0.0120
Roy's Greatest Root        3.28891482      6.58         4         8    0.0120

          MANOVA Test Criteria and Exact F Statistics for
          the Hypothesis of no Mode*Pressure Effect
          H = Type III SSCP Matrix for Mode*Pressure
                    E = Error SSCP Matrix

                         S=1    M=1    N=3

Statistic                      Value    F Value    Num DF    Den DF    Pr > F

Wilks' Lambda              0.44525446      2.49         4         8    0.1265
Pillai's Trace             0.55474554      2.49         4         8    0.1265
Hotelling-Lawley Trace     1.24590677      2.49         4         8    0.1265
Roy's Greatest Root        1.24590677      2.49         4         8    0.1265

Analysis of the Moldover et al. Data to Estimate R                    174
Quadratic Analysis with Proc GLM and Repeated Measurements
Note that WeightVar cannot be used in the standard way because it applies
at the "subject" level as opposed to the "observation" level.

The GLM Procedure
Repeated Measures Analysis of Variance

MANOVA Test Criteria and Exact F Statistics for the Hypothesis of no Mode*PressSq Effect
                    H = Type III SSCP Matrix for Mode*PressSq
                         E = Error SSCP Matrix

                         S=1    M=1    N=3

Statistic                      Value    F Value    Num DF    Den DF    Pr > F

Wilks' Lambda               0.53727789      1.72         4         8    0.2376
Pillai's Trace              0.46272211      1.72         4         8    0.2376
Hotelling-Lawley Trace      0.86123422      1.72         4         8    0.2376
Roy's Greatest Root         0.86123422      1.72         4         8    0.2376

Analysis of the Moldover et al. Data to Estimate R                                       175
Quadratic Analysis with Proc GLM and Repeated Measurements
Note that WeightVar cannot be used in the standard way because it applies
at the "subject" level as opposed to the "observation" level.

The GLM Procedure
Repeated Measures Analysis of Variance
Tests of Hypotheses for Between Subjects Effects

| Source | DF | Type III SS | Mean Square | F Value | Pr > F |
|--------|----|-----------| ------------|---------|--------|
| Pressure | 1 | 4663.761855 | 4663.761855 | 133438 | <.0001 |
| PressSq | 1 | 80.252524 | 80.252524 | 2296.15 | <.0001 |
| Error | 11 | 0.384460 | 0.034951 | | |

Analysis of the Moldover et al. Data to Estimate R                                    176
Quadratic Analysis with Proc GLM and Repeated Measurements
Note that WeightVar cannot be used in the standard way because it applies
at the "subject" level as opposed to the "observation" level.

The GLM Procedure
Repeated Measures Analysis of Variance
Univariate Tests of Hypotheses for Within Subject Effects

| Source | DF | Type III SS | Mean Square | F Value | Pr > F |
|--------|-----|-------------|-------------|---------|--------|
| Mode | 4 | 0.60138197 | 0.15034549 | 11.29 | <.0001 |
| Mode*Pressure | 4 | 0.35972111 | 0.08993028 | 6.75 | 0.0003 |
| Mode*PressSq | 4 | 0.27565992 | 0.06891498 | 5.18 | 0.0017 |
| Error(Mode) | 44 | 0.58591874 | 0.01331633 | | |

| | Adj Pr > F | |
|--------|-------|-------|
| Source | G - G | H - F |
| Mode | 0.0026 | 0.0009 |
| Mode*Pressure | 0.0146 | 0.0079 |
| Mode*PressSq | 0.0302 | 0.0195 |
| Error(Mode) | | |

| | |
|--------|--------|
| Greenhouse-Geisser Epsilon | 0.3332 |
| Huynh-Feldt Epsilon | 0.4308 |

Analysis of the Moldover et al. Data to Estimate R     177
Duplicate the Quadratic Analysis (earlier done with Proc REG) with Proc GLM

The GLM Procedure

Number of Observations Read     70
Number of Observations Used     70

Analysis of the Moldover et al. Data to Estimate R                          178
Duplicate the Quadratic Analysis (earlier done with Proc REG) with Proc GLM

The GLM Procedure

Dependent Variable: SpeedSq   SpeedSq (m^2/s^2)

Weight: WeightVar   Weight Variable

| Source | DF | Sum of Squares | Mean Square | F Value | Pr > F |
|--------|-----|------|------|------|------|
| Model | 2 | 47749338.65 | 23874669.32 | 1.893E7 | <.0001 |
| Error | 67 | 84.48 | 1.26 | | |
| Corrected Total | 69 | 47749423.13 | | | |

| R-Square | Coeff Var | Root MSE | SpeedSq Mean |
|--------|------|------|------|
| 0.999998 | 0.001184 | 1.122905 | 94822.73 |

| Source | DF | Type I SS | Mean Square | F Value | Pr > F |
|--------|-----|------|------|------|------|
| Pressure | 1 | 47715919.56 | 47715919.56 | 3.784E7 | <.0001 |
| PressSq | 1 | 33419.08 | 33419.08 | 26503.8 | <.0001 |

| Source | DF | Type III SS | Mean Square | F Value | Pr > F |
|--------|-----|------|------|------|------|
| Pressure | 1 | 1751660.758 | 1751660.758 | 1389198 | <.0001 |
| PressSq | 1 | 33419.081 | 33419.081 | 26503.8 | <.0001 |

| Parameter | Estimate | Standard Error | t Value | Pr > |t| | 95% Confidence Limits | |
|--------|------|------|------|------|------|------|
| Intercept | 94756.25089 | 0.02247974 | 4215184 | <.0001 | 94756.20602 | 94756.29576 |
| Pressure | 0.00022 | 0.00000019 | 1178.64 | <.0001 | 0.00022 | 0.00022 |
| PressSq | 0.00000 | 0.00000000 | 162.80 | <.0001 | 0.00000 | 0.00000 |

Analysis of the Moldover et al. Data to Estimate R                                    179
Duplicate the Quadratic Analysis with Proc Mixed
Intercept is almost identical to value from Proc REG except for last digit.
Standard errors of intercept differ in the fourth digit.
The SINGRES option of the model statement is needed to yield non-zero
estimates of the standard errors of the Pressure and Pressure-squared parameters.
The NOPROFILE options of the proc mixed statement is required for the same
standard error of b0 as with NLIN and REG.

The Mixed Procedure

                        Model Information

Data Set                         WORK.MERGED
Dependent Variable               SpeedSq
Weight Variable                  WeightVar
Covariance Structure             Variance Components
Estimation Method                REML
Residual Variance Method         Parameter
Fixed Effects SE Method          Model-Based
Degrees of Freedom Method        Between-Within

                 Dimensions

Covariance Parameters             1
Columns in X                      3
Columns in Z                      0
Subjects                         70
Max Obs Per Subject               1

              Number of Observations

Number of Observations Read           70
Number of Observations Used           70
Number of Observations Not Used        0

                Iteration History

Iteration     Evaluations     -2 Res Log Like        Criterion

        0              1         -96.70702346
        1              1         -97.13279808     0.00000000

Analysis of the Moldover et al. Data to Estimate R                              180
Duplicate the Quadratic Analysis with Proc Mixed
Intercept is almost identical to value from Proc REG except for last digit.
Standard errors of intercept differ in the fourth digit.
The SINGRES option of the model statement is needed to yield non-zero
estimates of the standard errors of the Pressure and Pressure-squared parameters.
The NOPROFILE options of the proc mixed statement is required for the same
standard error of b0 as with NLIN and REG.

The Mixed Procedure

                        Convergence criteria met.

Covariance Parameter
      Estimates

Cov Parm     Estimate

Residual       1.2609

            Fit Statistics

-2 Res Log Likelihood          -97.1
AIC (smaller is better)        -95.1
AICC (smaller is better)       -95.1
BIC (smaller is better)        -92.9

   Null Model Likelihood Ratio Test

   DF     Chi-Square      Pr > ChiSq

    0        0.43          1.0000

                 Solution for Fixed Effects

                              Standard
Effect              Estimate     Error     DF    t Value    Pr > |t|    Alpha

Intercept             94756     0.02248     67    4215184     <.0001      0.05
Pressure           0.000224    1.904E-7     67    1178.64     <.0001      0.05
Pressure*Pressure   5.48E-11   3.37E-13     67     162.80     <.0001      0.05

Analysis of the Moldover et al. Data to Estimate R                181
Duplicate the Quadratic Analysis with Proc Mixed
Intercept is almost identical to value from Proc REG except for last digit.
Standard errors of intercept differ in the fourth digit.
The SINGRES option of the model statement is needed to yield non-zero
estimates of the standard errors of the Pressure and Pressure-squared parameters.
The NOPROFILE options of the proc mixed statement is required for the same
standard error of b0 as with NLIN and REG.

The Mixed Procedure

       Solution for Fixed Effects

Effect                  Lower       Upper

Intercept                94756       94756
Pressure             0.000224    0.000225
Pressure*Pressure    5.41E-11    5.55E-11

         Covariance Matrix for Fixed Effects

Row    Effect               Col1        Col2        Col3

  1    Intercept         0.000505      -3.9E-9     6.21E-15
  2    Pressure           -3.9E-9     3.63E-14    -626E-22
  3    Pressure*Pressure 6.21E-15    -626E-22     1.13E-25

        Type 3 Tests of Fixed Effects

                   Num     Den
Effect              DF      DF    F Value    Pr > F

Pressure             1      67    1389198    <.0001
Pressure*Pressure    1      67    26503.8    <.0001

Analysis of the Moldover et al. Data to Estimate R　　　　182
Duplicate the Quadratic Analysis with Proc Mixed
Intercept is almost identical to value from Proc REG except for last digit.
Standard errors of intercept differ in the fourth digit.
The SINGRES option of the model statement is needed to yield non-zero
estimates of the standard errors of the Pressure and Pressure-squared parameters.
The NOPROFILE options of the proc mixed statement is required for the same
standard error of b0 as with NLIN and REG.

| Obs | Effect | Estimate | StdErr | DF | tValue | Probt |
|---|---|---|---|---|---|---|
| 1 | Intercept | 9.4756251E+04 | 2.25E-02 | 67 | 4215184.24 | <.0001 |
| 2 | Pressure | 2.2446960E-04 | 1.90E-07 | 67 | 1178.64 | <.0001 |
| 3 | Pressure*Pressure | 5.4783954E-11 | 3.37E-13 | 67 | 162.80 | <.0001 |

Analysis of the Moldover et al. Data to Estimate R
Duplicate the Quadratic Analysis with Proc Mixed
Intercept is almost identical to value from Proc REG except for last digit.
Standard errors of intercept differ in the fourth digit.
The SINGRES option of the model statement is needed to yield non-zero
estimates of the standard errors of the Pressure and Pressure-squared parameters.
The NOPROFILE options of the proc mixed statement is required for the same
standard error of bO as with NLIN and REG.

| Estimate of Intercept from Quadratic Model via Mixed | Estimate of Standard Error of Intercept from Quadratic Model via Mixed | Estimate of R from Quadratic Model via Mixed |
|---|---|---|
| 9.4756250893325000E+04 | 2.2479741205484000E-02 | 8.3144773564461000E+00 |

Analysis of the Moldover et al. Data to Estimate R                                  184
Perform a Repeated Measurements Analysis with Moldover et al. Model and Proc Mixed
The MAXFUNC, CONVH, and SINGRES options are needed to obtain a solution

The Mixed Procedure

### Model Information

| | |
|---|---|
| Data Set | WORK.MERGED |
| Dependent Variable | SpeedSq |
| Weight Variable | WeightVar |
| Covariance Structure | Compound Symmetry |
| Subject Effect | Pressure |
| Estimation Method | REML |
| Residual Variance Method | Profile |
| Fixed Effects SE Method | Model-Based |
| Degrees of Freedom Method | Between-Within |

### Dimensions

| | |
|---|---|
| Covariance Parameters | 2 |
| Columns in X | 5 |
| Columns in Z | 0 |
| Subjects | 14 |
| Max Obs Per Subject | 5 |

### Number of Observations

| | |
|---|---|
| Number of Observations Read | 70 |
| Number of Observations Used | 70 |
| Number of Observations Not Used | 0 |

### Iteration History

| CovP1 | CovP2 | Iteration | Evaluations | -2 Res Log Like | Criterion |
|---|---|---|---|---|---|
| 0 | 1.2398 | 0 | 1 | -39.53061925 | |
| 0.3075 | 1.0124 | 1 | 2 | -43.86960122 | 0.00000037 |
| 0.3086 | 1.0112 | 2 | 1 | -43.91983693 | 0.00000004 |

Convergence criteria met.

Perform a Repeated Measurements Analysis with Moldover et al. Model and Proc Mixed
The MAXFUNC, CONVH, and SINGRES options are needed to obtain a solution

The Mixed Procedure

### Estimated R Matrix for Subject 1/Weighted by WeightVar

| Row | Col1 | Col2 | Col3 | Col4 | Col5 |
|-----|--------|---------|----------|----------|----------|
| 1 | 2.0111 | 0.1629 | 0.08508 | 0.05420 | 0.03910 |
| 2 | 0.1629 | 0.2414 | 0.02948 | 0.01878 | 0.01355 |
| 3 | 0.08508 | 0.02948 | 0.06585 | 0.009808 | 0.007075 |
| 4 | 0.05420 | 0.01878 | 0.009808 | 0.02672 | 0.004507 |
| 5 | 0.03910 | 0.01355 | 0.007075 | 0.004507 | 0.01390 |

### Estimated R Matrix for Subject 2/Weighted by WeightVar

| Row | Col1 | Col2 | Col3 | Col4 | Col5 |
|-----|---------|---------|---------|---------|---------|
| 1 | 0.1436 | 0.01271 | 0.007680 | 0.005846 | 0.005029 |
| 2 | 0.01271 | 0.02057 | 0.002907 | 0.002213 | 0.001903 |
| 3 | 0.007680 | 0.002907 | 0.007514 | 0.001337 | 0.001150 |
| 4 | 0.005846 | 0.002213 | 0.001337 | 0.004354 | 0.000876 |
| 5 | 0.005029 | 0.001903 | 0.001150 | 0.000876 | 0.003222 |

### Estimated R Matrix for Subject 3/Weighted by WeightVar

| Row | Col1 | Col2 | Col3 | Col4 | Col5 |
|-----|---------|---------|---------|---------|---------|
| 1 | 0.1433 | 0.01269 | 0.007668 | 0.005839 | 0.005023 |
| 2 | 0.01269 | 0.02054 | 0.002903 | 0.002210 | 0.001901 |
| 3 | 0.007668 | 0.002903 | 0.007505 | 0.001336 | 0.001149 |
| 4 | 0.005839 | 0.002210 | 0.001336 | 0.004351 | 0.000875 |
| 5 | 0.005023 | 0.001901 | 0.001149 | 0.000875 | 0.003220 |

### Estimated R Matrix for Subject 4/Weighted by WeightVar

| Row | Col1 | Col2 | Col3 | Col4 | Col5 |
|-----|---------|---------|---------|---------|---------|
| 1 | 0.03241 | 0.003383 | 0.002467 | 0.002158 | 0.002025 |
| 2 | 0.003383 | 0.006459 | 0.001101 | 0.000963 | 0.000904 |
| 3 | 0.002467 | 0.001101 | 0.003435 | 0.000703 | 0.000659 |
| 4 | 0.002158 | 0.000963 | 0.000703 | 0.002629 | 0.000577 |
| 5 | 0.002025 | 0.000904 | 0.000659 | 0.000577 | 0.002314 |

Analysis of the Moldover et al. Data to Estimate R                                186
Perform a Repeated Measurements Analysis with Moldover et al. Model and Proc Mixed
The MAXFUNC, CONVH, and SINGRES options are needed to obtain a solution

The Mixed Procedure

### Estimated R Matrix for Subject 5/Weighted by WeightVar

| Row | Col1 | Col2 | Col3 | Col4 | Col5 |
|-----|------|------|------|------|------|
| 1 | 0.01272 | 0.001609 | 0.001335 | 0.001244 | 0.001204 |
| 2 | 0.001609 | 0.003722 | 0.000722 | 0.000673 | 0.000651 |
| 3 | 0.001335 | 0.000722 | 0.002563 | 0.000559 | 0.000541 |
| 4 | 0.001244 | 0.000673 | 0.000559 | 0.002227 | 0.000504 |
| 5 | 0.001204 | 0.000651 | 0.000541 | 0.000504 | 0.002086 |

### Estimated R Matrix for Subject 14/Weighted by WeightVar

| Row | Col1 | Col2 | Col3 | Col4 | Col5 |
|-----|------|------|------|------|------|
| 1 | 0.001986 | 0.000455 | 0.000452 | 0.000451 | 0.000451 |
| 2 | 0.000455 | 0.001903 | 0.000443 | 0.000442 | 0.000441 |
| 3 | 0.000452 | 0.000443 | 0.001883 | 0.000439 | 0.000439 |
| 4 | 0.000451 | 0.000442 | 0.000439 | 0.001875 | 0.000438 |
| 5 | 0.000451 | 0.000441 | 0.000439 | 0.000438 | 0.001872 |

### Covariance Parameter Estimates

| Cov Parm | Subject | Estimate |
|----------|---------|----------|
| CS | Pressure | 0.3086 |
| Residual | | 1.0112 |

### Fit Statistics

| | |
|---|---|
| -2 Res Log Likelihood | -43.9 |
| AIC (smaller is better) | -39.9 |
| AICC (smaller is better) | -39.7 |
| BIC (smaller is better) | -38.6 |

Analysis of the Moldover et al. Data to Estimate R                                187
Perform a Repeated Measurements Analysis with Moldover et al. Model and Proc Mixed
The MAXFUNC, CONVH, and SINGRES options are needed to obtain a solution

The Mixed Procedure

Null Model Likelihood Ratio Test

| DF | Chi-Square | Pr > ChiSq |
|----|-----------|-----------|
| 1  | 4.39      | 0.0362    |

Solution for Fixed Effects

| Effect | Estimate | Standard Error | DF | t Value | Pr > \|t\| | Alpha |
|--------|----------|----------------|-----|---------|-----------|-------|
| Intercept | 94756 | 0.1674 | 9 | 566039 | <.0001 | 0.05 |
| Pressure | 0.000223 | 1.575E-6 | 9 | 141.51 | <.0001 | 0.05 |
| Pressure*Pressure | 6.08E-11 | 5.26E-12 | 9 | 11.55 | <.0001 | 0.05 |
| Pressu*Pressu*Pressu | -669E-20 | 5.6E-18 | 9 | -1.20 | 0.2624 | 0.05 |
| PressInv | -3655.96 | 5233.62 | 9 | -0.70 | 0.5025 | 0.05 |

Solution for Fixed Effects

| Effect | Lower | Upper |
|--------|-------|-------|
| Intercept | 94756 | 94757 |
| Pressure | 0.000219 | 0.000226 |
| Pressure*Pressure | 4.89E-11 | 7.27E-11 |
| Pressu*Pressu*Pressu | -194E-19 | 5.97E-18 |
| PressInv | -15495 | 8183.31 |

Covariance Matrix for Fixed Effects

| Row | Effect | Col1 | Col2 | Col3 | Col4 | Col5 |
|-----|--------|------|------|------|------|------|
| 1 | Intercept | 0.02802 | -2.56E-7 | 8.09E-13 | -813E-21 | -833.28 |
| 2 | Pressure | -2.56E-7 | 2.48E-12 | -816E-20 | 8.42E-24 | 0.007134 |
| 3 | Pressure*Pressure | 8.09E-13 | -816E-20 | 2.77E-23 | -292E-31 | -2.18E-8 |
| 4 | Pressu*Pressu*Pressu | -813E-21 | 8.42E-24 | -292E-31 | 3.13E-35 | 2.14E-14 |
| 5 | PressInv | -833.28 | 0.007134 | -2.18E-8 | 2.14E-14 | 27390787 |

Perform a Repeated Measurements Analysis with Moldover et al. Model and Proc Mixed
The MAXFUNC, CONVH, and SINGRES options are needed to obtain a solution

The Mixed Procedure

### Type 3 Tests of Fixed Effects

| Effect | Num DF | Den DF | F Value | Pr > F |
|--------|--------|--------|---------|--------|
| Pressure | 1 | 9 | 20023.8 | <.0001 |
| Pressure*Pressure | 1 | 9 | 133.45 | <.0001 |
| Pressu*Pressu*Pressu | 1 | 9 | 1.43 | 0.2624 |
| PressInv | 1 | 9 | 0.49 | 0.5025 |

Perform a Repeated Measurements Analysis with Moldover et al. Model and Proc Mixed
The MAXFUNC, CONVH, and SINGRES options are needed to obtain a solution

| Obs | Effect | Estimate | StdErr | DF | tValue | Probt |
|---|---|---|---|---|---|---|
| 1 | Intercept | 9.4756396E+04 | 1.67E-01 | 9 | 566039.11 | <.0001 |
| 2 | Pressure | 2.2282527E-04 | 1.57E-06 | 9 | 141.51 | <.0001 |
| 3 | Pressure*Pressure | 6.0810050E-11 | 5.26E-12 | 9 | 11.55 | <.0001 |
| 4 | Pressu*Pressu*Pressu | -6.6933106E-18 | 5.60E-18 | 9 | -1.20 | 0.2624 |
| 5 | PressInv | -3.6559645E+03 | 5.23E+03 | 9 | -0.70 | 0.5025 |

Analysis of the Moldover et al. Data to Estimate R
Perform a Repeated Measurements Analysis with Moldover et al. Model and Proc Mixed
The MAXFUNC, CONVH, and SINGRES options are needed to obtain a solution

| Estimate of Intercept from Rep. Meas. Moldover et al. Model | Estimate of Standard Error of Intercept from Rep. Meas. Moldover et al. Model | Estimate of R from Rep. Meas. Moldover et al. Model |
|---|---|---|
| 9.4756396154510000E+04 | 1.6740256007757000E-01 | 8.3144901025269000E+00 |

Analysis of the Moldover et al. Data to Estimate R                    191
Perform a Repeated Measurements Quadratic Analysis with Proc Mixed
without weighting and with a GROUP variable.
The MAXFUNC, MAXITER, CONVH, and SINGRES options are needed to obtain a solution

The Mixed Procedure

              Model Information

Data Set                      WORK.MERGED2
Dependent Variable            SpeedSq
Covariance Structure          Compound Symmetry
Subject Effect                Pressure
Group Effect                  GroupVar
Estimation Method             REML
Residual Variance Method      None
Fixed Effects SE Method       Model-Based
Degrees of Freedom Method     Between-Within

              Dimensions

Covariance Parameters              6
Columns in X                       3
Columns in Z                       0
Subjects                          14
Max Obs Per Subject                5

           Number of Observations

Number of Observations Read        70
Number of Observations Used        70
Number of Observations Not Used     0

                    Parameter Search

   CovP1      CovP2      CovP3      CovP4      CovP5      CovP6         Res Log Like

  0.2500    0.03000    0.01695         0    0.001970   0.000977            44.3284

   Parameter
    Search

-2 Res Log Like

      -88.6568

Analysis of the Moldover et al. Data to Estimate R                    192
Perform a Repeated Measurements Quadratic Analysis with Proc Mixed
without weighting and with a GROUP variable.
The MAXFUNC, MAXITER, CONVH, and SINGRES options are needed to obtain a solution

The Mixed Procedure

### Iteration History

| CovP1 | CovP2 | CovP3 | CovP4 | CovP5 | CovP6 | Iteration | Evaluations |
|---|---|---|---|---|---|---|---|
| 0.2500 | 0.03000 | 0.01664 | -0.00180 | 0.001970 | 0.000980 | 1 | 2 |

### Iteration History

| -2 Res Log Like | Criterion |
|---|---|
| -89.91849534 | 4230.6883079 |

### Iteration History

| CovP1 | CovP2 | CovP3 | CovP4 | CovP5 | CovP6 | Iteration | Evaluations |
|---|---|---|---|---|---|---|---|
| 0.2500 | 0.03000 | 0.01641 | -0.00304 | 0.001970 | 0.000998 | 2 | 1 |

### Iteration History

| -2 Res Log Like | Criterion |
|---|---|
| -91.74545784 | 35315.000264 |

### Iteration History

| CovP1 | CovP2 | CovP3 | CovP4 | CovP5 | CovP6 | Iteration | Evaluations |
|---|---|---|---|---|---|---|---|
| 0.2500 | 0.03000 | 0.01639 | -0.00315 | 0.001970 | 0.000994 | 3 | 2 |

### Iteration History

| -2 Res Log Like | Criterion |
|---|---|
| -91.98721123 | 0.00033543 |

### Iteration History

| CovP1 | CovP2 | CovP3 | CovP4 | CovP5 | CovP6 | Iteration | Evaluations |
|---|---|---|---|---|---|---|---|
| 0.2529 | 0.03703 | 0.01702 | -0.00331 | 0.001970 | 0.000978 | 4 | 1 |

Perform a Repeated Measurements Quadratic Analysis with Proc Mixed
without weighting and with a GROUP variable.
The MAXFUNC, MAXITER, CONVH, and SINGRES options are needed to obtain a solution

The Mixed Procedure

     Iteration History

-2 Res Log Like     Criterion

   -92.09305408     0.00005206

Convergence criteria met.

Estimated R Matrix for Subject 1

| Row | Col1 | Col2 | Col3 | Col4 | Col5 |
|-----|------|------|------|------|------|
| 1 | 0.2899 | 0.03703 | 0.03703 | 0.03703 | 0.03703 |
| 2 | 0.03703 | 0.2899 | 0.03703 | 0.03703 | 0.03703 |
| 3 | 0.03703 | 0.03703 | 0.2899 | 0.03703 | 0.03703 |
| 4 | 0.03703 | 0.03703 | 0.03703 | 0.2899 | 0.03703 |
| 5 | 0.03703 | 0.03703 | 0.03703 | 0.03703 | 0.2899 |

Estimated R Matrix for Subject 2

| Row | Col1 | Col2 | Col3 | Col4 | Col5 |
|-----|------|------|------|------|------|
| 1 | 0.01371 | -0.00331 | -0.00331 | -0.00331 | -0.00331 |
| 2 | -0.00331 | 0.01371 | -0.00331 | -0.00331 | -0.00331 |
| 3 | -0.00331 | -0.00331 | 0.01371 | -0.00331 | -0.00331 |
| 4 | -0.00331 | -0.00331 | -0.00331 | 0.01371 | -0.00331 |
| 5 | -0.00331 | -0.00331 | -0.00331 | -0.00331 | 0.01371 |

Estimated R Matrix for Subject 3

| Row | Col1 | Col2 | Col3 | Col4 | Col5 |
|-----|------|------|------|------|------|
| 1 | 0.01371 | -0.00331 | -0.00331 | -0.00331 | -0.00331 |
| 2 | -0.00331 | 0.01371 | -0.00331 | -0.00331 | -0.00331 |
| 3 | -0.00331 | -0.00331 | 0.01371 | -0.00331 | -0.00331 |
| 4 | -0.00331 | -0.00331 | -0.00331 | 0.01371 | -0.00331 |

Analysis of the Moldover et al. Data to Estimate R                                        194
Perform a Repeated Measurements Quadratic Analysis with Proc Mixed
without weighting and with a GROUP variable.
The MAXFUNC, MAXITER, CONVH, and SINGRES options are needed to obtain a solution

The Mixed Procedure

### Estimated R Matrix for Subject 3

| Row | Col1 | Col2 | Col3 | Col4 | Col5 |
|-----|------|------|------|------|------|
| 5 | -0.00331 | -0.00331 | -0.00331 | -0.00331 | 0.01371 |

### Estimated R Matrix for Subject 4

| Row | Col1 | Col2 | Col3 | Col4 | Col5 |
|-----|------|------|------|------|------|
| 1 | 0.002948 | 0.000978 | 0.000978 | 0.000978 | 0.000978 |
| 2 | 0.000978 | 0.002948 | 0.000978 | 0.000978 | 0.000978 |
| 3 | 0.000978 | 0.000978 | 0.002948 | 0.000978 | 0.000978 |
| 4 | 0.000978 | 0.000978 | 0.000978 | 0.002948 | 0.000978 |
| 5 | 0.000978 | 0.000978 | 0.000978 | 0.000978 | 0.002948 |

### Estimated R Matrix for Subject 5

| Row | Col1 | Col2 | Col3 | Col4 | Col5 |
|-----|------|------|------|------|------|
| 1 | 0.002948 | 0.000978 | 0.000978 | 0.000978 | 0.000978 |
| 2 | 0.000978 | 0.002948 | 0.000978 | 0.000978 | 0.000978 |
| 3 | 0.000978 | 0.000978 | 0.002948 | 0.000978 | 0.000978 |
| 4 | 0.000978 | 0.000978 | 0.000978 | 0.002948 | 0.000978 |
| 5 | 0.000978 | 0.000978 | 0.000978 | 0.000978 | 0.002948 |

### Estimated R Matrix for Subject 14

| Row | Col1 | Col2 | Col3 | Col4 | Col5 |
|-----|------|------|------|------|------|
| 1 | 0.002948 | 0.000978 | 0.000978 | 0.000978 | 0.000978 |
| 2 | 0.000978 | 0.002948 | 0.000978 | 0.000978 | 0.000978 |
| 3 | 0.000978 | 0.000978 | 0.002948 | 0.000978 | 0.000978 |
| 4 | 0.000978 | 0.000978 | 0.000978 | 0.002948 | 0.000978 |
| 5 | 0.000978 | 0.000978 | 0.000978 | 0.000978 | 0.002948 |

Perform a Repeated Measurements Quadratic Analysis with Proc Mixed
without weighting and with a GROUP variable.
The MAXFUNC, MAXITER, CONVH, and SINGRES options are needed to obtain a solution

The Mixed Procedure

       Covariance Parameter Estimates

Cov Parm       Subject      Group       Estimate

Variance       Pressure     Group 1      0.2529
CS             Pressure     Group 1      0.03703
Variance       Pressure     Group 2      0.01702
CS             Pressure     Group 2     -0.00331
Variance       Pressure     Group 3      0.001970
CS             Pressure     Group 3      0.000978

            Fit Statistics

-2 Res Log Likelihood          -92.1
AIC (smaller is better)        -80.1
AICC (smaller is better)       -78.7
BIC (smaller is better)        -76.3

    PARMS Model Likelihood Ratio Test

    DF     Chi-Square      Pr > ChiSq

     6        3.44            0.7524

              Solution for Fixed Effects

                          Standard
Effect            Estimate      Error      DF    t Value    Pr > |t|    Alpha

Intercept         94756       0.01398      11    6779382    <.0001      0.05

        Solution for Fixed Effects

Effect              Lower       Upper

Intercept          94756       94756

Analysis of the Moldover et al. Data to Estimate R                196
Perform a Repeated Measurements Quadratic Analysis with Proc Mixed
without weighting and with a GROUP variable.
The MAXFUNC, MAXITER, CONVH, and SINGRES options are needed to obtain a solution

The Mixed Procedure

### Solution for Fixed Effects

| Effect | Estimate | Standard Error | DF | t Value | Pr > \|t\| | Alpha |
|---|---|---|---|---|---|---|
| Pressure | 0.000224 | 2.277E-7 | 11 | 984.97 | <.0001 | 0.05 |
| Pressure*Pressure | 5.51E-11 | 4.77E-13 | 11 | 115.66 | <.0001 | 0.05 |

### Solution for Fixed Effects

| Effect | Lower | Upper |
|---|---|---|
| Pressure | 0.000224 | 0.000225 |
| Pressure*Pressure | 5.41E-11 | 5.62E-11 |

### Covariance Matrix for Fixed Effects

| Row | Effect | Col1 | Col2 | Col3 |
|---|---|---|---|---|
| 1 | Intercept | 0.000195 | -2.82E-9 | 5.5E-15 |
| 2 | Pressure | -2.82E-9 | 5.18E-14 | -106E-21 |
| 3 | Pressure*Pressure | 5.5E-15 | -106E-21 | 2.27E-25 |

### Type 3 Tests of Fixed Effects

| Effect | Num DF | Den DF | F Value | Pr > F |
|---|---|---|---|---|
| Pressure | 1 | 11 | 970169 | <.0001 |
| Pressure*Pressure | 1 | 11 | 13376.8 | <.0001 |

Perform a Repeated Measurements Quadratic Analysis with Proc Mixed
without weighting and with a GROUP variable.
The MAXFUNC, MAXITER, CONVH, and SINGRES options are needed to obtain a solution

| Obs | Effect | Estimate | StdErr | DF | tValue | Probt |
|-----|--------|----------|--------|-----|--------|-------|
| 1 | Intercept | 9.4756287E+04 | 1.40E-02 | 11 | 6779381.82 | <.0001 |
| 2 | Pressure | 2.2422904E-04 | 2.28E-07 | 11 | 984.97 | <.0001 |
| 3 | Pressure*Pressure | 5.5139933E-11 | 4.77E-13 | 11 | 115.66 | <.0001 |

Analysis of the Moldover et al. Data to Estimate R
Perform a Repeated Measurements Quadratic Analysis with Proc Mixed
without weighting and with a GROUP variable.
The MAXFUNC, MAXITER, CONVH, and SINGRES options are needed to obtain a solution

| Estimate of Intercept from GROUPED Rep. Meas. Quadratic Model | Estimate of Standard Error of Intercept from GROUPED Rep. Meas. Quadratic Model | Estimate of R from GROUPED Rep. Meas. Quadratic Model |
|---|---|---|
| 9.4756287274020000E+04 | 1.3977127966840000E-02 | 8.3144805487048000E+00 |

Analysis of the Moldover et al. Data to Estimate R                                199
Repeat the quadratic fit, with a rescaled version of SpeedSq in which
SpeedSqResc = SpeedSq - 94,808.35, where 94,808.35 is the rough average
of the SpeedSq values.  OMIT THE WEIGHTING.

The REG Procedure
Model: MODEL1
Dependent Variable: SpeedSqResc Rescaled Speed-Squared = SpeedSq - 94,808.35

Number of Observations Read          70
Number of Observations Used          70

### Analysis of Variance

| Source | DF | Sum of Squares | Mean Square | F Value | Pr > F |
|--------|----|----|----|----|----|
| Model | 2 | 108023 | 54012 | 2234623 | <.0001 |
| Error | 67 | 1.61942 | 0.02417 | | |
| Corrected Total | 69 | 108025 | | | |

| | | | | |
|--------|--------|--------|--------|--------|
| Root MSE | 0.15547 | R-Square | 1.0000 | |
| Dependent Mean | 0.00399 | Adj R-Sq | 1.0000 | |
| Coeff Var | 3900.63944 | | | |

### Parameter Estimates

| Variable | Label | DF | Parameter Estimate | Standard Error | t Value |
|----------|-------|----|----|----|----|
| Intercept | Intercept | 1 | -51.95381 | 0.04865 | -1068.0 |
| Pressure | Pressure (Pa) | 1 | 0.00022334 | 5.084336E-7 | 439.26 |
| PressSq | Pressure Squared | 1 | 5.65996E-11 | 9.82259E-13 | 57.62 |

### Parameter Estimates

| Variable | Label | DF | Pr > |t| |
|----------|-------|----|----|
| Intercept | Intercept | 1 | <.0001 |
| Pressure | Pressure (Pa) | 1 | <.0001 |
| PressSq | Pressure Squared | 1 | <.0001 |

Analysis of the Moldover et al. Data to Estimate R
Repeat the quadratic fit, with a rescaled version of SpeedSq in which
SpeedSqResc = SpeedSq - 94,808.35, where 94,808.35 is the rough average
of the SpeedSq values.  OMIT THE WEIGHTING.

Here is a more accurate table of the values of the parameters:

| Variable | DF | Parameter Estimate | Standard Error | t | p |
|---|---|---|---|---|---|
| Intercept | 1 | -5.1953812E+01 | 4.86E-02 | -1068.0 | <.0001 |
| Pressure | 1 | 2.2333687E-04 | 5.08E-07 | 439.3 | <.0001 |
| PressSq | 1 | 5.6599632E-11 | 9.82E-13 | 57.6 | <.0001 |

Analysis of the Moldover et al. Data to Estimate R
Repeat the quadratic fit, with a rescaled version of SpeedSq in which
SpeedSqResc = SpeedSq - 94,808.35, where 94,808.35 is the rough average
of the SpeedSq values.  OMIT THE WEIGHTING.

Here is the estimate of R from the unweighted quadratic regression.

       Estimate of R from
      Unweighted Quadratic
              Regression
    ────────────────────────
    8.3144901054809000E+00

Analysis of the Moldover et al. Data to Estimate R
Rerun Repeated Measurements Analysis to get the RMS Average Weighted Residual
Note that some of the relevant output is in the SAS log.

The Mixed Procedure

### Model Information

| | |
|---|---|
| Data Set | WORK.MERGED |
| Dependent Variable | SpeedSq |
| Weight Variable | WeightVar |
| Covariance Structure | Compound Symmetry |
| Subject Effect | Pressure |
| Estimation Method | REML |
| Residual Variance Method | Profile |
| Fixed Effects SE Method | Model-Based |
| Degrees of Freedom Method | Between-Within |

### Dimensions

| | |
|---|---|
| Covariance Parameters | 2 |
| Columns in X | 3 |
| Columns in Z | 0 |
| Subjects | 14 |
| Max Obs Per Subject | 5 |

### Number of Observations

| | |
|---|---|
| Number of Observations Read | 70 |
| Number of Observations Used | 70 |
| Number of Observations Not Used | 0 |

### Iteration History

| Iteration | Evaluations | -2 Res Log Like | Criterion |
|---|---|---|---|
| 0 | 1 | -96.87722971 | |
| 1 | 2 | -101.81022097 | 0.00000000 |

Convergence criteria met.

Rerun Repeated Measurements Analysis to get the RMS Average Weighted Residual
Note that some of the relevant output is in the SAS log.

The Mixed Procedure

### Estimated R Matrix for Subject 1/Weighted by WeightVar

| Row | Col1 | Col2 | Col3 | Col4 | Col5 |
|---|---|---|---|---|---|
| 1 | 2.0005 | 0.1639 | 0.08558 | 0.05452 | 0.03933 |
| 2 | 0.1639 | 0.2401 | 0.02965 | 0.01889 | 0.01363 |
| 3 | 0.08558 | 0.02965 | 0.06551 | 0.009866 | 0.007116 |
| 4 | 0.05452 | 0.01889 | 0.009866 | 0.02658 | 0.004533 |
| 5 | 0.03933 | 0.01363 | 0.007116 | 0.004533 | 0.01383 |

### Estimated R Matrix for Subject 2/Weighted by WeightVar

| Row | Col1 | Col2 | Col3 | Col4 | Col5 |
|---|---|---|---|---|---|
| 1 | 0.1429 | 0.01278 | 0.007725 | 0.005881 | 0.005059 |
| 2 | 0.01278 | 0.02046 | 0.002924 | 0.002226 | 0.001915 |
| 3 | 0.007725 | 0.002924 | 0.007474 | 0.001345 | 0.001157 |
| 4 | 0.005881 | 0.002226 | 0.001345 | 0.004332 | 0.000881 |
| 5 | 0.005059 | 0.001915 | 0.001157 | 0.000881 | 0.003205 |

### Estimated R Matrix for Subject 3/Weighted by WeightVar

| Row | Col1 | Col2 | Col3 | Col4 | Col5 |
|---|---|---|---|---|---|
| 1 | 0.1426 | 0.01276 | 0.007714 | 0.005873 | 0.005053 |
| 2 | 0.01276 | 0.02043 | 0.002920 | 0.002223 | 0.001913 |
| 3 | 0.007714 | 0.002920 | 0.007466 | 0.001344 | 0.001156 |
| 4 | 0.005873 | 0.002223 | 0.001344 | 0.004328 | 0.000880 |
| 5 | 0.005053 | 0.001913 | 0.001156 | 0.000880 | 0.003203 |

### Estimated R Matrix for Subject 4/Weighted by WeightVar

| Row | Col1 | Col2 | Col3 | Col4 | Col5 |
|---|---|---|---|---|---|
| 1 | 0.03224 | 0.003403 | 0.002481 | 0.002171 | 0.002037 |
| 2 | 0.003403 | 0.006425 | 0.001108 | 0.000969 | 0.000909 |
| 3 | 0.002481 | 0.001108 | 0.003417 | 0.000707 | 0.000663 |
| 4 | 0.002171 | 0.000969 | 0.000707 | 0.002616 | 0.000580 |
| 5 | 0.002037 | 0.000909 | 0.000663 | 0.000580 | 0.002302 |

Analysis of the Moldover et al. Data to Estimate R                            204
Rerun Repeated Measurements Analysis to get the RMS Average Weighted Residual
Note that some of the relevant output is in the SAS log.

The Mixed Procedure

### Estimated R Matrix for Subject 5/Weighted by WeightVar

| Row | Col1    | Col2     | Col3     | Col4     | Col5     |
|-----|---------|----------|----------|----------|----------|
| 1   | 0.01265 | 0.001618 | 0.001343 | 0.001251 | 0.001211 |
| 2   | 0.001618| 0.003703 | 0.000726 | 0.000677 | 0.000655 |
| 3   | 0.001343| 0.000726 | 0.002550 | 0.000562 | 0.000544 |
| 4   | 0.001251| 0.000677 | 0.000562 | 0.002215 | 0.000507 |
| 5   | 0.001211| 0.000655 | 0.000544 | 0.000507 | 0.002075 |

### Estimated R Matrix for Subject 14/Weighted by WeightVar

| Row | Col1     | Col2     | Col3     | Col4     | Col5     |
|-----|----------|----------|----------|----------|----------|
| 1   | 0.001976 | 0.000457 | 0.000455 | 0.000454 | 0.000453 |
| 2   | 0.000457 | 0.001893 | 0.000445 | 0.000444 | 0.000444 |
| 3   | 0.000455 | 0.000445 | 0.001873 | 0.000442 | 0.000442 |
| 4   | 0.000454 | 0.000444 | 0.000442 | 0.001866 | 0.000441 |
| 5   | 0.000453 | 0.000444 | 0.000442 | 0.000441 | 0.001862 |

### Covariance Parameter Estimates

| Cov Parm | Subject  | Estimate | Standard Error | Z Value | Pr Z   | Alpha | Lower   | Upper  |
|----------|----------|----------|----------------|---------|--------|-------|---------|--------|
| CS       | Pressure | 0.3104   | 0.2201         | 1.41    | 0.1585 | 0.05  | -0.1210 | 0.7418 |
| Residual |          | 1.0024   | 0.1902         | 5.27    | <.0001 | 0.05  | 0.7136  | 1.5113 |

### Fit Statistics

| | |
|---|---|
| -2 Res Log Likelihood    | -101.8 |
| AIC (smaller is better)  | -97.8  |
| AICC (smaller is better) | -97.6  |
| BIC (smaller is better)  | -96.5  |

Analysis of the Moldover et al. Data to Estimate R     205
Rerun Repeated Measurements Analysis to get the RMS Average Weighted Residual
Note that some of the relevant output is in the SAS log.

The Mixed Procedure

  Null Model Likelihood Ratio Test

   DF     Chi-Square      Pr > ChiSq

    1         4.93           0.0263

              Solution for Fixed Effects

                           Standard
Effect              Estimate       Error     DF    t Value    Pr > |t|    Alpha

Intercept             94756       0.02956     11   3205167     <.0001      0.05
Pressure            0.000225      2.585E-7     11    868.62     <.0001      0.05
Pressure*Pressure   5.47E-11      4.63E-13     11    118.08     <.0001      0.05

          Solution for Fixed Effects

Effect               Lower       Upper

Intercept             94756       94756
Pressure            0.000224    0.000225
Pressure*Pressure   5.37E-11    5.57E-11

          Covariance Matrix for Fixed Effects

Row    Effect                Col1        Col2        Col3

  1    Intercept          0.000874     -6.92E-9     1.11E-14
  2    Pressure           -6.92E-9      6.68E-14    -117E-21
  3    Pressure*Pressure   1.11E-14    -117E-21     2.15E-25

          Type 3 Tests of Fixed Effects

                   Num     Den
Effect              DF      DF    F Value    Pr > F

Pressure             1      11    754498     <.0001
Pressure*Pressure    1      11    13943.3    <.0001

Rerun Standard Regression with MIXED to get the RMS Average Weighted Residual
Note that some of the relevant output is in the SAS log.

The Mixed Procedure

### Model Information

| | |
|---|---|
| Data Set | WORK.MERGED |
| Dependent Variable | SpeedSq |
| Weight Variable | WeightVar |
| Covariance Structure | Diagonal |
| Estimation Method | REML |
| Residual Variance Method | Profile |
| Fixed Effects SE Method | Model-Based |
| Degrees of Freedom Method | Residual |

### Dimensions

| | |
|---|---|
| Covariance Parameters | 1 |
| Columns in X | 3 |
| Columns in Z | 0 |
| Subjects | 1 |
| Max Obs Per Subject | 70 |

### Number of Observations

| | |
|---|---|
| Number of Observations Read | 70 |
| Number of Observations Used | 70 |
| Number of Observations Not Used | 0 |

### Covariance Parameter Estimates

| Cov Parm | Estimate | Standard Error | Z Value | Pr > Z | Alpha | Lower | Upper |
|---|---|---|---|---|---|---|---|
| Residual | 1.2614 | 0.2179 | 5.79 | <.0001 | 0.05 | 0.9235 | 1.8269 |

### Fit Statistics

| | |
|---|---|
| -2 Res Log Likelihood | -96.9 |
| AIC (smaller is better) | -94.9 |
| AICC (smaller is better) | -94.8 |
| BIC (smaller is better) | -92.7 |

Analysis of the Moldover et al. Data to Estimate R     207
Rerun Standard Regression with MIXED to get the RMS Average Weighted Residual
Note that some of the relevant output is in the SAS log.

The Mixed Procedure

### Solution for Fixed Effects

| Effect | Estimate | Standard Error | DF | t Value | Pr > \|t\| | Alpha |
|--------|----------|----------------|----|---------|-----------|-------|
| Intercept | 94756 | 0.02248 | 68 | 4214334 | <.0001 | 0.05 |
| Pressure | 0.000224 | 1.905E-7 | 68 | 1178.40 | <.0001 | 0.05 |
| Pressure*Pressure | 5.48E-11 | 3.37E-13 | 68 | 162.77 | <.0001 | 0.05 |

### Solution for Fixed Effects

| Effect | Lower | Upper |
|--------|-------|-------|
| Intercept | 94756 | 94756 |
| Pressure | 0.000224 | 0.000225 |
| Pressure*Pressure | 5.41E-11 | 5.55E-11 |

### Covariance Matrix for Fixed Effects

| Row | Effect | Col1 | Col2 | Col3 |
|-----|--------|------|------|------|
| 1 | Intercept | 0.000506 | -3.9E-9 | 6.22E-15 |
| 2 | Pressure | -3.9E-9 | 3.63E-14 | -626E-22 |
| 3 | Pressure*Pressure | 6.22E-15 | -626E-22 | 1.13E-25 |

### Type 3 Tests of Fixed Effects

| Effect | Num DF | Den DF | F Value | Pr > F |
|--------|--------|--------|---------|--------|
| Pressure | 1 | 68 | 1388638 | <.0001 |
| Pressure*Pressure | 1 | 68 | 26493.1 | <.0001 |

Analysis of the Moldover et al. Data to Estimate R     208
Rerun Standard Regression with MIXED to get the RMS Average Weighted Residual
Note that some of the relevant output is in the SAS log.

The Mixed Procedure

Influence Diagnostics

| Deleted Obs. Index | Observed Value | Predicted Value | Residual | Leverage | PRESS Residual | Internally Studentized Residual | RMSE without deleted obs |
|---|---|---|---|---|---|---|---|
| 1 | 94,763.099 | 94762 | 1.1121 | 0.00018 | 1.1123 | 0.8022 | 1.12616 |
| 2 | 94,762.457 | 94762 | 0.4701 | 0.00146 | 0.4708 | 0.9794 | 1.12348 |
| 3 | 94,761.858 | 94762 | -0.1289 | 0.00535 | -0.1295 | -0.5150 | 1.12937 |
| 4 | 94,762.258 | 94762 | 0.2711 | 0.0132 | 0.2748 | 1.7078 | 1.10670 |
| 5 | 94,761.909 | 94762 | -0.0779 | 0.0253 | -0.0799 | -0.6841 | 1.12765 |
| 6 | 94,767.836 | 94768 | 0.2208 | 0.00162 | 0.2211 | 0.5963 | 1.12860 |
| 7 | 94,767.576 | 94768 | -0.0392 | 0.0113 | -0.0397 | -0.2815 | 1.13094 |
| 8 | 94,767.494 | 94768 | -0.1212 | 0.0309 | -0.1251 | -1.4533 | 1.11363 |
| 9 | 94,767.586 | 94768 | -0.0292 | 0.0534 | -0.0309 | -0.4659 | 1.12977 |
| 10 | 94,767.750 | 94768 | 0.1348 | 0.0721 | 0.1452 | 2.5209 | 1.07661 |
| 11 | 94,767.808 | 94768 | 0.1872 | 0.00162 | 0.1875 | 0.5063 | 1.12944 |
| 12 | 94,767.572 | 94768 | -0.0488 | 0.0113 | -0.0493 | -0.3500 | 1.13057 |
| 13 | 94,767.564 | 94768 | -0.0568 | 0.0310 | -0.0586 | -0.6808 | 1.12769 |
| 14 | 94,767.723 | 94768 | 0.1022 | 0.0534 | 0.1080 | 1.6295 | 1.10896 |
| 15 | 94,767.528 | 94768 | -0.0928 | 0.0721 | -0.1000 | -1.7358 | 1.10587 |
| 16 | 94,773.462 | 94773 | 0.0319 | 0.00467 | 0.0320 | 0.1816 | 1.13133 |
| 17 | 94,773.355 | 94773 | -0.0751 | 0.0234 | -0.0769 | -0.9673 | 1.12368 |
| 18 | 94,773.430 | 94773 | -0.0001 | 0.0441 | -0.0001 | -0.0019 | 1.13161 |
| 19 | 94,773.390 | 94773 | -0.0401 | 0.0576 | -0.0426 | -0.8241 | 1.12586 |
| 20 | 94,773.397 | 94773 | -0.0331 | 0.0654 | -0.0354 | -0.7282 | 1.12712 |
| 21 | 94,779.334 | 94779 | 0.0269 | 0.00807 | 0.0271 | 0.2446 | 1.13110 |
| 22 | 94,779.347 | 94779 | 0.0399 | 0.0276 | 0.0410 | 0.6776 | 1.12772 |
| 23 | 94,779.360 | 94779 | 0.0529 | 0.0400 | 0.0551 | 1.0899 | 1.12153 |
| 24 | 94,779.230 | 94779 | -0.0771 | 0.0461 | -0.0809 | -1.7121 | 1.10658 |
| 25 | 94,779.294 | 94779 | -0.0131 | 0.0492 | -0.0138 | -0.3019 | 1.13084 |
| 26 | 94,779.808 | 94780 | 0.2394 | 0.00821 | 0.2414 | 2.2167 | 1.08932 |
| 27 | 94,779.590 | 94780 | 0.0214 | 0.0276 | 0.0220 | 0.3674 | 1.13047 |
| 28 | 94,779.648 | 94780 | 0.0794 | 0.0397 | 0.0827 | 1.6442 | 1.10854 |
| 29 | 94,779.603 | 94780 | 0.0344 | 0.0456 | 0.0361 | 0.7657 | 1.12665 |
| 30 | 94,779.585 | 94780 | 0.0164 | 0.0486 | 0.0173 | 0.3779 | 1.13040 |
| 31 | 94,791.066 | 94791 | -0.0379 | 0.0146 | -0.0384 | -0.5677 | 1.12888 |
| 32 | 94,791.065 | 94791 | -0.0389 | 0.0280 | -0.0400 | -0.8120 | 1.12603 |
| 33 | 94,791.076 | 94791 | -0.0279 | 0.0326 | -0.0288 | -0.6297 | 1.12825 |
| 34 | 94,791.088 | 94791 | -0.0159 | 0.0345 | -0.0164 | -0.3689 | 1.13046 |
| 35 | 94,791.125 | 94791 | 0.0211 | 0.0354 | 0.0219 | 0.4981 | 1.12951 |
| 36 | 94,803.326 | 94803 | 0.0159 | 0.0250 | 0.0164 | 0.2981 | 1.13086 |

Rerun Standard Regression with MIXED to get the RMS Average Weighted Residual
Note that some of the relevant output is in the SAS log.

The Mixed Procedure

Influence Diagnostics

| Deleted Obs. Index | Externally Studentized Residual | Cook's D | DFFITS | COVRATIO | Restricted Likelihood Distance |
|---|---|---|---|---|---|
| 1 | 0.8001 | 0.00004 | 0.01059 | 1.0165 | 0.0011 |
| 2 | 0.9791 | 0.00047 | 0.03743 | 1.0033 | 0.0014 |
| 3 | -0.5122 | 0.00048 | -0.03756 | 1.0393 | 0.0055 |
| 4 | 1.7332 | 0.01299 | 0.20033 | 0.9276 | 0.0695 |
| 5 | -0.6813 | 0.00406 | -0.10986 | 1.0510 | 0.0142 |
| 6 | 0.5935 | 0.00019 | 0.02389 | 1.0312 | 0.0037 |
| 7 | -0.2795 | 0.00030 | -0.02988 | 1.0544 | 0.0073 |
| 8 | -1.4657 | 0.02247 | -0.26184 | 0.9806 | 0.0783 |
| 9 | -0.4631 | 0.00408 | -0.10996 | 1.0944 | 0.0167 |
| 10 | 2.6298 | 0.16464 | 0.73316 | 0.8361 | 0.7843 |
| 11 | 0.5034 | 0.00014 | 0.02028 | 1.0359 | 0.0046 |
| 12 | -0.3477 | 0.00047 | -0.03719 | 1.0523 | 0.0072 |
| 13 | -0.6781 | 0.00493 | -0.12118 | 1.0573 | 0.0169 |
| 14 | 1.6504 | 0.04992 | 0.39194 | 0.9789 | 0.1754 |
| 15 | -1.7629 | 0.07807 | -0.49151 | 0.9821 | 0.2741 |
| 16 | 0.1803 | 0.00005 | 0.01235 | 1.0495 | 0.0072 |
| 17 | -0.9668 | 0.00748 | -0.14976 | 1.0270 | 0.0225 |
| 18 | -0.0019 | 0.00000 | -0.00041 | 1.0944 | 0.0075 |
| 19 | -0.8221 | 0.01383 | -0.20317 | 1.0766 | 0.0421 |
| 20 | -0.7256 | 0.01237 | -0.19195 | 1.0930 | 0.0385 |
| 21 | 0.2429 | 0.00016 | 0.02191 | 1.0518 | 0.0072 |
| 22 | 0.6749 | 0.00434 | 0.11363 | 1.0538 | 0.0151 |
| 23 | 1.0914 | 0.01651 | 0.22289 | 1.0328 | 0.0500 |
| 24 | -1.7377 | 0.04720 | -0.38194 | 0.9590 | 0.1757 |
| 25 | -0.2998 | 0.00157 | -0.06819 | 1.0958 | 0.0109 |
| 26 | 2.2855 | 0.01356 | 0.20798 | 0.8393 | 0.1710 |
| 27 | 0.3650 | 0.00128 | 0.06150 | 1.0694 | 0.0094 |
| 28 | 1.6658 | 0.03730 | 0.33891 | 0.9628 | 0.1380 |
| 29 | 0.7633 | 0.00934 | 0.16683 | 1.0676 | 0.0291 |
| 30 | 0.3754 | 0.00243 | 0.08484 | 1.0926 | 0.0128 |
| 31 | -0.5648 | 0.00159 | -0.06878 | 1.0464 | 0.0082 |
| 32 | -0.8099 | 0.00633 | -0.13746 | 1.0448 | 0.0198 |
| 33 | -0.6268 | 0.00445 | -0.11507 | 1.0623 | 0.0160 |
| 34 | -0.3665 | 0.00162 | -0.06924 | 1.0769 | 0.0104 |
| 35 | 0.4953 | 0.00303 | 0.09483 | 1.0725 | 0.0133 |
| 36 | 0.2961 | 0.00076 | 0.04741 | 1.0687 | 0.0085 |

Analysis of the Moldover et al. Data to Estimate R                                210
Rerun Standard Regression with MIXED to get the RMS Average Weighted Residual
Note that some of the relevant output is in the SAS log.

The Mixed Procedure

Influence Diagnostics

| Deleted Obs. Index | Observed Value | Predicted Value | Residual | Leverage | PRESS Residual | Internally Studentized Residual | RMSE without deleted obs |
|---|---|---|---|---|---|---|---|
| 37 | 94,803.321 | 94803 | 0.0109 | 0.0356 | 0.0113 | 0.2456 | 1.13110 |
| 38 | 94,803.308 | 94803 | -0.0021 | 0.0385 | -0.0021 | -0.0480 | 1.13159 |
| 39 | 94,803.320 | 94803 | 0.0099 | 0.0396 | 0.0104 | 0.2358 | 1.13114 |
| 40 | 94,803.362 | 94803 | 0.0519 | 0.0402 | 0.0541 | 1.2406 | 1.11853 |
| 41 | 94,815.834 | 94816 | -0.0213 | 0.0359 | -0.0221 | -0.4416 | 1.12996 |
| 42 | 94,815.768 | 94816 | -0.0873 | 0.0443 | -0.0913 | -2.0235 | 1.09648 |
| 43 | 94,815.831 | 94816 | -0.0243 | 0.0465 | -0.0254 | -0.5765 | 1.12880 |
| 44 | 94,815.814 | 94816 | -0.0413 | 0.0473 | -0.0433 | -0.9896 | 1.12331 |
| 45 | 94,815.865 | 94816 | 0.0097 | 0.0477 | 0.0102 | 0.2347 | 1.13114 |
| 46 | 94,828.439 | 94829 | -0.0671 | 0.0401 | -0.0699 | -1.4699 | 1.11321 |
| 47 | 94,828.454 | 94829 | -0.0521 | 0.0461 | -0.0546 | -1.2272 | 1.11882 |
| 48 | 94,828.471 | 94829 | -0.0351 | 0.0476 | -0.0369 | -0.8405 | 1.12563 |
| 49 | 94,828.487 | 94829 | -0.0191 | 0.0482 | -0.0201 | -0.4604 | 1.12982 |
| 50 | 94,828.516 | 94829 | 0.0099 | 0.0484 | 0.0104 | 0.2388 | 1.13113 |
| 51 | 94,841.578 | 94842 | 0.0265 | 0.0370 | 0.0275 | 0.5961 | 1.12860 |
| 52 | 94,841.552 | 94842 | 0.0005 | 0.0407 | 0.0005 | 0.0114 | 1.13161 |
| 53 | 94,841.563 | 94842 | 0.0115 | 0.0417 | 0.0120 | 0.2750 | 1.13097 |
| 54 | 94,841.577 | 94842 | 0.0255 | 0.0420 | 0.0266 | 0.6132 | 1.12843 |
| 55 | 94,841.587 | 94842 | 0.0355 | 0.0422 | 0.0370 | 0.8557 | 1.12541 |
| 56 | 94,855.267 | 94855 | 0.0976 | 0.0348 | 0.1011 | 2.2344 | 1.08863 |
| 57 | 94,855.194 | 94855 | 0.0246 | 0.0374 | 0.0255 | 0.5841 | 1.12872 |
| 58 | 94,855.196 | 94855 | 0.0266 | 0.0380 | 0.0276 | 0.6371 | 1.12818 |
| 59 | 94,855.214 | 94855 | 0.0446 | 0.0383 | 0.0464 | 1.0721 | 1.12186 |
| 60 | 94,855.205 | 94855 | 0.0356 | 0.0384 | 0.0370 | 0.8571 | 1.12539 |
| 61 | 94,868.238 | 94868 | 0.0750 | 0.0509 | 0.0790 | 1.7519 | 1.10539 |
| 62 | 94,868.142 | 94868 | -0.0210 | 0.0538 | -0.0222 | -0.5045 | 1.12946 |
| 63 | 94,868.116 | 94868 | -0.0470 | 0.0546 | -0.0497 | -1.1375 | 1.12063 |
| 64 | 94,868.163 | 94868 | 0.0000 | 0.0548 | 0.0000 | 0.0005 | 1.13161 |
| 65 | 94,868.147 | 94868 | -0.0160 | 0.0550 | -0.0169 | -0.3885 | 1.13033 |
| 66 | 94,883.396 | 94883 | 0.0469 | 0.123 | 0.0535 | 1.1502 | 1.12038 |
| 67 | 94,883.346 | 94883 | -0.0031 | 0.129 | -0.0035 | -0.0774 | 1.13156 |
| 68 | 94,883.321 | 94883 | -0.0281 | 0.130 | -0.0323 | -0.7097 | 1.12735 |
| 69 | 94,883.320 | 94883 | -0.0291 | 0.131 | -0.0335 | -0.7367 | 1.12702 |
| 70 | 94,883.270 | 94883 | -0.0791 | 0.131 | -0.0910 | -2.0057 | 1.09711 |

Analysis of the Moldover et al. Data to Estimate R                          211
Rerun Standard Regression with MIXED to get the RMS Average Weighted Residual
Note that some of the relevant output is in the SAS log.

The Mixed Procedure

Influence Diagnostics

| Deleted Obs. Index | Externally Studentized Residual | Cook's D | DFFITS | COVRATIO | Restricted Likelihood Distance |
|---|---|---|---|---|---|
| 37 | 0.2438 | 0.00074 | 0.04685 | 1.0818 | 0.0089 |
| 38 | -0.0477 | 0.00003 | -0.00953 | 1.0879 | 0.0076 |
| 39 | 0.2341 | 0.00076 | 0.04755 | 1.0866 | 0.0090 |
| 40 | 1.2457 | 0.02146 | 0.25480 | 1.0165 | 0.0672 |
| 41 | -0.4390 | 0.00242 | -0.08471 | 1.0757 | 0.0121 |
| 42 | -2.0727 | 0.06333 | -0.44648 | 0.9060 | 0.2779 |
| 43 | -0.5736 | 0.00540 | -0.12661 | 1.0809 | 0.0194 |
| 44 | -0.9894 | 0.01620 | -0.22043 | 1.0506 | 0.0486 |
| 45 | 0.2330 | 0.00092 | 0.05215 | 1.0958 | 0.0095 |
| 46 | -1.4830 | 0.03012 | -0.30329 | 0.9878 | 0.1026 |
| 47 | -1.2319 | 0.02427 | -0.27086 | 1.0244 | 0.0754 |
| 48 | -0.8387 | 0.01176 | -0.18744 | 1.0640 | 0.0358 |
| 49 | -0.4577 | 0.00358 | -0.10295 | 1.0887 | 0.0153 |
| 50 | 0.2371 | 0.00097 | 0.05349 | 1.0966 | 0.0096 |
| 51 | 0.5932 | 0.00455 | 0.11624 | 1.0691 | 0.0167 |
| 52 | 0.0113 | 0.00000 | 0.00233 | 1.0906 | 0.0075 |
| 53 | 0.2731 | 0.00110 | 0.05695 | 1.0879 | 0.0097 |
| 54 | 0.6103 | 0.00550 | 0.12783 | 1.0738 | 0.0193 |
| 55 | 0.8539 | 0.01076 | 0.17927 | 1.0568 | 0.0327 |
| 56 | 2.3052 | 0.06003 | 0.43783 | 0.8592 | 0.3249 |
| 57 | 0.5812 | 0.00442 | 0.11454 | 1.0703 | 0.0164 |
| 58 | 0.6342 | 0.00535 | 0.12609 | 1.0678 | 0.0186 |
| 59 | 1.0733 | 0.01525 | 0.21411 | 1.0327 | 0.0460 |
| 60 | 0.8554 | 0.00978 | 0.17092 | 1.0525 | 0.0298 |
| 61 | 1.7800 | 0.05491 | 0.41240 | 0.9577 | 0.2044 |
| 62 | -0.5016 | 0.00483 | -0.11967 | 1.0931 | 0.0185 |
| 63 | -1.1401 | 0.02489 | -0.27387 | 1.0436 | 0.0757 |
| 64 | 0.0005 | 0.00000 | 0.00012 | 1.1069 | 0.0075 |
| 65 | -0.3860 | 0.00293 | -0.09311 | 1.0996 | 0.0141 |
| 66 | 1.1530 | 0.06210 | 0.43268 | 1.1242 | 0.1880 |
| 67 | -0.0768 | 0.00030 | -0.02953 | 1.2006 | 0.0083 |
| 68 | -0.7070 | 0.02513 | -0.27356 | 1.1758 | 0.0767 |
| 69 | -0.7341 | 0.02721 | -0.28472 | 1.1745 | 0.0827 |
| 70 | -2.0532 | 0.20216 | -0.79723 | 0.9998 | 0.7104 |

Rerun the Quadratic Regression to get the RMS Average Weighted Residual for Checking
Note that some of the relevant output is in the SAS log.

The REG Procedure
Model: MODEL1
Dependent Variable: SpeedSqResc Rescaled Speed-Squared = SpeedSq - 94,808.35

Number of Observations Read          70
Number of Observations Used          70

Weight: WeightVar Weight Variable

### Analysis of Variance

| Source | DF | Sum of Squares | Mean Square | F Value | Pr > F |
|--------|----|----|----|----|----|
| Model | 2 | 47749339 | 23874669 | 1.893E7 | <.0001 |
| Error | 67 | 84.48133 | 1.26092 | | |
| Corrected Total | 69 | 47749423 | | | |

| | | | | |
|--------|----|----|----|----|
| Root MSE | 1.12290 | R-Square | 1.0000 | |
| Dependent Mean | 14.38129 | Adj R-Sq | 1.0000 | |
| Coeff Var | 7.80809 | | | |

### Parameter Estimates

| Variable | Label | DF | Parameter Estimate | Standard Error | t Value |
|----------|-------|----|----|----|----|
| Intercept | Intercept | 1 | -52.09911 | 0.02248 | -2317.6 |
| Pressure | Pressure (Pa) | 1 | 0.00022447 | 1.904476E-7 | 1178.64 |
| PressSq | Pressure Squared | 1 | 5.4784E-11 | 3.36511E-13 | 162.80 |

### Parameter Estimates

| Variable | Label | DF | Pr > |t| | Variance Inflation |
|----------|-------|----|----|----|
| Intercept | Intercept | 1 | <.0001 | 0 |
| Pressure | Pressure (Pa) | 1 | <.0001 | 21.15237 |
| PressSq | Pressure Squared | 1 | <.0001 | 21.15237 |

Analysis of the Moldover et al. Data to Estimate R                          213
Rerun the Quadratic Regression to get the RMS Average Weighted Residual for Checking
Note that some of the relevant output is in the SAS log.

The REG Procedure
Model: MODEL1
Dependent Variable: SpeedSqResc Rescaled Speed-Squared = SpeedSq - 94,808.35

Collinearity Diagnostics

| Number | Eigenvalue | Condition Index | Intercept | Pressure | PressSq |
|--------|------------|-----------------|-----------|----------|---------|
| 1 | 2.74044 | 1.00000 | 0.00815 | 0.00139 | 0.00244 |
| 2 | 0.25305 | 3.29081 | 0.17774 | 0.00099369 | 0.02849 |
| 3 | 0.00651 | 20.52008 | 0.81411 | 0.99761 | 0.96907 |

Collinearity Diagnostics (intercept adjusted)

| Number | Eigenvalue | Condition Index | Pressure | PressSq |
|--------|------------|-----------------|----------|---------|
| 1 | 1.97608 | 1.00000 | 0.01196 | 0.01196 |
| 2 | 0.02392 | 9.08831 | 0.98804 | 0.98804 |

# CONTENTS